\documentclass[a4paper,12pt]{article}
\usepackage{jheppub,esint,psfrag,bm}
\usepackage{booktabs,array,graphicx,blkarray}
\usepackage{xcolor,tikz}
\usetikzlibrary{decorations.pathmorphing,calc}
\newcommand{\bgleg}[2]{\draw[thick,shorten >=2.6pt] (0,0)--(#1,#2);
\draw[thick] (#1,#2) circle (2.6pt);
\draw (#1,#2)++(-1.84pt,-1.84pt)--++(3.68pt,3.68pt); \draw (#1,#2)++(-1.84pt,1.84pt)--++(3.68pt,-3.68pt);}
\usepackage{longtable,pdflscape,url}
\usepackage[hypcap=false]{caption}
\newcommand{\Atlas}[1]{\textbf{#1}}
\newcommand{\FFgraph}[2]{\texorpdfstring{\ensuremath{\mathrm{FF}^{(#1)}_{#2}}}{FF(#1)-#2}}
\newcommand{\SEgraph}[2]{\texorpdfstring{\ensuremath{\mathrm{SE}^{(#1)}_{#2}}}{SE(#1)-#2}}
\newcommand{\tadgraph}[2]{\texorpdfstring{\ensuremath{\mathrm{tad}^{(#1)}_{#2}}}{tad(#1)-#2}}
\newcommand{\WalkingResult}[1]{#1}
\usepackage[utf8]{inputenc}

\usepackage{epsfig,amssymb,amsmath,psfrag,subfigure,rotate,color,wasysym,xcolor}
\usepackage[bbgreekl]{mathbbol}

\allowdisplaybreaks

\DeclareSymbolFontAlphabet{\mathbbm}{bbold}
\DeclareSymbolFontAlphabet{\mathbb}{AMSb}%
\newcommand{\SU}{{\rm SU}}
\newcommand{\su}{{\rm su}}
\newcommand{\U}{{\rm U}}
\newcommand{\dd}{\mathrm d}
\newcommand{\cN}{\mathcal N}

\newcommand{\gym}{g_{\rm\scriptscriptstyle YM}}
\newcommand{\Ds}{D_s}
\newcommand{\tr}{\operatorname{tr}}
\newcommand{\ii}{\mathrm{i}}
\newcommand{\hI}{\widehat I}
\newcommand{\hJ}{\widehat J}
\newcommand{\cD}{\mathcal{D}}
\newcommand{\unit}{\mathbb{I}}

\def\XXint#1#2#3{{\setbox0=\hbox{$#1{#2#3}{\int}$ }
\vcenter{\hbox{$#2#3$ }}\kern-.6\wd0}}

\def \be  {\begin{equation}}
\def \ee  {\end{equation}}
\def \ba  {\begin{eqnarray}}
\def \ea  {\end{eqnarray}}
\def \baa {\begin{eqnarray*}}
\def \eaa {\end{eqnarray*}}

\def \lab #1 {\label{#1}}

\newcommand\re[1]{(\ref{#1})}
\def\d{\hbox{{d}\kern-.20em\hbox{l}}}

\def \matrix #1 {\left(\begin{array}{cc} #1 \end{array}\right)}

\def \tr {\mathop{\rm tr}\nolimits}
\def \e  {\mathop{\rm e}\nolimits}

\newcommand \vev [1] {\langle{#1}\rangle}

\newcommand{\ft}[2]{{\textstyle\frac{#1}{#2}}}

\def\1{\mathbbm{1}}

\title{Walking Sudakov with running coupling:\\ adjoint scalar YM. I}

\author[a]{A.V.~Belitsky,}
\author[b]{L.V.~Bork,}
\author[b]{A.A.~Zhukov}
\affiliation[a] {Department of Physics, Arizona State University,  Tempe, AZ 85287-1504, USA}  
\affiliation[b]{Dukhov Automatics Research Institute, 127055 Moscow, Russia}

\abstract{
We study the Sudakov form factor of the Lagrangian in Yang-Mills theory
coupled to adjoint scalars, with masses generated by spontaneous symmetry
breaking. Dimensional reduction provides a common description of the
vector and scalar sectors, while analytic continuation of the background
allows an infrared hierarchy between the internal masses and external
virtualities. We determine the one-loop effective potential and distinguish
its null stationary branch from the Coleman-Weinberg branch. At a generic
background, a holding source cancels the one-point function and permits
the walking hierarchy to be varied independently of the stationarity
condition. In this prescription we calculate the planar form factor
through two loops at leading order on the walking trajectory.
The calculation combines LSZ pole and residue corrections with action
counterterms and composite-operator mixing. We retain the independent
mass and subtraction-scale logarithms and organize the result as local
scale integrals over the running coupling. Two double-logarithmic walking
anomalous dimensions connect the cusp and octagon boundary data, and the 
maximally transcendental walking term agrees with the maximally supersymmetric 
result. The strictly off-shell form factor on the null branch is treated in a
companion paper. Our consideration is a prototypical environment for studying 
off-shell observables in pure gluodynamics but in a controlled gauge-invariant setting.
}

\begin{document}

\maketitle
\flushbottom
\renewcommand{\floatpagefraction}{0.8}
\setcounter{footnote} 0

\section{Introduction}

Infrared singularities of massless scattering amplitudes have been at the forefront of perturbative studies since the
inception of gauge theories. The latter are the backbone of the Standard Model of Particle Physics, describing the
electroweak and strong forces. Depending on the sector, its gauge symmetry can be broken or not and, as a
consequence, spectra of particle excitations can be massive or massless. With the discovery of the Higgs boson~\cite{ATLAS:2012yve,CMS:2012qbp}, the
underlying mechanism for mass generation was experimentally confirmed to be governed by spontaneous symmetry
breaking~\cite{Englert:1964et,Higgs:1964pj,Guralnik:1964eu}. Depending on the generated hierarchy of mass scales, the nature of infrared divergences can change. Naively,
since the latter emerge from copious emission of soft/collinear gauge bosons, this cloud is insensitive to the quantum
numbers of the parent emitters, hence the well-known {\sl universality} property touted in the literature~\cite{Sudakov:1954sw,Collins:1989gx,Catani:1998bh,Sterman:2002qn,Becher:2009qa,Agarwal:2021ais}. However,
their qualitative form depends on whether external states are heavier or lighter than intermediate ones, and so the
universality could be lost, at least in its original sense.

The question of infrared universality received close attention in recent years within the context of maximally supersymmetric 
gauge theory ($\cN = 4$ sYM) on the Coulomb branch~\cite{Alday:2009zm,Henn:2010ir,Henn:2011by,Caron-Huot:2021usw,Belitsky:2022itf,Bork:2022vat,Belitsky:2023ssv,Belitsky:2024dcf,Belitsky:2024agy}. The latter is characterized by non-vanishing moduli of some of the
scalar fields populating its Lagrangian. Two polar opposite cases were addressed first: (i) when some of the states
propagating in loops admit masses, (ii) when only external states are massive. We will loosely refer to these two
situations as on- and off-shell, respectively. The former was merely regarded as an alternative to regulating massless
gauge theories with dimensional continuation~\cite{Alday:2009zm,Henn:2010ir,Henn:2011by}, while the latter served as a
prototype for off-shell scattering~\cite{Caron-Huot:2021usw,Belitsky:2022itf,Belitsky:2023ssv}.
Unsurprisingly, the double-logarithmic, or Sudakov, asymptotics in the former case were found to be driven by the very same
infrared exponent known for decades, the cusp anomalous dimension~\cite{Polyakov:1980ca,Korchemsky:1987wg,Korchemsky:1988hd,BES}. It is the latter setup that was out of the ordinary.
Namely, instead of the cusp, the octagon anomalous dimension took center stage~\cite{Caron-Huot:2021usw,Belitsky:2022itf}.
The latter first emerged in studies of the null limit of polarized four-point correlation functions of infinitely heavy
half-BPS operators~\cite{Coronado:2018ypq,Coronado:2018cxj,Kostov:2019auq,Belitsky:2019fan,Belitsky:2020qzm}, an object
apparently far removed from amplitudes. The cusp and octagon anomalous dimensions are different functions of the
't Hooft coupling, starting with their transcendental structure order-by-order in coupling and culminating in a stark
difference in their strong-coupling behavior.

From the point of view of the effective field theory of momentum modes underlying the dynamics of the above two
extremes, it appeared that the passage from the on- to off-shell kinematics is non-analytic in ratios of mass
parameters involved. While the two regimes indeed share the hard and collinear modes\footnote{We rely here on the
infrared counting within the Method of Regions~\cite{Beneke:1997zp}. So a knowledgeable reader would acknowledge the absence of soft modes,
which yield scaleless integrals that vanish.}, the off-shell case possesses additional, ultra-soft modes. These exist
only when excitations propagating in loops are strictly massless. However, it turned out that the collinear modes can
pick up the torch and `compensate' for the lack of the ultra-soft degrees of freedom. In a certain double scaling limit,
one can then safely navigate between the on- and off-shell regimes. The corresponding {\sl walking} infrared anomalous
dimension~\cite{Alday:2026walk,Calisto:2026xfb} depends on a new scaling variable (in addition to the 't Hooft coupling) and interpolates between the cusp and
octagon. Then the universality alluded to before is restored once one accepts it as the true Sudakov exponent.

The question of how the above considerations get modified in non-conformal gauge theories is addressed in the present
paper. As a small step away from $\cN = 4$ sYM, we eliminate fermions from the latter and consider Yang-Mills fields coupled
to massless adjoint scalars. This model is quite rich and allows us to study a plethora of topics that shape the
following presentation. Namely, in the next section, we obtain the model as a dimensional reduction of
$\Ds$-dimensional pure Yang-Mills theory down to $d$. Then in Sect.\ \ref{sec:vacua}, we introduce a pattern of
spontaneous gauge symmetry breaking to an analytically continued complex background. We quantize the theory
in the generalized $R_\xi$ gauges in Sect.\ \ref{sec:Quantization} and discuss the renormalization of the scalar potential,
which is not protected by any symmetry, in Sect.\ \ref{sec:Renormalization}. We then land on two stationary branches of 
the model, the null and Coleman-Weinberg branches, that we dub A and B, respectively. We elucidate their
quantum stability in Sect.\ \ref{sec:QuantumStability}, using the one-loop effective potential: branch B is a genuine 
minimum, while branch A is a stationary point sitting on top of a hill. The subsequent analysis is focused on the simplest
infrared-sensitive observable in the model, the Sudakov form factor of the model's Lagrangian. The branch-A calculation,
which corresponds to the strictly off-shell regime and determines the octagon anomalous dimension, is presented in a
companion paper. Here we calculate the walking form factor through two loops and explain its relation to branch B,
which naturally embraces this regime. The calculation is performed in the held-background prescription, allowing
arbitrary interior walking trajectories. The leading double-logarithmic behavior is insensitive to this background
choice, whereas single logarithms and constants retain its dependence. In Sect.\ \ref{sec:integral}, we provide an integral representation for the walking Sudakov
exponent with running coupling. Finally, we conclude.

\section{Adjoint scalar Yang-Mills from dimensional reduction}\label{sec:model}

Dimensional reduction fixes the classical relation between the gauge and
scalar interactions and provides the notation used throughout the
calculation. We first derive this common Lagrangian, then specify the
continuation of the field multiplicities needed in dimensional
regularization. Our starting point is the pure Yang-Mills Lagrangian
\begin{equation}
\label{eq:LDs}
\mathcal L_{\Ds}=-\tfrac12\tr\bigl(F_{MN}F^{MN}\bigr)
\, ,\qquad
F_{MN}=\partial_MA_N-\partial_NA_M-\ii\gym[A_M,A_N]
\, ,
\end{equation}
Here the field strength is built from the matrix-valued gauge field
\begin{equation}
A_M=\sum_AA_M^A\,T^A
\, ,
\end{equation}
with the Hermitian traceless generators $T^A$ in the fundamental representation of the group $\SU(N+2)$, normalized as
$\tr(T^AT^B)=\tfrac12\delta^{AB}$. The index $M$ runs over $\Ds$ spacetime dimensions, $M=0,\dots,\Ds-1$, and we use a
mostly minus metric, $\eta_{MN} = \mbox{diag}(1, -1, \dots, -1)$.

Anticipating the use of dimensional regularization for loop calculations, we split the
index\footnote{The hat is used to disambiguate an extra-dimensional label from a spacetime value with the same number.}
$M=(\mu,\hI)$ into the physical spacetime values $\mu=0,\dots,d-1$ and the remaining values along the extra directions
$\hI$, $I=1,\dots, n$, which enumerate $n \equiv \Ds-d$ scalars. The metric tensor is decomposed accordingly,
\begin{equation}\label{eq:metric}
\eta_{\mu\nu}
=
g_{\mu\nu}
\,,\qquad
\eta_{\hI\hJ}=-\delta_{IJ}
\, ,
\qquad \eta_{\mu\hI}=0
\, .
\end{equation}
Dimensional reduction eliminates the dependence on the extra-dimensional coordinates, i.e., $\partial_{\hI}\equiv0$,
and renames the corresponding components of the gauge field as scalars,
\begin{equation}\label{eq:reduction}
A_M = (A_\mu, \Phi_I)
\, ,\qquad
A_{\hI} \equiv \Phi_I
\, .
\end{equation}
The field strength then decomposes into the $d$-dimensional field strength, the covariant derivatives of the
scalars, and their commutators,
\begin{equation}
F_{\mu\nu}=\partial_\mu A_\nu-\partial_\nu A_\mu-\ii\gym[A_\mu,A_\nu]
\, ,\qquad
F_{\mu\hI}=\cD_\mu\Phi_I
\, ,\qquad
F_{\hI\hJ}=-\ii\gym[\Phi_I,\Phi_J],
\end{equation}
with $\cD_\mu\Phi_I \equiv \partial_\mu\Phi_I-\ii\gym[A_\mu,\Phi_I]$. Inserting this into Eq.~\eqref{eq:LDs} with the
metric \eqref{eq:metric} gives the classical Lagrangian of adjoint scalar Yang-Mills theory in $d$ dimensions,
\begin{equation}\label{eq:Lcl}
\mathcal L_{\rm cl}=-\tfrac12\tr\bigl(F_{\mu\nu}F^{\mu\nu}\bigr)
+
\sum_{I=1}^{n}\tr\bigl(\cD_\mu\Phi_I\,\cD^\mu\Phi_I\bigr)
-U_{\rm cl}(\Phi)
\, ,
\end{equation}
with the scalars interacting via a double-commutator potential,
\begin{equation}
U_{\rm cl}(\Phi)=-\tfrac{1}{2}\gym^2 \sum_{I,J = 1}^{n}\tr\bigl([\Phi_I,\Phi_J]^2\bigr)\ \ge0 .
\label{eq:Ucl}
\end{equation}
All couplings, cubic and quartic, are inherited from the single gauge coupling $\gym$: the model has no independent
scalar self-coupling at the classical level. This rigidity is what makes the above dimensional-reduction picture useful.
Quantum mechanically, however, it will be a different story altogether. Notice that in the spirit of Coleman and
Weinberg~\cite{Coleman:1973jx}, we do not introduce a negative mass term, nor any mass term at all. In $d=4$ the classical theory then contains
no dimensionful parameters and is therefore classically scale (in fact, conformally) invariant.

The local gauge symmetry of Eq.~\eqref{eq:LDs} restricted to $x^{\hI}$-independent transformations is the
$d$-dimensional $\SU(N+2)$ gauge symmetry under which all fields are adjoint. The rotations of the extra directions
become a global flavor symmetry ${\rm O} (n)$ acting on the index~$I$.

For four-dimensional physics, $d=4$, the model contains $\Ds-4$ real adjoint scalars; for $\Ds=10$ these are the six
scalars of $\cN=4$ sYM theory without its fermions. Indeed, $\cN=4$ sYM was originally constructed precisely in this
way, by dimensional reduction of ten-dimensional $\cN=1$ sYM~\cite{Brink:1976bc,Gliozzi:1976qd}. We will not set $\Ds$ to any particular value; it will be an arbitrary 
parameter. The model is ultraviolet divergent and the coupling constant $\gym$ runs. Moreover, quantum mechanically 
the double-commutator structure \re{eq:Ucl} is not preserved: renormalization generates all symmetry-allowed quartic 
scalar operators, which run independently of $\gym$. In loop calculations, the momentum integrals are continued to 
$d=4-2\epsilon$ while $\Ds$ is held fixed and integral. The scalar index then formally runs over $n = \Ds-d$ values, such 
that sums are defined by $\sum_I\delta_{II}=n$, with $\sum_\mu\delta^\mu{}_\mu=d$, and the total number of field 
components is always $\Ds$. This is a {\sl dimensional-reduction}-type scheme~\cite{Siegel:1979wq,Bern:1991aq}, natural here, since the model
itself is a dimensional reduction. At the classical level it plays no role, 
and $d=4$ may be used throughout.

\section{Classical vacua and mass spectra}
\label{sec:vacua}

The infrared scales of the form factor arise from the choice of scalar
background, rather than from explicit mass terms in the Lagrangian.
We therefore identify the classical vacua, choose the symmetry-breaking
pattern relevant to the external charged vectors, and derive the masses
of the fields that propagate in their loops.

The potential $U_{\rm cl}$ is a sum of squares and vanishes if and only if $[\Phi_I,\Phi_J]=0$ for all $I,J$. Constant
scalar fields $\Phi_I=V_I$, for every $I = 1, \dots, n$, with
\begin{equation}
\label{eq:moduli}
V_I^\dagger=V_I
\,, \qquad
\tr V_I=0
\, ,\qquad [V_I,V_J]=0 ,
\end{equation}
are therefore classical vacua. They form the classical moduli space $\mathcal M_0$,
\begin{equation}\label{eq:M0}
\mathcal M_0=\Bigl\{(V_1,\dots,V_n)\in {\su}(N+2)^{\,n}:\ [V_I,V_J]=0\ \ \forall I,J\Bigr\}\Big/\SU(N+2)
\, ,
\end{equation}
a continuous family of commuting traceless Hermitian color matrices, up to constant gauge transformations
$V_I \to U V_I U^\dagger$, $U \in \SU(N+2)$. The transformations are constant because the residual gauge transformations
preserving $A_\mu=0$ are those with $\partial_\mu U=0$. The flavor rotations $V_I\to R_{IJ}V_J$, $R\in {\rm O}(n)$, by
contrast, are a physical symmetry acting on $\mathcal M_0$ and are not divided out. From the higher-dimensional vantage
point of the previous section, the background is a constant gauge field in the extra directions with vanishing field
strength, $F_{\hI\hJ}=0$. This flat-connection condition implies that it possesses no curvature and, thus, it costs no
energy to move along $\mathcal M_0$. It is a classically flat direction.

Instead of the above color-matrix representation, it is instructive to look at the vacuum from the flavor-space vector
perspective. Commuting Hermitian matrices can be diagonalized simultaneously, so we can choose a gauge where
$V_I = {\rm diag} (V_{I;11}, \dots, V_{I;N+2,N+2})$. Collecting their eigenvalues into flavor vectors\footnote{We multiply the
eigenvalues by the Yang-Mills coupling to obtain mass-normalized quantities, which is how they enter the Lagrangian.}
\begin{align}
x_k= \gym (V_{1;kk},\dots,V_{n;kk})\in\mathbb R^n
\, ,
\label{FlavorX}
\end{align} 
we see that every point in the moduli space has such a representative, and the only freedom left is the permutation
of colors, i.e., the Weyl group $S_{N+2}$,
\begin{equation}\label{eq:M0x}
\mathcal M_0
\cong
\Bigl\{(x_1,\dots,x_{N+2})\in(\mathbb R^{n})^{N+2}:\ \textstyle\sum_k x_k=0\Bigr\}\Big/S_{N+2}
\, .
\end{equation}
In fact, physical quantities depend only on the differences $x_i-x_j$, i.e., we effectively have translational
invariance in flavor space. The reason for this is easy to understand. The fields in the classical Lagrangian
\re{eq:Lcl}-\re{eq:Ucl} solely appear through commutators, which are blind to the shifts $V_I\to V_I+a_I{\mathbb I}_{N+2}$. In
the $\U(N+2)$ case, this shift would act only on the decoupled trace mode, while in the $\SU(N+2)$ case, it is removed by
the tracelessness condition, $\sum_kx_k=0$, which fixes the `center of mass'. In the geometric brane picture~\cite{Witten:1995im}, this
becomes almost obvious: only the relative positions of the $N+2$ branes in the transverse flavor space matter, and
their center of mass is the decoupled $\U(1)$. In this picture, the off-diagonal matrix entries correspond to open
strings stretched between pairs of branes, whose masses are proportional to the brane separations; this is the
origin of the mass formula derived below. Thus, equivalently, $\mathcal M_0\cong(\mathbb R^n)^{N+2}/(\mathbb R^n\times S_{N+2})$. 
The dimension of the moduli space is $\dim_{\mathbb R}\mathcal M_0=n(N+1)$. 

\subsection{Pattern of spontaneous symmetry breaking}

We select two distinguished color directions from the common block.
This isolates the charged vector states used as external particles and
keeps their couplings to the large-$N$ internal sector explicit.

Choosing a point $V_I$ of $\mathcal M_0$ and expanding the scalar fields around it,
\begin{equation}\label{eq:shift}
\Phi_I = V_I + \chi_I
\, ,
\end{equation}
spontaneously breaks the gauge symmetry down to the subgroup commuting with all $V_I = \vev{\Phi_I}$. In this paper, we are
interested in a very particular pattern of vacuum expectation values (vevs) $\vev{\Phi_I}$.

Namely, let us order the color indices as $(a,b,3,\dots,N+2)$, with\footnote{We prefer letters rather than the numerical
indices 1 and 2 in order to emphasize their special role and to avoid confusion with other numerical labels, for instance,
flavor ones.} $a=1$ and $b=2$ distinguished from the rest, $r,s,t=3,\dots,N+2$, which form a common square block of size $N$. We
let only the first two scalar flavors, $I=1,2$, acquire nonvanishing vevs and introduce in this manner a two-parameter family of
classical vacua,
\begin{align}
\gym V_1&=\tfrac{1}{2} m_0\operatorname{diag}(1,-1,0,\dots,0),\label{eq:V1}\\
\gym V_2&=\tfrac{1}{2} m_0 z \big[\operatorname{diag}(1,1,0,\dots,0) - \tfrac{2}{N+2}\,\mathbb{I}_{N+2}\big]
\, , \qquad 
V_{I>2}=0 \, .
\label{eq:V2}
\end{align}
Here $m_0>0$ and $z$ are, respectively, dimensionful and dimensionless real (for now) parameters. The matrices satisfy Eq.~\eqref{eq:moduli}, 
so for every $z$ the family lies in $\mathcal M_0$ and $U_{\rm cl}(V)=0$. Obviously, the setup requires at least two scalars, so
$\Ds\ge6$ for $d=4$. The common block of $V_2$ carries the entry $\gym(V_2)_{rr}=-m_0z/(N+2)$: it ensures tracelessness, but is
suppressed at large $N$, a limit to be taken shortly.

The three diagonal blocks of the background are pairwise distinct for every $z$. The gauge group is therefore broken down to
\begin{equation}\label{eq:breaking}
\SU(N+2)\longrightarrow {\rm S}\bigl(\U(1)_a\times \U(1)_b\times \U(N)\bigr),\qquad
U=\operatorname{diag}\bigl(\e^{\ii\theta_a},\e^{\ii\theta_b},U_N\bigr),\quad \e^{\ii(\theta_a+\theta_b)}\det U_N=1 
\, .
\end{equation}
The pattern is the same for all $z$. In flavor space, the background is characterized by the points $x_a$, $x_b$ and $N$ coincident $x_r$. 
What $z$ controls is the size of the breaking in the directions connecting $a$ and $b$ to the common block, as can be seen from
the mass-normalized flavor vectors
\begin{align}
\label{eq:vc}
v_c \equiv x_c - x_r
\, , \qquad
v_a = \tfrac{1}{2} m_0\, (1, z, 0,\dots,0)
\, , \qquad
v_b = \tfrac{1}{2} m_0\, (-1, z, 0,\dots,0)
\, .
\end{align}
In the geometric picture, $v_a$ and $v_b$ are the separations of the branes $a$ and $b$ from the stack of $N$ coincident
branes $r$.

Field components, collectively denoted $X_{ij} = ( A_\mu, \chi_I)_{ij}$, carry the color-index pair $(ij)$ and acquire masses
through the interaction terms. For diagonal $V_I$, the adjoint action is multiplicative (no summation over $i,j$ is implied),
\begin{equation}
\label{eq:adjoint}
\gym[V_I,X]_{ij}
=
\kappa_{ij,I}\,X_{ij}
\, ,\qquad
\kappa_{ij}
\equiv
x_i - x_j
\, ,
\end{equation}
where $\kappa_{ij}$ is the mass vector of the pair $(ij)$; it lives in flavor space. For the family of classical vacua
\re{eq:V1}-\re{eq:V2},
\begin{equation}\label{eq:kappas}
\kappa_{ab}
=
v_a-v_b
=(m_0,0,\dots,0)
\, ,\qquad 
\kappa_{ar} = v_a
\, ,\qquad 
\kappa_{br}=v_b
\, , \qquad
\kappa_{rs} = \kappa_{cc}=0 
\, .
\end{equation}
This confirms the expectation that only translationally invariant combinations of the $x$-vectors enter the spectrum.

\subsection{Scalar and gauge mass matrices}
\label{sect:Massed}

The quadratic action determines both the scalar spectrum and the masses acquired by the gauge fields. We derive these together before separating the physical scalar modes from the Goldstone directions. We begin with the mass matrix of the scalars $\chi_I$. For a diagonal background, the expansion \re{eq:shift} of
$U_{\rm cl}$ around it starts at second order,
\begin{equation}
U_{\rm cl}(V+\chi)
=
\tfrac12\,\chi\cdot H \cdot\chi+O(\chi^3)
\, ,
\end{equation}
where $H$ is the Hessian
\begin{equation}
\chi\cdot H\cdot\chi
=
-
\gym^2 \sum_{I,J = 1}^n \tr \big([V_I,\chi_J] - [V_J,\chi_I]\big)^2
\, .
\end{equation}
Collecting the flavor components of one $(ij)$-matrix entry into the flavor vector $\chi_{ij}\equiv(\chi_{I,ij})_I$, the
quadratic form becomes a sum over ordered color pairs,
\begin{equation}\label{eq:Hform}
\tfrac12\,\chi\cdot H\cdot\chi=\sum_{i\neq j}\chi_{ji}^{\rm T}\,M_{ij}\,\chi_{ij}
\, ,
\end{equation}
where $M_{ij}$ is an $n \times n$ matrix in flavor space,
\begin{equation}
\label{eq:Mijdef}
M_{ij}\equiv\kappa_{ij}^2\,\unit_{n}-\kappa_{ij}\otimes\kappa_{ij}
\, .
\end{equation}
Here $\unit_{n}$ is the unit matrix in flavor space, $\kappa_{ij}^2 = \sum_I\kappa_{ij,I}\kappa_{ij,I}$, and 
$\kappa\otimes\kappa$ is the direct product, $(\kappa\otimes\kappa)_{IJ} = \kappa_I\kappa_J$, acting 
as $(\kappa\otimes\kappa) \chi=\kappa\,(\kappa\cdot \chi)$. All products of vev-vectors are bilinear and 
$M_{ji}=M_{ij}$ because $\kappa_{ji}=-\kappa_{ij}$. Three properties follow directly from Eq.~\eqref{eq:Hform}:
\begin{itemize}
\item the diagonal entries $\chi_{ii}$ do not appear: $H$ annihilates the diagonal (Cartan) sector;
\item different color pairs do not mix: $H$ maps the sector of the pair $\{i,j\}$ into itself;
\item inside a color pair with $\kappa_{ij}^2\neq0$, we can decompose the field
\begin{align}
\chi_{ij} = \chi_{ij}^{||} + \chi_{ij}^\perp
\, , \qquad
\chi_{ij}^{||} = \kappa_{ij} (\kappa_{ij}  \cdot \chi_{ij})/\kappa_{ij}^2
\, ,
\end{align}
into its components along and transverse to $\kappa_{ij}$, such that
\begin{align}
M_{ij} \kappa_{ij}=0
\, , \qquad
M_{ij} \chi^\perp_{ij} =\kappa_{ij}^2\, \chi^\perp_{ij}
\, .
\end{align}
\end{itemize}
The last property implies that the component along $\kappa_{ij}$ is the would-be Goldstone boson. It is eaten by the vector
$(A_\mu)_{ij}$, which acquires the same mass $m_{ij}$ as the transverse scalars,
\begin{align}
\label{eq:masses}
m_{ij}^2 = \kappa_{ij}^2
\, .
\end{align}
The remaining $n-1$ scalars are physical with mass $m_{ij}$.
Each massive complex pair $(ij),(ji)$ thus carries
\begin{equation}\label{eq:dof}
2\Bigl[\underbrace{(d-1)}_{\text{massive vector}}+\underbrace{(n - 1)}_{\text{physical scalars}}\Bigr]=2(\Ds-2)
\end{equation}
degrees of freedom, independently of $d$. This matches the count for a massless gauge field in $\Ds$ dimensions. For pairs with
$\kappa_{ij}=0$, i.e., the common block and the diagonal components, all $n$ scalars are massless, and the vector is a
massless gauge field.

To summarize, ordering the scalar field vector $\chi_{I, ij}$ by the color sectors,
\begin{equation}
\chi 
=
\big(
\chi_{\rm diag}\ ;\ \chi_{ab}\ ;\ \chi_{a3},\dots,\chi_{a,N+2}\ ;\ \chi_{b3},\dots,\chi_{b,N+2}\ ;\ \{\chi_{rs}\}_{3\le r<s\le N+2}
\big)
\, ,
\end{equation}
the mass matrix takes a block-diagonal form\footnote{Off-diagonal entries are listed with $i<j$, since
$\chi_{I,ji}=\chi_{I,ij}^*$.}
\begingroup
\makeatletter
\setlength{\BA@colsep}{4pt}
\makeatother
\renewcommand{\arraystretch}{1.3}
\setlength{\aboverulesep}{0pt}
\setlength{\belowrulesep}{0pt}
\setlength{\cmidrulesep}{0pt}
\begin{equation}
\label{eq:Hblock}
M
=
\begin{blockarray}{ccccc@{\quad}l}
n(N{+}1) & n & nN & nN & \tfrac12nN(N{-}1) & \\
\begin{block}{(c|c|c|c|c)@{\quad}l}
\mathbf 0_{\rm diag} & 0 & 0 & 0 & 0 & n(N{+}1)\quad\text{diagonal}\\ \cmidrule[.4pt](l{5pt}r{7pt}){1-5}
0 & M_{ab} & 0 & 0 & 0 & n\quad\text{pair }ab\\ \cmidrule[.4pt](l{5pt}r{7pt}){1-5}
0 & 0 & \unit_N \otimes M_{ar} & 0 & 0 & nN\quad\text{pairs }ar\\ \cmidrule[.4pt](l{5pt}r{7pt}){1-5}
0 & 0 & 0 & \unit_N \otimes M_{br} & 0 & nN\quad\text{pairs }br\\ \cmidrule[.4pt](l{5pt}r{7pt}){1-5}
0 & 0 & 0 & 0 & \mathbf 0_{rs} & \tfrac12nN(N{-}1)\quad\text{pairs }rs\\
\end{block}
\end{blockarray}
\end{equation}
\endgroup
with $\unit_N$ the $N\times N$ unit matrix in the bulk color index. The off-diagonal blocks vanish, and $\mathbf 0_{\rm diag}$, 
$\mathbf 0_{rs}$ are zero matrices of the indicated sizes. The nonzero blocks are:
\begin{itemize}
\item $M_{ab}=m_0^2\,\unit_n-\kappa_{ab}\otimes\kappa_{ab}$, with $\kappa_{ab}=v_a-v_b$;
\item $M_{cr}=m_z^2\,\unit_n-v_c\otimes v_c$ for $c = a,b$, with
\begin{align}
\label{eq:mz}
m_z^2 = v_a^2 = v_b^2 =
\tfrac{1}{4}m_0^2(1+z^2)
\, ;
\end{align}
these are the same for every bulk color $r$, hence the factor $\unit_N\otimes$;
\item $M_{rs}=0$, since $\kappa_{rs}=0$.
\end{itemize}
This explicit representation makes it easy to check the number of real field components:
$n(N+1)+2\bigl[n+2nN+\tfrac12nN(N-1)\bigr]=n\bigl[(N+2)^2-1\bigr]$, the real dimension of $n$ copies of
${\su}(N+2)$; the factor $2$ counts the real and imaginary parts of each complex entry with $i<j$. Written over $i<j$, the
quadratic form \eqref{eq:Hform} becomes $2\sum_{i<j}\chi_{ij}^{\dagger}M_{ij}\chi_{ij}$, and the kinetic term
$\tr(\partial\chi_I)^2=\sum_{i,j}\partial\chi_{I,ij}\,\partial\chi_{I,ji}=2\sum_{i<j}|\partial\chi_{I,ij}|^2+\dots$ carries
the same factor $2$. The physical masses squared in a pair are therefore the eigenvalues of $M_{ij}$.

It is natural to pass to physical fields with canonically normalized kinetic terms. The scalars are split by Eq.~\eqref{eq:Mijdef} 
into one Goldstone direction and $n-1$ physical directions in each charged color pair:
\begin{itemize}
\item in the pair $ab$, the Goldstone direction is along $\kappa_{ab}\propto(1,0,\dots,0)$, i.e., $\chi_{1,ab}$ and
$\chi_{1,ba}$, and the physical scalars are $\phi^+_I = \sqrt2\, \chi_{I,ab},\ \phi^-_I = \sqrt2\,\chi_{I,ba}$ with $I = 2, \dots, n$;
\item in the pairs $cr$, $c=a,b$, the Goldstone direction is along $v_c\propto(\pm1,z,0,\dots,0)$, i.e., $v_c\cdot\chi_{cr}$. The 
physical scalars are $\sqrt2\,w_c\cdot\chi_{cr}$ and $\sqrt2\,\chi_{I,cr}$ with $I=3,\dots,n$, collectively denoted $(\phi_c)_r$, 
together with their conjugates $(\bar\phi_c)_r$. Here
\begin{equation}
w_a=\frac{(-z,1,0,\dots,0)}{\sqrt{1+z^2}},\qquad w_b=\frac{(z,1,0,\dots,0)}{\sqrt{1+z^2}},\qquad w_c\cdot v_c=0,\qquad w_c^2=1
\, .
\end{equation}
Notice that this normalization degenerates at the null points $z=\pm\ii$, where $v_c^2=0$;
\item the traceless part of the common $N \times N$ block,
\begin{equation}
(\phi_I)_{rs}=(\chi_I)_{rs}-\frac{\delta_{rs}}{N}\sum_{t=3}^{N+2}(\chi_I)_{tt}
\, ,
\end{equation}
with all $n$ flavors massless;
\item the Cartan components $\sigma^{(1)}_I=(\chi_I)_{aa}-(\chi_I)_{bb}$ and
$\sigma^{(2)}_I=\sqrt{\tfrac{N+2}{N}}\,[(\chi_I)_{aa}+(\chi_I)_{bb}]$, which are massless.
\end{itemize}

A few comments are in order. The Cartan scalars and the diagonal components of the common block are the fluctuations along 
$\mathcal M_0$. They are massless because the classical potential is flat in these directions. Whether this flatness survives 
quantum corrections is not a classical question and will be discussed below.

The charges $(q_a,q_b)$ under the two Abelian phases and the $\U(N)$ representations do not depend on $z$. 

The masses obey the vector identity $\kappa_{ab}=\kappa_{ar}-\kappa_{br}$, so the squared masses are related through the bilinear 
products, e.g., $m_0^2 = 2m_z^2-2v_a\cdot v_b$. The masses of all physical states follow from Eqs.~\eqref{eq:kappas},
\eqref{eq:masses} and \eqref{eq:mz}; they are listed in Table~\ref{tab:spectrum}, together with those of the vector fields
introduced below.

\begin{table}[t]
\begin{center}
\begin{tabular}{l|lccccc}
\toprule
\multicolumn{2}{c}{States} & & & {Properties} & & \\
\cmidrule(lr){1-2}
\cmidrule(lr){3-7}
gauge & scalar & $q_a$ & $q_b$ & $\U(N)$ & mass$^2$ & Goldstone direction\\
\midrule
$W^\pm$ & $\phi^\pm$ ($n-1$) & $\pm1$ & $\mp1$ & $\mathbf 1$ & $m_0^2$ & $\kappa_{ab}$\\
$w_a,\bar w_a$ & $\phi_a,\bar\phi_a$ ($n-1$) & $\pm1$ & $0$ & $\overline{\mathbf N},\mathbf N$ & $m_z^2$ & $v_a$\\
$w_b,\bar w_b$ & $\phi_b,\bar\phi_b$ ($n-1$) & $0$ & $\pm1$ & $\overline{\mathbf N},\mathbf N$ & $m_z^2$ & $v_b$\\
$g_{rs}$ & $\phi_{rs}$ ($n$) & $0$ & $0$ & adj $\SU(N)$ & $0$ & -\\
$B^{(1,2)}$ & $\sigma^{(1,2)}_I$ ($n$) & $0$ & $0$ & $\mathbf 1$ & $0$ & -\\
\bottomrule
\end{tabular}
\end{center}
\caption{Spectrum of the classical vacua \re{eq:V1}-\re{eq:V2}. Gauge and scalar states are separated by the vertical
line; the number of scalar flavors in each multiplet is given in parentheses. In the charged multiplets, the scalar along
the Goldstone direction is eaten by the vector.}
\label{tab:spectrum}
\end{table}

For the vector fields, the decomposition into the three diagonal blocks is the same as for the scalars. In color-matrix
form,
\begin{equation}
A_\mu=\begin{pmatrix}
(A_\mu)_{aa}&W_\mu^+/\sqrt2&(w_{a,\mu})_s/\sqrt2\\
W_\mu^-/\sqrt2&(A_\mu)_{bb}&(w_{b,\mu})_s/\sqrt2\\
(\bar w_{a,\mu})_r/\sqrt2&(\bar w_{b,\mu})_r/\sqrt2&(A_\mu)_{rs}
\end{pmatrix}
\, ,
\label{eq:42}
\end{equation}
where the three blocks have sizes $1,1,N$, respectively, and the normalizations of $W^\pm$, $w_c$ and $\bar w_c$ are the
same as for the charged scalars. The diagonal components obey $(A_\mu)_{aa}+(A_\mu)_{bb}+\sum_r(A_\mu)_{rr}=0$. The two
diagonal components not contained in the traceless part of the common block are the canonically normalized singlets
\begin{equation}
B^{(1)}_\mu=(A_\mu)_{aa}-(A_\mu)_{bb},\qquad B^{(2)}_\mu=\sqrt{\tfrac{N+2}{N}}\bigl[(A_\mu)_{aa}+(A_\mu)_{bb}\bigr]
\, .
\end{equation}
The traceless part of the lower-right $N \times N$ block is the gluon field of the unbroken $\SU(N)$ group,
\begin{equation}
(g_\mu)_{rs}=(A_\mu)_{rs}-\frac{\delta_{rs}}N\sum_{t=3}^{N+2}(A_\mu)_{tt}
\, .
\label{eq:43}
\end{equation}
The masses of all these fields are given in Table~\ref{tab:spectrum}.

\subsection{Extra-dimensional perspective on masses}
\label{subsec:ExtraDimMasses}

In the dimensionally reduced theory with spontaneously broken symmetry, the background enters only through the replacement
$\partial_{\hI} \to -\ii\gym[V_I,\cdot\,]$ in the extra directions. Acting on a component with colors $(ij)$, this is multiplication 
by $-\ii\kappa_{ij,I}$. Thus, the vector $\kappa_{ij}$ plays the role of a momentum in the extra dimensions. A field component
with $d$-momentum $p$ and colors $(ij)$ therefore carries the $\Ds$-dimensional momentum
\begin{equation}\label{eq:Pij}
P_{ij}=(p_\mu, \kappa_{ij,I}) 
\, ,\qquad 
P_{ij}^2=p^2-\kappa_{ij}^2
\, ,
\end{equation}
and the massless $\Ds$-dimensional wave equation $P^2=0$ becomes the $d$-dimensional mass shell condition
\begin{equation}
\label{eq:mass}
p^2=m_{ij}^2
\, , \qquad 
m_{ij}^2\equiv\kappa_{ij}^2
\, .
\end{equation}
This is the Higgs mechanism seen from $\Ds$ dimensions. A constant flat connection in the extra directions shifts the
extra-dimensional momentum of each color component, exactly as a Wilson line on a compact direction would; the mass of
a component is the length of its shift. Mass generation from extra dimensions goes back to Scherk and
Schwarz~\cite{Scherk:1979zr}, and symmetry breaking by flat extra-dimensional connections is the Hosotani
mechanism~\cite{Hosotani:1983xw}. In both cases the flat directions are classically degenerate, and it is the one-loop
potential that decides which of them is selected, a theme to which we return in Sect.~\ref{sec:QuantumStability}. The mass relation $\kappa_{ab}=\kappa_{ar}-\kappa_{br}$ quoted above is then
momentum conservation in the extra directions at the cubic vertex $(ab) \to(ar) + (rb)$. This point of view will
significantly simplify the Feynman rules introduced in the next section, since masses, vertices and propagators are all
Yang-Mills objects in $\Ds$ dimensions, evaluated at special values of the extra-dimensional ``momenta.''

Let us finally point out that this is the point of view adopted in previous studies of the Coulomb branch of $\cN = 4$
sYM~\cite{Alday:2009zm,Caron-Huot:2010nes,Caron-Huot:2021usw,Belitsky:2022itf}, where the theory was obtained by a
generalized dimensional reduction of ten-dimensional $\cN = 1$ sYM down to four dimensions, keeping the
extra-dimensional components of the ten-dimensional momenta nonvanishing.

\subsection{Complexification}

To realize the walking hierarchy, as well as the strictly off-shell kinematics considered in the companion paper,
we lift the reality condition on $z$ and allow it to take imaginary or complex values. To
this end, the flavor space $\mathbb R^n$ is complexified to $\mathbb C^n$, and products of flavor vectors are defined
bilinearly, without complex conjugation,
\begin{align}
\label{eq:bilinear}
v\cdot w\equiv\sum_Iv_Iw_I
\, . 
\end{align}

For complex $z$, we drop the Hermiticity condition and work with commuting elements of ${\rm sl}(N+2,\mathbb C)^n$
modulo global ${\rm SL}(N+2,\mathbb C)$ conjugations. Commuting non-Hermitian matrices need not be simultaneously
diagonalizable, so we restrict ourselves to the diagonal slice: complex points $x_k\in\mathbb C^n$ with $\sum_kx_k=0$,
modulo the Weyl permutations $S_{N+2}$. This slice contains the continued complex-$z$ family.

The bilinear product \eqref{eq:bilinear} admits nonzero vectors $v$ with $v \cdot v=0$. For our family they occur at the
{\sl null} points $z=\pm \ii$, which will play a central role: there the Goldstone directions $v_c$, $c = a,b$, are nonzero
but $v_c^2 =0$, while $(v_a-v_b)^2=m_0^2$ remains unchanged. In contrast to real $z$, where $m_z^2\ge m_0^2/4 > 0$, the
continued background is not Hermitian. The continued theory is defined by its Feynman rules, with the bilinear products
\eqref{eq:bilinear} and the standard $\ii0$ prescription, rather than as a real vacuum. We will show below that the null
background is nevertheless a stationary point, albeit one sitting on top of a hill.

The multiplets $w_c,\bar{w}_c$ become massless at the null points, while $W^\pm$ keeps its mass. This realizes off-shell
external legs with massless internal lines in the leading planar loops. For certain imaginary $z\ne\pm\ii$, we will find
that the theory resides in a true vacuum. Continuing the external virtuality to Euclidean values, $m_0^2\to-M_E^2$ with
$M_E^2>0$, will then give access to the walking regime.

Having mentioned loops, we now turn to the quantization of the classical action \re{eq:Lcl}-\re{eq:Ucl}.

\section{Gauge fixing and quantization}
\label{sec:Quantization}

It is straightforward to quantize adjoint scalar Yang-Mills theory in the component form. However, dimensional reduction of the
$D_s$-dimensional gauge field in terms of vectors and scalars, as well as their subsequent decomposition into color blocks that 
acquire different masses upon spontaneous symmetry breaking, leads to an untamed proliferation of field components and Feynman rules that one has to follow through. This is
obvious from Table~\ref{tab:spectrum}. To tackle masses,
we will recycle the higher-dimensional perspective in a moment, as was advocated in Sect.\ \ref{subsec:ExtraDimMasses}. 
But to start with, it is more economical to step back and treat both the vector and scalar fluctuations on top of the 
spontaneously broken vacuum as the components of a single $\Ds$-dimensional {\sl uplifted} gauge field $X_{M; ij}$,
with $M = (\mu,\hI)$,
\begin{equation}
\label{eq:X}
X_{\mu;ij}
=
(A_\mu)_{ij}
\, ,\qquad 
X_{\hI;ij}=(\chi_I)_{ij}
\, ,
\end{equation}
so that the full field is $A_M = \mathcal V_M + X_M$ with background $\mathcal V_\mu=0$, $\mathcal V_{\hI}=V_I$. For fixed 
$M$, $X_M$ is a traceless Hermitian $(N+2) \times(N+2)$ matrix. Indices are raised and lowered with $\eta$ of Eq.~\eqref{eq:metric}, 
and the traces are
\begin{equation}
\sum_\mu\delta^\mu{}_\mu=d
\, ,\qquad 
\sum_I\delta_{II}=n
\, , \qquad 
\eta^{MN}\eta_{NM} = \Ds 
\, .
\end{equation}

The symbol $X$ alone does not fix the mass or color representation of a component; the color pair $(ij)$ does, through $\kappa_{ij}$.
In section \ref{subsec:ExtraDimMasses}, we combined the $d$-dimensional momentum and mass flow into a single {\sl uplifted} 
momentum $P$. Its origin obviously stems from the background covariant derivative of the reduced theory (we reserve the
calligraphic $\cD$ for the covariant derivative containing the full gauge field),
\begin{equation}
D_MX\equiv\partial_MX-\ii\gym[\mathcal V_M,X]
\, ,\qquad 
D_\mu=\partial_\mu
\, ,\qquad 
D_{\hI}X=-\ii\gym[V_I,X]
\, .
\end{equation}
It acts on a component with momentum $p$ (convention $\partial_\mu\to-\ii p_\mu$) and colors $(ij)$ as multiplication by
$-\ii P_{M;ij}$, with
\begin{equation}
\label{eq:P}
P_{\mu;ij}(p)=p_\mu
\, ,\qquad 
P_{\hI;ij}(p)=\kappa_{ij,I}
\, ,\qquad
P_{M;ij}^2\equiv\eta^{MN}P_{M;ij}P_{N;ij}=p^2-m_{ij}^2 
\, .
\end{equation}
Reversing the momentum and exchanging the colors flips both parts, $P_{M;ji}(-p)=-P_{M;ij}(p)$. Only $p_\mu$ is
integrated over in loops; $\kappa_{ij}$ is a fixed external vector.

\subsection{Uplifted classical action}

The uplifted formulation packages the scalar and vector interactions
into tensors covariant with respect to the background. Expanding the action in
this form makes the quadratic operator and the cubic and quartic
vertices available without repeating the component decomposition.

Because the background is a flat connection, $[\mathcal V_M,\mathcal V_N]=0$, the field strength expanded about it is
\begin{equation}
F_{MN}=D_MX_N-D_NX_M-\ii\gym[X_M,X_N],
\end{equation}
and the action density \eqref{eq:LDs}, 
\begin{align}
{\mathcal L}_{\rm cl} = \mathcal L^{(2)} + \mathcal L^{(3)} + \mathcal L^{(4)}
\, ,
\end{align}
decomposes into
\begin{align}
\mathcal L^{(2)}&=\tr\bigl(D_MX_N\,D^NX^M - D_MX_N\,D^MX^N\bigr)
\, ,
\label{eq:L2}\\[1mm]
\mathcal L^{(3)}&=2\ii\gym\,\tr\bigl(D_MX_N\,[X^M,X^N]\bigr)
\, ,
\label{eq:L3}\\[1mm]
\mathcal L^{(4)}&=\tfrac{1}{2}\gym^2 \,\tr\bigl([X_M,X_N][X^M,X^N]\bigr)
\, .
\label{eq:L4}
\end{align}
There are no terms linear in $X$, since the background is a classical solution. The background enters only through
$D$, i.e., through the uplifted momenta \eqref{eq:P}.

Restricting the indices of Eqs.~\eqref{eq:L2}-\eqref{eq:L4} reproduces every term of the reduced Lagrangian
\eqref{eq:Lcl}. The quadratic term \eqref{eq:L2} contains the masses of Sect.~\ref{sect:Massed} through
$D_{\hI}\to-\ii\kappa_{ij,I}$. The cubic and quartic terms are the $\Ds$-dimensional Yang-Mills interactions, with
the background entering the cubic couplings only through the extra-dimensional components of the uplifted momenta.

\subsection{Generalized $R_\xi$ gauge and ghosts}
\label{sec:gauge}

To remove the redundant unphysical degrees of freedom from quantum loops, we choose the generalized $R_\xi$
gauge\footnote{The conventional gauge parameter $\xi$ corresponds to $\alpha$ in our notation.}
\begin{equation}
\label{eq:Gc}
\mathcal L_{\rm gf}
=
-
\frac1\alpha\tr\bigl(\mathcal G_c^2\bigr)
\, , \qquad
\mathcal G_c 
=
\partial^\mu A_\mu+\ii c\,\gym\sum_I[V_I,\chi_I]
\, .
\end{equation}
It is a two-parameter $(\alpha, c)$ family of gauges. The Faddeev-Popov procedure~\cite{Faddeev:1967fc} turns this
choice into a ghost Lagrangian, discussed below. In the uplifted form, using Eq.~\re{eq:adjoint},
\begin{equation}
(\mathcal G_c)_{ij} \stackrel{{\rm Fourier}}{\longrightarrow} -\ii\,Q^M{}_{;ij}X_{M;ij}
\, ,
\end{equation}
where the gauge-fixing vector $Q$ differs from the uplifted momentum $P$ only by the factor $c$ in its
extra-dimensional part,
\begin{equation}
\label{eq:Q}
Q_{M;ij}
=
(p_\mu, c\,\kappa_{ij,I})
\, ,\qquad
Q^M{}_{;ij}=(p^\mu,-c\,\kappa_{ij,I})
\, .
\end{equation}
The invariants built from $P$ and $Q$ are
\begin{equation}
\label{eq:invPQ}
P^2=p^2-m_{ij}^2
\, ,\qquad
P\cdot Q=p^2-c\,m_{ij}^2
\, ,\qquad
Q^2=p^2-c^2m_{ij}^2
\, ,
\end{equation}
so that $c$ controls where the unphysical poles of the propagators sit.
Several values of $c$ are special:
\begin{itemize}
\item \textbf{$c = 0$} [Fermi gauges]: the scalar term disappears, and the condition is $\partial^\mu A_\mu = 0$. The
vector-Goldstone mixing then survives in full, and the ghosts are massless.
\item \textbf{$c=\alpha$} [$R_\xi$ gauges]: the vector-Goldstone mixing cancels, and the unphysical Goldstone and ghost
poles sit at $p^2=\alpha\,m_{ij}^2$. This is the gauge-fixing term introduced by 't~Hooft~\cite{tHooft:1971rn} for
$\alpha=1$, the 't~Hooft-Feynman gauge, and extended to arbitrary $\alpha$ by Fujikawa, Lee and
Sanda~\cite{Fujikawa:1972fe}. In these gauges, unitarity and renormalizability are simultaneously manifest:
the $\alpha$-dependent poles cancel in physical quantities.
\item \textbf{$c=1$} [background-covariant gauge]: $\mathcal G_1=D^MX_M$ is the Lorenz condition of the
$\Ds$-dimensional theory in the background, i.e., the background-field gauge~\cite{DeWitt:1967ub,Abbott:1980hw,Abbott:1981ke}
of the parent Yang-Mills theory. The uplifted gauge-fixing and momentum vectors coincide, $Q=P$, and the gauge-fixed
quadratic action is a function of $P$ alone. All propagator poles, including the unphysical ones, sit at $P^2=0$, i.e.,
at the physical masses $p^2=m_{ij}^2$. The vector-Goldstone mixing, $\propto 1-1/\alpha$, does not cancel unless
$\alpha=1$; in the uplifted language it is simply the longitudinal part $\propto P_MP_N$ of the $\Ds$-dimensional
propagator.
\end{itemize}
At $\alpha=c=1$ all three choices meet in the 't~Hooft-Feynman gauge of the parent $\Ds$-dimensional theory. In what
follows we use $c=1$, which keeps the uplifted $\Ds$-dimensional covariance manifest. The general Feynman rules retain
arbitrary $\alpha$; the walking form-factor calculation specializes to $\alpha=1$. The gauge-parameter analysis of the
branch-A form factor is presented in the companion paper. Let us stress that the effective potential discussed in
Sect.~\ref{sec:QuantumStability} is gauge dependent off shell; by the Nielsen identity~\cite{Nielsen:1975fs}, the
location of its stationary points and the vacuum energy there are not.

Finally, the gauge-fixed Lagrangian needs to be supplemented by Faddeev-Popov ghosts. They arise from the gauge variation of 
$\mathcal G_c$. With anticommuting adjoint ghosts $\omega,\bar\omega$, we have
\begin{equation}
\label{eq:Lgh}
\mathcal L_{\rm gh}
=
-2\tr\Bigl\{\bar\omega\Bigl[\partial^\mu \cD_\mu\omega+c\,\gym^2\sum_I\bigl[V_I,[V_I+\chi_I,\omega]\bigr]\Bigr]\Bigr\}
\, .
\end{equation}
The ghosts couple to the vector fields through the covariant derivative, $\cD_\mu\omega=\partial_\mu\omega-\ii\gym[A_\mu,\omega]$, 
and to the scalars through the background, with strength proportional to $c$. Their quadratic part generates the mass $c\,m_{ij}^2$ for 
the ghost component with $(ij)$-colors. For our value of choice, $c=1$, Eq.~\eqref{eq:Lgh} becomes 
\begin{align}
\mathcal L_{\rm gh} 
= 
- 2\tr(\bar\omega\,D^M\cD_M\omega)
\, ,
\end{align} 
the ghost Lagrangian of the $\Ds$-dimensional Lorenz gauge, with $\cD_M$ the full covariant derivative in the background,
\begin{equation}
\label{eq:fullD}
\cD_M\omega\equiv\partial_M\omega-\ii\gym[\mathcal V_M+X_M,\omega]
\, ,
\end{equation}
i.e., $\cD_\mu\omega=\partial_\mu\omega-\ii\gym[A_\mu,\omega]$ and $\cD_{\hI}\omega=-\ii\gym[V_I+\chi_I,\omega]$.
It reduces to the background derivative $D_M$ when the fluctuations are switched off, $X_M\to0$. Using
$\eta^{\hI\hJ}=-\delta_{IJ}$, one checks that $D^M\cD_M\omega=\partial^\mu\cD_\mu\omega+\gym^2\sum_I[V_I,[V_I+\chi_I,\omega]]$,
reproducing Eq.~\eqref{eq:Lgh} at $c=1$.

The total Lagrangian to be used in perturbative calculations is the sum of all of the above
\begin{equation}
\label{TotalQuantumL}
\mathcal L=\mathcal L_{\rm cl}+\mathcal L_{\rm gf}+\mathcal L_{\rm gh}
\, .
\end{equation}

\section{Feynman rules}
\label{sec:rules}

Let us proceed by deriving the propagators and vertices for the Lagrangian \re{TotalQuantumL}. Along the way, we will advocate
for the choice $c=1$ alluded to in the previous section.

The classical quadratic term \re{eq:L2} contains the mixing between the vector and the scalars along $\kappa_{ij}$,
\begin{equation*}
2\ii\gym\sum_I\tr\bigl(\partial^\mu A_\mu[V_I,\chi_I]\bigr) .
\end{equation*}
The gauge-fixing term \re{eq:Gc} adds to it $-(2\ii c\gym/\alpha)\sum_I\tr(\partial^\mu A_\mu[V_I,\chi_I])$, together with a
Goldstone mass term $(c^2\gym^2/\alpha)\tr\bigl(\sum_I[V_I,\chi_I]\bigr)^2$. The total mixing, therefore, becomes
\begin{equation}
\label{eq:mix}
\mathcal L^{(2)}_{\rm mix}=2\ii\gym\Bigl(1-\frac c\alpha\Bigr)\sum_I\tr\bigl(\partial^\mu A_\mu\,[V_I,\chi_I]\bigr)
\, .
\end{equation}
Per generator, the full quadratic kernel of $X$ in momentum space is
\begin{equation}
\label{eq:kernel}
\mathcal L^{(2)}+\mathcal L_{\rm gf} \stackrel{{\rm Fourier}}{\longrightarrow} \tfrac12\,X_M\,K^{MN}X_N
\, ,
\qquad
K^{MN}=-P^2\eta^{MN}+P^MP^N-\frac1\alpha\,Q^MQ^N
\, .
\end{equation}
The first two terms are the transverse $\Ds$-dimensional Yang-Mills kernel, and the last term is the gauge fixing.
Inverting \eqref{eq:kernel} gives, for matrix components,
\begin{multline}
\label{eq:propgen}
\bigl\langle X_{M;ij}(p)X_{N;kl}(-p)\bigr\rangle_0=\frac{\ii}{2}
\Bigl(\delta_{il}\delta_{jk}-\frac{\delta_{ij}\delta_{kl}}{N+2}\Bigr)
\\
\times\left[
-
\frac{\eta_{MN}}{P^2}
-
\frac{\alpha P^2+Q^2}{P^2(P\!\cdot\!Q)^2}\,P_MP_N
+
\frac{P_MQ_N+Q_MP_N}{P^2\,(P\!\cdot\!Q)}
\right] \, ,
\end{multline}
with $P,Q$ evaluated for the color pair $(ij)$ and their invariants given in Eq.~\re{eq:invPQ}. The color factor is the
$\SU(N+2)$ completeness relation $\sum_AT^A_{ij}T^A_{kl}$; the first field carries momentum $p$, the second $-p$.

As usual, one is tempted to set $c = \alpha$ in Eq.~\re{eq:mix} to get rid of the mixing. However, for any $c\neq1$ the
propagator \re{eq:propgen} develops unphysical poles at $P\cdot Q=0$, i.e., at $p^2=c\,m_{ij}^2$, in addition to the
physical poles at $p^2=m_{ij}^2$; for $c=\alpha$ they sit at $p^2=\alpha\,m_{ij}^2$ and coincide with the physical ones only
in the 't~Hooft-Feynman gauge $\alpha=c=1$. Keeping $\alpha$ arbitrary makes the gauge-parameter dependence of the general
Feynman rules explicit, as needed for the branch-A analysis in the companion paper. With $\alpha$-dependent unphysical
thresholds, such an analysis would be considerably more cumbersome. We are thus naturally led to the value
$c = 1$ with arbitrary $\alpha$: then the gauge parameter enters only the numerator of a double pole at the physical mass,
$Q=P$, and every Feynman rule is that of $\Ds$-dimensional Yang-Mills theory with the momenta replaced by the uplifted
momenta \eqref{eq:P}. This enormously simplifies the construction of loop integrands.

Since the pure Yang-Mills Feynman rules are textbook material, we list them for the reader's convenience in
Appendix~\ref{App:FeynRules} rather than in the main text.

\section{Renormalized scalar potential}
\label{sec:Renormalization}

Before turning to the quantum fate of the classical vacua, we have to specify what the scalar potential of the quantum
theory is. At tree level, dimensional reduction fixes it uniquely; it is the double commutator \re{eq:Ucl}, with no
independent coupling. This structure is not protected by any symmetry of the reduced theory. The flavor symmetry
${\rm O}(n)$ and the gauge symmetry $\SU(N+2)$ allow a larger set of quartic invariants, and in the absence of supersymmetry
loop corrections generate all of them. This is a familiar feature of gauge theories with scalars~\cite{Cheng:1973nv,
Machacek:1983tz,Machacek:1984zw}, and of non-supersymmetric large-$N$ theories in particular, where the generated double-trace
couplings play an important role~\cite{Tseytlin:1999ii,Adams:2001jb,Dymarsky:2005uh,Pomoni:2008de}.

\subsection{Quartic operator basis}

Renormalization need not preserve the single scalar interaction inherited
from dimensional reduction. We first identify the complete quartic basis
allowed by gauge and flavor symmetry, so that every local ultraviolet
divergence can be assigned an independent counterterm. The basis contains
two single-trace and two double-trace structures,
\begin{align}
\label{eq:Rbasis}
O_{c}
=A-B
\, ,\quad
O_s
=A+B
\, ,\quad
O_3
=\Bigl(\sum_I\tr\Phi_I^2\Bigr)^2
\, ,\quad
O_4
=
\sum_{I,J}\bigl(\tr\Phi_I\Phi_J\bigr)^2
\, , \nonumber
\end{align}
with two flavor-ordered combinations
\begin{align}
A
=\sum_{I,J}\tr(\Phi_I\Phi_I\Phi_J\Phi_J)
\, ,\qquad
B=\sum_{I,J}\tr(\Phi_I\Phi_J\Phi_I\Phi_J)
\end{align}
The classical potential is $U_{\rm cl}=\gym^2O_c$, since $O_c=-\tfrac12\sum_{I,J}\tr[\Phi_I,\Phi_J]^2$. The combination $O_s$ is
the symmetric single-trace partner of the commutator structure; it does not vanish on commuting configurations and is,
therefore, the operator that can lift the classical moduli space. The renormalized potential reads\footnote{The square brackets
stand for renormalized operators that generate finite Green functions with elementary fields, i.e., $[O] = O + {\rm counterterms}$.}
\begin{equation}
\label{eq:UR}
U_R(\Phi)=\lambda_{c,R}\,[O_c]+\lambda_{s,R}\,[O_s]+\frac{\lambda_{3,R}}{N}\,[O_3]+\frac{\lambda_{4,R}}{N}\,[O_4]
\, ,
\end{equation}
where the explicit factors of $1/N$ in front of the double-trace operators make all four terms contribute at the same
order in the 't~Hooft limit to the four-point functions they generate. We define the theory by the renormalization
conditions
\begin{equation}
\label{eq:Rcond}
\lambda_{c,R}(\mu_*)=\gym^2
\, ,\qquad
\lambda_{s,R}(\mu_*)=\lambda_{3,R}(\mu_*)=\lambda_{4,R}(\mu_*)=0
\, ,
\end{equation}
at a renormalization scale $\mu_*$. This exhibits the fact that the theory coincides with the dimensionally reduced Yang-Mills 
theory at $\mu_*$, and all deviations from the double-commutator structure are generated radiatively. The scale $\mu_*$ is 
the only dimensionful parameter of the quantum theory.

\subsection{Four-scalar function at one loop}

The four-scalar function determines which of the symmetry-allowed
interactions are generated by quantum fluctuations. We extract its local
one-loop pole and match it to the quartic basis, including the external
scalar wave-function renormalization. Its divergent graphs,
with four external scalars $\chi$ and internal scalar (solid) and gauge (wavy) lines, are
\begin{equation}
\label{eq:Rgraphs}
\begin{tikzpicture}[baseline=-3pt,scale=1.25,
gauge/.style={thick,decorate,decoration={snake,amplitude=0.55mm,segment length=1.8mm}}]
\begin{scope}[shift={(0,0)}]
\draw[thick] (-1,0.5)--(-0.5,0); \draw[thick] (-1,-0.5)--(-0.5,0);
\draw[thick] (0.5,0)--(1,0.5); \draw[thick] (0.5,0)--(1,-0.5);
\draw[thick] (0,0) circle (0.5);
\fill (-0.5,0) circle (1.3pt); \fill (0.5,0) circle (1.3pt);
\node[font=\small] at (0,-0.95) {(a)};
\end{scope}
\begin{scope}[shift={(2.6,0)}]
\draw[thick] (-1,0.5)--(-0.5,0); \draw[thick] (-1,-0.5)--(-0.5,0);
\draw[thick] (0.5,0)--(1,0.5); \draw[thick] (0.5,0)--(1,-0.5);
\draw[gauge] (0.5,0) arc (0:180:0.5);
\draw[gauge] (-0.5,0) arc (180:360:0.5);
\fill (-0.5,0) circle (1.3pt); \fill (0.5,0) circle (1.3pt);
\node[font=\small] at (0,-0.95) {(b)};
\end{scope}
\begin{scope}[shift={(5.2,0)}]
\draw[thick] (-1.05,0.5)--(-0.55,0); \draw[thick] (-1.05,-0.5)--(-0.55,0);
\draw[thick] (0.4,0.5)--(0.95,0.75); \draw[thick] (0.4,-0.5)--(0.95,-0.75);
\draw[gauge] (-0.55,0)--(0.4,0.5);
\draw[gauge] (-0.55,0)--(0.4,-0.5);
\draw[thick] (0.4,0.5)--(0.4,-0.5);
\fill (-0.55,0) circle (1.3pt); \fill (0.4,0.5) circle (1.3pt); \fill (0.4,-0.5) circle (1.3pt);
\node[font=\small] at (0,-0.95) {(c)};
\end{scope}
\begin{scope}[shift={(7.8,0)}]
\draw[thick] (-0.85,0.75)--(-0.45,0.45); \draw[thick] (0.85,0.75)--(0.45,0.45);
\draw[thick] (-0.85,-0.75)--(-0.45,-0.45); \draw[thick] (0.85,-0.75)--(0.45,-0.45);
\draw[thick] (-0.45,0.45)--(0.45,0.45); \draw[thick] (-0.45,-0.45)--(0.45,-0.45);
\draw[gauge] (-0.45,0.45)--(-0.45,-0.45);
\draw[gauge] (0.45,0.45)--(0.45,-0.45);
\fill (-0.45,0.45) circle (1.3pt); \fill (0.45,0.45) circle (1.3pt);
\fill (-0.45,-0.45) circle (1.3pt); \fill (0.45,-0.45) circle (1.3pt);
\node[font=\small] at (0,-0.95) {(d)};
\end{scope}
\end{tikzpicture}
\end{equation}
together with their crossed versions and the scalar wave-function renormalization on the external legs: (a) the scalar
bubble built from two quartic vertices of $U_{\rm cl}$, (b) the gauge bubble built from two seagull vertices
$\chi\chi A A$, (c) the mixed triangles with one seagull and two cubic vertices $\chi\chi A$, and (d) the gauge-scalar box.
In the uplifted language of Sect.~\ref{sec:Quantization}, they are simply the one-loop four-point graphs of the
$\Ds$-dimensional Yang-Mills theory restricted to external components along the extra directions. The color structures
of the divergent parts project onto the basis \re{eq:Rbasis}. Each graph produces several of these structures, and their
sum is not proportional to $O_c$: this is the non-closure of the double commutator under renormalization. With
the counterterm potential
\begin{equation}
\label{eq:RCT}
\Delta U_{\rm CT}=\gym^2\Bigl(h_c\,O_c+h_s\,O_s+\frac{h_w}{N}\,O_3+\frac{h_z}{N}\,O_4\Bigr)
\, ,
\end{equation}
the one-loop pole parts are, in terms of the 't~Hooft coupling
\begin{equation}
\label{eq:gR}
g^2\equiv\frac{\gym^2N}{16\pi^2}
\, ,
\end{equation}
held fixed in the planar limit $N\to\infty$,
\begin{equation}
\label{eq:Rpoles}
(h_c,h_s,h_w,h_z)_{\rm pole}=\frac{(\Ds-2)\,g^2}{\epsilon}\Bigl(\frac1{12},\frac14,\frac12,1\Bigr)
\, ,
\end{equation}
with the finite parts set to zero at $\mu_*$ in accord with Eq.~\re{eq:Rcond}. Equation \re{eq:Rpoles} implies that the
induced couplings run. For the one that matters most below,
\begin{equation}
\label{eq:betas}
\beta_s\equiv\frac{\dd}{\dd\log\mu^2}\,\frac{\lambda_{s,R}}{\gym^2}=\frac{(\Ds-2)\,g^2}{4}+O(g^4)
\, ,
\end{equation}
so that $\lambda_{s,R}(\mu)/\gym^2=\beta_s\log(\mu^2/\mu_*^2)$ at one loop. Note that $\beta_s$ is positive and has no
fixed point at $\lambda_s=0$: the double-commutator structure cannot be maintained at all scales, and $\mu_*$ is the
scale at which it holds exactly. A similar running of the double-trace couplings is known to obstruct conformality of
non-supersymmetric large-$N$ theories~\cite{Dymarsky:2005uh,Pomoni:2008de}.

\subsection{Descendants on the moduli space}

Expanding the quartic operators about the scalar background connects the four-point counterterms to the mass and insertion counterterms needed below. After the shift \re{eq:shift}, a local four-leg counterterm
produces descendants with fewer legs. Setting slots of the operator to the background ($\otimes$) generates the mass
(two-leg), tadpole (one-leg) and vacuum-energy (zero-leg) counterterms,
\begin{equation}
\label{eq:Rdesc}
\begin{tikzpicture}[baseline=-3pt]
\draw[thick] (-0.5,0.5)--(0.5,-0.5); \draw[thick] (-0.5,-0.5)--(0.5,0.5);
\draw[fill=white] (0,0) circle (0.16); \node[font=\scriptsize] at (0,0) {c};
\end{tikzpicture}
\;\to\;
\begin{tikzpicture}[baseline=-3pt]
\draw[thick] (-0.5,0.5)--(0,0); \draw[thick] (-0.5,-0.5)--(0,0);
\bgleg{0.5}{0.5} \bgleg{0.5}{-0.5}
\draw[fill=white] (0,0) circle (0.16); \node[font=\scriptsize] at (0,0) {c};
\end{tikzpicture}
\;\to\;
\begin{tikzpicture}[baseline=-3pt]
\draw[thick] (-0.7,0)--(0,0);
\bgleg{0.5}{0.5} \bgleg{0.5}{-0.5} \bgleg{0.68}{0}
\draw[fill=white] (0,0) circle (0.16); \node[font=\scriptsize] at (0,0) {c};
\end{tikzpicture}
\;\to\;
\begin{tikzpicture}[baseline=-3pt]
\bgleg{0.5}{0.5} \bgleg{0.5}{-0.5} \bgleg{-0.5}{0.5} \bgleg{-0.5}{-0.5}
\draw[fill=white] (0,0) circle (0.16); \node[font=\scriptsize] at (0,0) {c};
\end{tikzpicture}
\, .
\end{equation}
No new counterterms are required for the vacuum energy and for the one-point function: they are the descendants of the
four-point counterterms \re{eq:RCT}. That this is consistent is a nontrivial check. On any commuting traceless background,
the masses $m_{kl}^2=(x_k-x_l)^2$ of Sect.~\ref{sec:vacua} obey the identity
\begin{equation}
\label{eq:Rquarticid}
\sum_{k\neq l}m_{kl}^4=\gym^4\bigl[(N+2)\,O_s(V)+2\,O_3(V)+4\,O_4(V)\bigr]
\, ,\qquad O_c(V)=0
\, ,
\end{equation}
which expresses the sum of the fourth powers of the masses, i.e., the quantity that controls the divergence of the vacuum
energy, through the operators of the basis. In Sect.~\ref{sec:QuantumStability} we will see that the zero-leg descendant
of Eq.~\re{eq:RCT}, with the pole parts \re{eq:Rpoles}, cancels the divergence of the one-loop vacuum energy exactly. Note
also that, while the individual operators in Eq.~\re{eq:Rquarticid} are not invariant under the flavor translations
$V_I\to V_I+a_I\mathbb I$ of Sect.~\ref{sec:vacua}, the combination on the right-hand side is.

\subsection{One-point vertices}

The one-point descendants determine how the same counterterms enter the
background equation of motion. Computing them explicitly makes the
renormalization of the tadpole consistent with the four-scalar
calculation, rather than introducing a separate prescription for the
vacuum force. On the commuting backgrounds of the family 
\re{eq:V1}-\re{eq:V2}, with a degenerate common block and $m_a^2=m_b^2=m_z^2$, the gradients 
$\widehat U_{X;aI}\equiv\partial O_X/\partial v_{aI}$ are, exactly in $N$,
\begin{align}
\label{eq:RVX}
\widehat U_{c;aI}
&
=0
\, , \nonumber\\
\widehat U_{s;aI}
&
=
8\Bigl[v_{aI}m_z^2-\frac{S_I}{N+2}\Bigl(m_z^2-\frac{S^2}{N}\Bigr)\Bigr]
\, ,\nonumber\\
\widehat U_{3;aI}
&
=
4v_{aI}\Bigl(2m_z^2+\frac{S^2}{N}\Bigr)
\, ,\nonumber\\
\widehat U_{4;aI}
&
=
4\Bigl[v_{aI}m_z^2+v_{bI}\,\chi+\frac{S_I}{N}\bigl(m_z^2+\chi\bigr)\Bigr]
\, ,
\end{align}
where $S_I=(v_a+v_b)_I$ and $\chi=v_a\cdot v_b$; the $b$-channel follows by $a\leftrightarrow b$. Three features of these
vertices determine which counterterms can move the vacuum; we spell them out in Sect.~\ref{sec:QuantumStability}. First,
the classical structure $O_c$ is quadratic in commutators, so that its gradient $\propto[\Phi_J,[\Phi_I,\Phi_J]]$ vanishes
identically on $\mathcal M_0$. Second, $O_s$ has an unsuppressed gradient along $v_a$. Third, the double-trace gradients
carry the explicit $1/N$ of Eq.~\re{eq:UR}, which is not compensated by an index loop, and $\widehat U_4$ has a component
along $v_b$, i.e., transverse to $v_a$.

\section{Quantum stationary points}
\label{sec:QuantumStability}

The classical moduli space $\mathcal M_0$ of Sect.~\ref{sec:vacua} is flat: every commuting background costs no energy. This
degeneracy is an artifact of the classical approximation. In the absence of supersymmetry, quantum fluctuations generate a
potential on $\mathcal M_0$ that selects distinguished points and determines whether they are stable. A consistent Sudakov calculation must either expand about a stationary background or include the source that holds a chosen background fixed. We therefore determine the quantum force before computing the observable. In this section, using the renormalized potential of Sect.~\ref{sec:Renormalization}, we determine the one-loop effective
potential on $\mathcal M_0$ and its stationary points within the family \re{eq:V1}-\re{eq:V2}. Two branches emerge: the null branch A at
$z=\pm\ii$ and the Coleman-Weinberg branch B, whose infrared hierarchies are distinguished below.

\subsection{One-loop effective potential}

Quantum fluctuations lift the classical moduli space through the
mass-dependent vacuum energy. We obtain this contribution from the
quadratic fluctuation operator, keeping the vector, scalar and ghost
multiplicities together so that the background force can subsequently
be renormalized with the same scalar counterterms.

For constant scalar fields, the effective action reduces to 
\begin{align}
\Gamma = - \int\dd^dx\,U_{\rm eff}(\Phi)
\, ,
\end{align}
with the effective potential $U_{\rm eff}$ admitting the loop expansion
\begin{equation}
\label{eq:QSloop}
U_{\rm eff}(\Phi)=U_{\rm cl} (\Phi)+\hbar\,U_1(\Phi)+\hbar^2\,U_2(\Phi)+\dots
\, .
\end{equation}
Here, $U_{\rm cl}$ is the tree-level potential of the quantum theory, i.e., the renormalized potential \re{eq:UR} with the renormalized couplings at 
the scale $\mu_*$. By the renormalization conditions \re{eq:Rcond}, $U_{\rm cl} (\Phi)=\gym^2\,O_c(\Phi)$, so that $U_{\rm cl} (V)=0$ at 
every point $V$ of the classical moduli space, in particular on the family \re{eq:V1}-\re{eq:V2}. The power of $\hbar$ counts loops: with $\hbar$ 
restored, each propagator carries $\hbar$ and each vertex $\hbar^{-1}$. In units $\hbar=1$, the same bookkeeping is carried by the 't~Hooft 
coupling \re{eq:gR}: after the rescaling $\Phi=\tilde\Phi/\gym$ the action is $\tilde S/\gym^2$, and in the planar limit each loop also produces a 
free color-index loop, $N$, and a factor $1/(16\pi^2)$ from the momentum measure. Hence ``order $\hbar$'' and ``order $g^2$'' are
interchangeable below. The higher terms $U_{L\geq1}$ include the counterterms of Sect.~\ref{sec:Renormalization} at the corresponding order.

Since $U_{\rm cl} (V)=0$ on $\mathcal M_0$, the fate of the classical vacua is decided by $U_1$. Writing the effective action
as $\Gamma=S+\hbar\,\Gamma_1+O(\hbar^2)$, so that $\Gamma_1=-\int\dd^dx\,U_1$ for constant fields, the one-loop term is given by
the Gaussian integral over the fluctuations of the uplifted field and of the ghosts,
\begin{equation}
\label{eq:QSGamma1}
\Gamma_1 = \frac{\ii}{2}\operatorname{Tr}\ln K_X-\ii\operatorname{Tr}\ln K_{\rm gh}
\, ,
\end{equation}
where $K_X$ is the quadratic kernel \re{eq:kernel} and $K_{\rm gh}$ the ghost kernel following from Eq.~\re{eq:Lgh}. For a
constant diagonal background, both kernels are diagonal in momentum and in the color pair $(ij)$. At $c=1$, $K_X$ depends on
the uplifted momentum alone, with $\det_{MN}K_X\propto (P^2)^{\Ds}/\alpha$, while the ghost kernel is $-P^2$. The gauge
parameter thus cancels, and each complex color pair contributes $\Ds$ bosonic and two ghost factors of
$\ln(m_{ij}^2-p^2)$ so that we find after Wick rotation
\begin{equation}
\label{eq:QSU1det}
U_1=(\Ds-2)\sum_{i<j}J(m_{ij}^2)
\, ,\qquad
J(m^2)=\mu^{2\epsilon}\int\frac{\dd^dp_{\ell_y}}{(2\pi)^d}\ln\bigl(p_{\ell_y}^2+m^2\bigr)
\, .
\end{equation}
The factor $\Ds-2$ counts the physical polarizations per color pair, Eq.~\re{eq:dof}. Again, the dimensional-reduction
picture is manifest: the effective potential of the reduced theory is the vacuum energy of a massless $\Ds$-dimensional
gauge field whose extra-dimensional momenta are frozen at $\kappa_{ij}$.

Equation \re{eq:QSU1det} is the bosonic part of the supertrace ${\rm Str}\,J(\mathcal M^2)$ of the mass matrix. In a theory
with fermions in the same multiplets, e.g., in $\cN=4$ sYM at $\Ds=10$, the fermionic contribution $-8\sum_{i<j}J(m_{ij}^2)$
cancels the bosonic one exactly, and the moduli space stays flat to all orders. Here, nothing compensates the vacuum
energy of the background-dependent heavy states, so $\mathcal M_0$ is lifted already at one loop. This is the mechanism of
Coleman and Weinberg~\cite{Coleman:1973jx}, operating on a non-supersymmetric Coulomb branch.

Diagrammatically, $U_1$ is the sum of closed vacuum bubbles of $X$ and ghosts in the $V$-background, one for each color pair,
\begin{equation}
\label{eq:QSbubbles}
U_1\;=\;\sum_{i<j}\Biggl[\;
\begin{tikzpicture}[baseline=-3pt]
\draw[thick] (0,0) circle (0.42); \node[font=\scriptsize] at (0,0) {$X_{ij}$};

\end{tikzpicture}
\;+\;
\begin{tikzpicture}[baseline=-3pt]
\draw[thick,dashed] (0,0) circle (0.42); \node[font=\scriptsize] at (0,0) {$\omega_{ij}$};
\end{tikzpicture}\;\Biggr]
\, .
\end{equation}
Each bubble resums an arbitrary number of background insertions, i.e., two-point vertices $\propto\kappa_{ij}^2$, on a massless
loop. The force on the background is the one-point function, obtained by differentiating with respect to a diagonal
component $V_{cI}$. This attaches a zero-momentum external leg of the Cartan scalar $\chi_{cI}$ to the bubble, while the loop
stays closed. The leg couples through the cubic vertex $\chi_{cI}X_{cl}X_{lc}\propto\kappa_{cl,I}$, i.e., through the
extra-dimensional component of the uplifted three-point vertex \re{eq:V3}. Since $\partial_{m^2}\ln(p^2+m^2)=1/(p^2+m^2)$,
each bubble turns into a tadpole,
\begin{equation}
\label{eq:QStadpole}
T^{(1)}_{cI}\equiv-\frac{\partial U_1}{\partial V_{cI}}=
\begin{tikzpicture}[baseline=-3pt]
\draw[thick] (0,0) circle (0.38); \draw[thick] (-1.1,0) -- (-0.38,0);
\node[font=\scriptsize,anchor=south] at (-0.8,0) {$\chi_{cI}$}; \node[font=\scriptsize] at (0,0) {$X$};
\end{tikzpicture}
+
\begin{tikzpicture}[baseline=-3pt]
\draw[thick,dashed] (0,0) circle (0.38); \draw[thick] (-1.1,0) -- (-0.38,0);
\node[font=\scriptsize,anchor=south] at (-0.8,0) {$\chi_{cI}$}; \node[font=\scriptsize] at (0,0) {$\omega$};
\end{tikzpicture}
=\frac{2(\Ds-2)\gym}{16\pi^2}\sum_{l\neq c}\kappa_{cl,I}\,A_0(m_{cl}^2)
\, ,
\end{equation}
with the tadpole integral $A_0(m^2)=m^2\bigl(\tfrac1\epsilon+1-\log\tfrac{m^2}{\mu^2}\bigr)$. Only diagonal legs have a
one-point function: by color conservation, $\sum_\ell\kappa_\ell=0$, a charged leg cannot be attached to a vacuum bubble.
The force therefore always points along $\mathcal M_0$, and the quantum vacuum stays commuting.

For the external color $c=a$, the sum in Eq.~\re{eq:QStadpole} runs over $l=b$ and over the $N$ colors of the common block,
\begin{equation}
\sum_{l\neq a}\kappa_{al,I}\,A_0(m_{al}^2)=\underbrace{N\,v_{aI}\,A_0(m_z^2)}_{O(N)}
+\underbrace{\kappa_{ab,I}\,A_0(m_0^2)}_{O(N^0)}+O(N^{-1})
\, .
\end{equation}
In double-line notation, the external field $\chi_{aa}$ carries color $a$ on both lines. Attaching it cuts open the outer
color line $a$, which enters and leaves through the leg, while the inner color line still closes. For the common-block
colors, the closed inner line is a free index loop worth $N$,
\begin{equation}
\begin{tikzpicture}[baseline=-3pt]
\draw[thick] (170:0.48) arc (170:-170:0.48);
\draw[thick,blue] (0,0) circle (0.30);
\draw[thick] (-1.25,0.0834) -- (170:0.48); \draw[thick] (-1.25,-0.0834) -- (190:0.48);
\node[font=\scriptsize,anchor=south] at (-0.95,0.06) {$\chi_{aa}$};
\node[font=\scriptsize] at (0.62,0.42) {$a$}; \node[font=\scriptsize,blue] at (0,0) {$r$};
\end{tikzpicture}
\;\propto\;\sum_{r}1=N
\, ,
\qquad
\begin{tikzpicture}[baseline=-3pt]
\draw[thick] (170:0.48) arc (170:-170:0.48);
\draw[thick,red] (0,0) circle (0.30);
\draw[thick] (-1.25,0.0834) -- (170:0.48); \draw[thick] (-1.25,-0.0834) -- (190:0.48);
\node[font=\scriptsize,anchor=south] at (-0.95,0.06) {$\chi_{aa}$};
\node[font=\scriptsize] at (0.62,0.42) {$a$}; \node[font=\scriptsize,red] at (0,0) {$b$};
\end{tikzpicture}
\;\propto\;1
\, .
\end{equation}
At fixed $g^2$, the $N$ loops of the $w_a$ multiplet with mass $m_z$ dominate, while the loop of the $W$ multiplet with
mass $m_0$ is suppressed by $1/N$. The same counting applies to the $b$ channel. Thus, at leading planar order, the
quantum force is generated entirely by the states that will later propagate in the loops of the form factor.

\subsection{Coleman-Weinberg potential}

We now subtract the ultraviolet poles of the vacuum determinants.
Their finite logarithms then expose the dimensional-transmutation
scale and provide the renormalized force whose zeros distinguish the
stationary branches.

The bubbles, defined by Eq.\ \re{eq:QSU1det}, develop ultraviolet poles,
\begin{align}
J (m^2)
=
-\frac{m^4}{32\pi^2}\left(\frac1\epsilon+\frac32-\log\frac{m^2}{\mu^2}\right)
\, .
\end{align}
They are removed by the zero-leg descendants \re{eq:Rdesc} of the four-scalar counterterms of Sect.~\ref{sec:Renormalization}. 
Inserting the pole parts \re{eq:Rpoles} into the counterterm potential \re{eq:RCT} evaluated on the background and using the identity
\re{eq:Rquarticid}, one finds $\Delta U_{\rm CT}(V)|_{\rm pole}=-U_1|_{\rm pole}$, exactly for the double-trace operators and at
leading planar order for $O_s$. The subtraction of the vacuum bubbles is therefore not an extra prescription: it is fixed
by the independent four-point function, and its derivative is the one-leg cancellation of the pole of the tadpole
\re{eq:QStadpole}. After the subtraction at $\mu_*$,
\begin{equation}
J_R(m^2)=\frac{m^4}{32\pi^2}\Bigl(\log\frac{m^2}{\mu_*^2}-\frac32\Bigr)
\, .
\end{equation}
The constant $-\tfrac32$ is common to all components because, in the dimensional-reduction scheme, vector and
$\epsilon$-scalar components combine into a single multiplet.

At leading planar order, the renormalized one-loop potential on $\mathcal M_0$ is the Coleman-Weinberg potential of the
$w_a$ and $w_b$ multiplets,
\begin{equation}
\label{eq:QSU1}
U_1=K\sum_{c=a,b}m_c^4\Bigl(\log\frac{m_c^2}{\mu_*^2}-\frac32\Bigr)
\, ,\qquad
K=\frac{N(\Ds-2)}{32\pi^2}
\, ,\qquad
m_c^2=v_c\cdot v_c
\, ,
\end{equation}
and the renormalized force reads
\begin{equation}
\label{eq:QSforce}
T^{(1)}_{cI}=-\gym\frac{\partial U_1}{\partial v_{cI}}=\frac{2(\Ds-2)g^2}{\gym}\,v_{cI}\,m_c^2\bigl(1-L_c\bigr)
\, ,\qquad
L_c\equiv\log\frac{m_c^2-\ii0}{\mu_*^2}
\, .
\end{equation}
This is the renormalized version of the tadpole \re{eq:QStadpole}. In what follows, we drop the superscript and write
$T_{cI}$ whenever the loop order is clear from the context. In units of the natural scale $m_c^4/\gym^2$ of the classical quartic potential,
$U_1=\tfrac12(\Ds-2)\,g^2\,(m_c^4/\gym^2)\bigl(\log\tfrac{m_c^2}{\mu_*^2}-\tfrac32\bigr)$, i.e., it is a one-loop effect of
relative order $g^2$, in agreement with the counting below Eq.~\re{eq:QSloop}. At planar order, $U_1$ depends on the background only through
the two invariants $m_a^2$ and $m_b^2$. In accord with the remark at the end of Sect.~\ref{sec:gauge}, the force is
independent of the gauge parameter at this order.

\subsection{Two ways out}
\label{sec:TwoWays}

Away from special points of $\mathcal M_0$, the force \re{eq:QSforce} does not vanish, and the classical background is
not a quantum vacuum: the tadpole \re{eq:QStadpole} feeds into every loop calculation around it. There are two
consistent ways to deal with this.

\paragraph{Hold the background.}
One may add to the action a constant linear source for the Cartan scalars and tune it to the force,
\begin{equation}
\label{eq:held}
S\to S-\int\dd^dx\,\sum_{c,I}H_{cI}\,\chi_{cI}
\, ,\qquad
H_{cI}=T_{cI}(V)
\, ,
\end{equation}
order by order in perturbation theory. Since the one-loop force enters the effective action as $\int\dd^dx\,T_{cI}\chi_{cI}$,
the source cancels the tadpoles exactly, the background stays where it was put,
for any $z$, and the calculation proceeds as if the classical background $V$ were a vacuum. The price is that observables are matrix
elements of a {\sl source-deformed} theory. Since the source is a local, nonpropagating insertion attached to
%
zero-momentum lines, it affects single logarithms and constants of infrared observables but not their leading
double-logarithmic asymptotics. This held-background prescription is the one referred to in the Introduction.

We next examine which backgrounds admit the walking hierarchy without a
holding source. This complements the calculation in Sect.~\ref{sec:FF},
where the form factor is evaluated on a generic held background without
requiring it to be stationary in the absence of the source.

\paragraph{Move to a true stationary point.}
Alternatively, one may look for the points of $\mathcal M_0$ where the quantum force vanishes. To avoid ambiguity, we
distinguish three objects: the tree-level potential $U_{\rm cl}(\Phi)$, the classical background
$V\in\mathcal M_0$ of Sect.~\ref{sec:vacua}, and the quantum vacuum $\vev{\Phi}=V+\hbar V^{(1)}+O(\hbar^2)$, the point where the
effective potential \re{eq:QSloop} is stationary. Expanding the stationarity condition $\partial U_{\rm eff}/\partial\Phi=0$
around $V$, the tree-level term starts at order $\hbar$ through the classical Hessian $H$ of Sect.~\ref{sect:Massed}, and
one finds
\begin{equation}
\label{eq:QSvaccond}
H\cdot V^{(1)}=T^{(1)}(V)
\, .
\end{equation}
This is a linear equation for $V^{(1)}$, whose operator is fixed by the classical background and whose source is the
one-loop force. Its fate is decided by the block structure \re{eq:Hblock} of $H$. Decompose the correction into its
diagonal entries $\delta_{iI}$ and off-diagonal entries $o_{ij,I}$. By Eqs.~\re{eq:Hform} and \re{eq:Hblock},
\begin{align}
(H\cdot V^{(1)})_{I,ii}=0 
\, , \qquad
(H\cdot V^{(1)})_{I,ij}=2(M_{ij}\,o_{ij})_I
\end{align} 
for $i\neq j$. For a pair with $\kappa_{ij}^2\neq0$, the flavor vector $o_{ij}$ splits into a gauge component along $\kappa_{ij}$, 
annihilated by $M_{ij}$, and a transverse component $o^\perp_{ij}$, on which $M_{ij}$ acts as $m_{ij}^2$. The force, on the other 
hand, has diagonal entries only. Row by row, Eq.~\re{eq:QSvaccond} therefore states:
\begin{center}
\small
\begin{tabular}{>{\raggedright\arraybackslash}p{3.6cm}>{\raggedright\arraybackslash}p{3cm}>{\raggedright\arraybackslash}p{2.3cm}>{\raggedright\arraybackslash}p{4.6cm}}
\toprule
Entries of $V^{(1)}$ & action of $H$ & force & consequence\\
\midrule
$\delta_{aI}$, $\delta_{bI}$ (along $\mathcal M_0$) & $0$ & $T_{aI}$, $T_{bI}$ & $0=T_{cI}$: condition on $V$\\[2pt]
$\delta_{rI}$ (common block) & $0$ & $- \tfrac{T_{aI}+T_{bI}}{N}$ & satisfied once $T_a=T_b=0$\\[2pt]
$o^\parallel_{ij}$ (gauge) & $0$ & $0$ & undetermined, gauge fixed\\[2pt]
$o^\perp_{ab}$ (transverse) & $m_0^2\,o^\perp_{ab}$ & $0$ & $o^\perp_{ab}=0$\\[2pt]
$o^\perp_{ar}$, $o^\perp_{br}$ (transverse) & $m_z^2\,o^\perp$; nilpotent at $z=\pm\ii$ & $0$ & $o^\perp=0$; at $z=\pm\ii$ only $\kappa\cdot o=0$\\[2pt]
$o_{rs}$ (common block) & $0$ & $0$ & unbroken $\U(N)$ rotation\\
\bottomrule
\end{tabular}
\end{center}
Thus, at order $\hbar$, no off-diagonal correction is generated and the quantum vacuum stays commuting. The diagonal
rows, however, read $0\cdot\delta_{cI}=T_{cI}$, because $H$ annihilates the Cartan sector. They are not equations for
$V^{(1)}$ but solvability conditions on the classical background,
\begin{equation}
\label{eq:QSsolv}
T^{(1)}_{aI}(V)=T^{(1)}_{bI}(V)=0
\, .
\end{equation}
This is degenerate perturbation theory on a flat valley~\cite{Coleman:1973jx,Weinberg:1973am}: at one loop, quantum
corrections do not shift an arbitrary point of $\mathcal M_0$; they \textsl{select} the points where the quantum force
vanishes. The diagonal correction is fixed only at the next order, by $\partial^2U_1\cdot V^{(1)}_{\rm diag}=T^{(2)}_{\rm diag}$,
where the one-loop curvature supplies the stiffness absent at tree level. A point violating Eq.~\re{eq:QSsolv} has no
perturbative quantum vacuum nearby: regulating the vanishing diagonal block of $H$ by a mass $\lambda^2$ gives
$\delta V=T/\lambda^2\to\infty$. Such a point can be used only if it is held in place by the source \re{eq:held}.

The various directions in field space have a transparent geometric meaning. The diagonal entries are tangent to
$\mathcal M_0$. Among them, the radial direction $\delta\mathbf v=\varepsilon\,\mathbf v$, with $\mathbf v=(v_a,v_b)$, rescales all
masses at fixed $z$, while the angular directions change the ratio $m_a/m_b$ or the relative orientation $\chi=v_a\cdot v_b$.
The gauge directions $o_{ij}\propto\kappa_{ij}$ are infinitesimal gauge rotations of the background and are flat to all orders.
The transverse directions are normal to $\mathcal M_0$ and carry the tree-level stiffness $m_{ij}^2$. The quantum force acts
only along the valley floor, and its radial component,
\begin{equation}
\label{eq:QSradial}
\mathbf v\cdot\nabla_{\mathbf v}U_1=4K\sum_{c=a,b}m_c^4\bigl(L_c-1\bigr)
\, ,
\end{equation}
is what decides where on the floor the stationary points are.

\subsection{Counterterms moving the vacuum}
\label{sec:CTvacuum}

The loop force alone does not specify the background equation: finite
choices of the renormalized scalar couplings can also change it.
We distinguish this physical choice of theory from the pole
counterterms that are already fixed by ultraviolet finiteness.

Besides the loop term, the effective potential contains the counterterm potential \re{eq:RCT}, which acts on the vacuum
through the one-point vertices \re{eq:RVX},
\begin{equation}
T^{(1)}_{{\rm CT},cI}=-\frac{1}{\gym}\sum_{X=c,s,3,4}\hat h_X\,\widehat U_{X;cI}
\, ,\qquad
(\hat h_c,\hat h_s,\hat h_3,\hat h_4)=\Bigl(h_c,h_s,\frac{h_w}{N},\frac{h_z}{N}\Bigr)
\, .
\end{equation}
The pole parts of the $h_X$ cancel the ultraviolet pole of the loop tadpole \re{eq:QStadpole}; this cancellation is already
built into the renormalized force \re{eq:QSforce}. Only the finite parts $h^{\rm fin}_X$, i.e., the values of the induced
couplings at the scale where the vacuum is determined, can move the vacuum. Including them, the leading-planar
stationarity condition in the $a$ channel becomes
\begin{equation}
\label{eq:gencond}
\frac{v_{aI}m_z^2}{\gym}\Bigl[2(\Ds-2)g^2\bigl(1-L_z\bigr)-8h_s^{\rm fin}\Bigr]
-\frac{4}{\gym N}\Bigl[2h_w^{\rm fin}v_{aI}m_z^2+h_z^{\rm fin}\bigl(v_{aI}m_z^2+v_{bI}\chi\bigr)\Bigr]
+O\bigl(N^{-1}S\bigr)=0
\, ,
\end{equation}
with $L_z\equiv\log(m_z^2/\mu_*^2)$. Coupling by coupling:
\begin{itemize}
\item \textbf{$h_c$ has no effect}, to all orders in $1/N$ and for any finite part. The classical structure $O_c$ is
quadratic in commutators, so its gradient vanishes on every commuting background, $\widehat U_{c;aI}=0$. A change of
$\lambda_{c,R}=\gym^2(1+h_c)$ changes the curvature of the walls of the valley, i.e., the transverse masses, but never the
position along its floor.
\item \textbf{$h_s$ fixes the radial position} at leading planar order. From Eq.~\re{eq:gencond},
\begin{equation}
\label{eq:hs}
L_z=1-\frac{4h_s^{\rm fin}}{(\Ds-2)g^2}
\quad\Longleftrightarrow\quad
m_z^2={\rm e}\,\mu_*^2\exp\Bigl[-\frac{4h_s^{\rm fin}}{(\Ds-2)g^2}\Bigr]
\, .
\end{equation}
Since $h_s^{\rm fin}=O(g^2)$, the exponent is of order one: a finite part moves the vacuum by an $O(1)$ factor. With the
renormalization conditions \re{eq:Rcond}, $h_s^{\rm fin}(\mu_*)=0$ and $m_z^2={\rm e}\,\mu_*^2$.
\item \textbf{$h_w$ shifts the radial position at $O(1/N)$}, since its force is parallel to $v_a$.
\item \textbf{$h_z$ shifts the radial position at $O(1/N)$ and exerts an angular force.} Its vertex has a component
$\propto v_{bI}\,\chi$ along $v_b$. Together with the $O(N^0)$ loop of the $W$ multiplet, it lifts the planar flatness of the
relative orientation $\chi$ and therefore selects $z$ at finite $N$.
\end{itemize}
In short, at leading planar order the position of the vacuum is controlled by a single number, the value of the induced
single-trace coupling $\lambda_s$ at the scale of the vacuum.

\subsection{Branches A and B}

We next solve the renormalized stationarity condition within the chosen
background family. Its two branches have different infrared spectra;
their existence must be distinguished from the stability question,
which is decided by the Hessian and effective potential below.

\begin{figure}[t]
\begin{center}
\includegraphics[width=\textwidth]{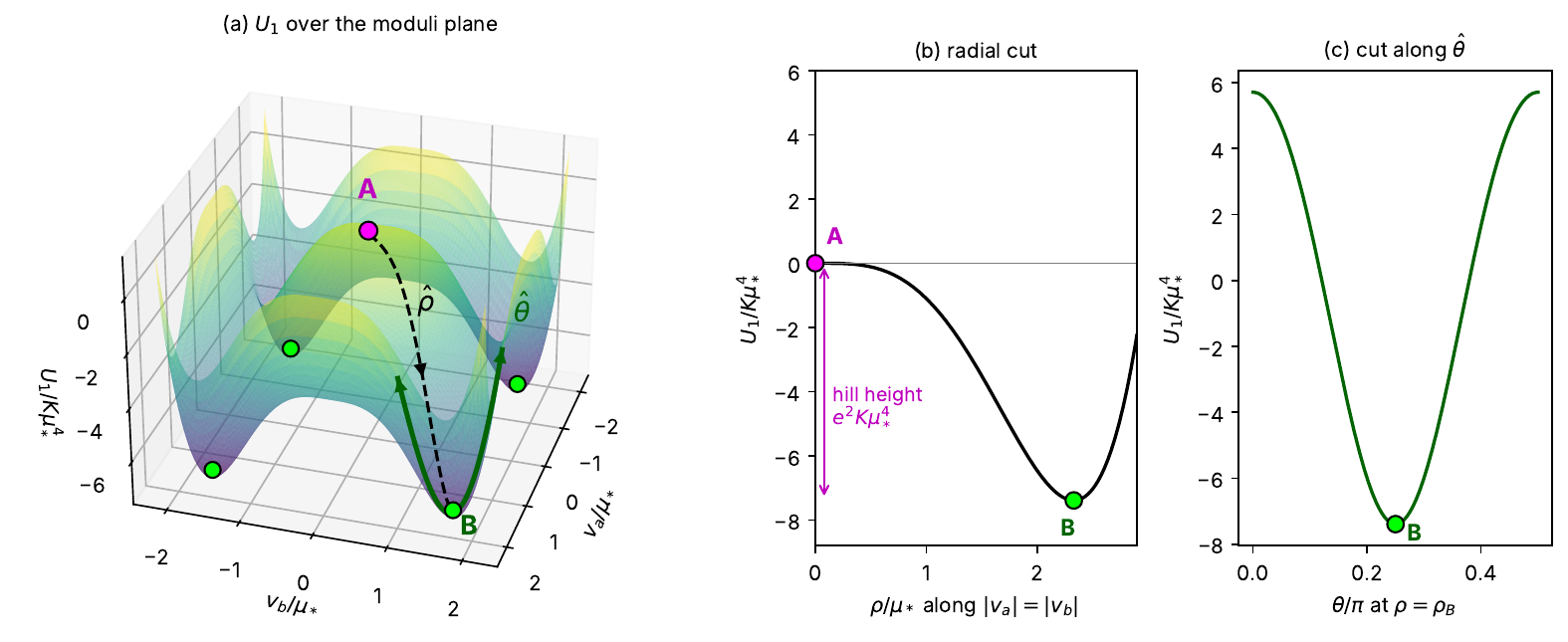}
\end{center}
\caption{The one-loop potential $U_1/(K\mu_*^4)=\sum_cf(v_c^2/\mu_*^2)$, $f(x)=x^2(\log x-\tfrac32)$, on the moduli plane,
where $U_{\rm cl}\equiv0$. (a) $U_1$ over signed $(v_a,v_b)$ along fixed flavor directions; it depends only on $v_a^2$ and
$v_b^2$. A (magenta) is a hilltop with $U_1=0$, surrounded by the four wells of branch B (lime) at $v_a^2=v_b^2={\rm e}\,\mu_*^2$.
Dashed: the roll-down path from A to B along $|v_a|=|v_b|$ (direction $\hat\rho$). Green arc: the channel-asymmetry direction
$\hat\theta$ at fixed $\rho^2=v_a^2+v_b^2$. (b) Radial cut: a hill of height ${\rm e}^2K\mu_*^4$ separates A from B. (c) Cut
along $\hat\theta$ at $\rho=\rho_B=\sqrt{2{\rm e}}\,\mu_*$, with $\tan\theta=|v_b|/|v_a|$: B is a minimum in this direction as well.
A appears here at real $|v|\to0$; within the $z$-family, it is reached at $z=\pm\ii$, where $v_c$ is null but nonzero.}
\label{fig:U1}
\end{figure}

With Eq.~\re{eq:QSforce}, the solvability conditions \re{eq:QSsolv} become $v_{cI}\,m_c^2(1-L_c)=0$. For $v_c\neq0$ there are
two disjoint sets of solutions,
\begin{equation}
\label{eq:QSbranches}
\mbox{branch A:}\quad m_a^2=m_b^2=0
\, ,\qquad\qquad
\mbox{branch B:}\quad m_a^2=m_b^2={\rm e}\,\mu_*^2
\, .
\end{equation}
Branch B is single valued in $m_c^2$: $\log(m_c^2/\mu_*^2)+2\pi\ii k=1$ yields the same $m_c^2$ on every sheet of the
logarithm. In the two-parameter family \re{eq:V1}-\re{eq:V2}, where $m_a^2=m_b^2=m_z^2=\tfrac14m_0^2(1+z^2)$,
\begin{itemize}
\item branch A is located at the null points $z=\pm\ii$, for any $m_0$;
\item branch B is the curve $m_0^2(1+z^2)=4{\rm e}\,\mu_*^2$, i.e., $z_B^2=4{\rm e}\,\mu_*^2/m_0^2-1$ at fixed $m_0$.
\end{itemize}
The two branches never meet: along branch B, $m_0$ diverges as $z\to\pm\ii$, because B keeps $m_z^2$ fixed and nonzero (see
also Table~\ref{tab:branches}). The
landscape is displayed in Fig.~\ref{fig:U1}. Branch A sits on top of a hill, branch B at the bottom of a well, and the
radial direction rolls from one to the other.

\subsubsection{The one-loop Hessian}

Whether a stationary point is a minimum is decided by the curvature of the effective potential along $\mathcal M_0$, where
the classical Hessian of Sect.~\ref{sect:Massed} vanishes identically. At one loop, the curvature is supplied by $U_1$. Since
the $a$-channel part of Eq.~\re{eq:QSU1} depends on $v_a$ only through $m_a^2=v_a\cdot v_a$, i.e., $U_1\supset K f(m_a^2)$ with
$f(x)=x^2\bigl(\log(x/\mu_*^2)-\tfrac32\bigr)$, the chain rule gives
\begin{align}
\label{eq:QSHess1}
\frac{\partial^2U_1}{\partial v_{cI}\partial v_{cJ}}
&=2K\,\delta_{IJ}\,f'(m_c^2)+4K\,v_{cI}v_{cJ}\,f''(m_c^2)
\nonumber\\
&=\frac{N(\Ds-2)}{8\pi^2}\Bigl[\delta_{IJ}\,m_c^2(L_c-1)+2\,v_{cI}v_{cJ}\,L_c\Bigr]
\, ,
\end{align}
with $f'(x)=2x(L-1)$ and $f''(x)=2L$, where $L=\log(x/\mu_*^2)$. The two terms have a transparent diagrammatic origin. The
second derivative of the effective potential is the zero-momentum two-point function of the Cartan scalar $\chi_{cI}$, which
at one loop receives two contributions,
\begin{equation}
\label{eq:QSHessdiag}
\frac{\partial^2U_1}{\partial v_{cI}\partial v_{cJ}}\ \propto\
\underbrace{\begin{tikzpicture}[baseline=-3pt]
\draw[thick] (0,0.42) circle (0.42);
\draw[thick] (-0.8,0)--(0,0)--(0.8,0);
\node[font=\scriptsize,anchor=north] at (-0.6,0) {$\chi_{cI}$};
\node[font=\scriptsize,anchor=north] at (0.6,0) {$\chi_{cJ}$};
\fill (0,0) circle (1.5pt);
\end{tikzpicture}}_{\text{seagull: }\ \delta_{IJ}\,A_0(m_c^2)}
\;+\;
\underbrace{\begin{tikzpicture}[baseline=-3pt]
\draw[thick] (0,0) circle (0.42);
\draw[thick] (-1.1,0)--(-0.42,0); \draw[thick] (0.42,0)--(1.1,0);
\node[font=\scriptsize,anchor=south] at (-0.8,0) {$\chi_{cI}$};
\node[font=\scriptsize,anchor=south] at (0.8,0) {$\chi_{cJ}$};
\fill (-0.42,0) circle (1.5pt); \fill (0.42,0) circle (1.5pt);
\node[font=\scriptsize] at (0,0.62) {$(cr)$};
\end{tikzpicture}}_{\text{bubble: }\ v_{cI}v_{cJ}\,B_0(0;m_c^2,m_c^2)}
\, .
\end{equation}
The seagull uses the quartic vertex $\chi\chi XX$ with both external legs at one point. Its loop is the tadpole
$A_0(m_c^2)$ of Eq.~\re{eq:QStadpole}, independent of the flavor directions, and after subtraction it gives the
$\delta_{IJ}\,m_c^2(L_c-1)$ term. The bubble uses two cubic vertices $\chi_{cI}X_{cr}X_{rc}$. On the background each of them is
proportional to the mass vector $\kappa_{cr}=v_c$ of the pair, i.e., to the extra-dimensional component of the uplifted
momentum, which produces the factor $v_{cI}v_{cJ}$. The two $(cr)$ propagators at zero external momentum form the massive
bubble
\begin{equation}
\label{eq:B00}
B_0(0;m^2,m^2)=\mu^{2\epsilon}\!\int\!\frac{\dd^dk_{\ell_y}}{\pi^{d/2}}\,\frac{1}{(k_{\ell_y}^2+m^2)^2}
=\frac1\epsilon-\log\frac{m^2}{\mu^2}+O(\epsilon)
\, ,
\end{equation}
whose subtracted part is $-L_c$. The correspondence with Eq.~\re{eq:QSHess1} is exact. Differentiating the vacuum
bubble $J(m^2)$ of Eq.~\re{eq:QSU1det} with respect to $m^2$ gives
\begin{equation}
\frac{\dd J}{\dd m^2}\ \xrightarrow{\ \rm subtracted\ }\ \frac{m^2(L-1)}{16\pi^2}
\, ,\qquad
\frac{\dd^2J}{\dd(m^2)^2}=-\frac{B_0(0;m^2,m^2)}{16\pi^2}\ \xrightarrow{\ \rm subtracted\ }\ \frac{L}{16\pi^2}=\frac{f''(m^2)}{32\pi^2}
\, ,
\end{equation}
i.e., the subtracted tadpole and the subtracted zero-momentum bubble, respectively.

Physically, $f''(m_c^2)$ measures how strongly the vacuum energy of the heavy states responds to a change of their mass.
It is the effective quartic self-coupling of the modulus induced by integrating out the $(cr)$ states. As for any
zero-momentum bubble, it grows logarithmically as these states become light, $L_c\to-\infty$ for $m_c^2\to0$, and at
$m_c^2=0$ the bubble $\int\dd^4k/k^4$ is infrared divergent. This behavior controls the character of both branches.

\subsubsection{Branch B: the Coleman-Weinberg vacuum}

The scale of branch B is not an input: it is generated by the running of the radiatively induced coupling $\lambda_s$. At
a general renormalization scale $\mu$, the tree-level $O_s$ term with $\lambda_{s,R}=\gym^2h_s^{\rm fin}$ contributes to the
force, and the stationarity condition \re{eq:gencond} takes the form
\begin{equation}
\label{eq:QSrg}
\frac{\lambda_{s,R}(\mu)}{\gym^2}=\frac{(\Ds-2)g^2}{4}\Bigl(1-\log\frac{m_z^2}{\mu^2}\Bigr)+O(g^4,N^{-1})
\, .
\end{equation}
The stationarity condition is renormalization-group consistent: both sides have the same $\mu$ dependence at this order,
governed by $\beta_s$ in Eq.~\re{eq:betas}, which is fixed by the pole of $h_s$.
The dimensionless coupling $\lambda_s$ is traded for the scale $\mu_*$ at which it vanishes,
and $m_z^2={\rm e}\,\mu_*^2$: this is dimensional transmutation. A different finite part of $h_s$ merely relabels $\mu_*$, see Eq.~\re{eq:hs}. 
Since $\lambda_s$ is itself of order $g^2$, the tree and one-loop terms are of the same order, and the minimum lies where the logarithm is of 
order one. The one-loop treatment is therefore self-consistent and free of large logarithms, as in the analysis in Ref.\ \cite{Gildener:1976ih}.

At B, $L_c=1$, so the seagull term of the Hessian \re{eq:QSHess1} vanishes and only the bubble survives,
\begin{equation}
\label{eq:QSHessB}
\frac{\partial^2U_1}{\partial v_{cI}\partial v_{cJ}}\ \stackrel{\rm B}{\longrightarrow}\ \frac{N(\Ds-2)}{4\pi^2}\,v_{cI}v_{cJ}
\, .
\end{equation}
Its only nonzero eigenvalue, $N(\Ds-2)m_z^2/(4\pi^2)>0$, belongs to the radial direction $\hat v_c$, while the directions
orthogonal to $v_c$ have zero curvature. Accordingly, the $2n$ diagonal moduli $(v_a,v_b)$ decompose as follows:
\begin{itemize}
\item \textbf{Two radial moduli}, one per channel ($\delta v_c\parallel v_c$), acquire a one-loop mass. The fluctuation
$\delta v_a=\gym\,\delta\chi_{aa}$ has the kinetic term $(\partial \delta\chi_{aa})^2$, the bulk adjustment required by tracelessness
being $O(1/N)$, so that the canonical field is $\varphi=\sqrt2\, \delta\chi_{aa}$ and
\begin{equation}
\label{eq:QSdilaton}
m_\varphi^2=\frac{\gym^2}{2}\,\frac{N(\Ds-2)}{4\pi^2}\,m_z^2=2(\Ds-2)\,g^2\,m_z^2=8\,\beta_s\,m_z^2
\, .
\end{equation}
The radial modulus is parametrically lighter than the $w_c$ multiplets, $m_\varphi^2/m_z^2=O(g^2)$. It is the
pseudo-dilaton of the classically scale-invariant theory: its mass is set by the $\beta$ function of the coupling that
lifts the flat direction, as in the original Coleman-Weinberg mechanism.
\item \textbf{$2n-3$ Goldstone bosons.} The two vectors $v_a,v_b$ of fixed length break the global flavor symmetry
${\rm O}(n)\to{\rm O}(n-2)$ for $z\neq0$, and ${\rm O}(n)\to{\rm O}(n-1)$, with $n-1$ Goldstone bosons, at $z=0$, where
$v_b=-v_a$. These directions are exactly flat to all orders.
\item \textbf{One relative-orientation modulus}, $\chi=v_a\cdot v_b$, equivalently $m_0^2=2m_z^2-2\chi$ at fixed $m_z$, is flat
at planar order, since $U_1$ depends only on $v_a^2$ and $v_b^2$. It is lifted at $O(N^0)$, see Sects.~\ref{sec:CTvacuum} and \ref{sec:beyondLO}.
\end{itemize}
The count is complete, $2n=2+(2n-3)+1$. The vacuum energy at B,
\begin{equation}
U_1\big|_{\rm B}=2K\,({\rm e}\,\mu_*^2)^2\Bigl(1-\frac32\Bigr)=-K\,{\rm e}^2\mu_*^4<0=U_1\big|_{\rm A}
\, ,
\end{equation}
lies below the classical value $U_{\rm cl}=0$ of the entire moduli space. The vacuum energy, the masses of the heavy
multiplets and the mass of the pseudo-dilaton are all controlled by the single generated scale $\mu_*$.

\subsubsection{Branch A: the null background}

At $z=\pm\ii$, the force vanishes because $m_z^2\log m_z^2\to0$. The classical background is stationary without any
correction, $V^{(1)}=0$ at this order, and no renormalization scale is selected: branch A exists for every $\mu_*$ and every
$m_0$. This is what makes it the natural background for the strictly off-shell Sudakov regime: massive external
$W$ bosons with massless planar internal lines, without any source or tuning.

Stationarity, however, does not imply stability. Along the deformations that keep the loop masses real, e.g.,
$z=\ii\zeta$ with $\zeta\to1^-$, for which $m_z^2=\tfrac14m_0^2(1-\zeta^2)>0$,
\begin{equation}
U_1-U_1\big|_{\rm A}=2K\,m_z^4\Bigl(\log\frac{m_z^2}{\mu_*^2}-\frac32\Bigr)<0
\quad
\mbox{for}\ 0<m_z^2<{\rm e}^{3/2}\mu_*^2
\, ,\qquad
\frac{\dd^2U_1}{\dd(m_z^2)^2}=4K\,L_z\to-\infty
\, .
\end{equation}
Branch A is a local maximum: the energy decreases in every such direction, quartically in the field, $\propto m_z^4\log
m_z^2$, with a logarithmically divergent negative curvature in $m_z^2$. The system rolls away from A, toward B whenever B
exists.

In the field coordinates, the first term of the Hessian \re{eq:QSHess1} vanishes at A, while the second does not,
\begin{equation}
\frac{\partial^2U_1}{\partial v_{aI}\partial v_{aJ}}\ \stackrel{\rm A}{\longrightarrow}\ \frac{N(\Ds-2)}{4\pi^2}\,L_a\,(v_a\otimes v_a)_{IJ}
\, ,\qquad L_a\to-\infty
\, .
\end{equation}
Only the bubble survives, and it is proportional to the matrix $P\equiv v_a\otimes v_a$. Since $v_a$ is null,
$P^2=(v_a\cdot v_a)\,P=0$: $P$ is nilpotent. Its characteristic polynomial is $\lambda^n$, so its only eigenvalue is zero.
A diagonalizable matrix whose eigenvalues all vanish is the zero matrix, whereas $P\neq0$; hence $P$ is not diagonalizable:
it has fewer independent eigenvectors than dimensions, and there is no eigenbasis in which the curvature could be read off
direction by direction. Equivalently, along a direction $\chi$ the curvature is $w\cdot\partial^2U_1\cdot w\propto L_a\,(v_a\cdot w)^2$,
which is complex for real $\chi$, because $v_a$ is complex, and diverges logarithmically. The quadratic test is therefore
inconclusive, and it is the potential itself that settles the question, as above.
The logarithmic divergence reflects the $w_a$ and $w_b$ multiplets becoming massless, so that the bubble
$B_0(0;m^2,m^2)$ is infrared divergent at $m=0$. Physically, branch A is a legitimate expansion point for perturbation
theory, being free of tadpoles, but it is an unstable equilibrium rather than a vacuum into which the system settles. The
infrared observables computed around it are well defined order by order in perturbation theory. The corresponding
branch-A Sudakov analysis is presented in the companion paper.

\subsection{Beyond leading order}
\label{sec:beyondLO}

The one-loop planar analysis leaves some background directions flat
and fixes the generated scale only to leading perturbative order.
We separate the effects of finite color number, higher-loop vacuum
forces, and gauge or subtraction conventions, indicating which
additional calculations each would require.

\paragraph{Finite $N$.}
At $O(N^0)$, the $W$-multiplet loop contributes $\propto m_0^4(\log m_0^2/\mu_*^2-\tfrac32)$ to $U_1$, and, as discussed
in Sect.~\ref{sec:CTvacuum}, the $O_4$ counterterm exerts an angular force. Both lift the planar flatness of the relative
orientation $\chi$. If the $W$-loop carries the same
Coleman-Weinberg structure, stationarity in $\chi$ would also require $m_0^2={\rm e}\,\mu_*^2$, i.e., $z^2=3$ within the family,
where all three masses are degenerate. Establishing this requires the complete one-point function at finite $N$.

\paragraph{Higher loops.}
The diagonal correction $V^{(1)}_\parallel$ is fixed at two loops by $\partial^2U_1\cdot V^{(1)}_\parallel=T^{(2)}$. Since
$\partial^2U_1=O(g^2)$ along the radial direction, the vacuum shift is of relative order $g^2$. Along the exact
Goldstone directions it is undetermined, as it must be. A shift of $m_z^2$ by a relative amount $O(g^2)$ changes the ratio $y$
of the internal mass to the external virtuality, Eq.~\re{eq:walkcond}, by the same relative amount and therefore affects only single logarithms of the Sudakov form factor, not its double-logarithmic
asymptotics.

\paragraph{Gauge and scheme.}
By the Nielsen identity~\cite{Nielsen:1975fs}, the location of the stationary points on $\mathcal M_0$ (up to field
redefinitions) and the vacuum energy there are gauge independent order by order. The number ${\rm e}$ in $m_z^2={\rm e}\,\mu_*^2$
reflects the constant $-\tfrac32$ of the dimensional-reduction scheme; in conventional dimensional regularization the
vector contribution carries a different constant. The physical content, a scale generated by dimensional transmutation
together with a light pseudo-dilaton of mass $m_\varphi^2=8\beta_s m_z^2$, is scheme independent.

\subsection{Stationary points in the complex $z$-plane}

The complex-$z$ description shows how the two stationary branches are
embedded in the same analytically continued background family.
Here, real slices are restrictions of the complexified background family
on which the effective potential is real, such as real $z$ or
$z=\ii\zeta$ with real $|\zeta|<1$, at fixed real $m_0>0$.
Statements about minima refer to variations within the specified slice.
We use the potential and its force to locate the branches, identify
the logarithmic cuts, and determine which real slices admit the mass
hierarchies needed for walking.
\label{sec:zplane}

\begin{figure}[t]
\begin{center}
\includegraphics[width=0.9\textwidth]{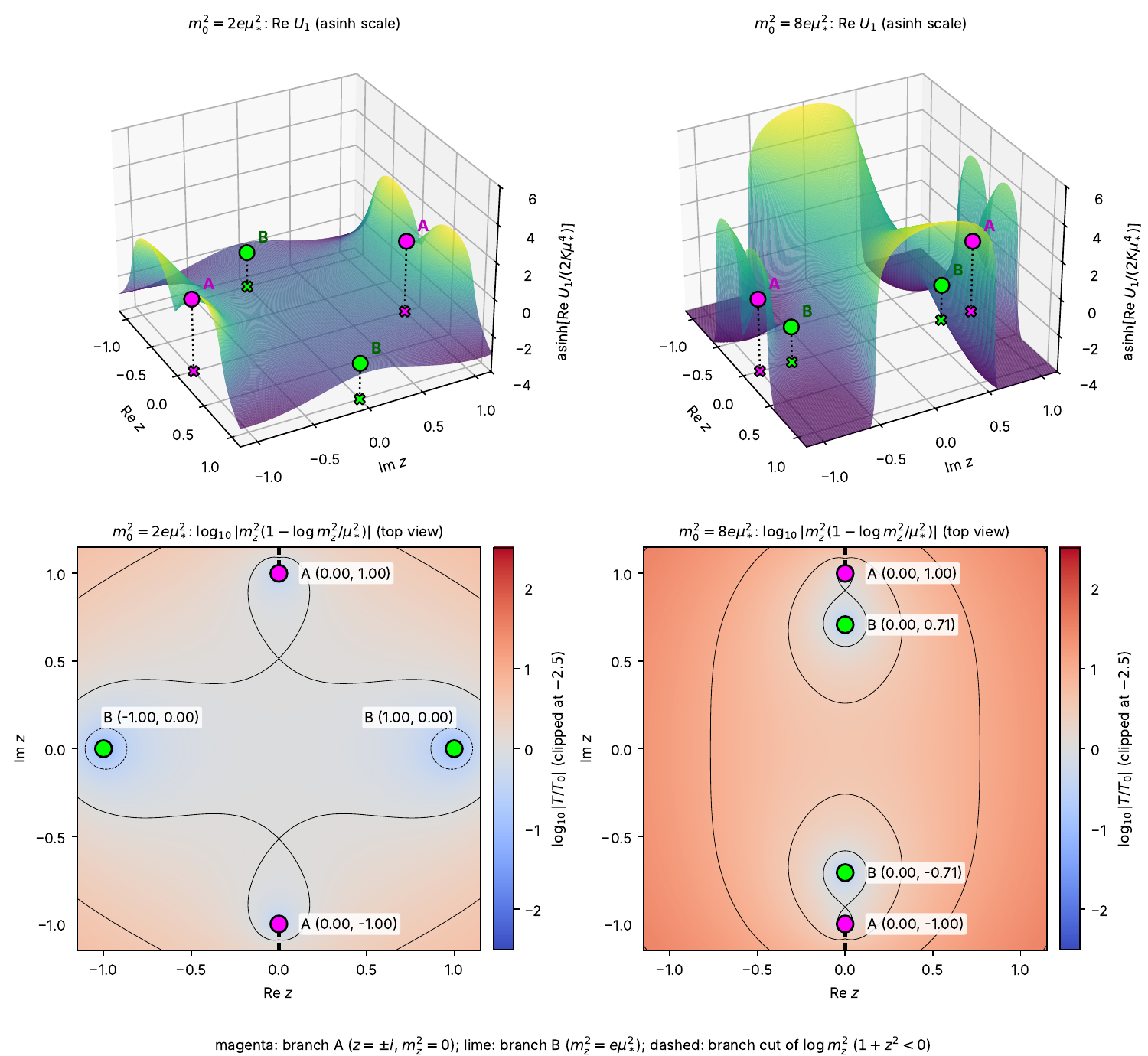}
\end{center}
\caption{Branches A (magenta, $z=\pm\ii$) and B (lime, $m_z^2={\rm e}\,\mu_*^2$) over the complex $z$-plane at fixed $m_0$:
$m_0^2=2{\rm e}\,\mu_*^2$ (left) and $m_0^2=8{\rm e}\,\mu_*^2$ (right). Top: ${\rm Re}\,U_1$ on an asinh scale, with the branch points
projected onto the plane (crosses). Bottom: top view of $\log_{10}|m_z^2(1-\log m_z^2/\mu_*^2)|$, which vanishes on both
branches, with the coordinates of the stationary points indicated. Dashed: the cut of $\log m_z^2$.}
\label{fig:z}
\end{figure}

Figure~\ref{fig:z} displays the potential and the force over the complex $z$-plane at fixed $m_0$. The force vanishes
exactly on the two branches. Branch B lies on the real axis ($z_B=\pm1$) for $m_0^2=2{\rm e}\,\mu_*^2$ and on the imaginary
axis ($z_B=\pm\ii/\sqrt2$) for $m_0^2=8{\rm e}\,\mu_*^2$. More generally:
\begin{itemize}
\item \textbf{Real $z$}: $m_z^2\ge m_0^2/4>0$, so branch A is unreachable. Branch B is reached on the real axis if and only if
$m_0^2\le4{\rm e}\,\mu_*^2$.
\item \textbf{Imaginary path $z=\ii\zeta$, $0\le\zeta\le1$}: the force vanishes at $\zeta=1$ (A) and, if
$m_0^2>4{\rm e}\,\mu_*^2$, also at $\zeta_B=\sqrt{1-4{\rm e}\,\mu_*^2/m_0^2}\in(0,1)$ (B). These are different points, and
$\zeta_B\to1$ only as $m_0\to\infty$. All masses are real and positive along this path, which is the one relevant for the
walking regime, see Sect.~\ref{sec:Realize}.
\item \textbf{$z=0$}: $\partial U_1/\partial z$ vanishes there because $\dd m_z^2/\dd z=\tfrac12m_0^2z=0$. This is
\textsl{not} a stationary point of $U_{\rm eff}$: the radial force is nonzero.
\item \textbf{Branch cut}: $\log m_z^2$ has a cut along $1+z^2<0$, i.e., along the imaginary axis beyond $|{\rm Im}\,z|=1$. It
ends at the points of branch A, which are branch points of $U_1\propto m_z^4\log m_z^2$.
\end{itemize}
For complex $z$, the background is not Hermitian. Stationary points are defined by the analytic continuation of the
stationarity condition, and the notion of a minimum acquires a meaning only on real slices.

\subsection{Realizing off-shell and walking regimes}
\label{sec:Realize}

The Sudakov regimes of the Introduction are distinguished by the hierarchy between the hard scale $Q^2$, the Euclidean
virtuality of the external $W$ bosons, $M_E^2$, and the mass $m_z^2$ of the planar internal states. In the notation of
Ref.~\cite{Alday:2026walk}, the walking regime is parametrized by
\begin{align}
\label{eq:walkcond}
w
\equiv
\frac{M_E^2}{Q^2}\to0
\, ,\qquad
y\equiv\frac{m_z^2}{M_E^2}
=
w^\eta
\, .
\end{align}
The interval $0\le\eta\le1$ interpolates between the cusp ($\eta=0$) and the octagon ($\eta=1$).\footnote{Our variable is $w=1/\omega_{\rm AABHZ}$, the inverse of the expansion variable in Ref.~\cite{Alday:2026walk}.} With the
external virtuality continued to Euclidean values, $p^2=-M_E^2$, as described in Sect.~\ref{sec:vacua}, and the Euclidean
dictionary $M_E^2=m_0^2$, the family \re{eq:V1}-\re{eq:V2} gives $y=m_z^2/m_0^2=(1+z^2)/4$. Which values of
$\eta$ are compatible with a stationary background?
\begin{itemize}
\item \textbf{Branch A}: $y\equiv0$, so formally $\eta=\infty$. Only the octagon plateau, $\eta\ge1$, is available. This is
the strictly off-shell regime: massive external $W$ bosons with massless planar internal lines, realized on a
tadpole-free background for every $m_0$, without any tuning.
\item \textbf{Branch B, real $z$}: $y=(1+z^2)/4\ge\tfrac14$, so $\eta\le0$. Only the cusp point $\eta=0$ and the window
$-1\le\eta\le0$ (with $z\to\infty$) are available, i.e., the on-shell side of the interpolation.
\item \textbf{Branch B, imaginary $z=\ii\zeta$, $0<\zeta<1$}: $y=(1-\zeta^2)/4\in(0,\tfrac14)$, and the full walking window
opens up. Since branch B pins $m_z^2={\rm e}\,\mu_*^2$, the walking trajectory is a $Q$-correlated family of vacua,
\begin{equation}
\label{eq:walkB}
1-\zeta^2=4\,w^{\eta}
\, ,\qquad
m_z^2={\rm e}\,\mu_*^2
\, ,\qquad
M_E^2={\rm e}\,\mu_*^2\,w^{-\eta}
\, ,\qquad
Q^2={\rm e}\,\mu_*^2\,w^{-1-\eta}
\, ,
\end{equation}
which realizes the hierarchy
\begin{equation*}
Q^2\gg M_E^2\gg m_z^2\gg M_E^4/Q^2
\end{equation*}
exactly for $0<\eta<1$. All masses along this trajectory are real and positive, while the vevs
$v_c=\tfrac12m_0(\pm1,\ii\zeta,0,\dots,0)$ are complex, i.e., the background is not Hermitian.
\end{itemize}
The walking regime is thus achievable on quantum-stationary backgrounds, but only on branch B with imaginary $z$, and
along a $Q$-correlated family of vacua: the vacuum pins $m_z$, and walking is produced by letting the external virtuality
grow as $M_E^2\propto Q^{2\eta/(1+\eta)}$. This is the same type of Euclidean continuation as the one employed for the
Coulomb branch of $\cN=4$ sYM in Ref.~\cite{Alday:2026walk}. Real $z$ allows only $\eta\le0$. In the held-background prescription of
Sect.~\ref{sec:TwoWays}, any trajectory is allowed, at the price of the source deformation; since the leading double
logarithms are insensitive to the source, both setups yield the same walking anomalous dimension.

The properties of the two branches established in this section, together with the Sudakov regimes they give access to,
are summarized in Table~\ref{tab:branches}.

\begin{table}[t]
\begin{center}
\small
\begin{tabular}{>{\raggedright\arraybackslash}p{3.2cm}>{\raggedright\arraybackslash}p{5.3cm}>{\raggedright\arraybackslash}p{5.3cm}}
\toprule
 & Branch A (null) & Branch B (Coleman-Weinberg)\\
\midrule
Condition & $v_c^2=0$, $v_c\neq0$ ($z=\pm\ii$) & $v_c^2={\rm e}\,\mu_*^2$\\
Background & complex, non-Hermitian & real for $m_0^2\le4{\rm e}\,\mu_*^2$, else $z$ imaginary\\
Scale selected & none (any $m_0$, $\mu_*$) & $m_z^2={\rm e}\,\mu_*^2$ (dimensional transmutation)\\
$U_1$ & $0$ & $-K{\rm e}^2\mu_*^4$\\
Curvature & nilpotent, infrared divergent & radial $m_\varphi^2=2(\Ds-2)g^2m_z^2$\\
Character & stationary, unstable (hilltop) & local minimum\\
Flat directions & - & $2n-3$ Goldstones (exact), $\chi$ (planar)\\
Sudakov regime & off-shell (octagon) & walking\\
\bottomrule
\end{tabular}
\end{center}
\caption{Summary of the two quantum stationary branches of the family \re{eq:V1}-\re{eq:V2} at one loop and leading planar
order.}
\label{tab:branches}
\end{table}

\section{Walking Sudakov form factor}
\label{sec:FF}

We now calculate the form factor in the walking hierarchy, where the
internal mass and the external virtualities provide distinct infrared
cutoffs. We first fix its operator and pole normalization, derive the
LSZ expansion, and then assemble the one- and two-loop contributions
before ultraviolet renormalization. The observable is defined by 
\begin{equation}
\label{eq:FFme}
\int d^d x \e^{- i q \cdot x}
\vev{W^+(p_1,\varepsilon_1)\,W^-(p_2,\varepsilon_2)|[G](x)|0}
=
(2 \pi)^d \delta^{(d)} (q + p_1 + p_2)
{\mathcal F}_G
\, ,
\end{equation}
Here $[G]$ is the renormalized insertion of minus twice the model's Lagrangian, 
\begin{equation}
\label{eq:Gop}
G=\tr\bigl(F_{\mu\nu}F^{\mu\nu}\bigr)-2\sum_I\tr\bigl(\cD_\mu\Phi_I\,\cD^\mu\Phi_I\bigr)
-\gym^2\sum_{I,J}\tr\bigl([\Phi_I,\Phi_J][\Phi_I,\Phi_J]\bigr)
\, ,
\end{equation}
The external states are two charged vector bosons $W^\pm$ of tree-level mass squared $m_0^2$, see Table~\ref{tab:spectrum}. This operator is conjugate 
to the gauge coupling. {\sl After the shift} \re{eq:shift}, $G$ acquires insertion vertices with two, three, and four dynamical fields, all of which 
have to be retained. The insertion carries the incoming momentum $q=-(p_1+p_2)$, with $s \equiv q^2$.

Three masses must be distinguished: the background parameter $m_0^2$, which appears in the Feynman rules of 
Sects.~\ref{sec:Quantization}-\ref{sec:rules}; the pole $M^2$ of the full vector propagator, which is the external shell; and the running mass 
scale introduced with the ultraviolet counterterms. Only $M^2$ is the physical external mass. On the complex background, the pole and residues 
may themselves be complex, and $\mathcal F_G$ is defined as the analytically continued pole-residue matrix element, continued from the 
Feynman boundary values. Because the insertion is the renormalized operator $[G]$, this matrix element
is ultraviolet finite after the action parameters have also been renormalized.
Regulator poles below belong only to auxiliary unsubtracted quantities.

Throughout this section, we work in the held-background prescription of Sect.~\ref{sec:TwoWays}: the classical background $V$ of the family 
\re{eq:V1}-\re{eq:V2} is kept fixed at generic $z$ by the source \re{eq:held}, which cancels the one-point function order by order. As explained 
there, this choice affects single logarithms and constants but not the leading double-logarithmic asymptotics, and it allows us to move freely 
along any walking trajectory. On the stationary branches A and B of Sect.~\ref{sec:QuantumStability} the source vanishes, and the held 
and source-free observables coincide. The branch-A form factor is studied separately in the companion paper. Here we
perform the walking calculation at $c=\alpha=1$, in the 't Hooft-Feynman gauge.

\subsection{LSZ reduction}
\label{sec:FFdef}

The charged external states are selected by the poles of their full
propagators. LSZ reduction relates the desired matrix element to the
amputated insertion while keeping the pole shifts and residues that
first mix with lower-loop vertices at two loops.

Let $\mathcal W_{1,\mu}$ and $\mathcal W_{2,\nu}$ be charged interpolating fields whose outgoing one-particle poles select the
two states in Eq.~\re{eq:FFme}. To organize the diagrammatic calculation,
first replace $[G]$ by the unsubtracted polynomial $G_0$, Eq.~\re{eq:Gop}
with unrenormalized fields and parameters, and consider the auxiliary correlator
\begin{equation}
\label{eq:FFgreen}
C_{G_0,\mu\nu}(p_1,p_2)= \ii \int\dd^dx\,\dd^dy\,{\rm e}^{\ii p_1\cdot x+\ii p_2\cdot y}\,
\vev{0|T\{\mathcal W_{1,\mu}(x)\,\mathcal W_{2,\nu}(y) G_0(0)\}|0}_{\rm c}
\, .
\end{equation}
We project the function by contracting both sides with polarization vectors $\varepsilon_i$ ($i=1,2$) of $W$'s. Of course, $C_{G_0,\mu\nu}$ 
possesses two independent tensor structures, but we eliminate the longitudinal one by imposing the conditions
\begin{equation}
\label{eq:FFproj}
\varepsilon_i \cdot p_j=0
\, ,\qquad
\mbox{for}
\qquad
i,j=1,2
\, .
\end{equation}
We strip the common transverse polarization product at this stage, and define
\begin{align}
\label{eq:FFtransverse-correlator}
C_J (s,t_1,t_2)
 \equiv
\frac{\varepsilon_1^\mu\varepsilon_2^\nu C_{G_0,\mu\nu}}{(\varepsilon_1\cdot\varepsilon_2)}
 \, .
\end{align}
The amputated vertex $\Gamma$, which will be the main object of our perturbative study below, is introduced via 
\begin{align}
\label{eq:FFfact}
C_J(s,t_1,t_2) =-\ii s D_T(t_1)D_T(t_2)\Gamma(s,t_1,t_2)
\, ,\qquad t_i=p_i^2
\, ,
\end{align}
by stripping down the two $W$-boson propagators $D_T$ to be defined momentarily. We also extracted the dimensionful
scale from it along with a phase, so that the resulting $\Gamma$ is dimensionless and real. Only the double-pole part contributes 
to the matrix element. At tree level it is
\begin{align}
\label{eq:FFtree-vertex}
\Gamma_0=1 - \frac{t_1+t_2-2m_0^2}{s}
\, .
\end{align}

The tensor decomposition defines the proper two-point function of the external vector fields
\begin{equation}
\label{eq:FFSigma}
\Pi_{\mu\nu}=\Bigl(g_{\mu\nu}-\frac{p_\mu p_\nu}{t}\Bigr)\Sigma_T(t)+\frac{p_\mu p_\nu}{t}\Sigma_L(t)
\, ,\qquad
\Sigma_T(t)=\frac{1}{(d-1)}\Bigl(g^{\mu\nu}-\frac{p^\mu p^\nu}{t}\Bigr) \Pi_{\mu\nu}
\, .
\end{equation}
External vector-scalar mixing is proportional to $p_\mu$ and drops out of the physical projection, while internal mixed
propagators are unaffected. The longitudinal projector belongs to the gauge sector and is {\sl not} the helicity-zero state of the
massive vector. The Dyson-summed transverse propagator, its pole and its residue are
\begin{gather}
D_T(t)=\frac{-\ii}{t-m_0^2-\Sigma_T(t)}
\, ,\qquad
D_T(t)\simeq\frac{-\ii Z}{t-M^2}
\, ,\nonumber\\
M^2=m_0^2+\Sigma_T(M^2)
\, ,\qquad
Z=\frac{1}{1-\Sigma^\prime_T(M^2)}
\, ,
\label{eq:FFpole}
\end{gather}
where $\Sigma^\prime (M^2) \equiv \partial_t\Sigma_T(t)|_{t = M^2}$ and the pole is the one continuously connected to $m_0^2$ as 
the coupling vanishes. Here $Z$ is a pole residue, not to be confused with the ultraviolet renormalization constant.

Allowing, for a moment, for different residues on the two legs, the pole part of Eq.~\re{eq:FFfact} is
\begin{align}
\label{eq:FFdoublepole}
C_J\big|_{\rm double\ pole}
&=\frac{(-\ii Z_1)(-\ii Z_2) (-\ii s) \Gamma(s,M_1^2,M_2^2)}
{(t_1-M_1^2)(t_2-M_2^2)}\, .
\end{align}
Multiplication by $\ii(t_i-M_i^2)/\sqrt{Z_i}$ removes the
pole on leg $i$ and divides out its field-to-state overlap~\cite{Fitzpatrick:2023lsz}, so that
\begin{align}
\label{eq:FFLSZ}
\mathcal F_{G_0,B}&=(\varepsilon_1\cdot\varepsilon_2)
\lim_{t_i\to M_i^2}\frac{\ii(t_1-M_1^2)}{\sqrt{Z_1}}
\frac{\ii(t_2-M_2^2)}{\sqrt{Z_2}}\, (-\ii)C_J\notag\\*
&=-s (\varepsilon_1\cdot\varepsilon_2) Z\, \Gamma(s,M^2,M^2)\, .
\end{align}
Thus Eqs.~\re{eq:FFtransverse-correlator}, \re{eq:FFfact} and \re{eq:FFLSZ} give, in order, the transverse scalar correlator, its
dimensionless amputated vertex and the auxiliary pole matrix element. The factor $Z^{-1/2}$ per leg of the unamputated Green 
function thus becomes $Z^{1/2}$ per leg multiplying the amputated vertex and gives access to the auxiliary matrix element of 
$G_0$, not yet the physical matrix element of $[G]$ in question. In the opposite-charge channels the poles and residues coincide, 
and for a complex residue $\sqrt{Z_+}\sqrt{Z_-}=Z$ is the analytic product continued from $Z=1$, not $|Z|$. Since $Z$ and 
$\Gamma$ are dimensionless, Eq.~\re{eq:FFLSZ} has mass dimension two. The vertex $\Gamma$ and the self-energy coefficients 
below refer to this unsubtracted calculation throughout.

The amputated vertex $\Gamma$ contains every connected graph with one insertion of $G_0$ that remains after the full propagators 
on the two external legs are stripped:
\begin{itemize}
\item self-energy subgraphs dressing an individual external leg are removed, since their effect is supplied by Eq.~\re{eq:FFpole};
\item self-energy insertions on internal lines are retained;
\item propagator bridges in the operator channel are retained: cutting a line carrying the $q$-momentum may separate the 
$G_0$-insertion from a component containing \textsl{both} external legs, which is not a self-energy of either leg;
\item in the two-point function, only strictly one-particle-irreducible insertions are kept, since reducible chains are
generated by the Dyson resummation.
\end{itemize}
Demanding strict one-particle irreducibility of the vertex would therefore remove genuine contributions, and retaining only
the quadratic insertion of $G_0$ would miss its cubic and quartic vertices.

\subsection{Perturbative expansion}

The vertex, pole position and residue must be expanded in the same
dimensionally regulated coupling before their contributions can be
combined. We state the loop normalization first, then derive the
coefficient-level LSZ formulas used in the diagrammatic calculation.
In $d=4-2\epsilon$ dimensions, the classical coupling of
Eq.~\re{eq:gR} is continued as 
\begin{align}
\label{eq:coupling-regulated}
g_\epsilon^2
&\equiv \e^{-\gamma_{\rm E}\epsilon}\mu^{-2\epsilon}
        \frac{N\gym^2}{(4\pi)^{d/2}}
 = \e^{-\gamma_{\rm E}\epsilon}(4\pi)^\epsilon\mu^{-2\epsilon}g^2
\, ,
\end{align}
such that the per-loop measure becomes
\begin{align}
\label{eq:FFloop-measure}
[\dd k]&\equiv 
\e^{\gamma_{\rm E}\epsilon}\mu^{2\epsilon} \frac{\dd^d k}{\ii\pi^{d/2}}\, .
\end{align}
The vertex and the transverse self-energy develop perturbative expansions in dimensionally regularized theory 
\begin{align}
\label{eq:FFexp0}
\Gamma&=\Gamma_0+g_\epsilon^2\Gamma_1+g_\epsilon^4\Gamma_2
       +O(g_\epsilon^6),\\
\Sigma_T&=g_\epsilon^2\sigma_1+g_\epsilon^4\sigma_2
       +O(g_\epsilon^6)\, .
\end{align}
Here $\Gamma_n$ is dimensionless by Eq.~\re{eq:FFfact}, whereas
$\sigma_n$ has mass dimension two by Eq.~\re{eq:FFpole}.

In the companion branch-A analysis, the expansion is organized around the tree mass $m_0^2$. Here the walking form
factor is defined at the pole $M^2$, and the displacement between these values generates the corrections displayed below.
The dimensionless pole shift is
\begin{align}
\label{eq:FFpole-shift-definition}
 \delta&\equiv\frac{M^2-m_0^2}{s}\, .
\end{align}
The pole equation and residue then become
\begin{align}
\label{eq:FFdelta}
s \delta&= g_\epsilon^2\sigma_1+
 g_\epsilon^4(\sigma_2+\sigma_1\sigma_1')
 +O(g_\epsilon^6),\\
\label{eq:FFZ}
Z&=1+g_\epsilon^2\sigma_1'
 +g_\epsilon^4\bigl[\sigma_2'+\sigma_1\sigma_1''
                   +(\sigma_1')^2\bigr]+O(g_\epsilon^6)\, ,
\end{align}
where the derivatives are evaluated at the unshifted pole,
\begin{align}
\label{eq:FFself-energy-derivatives}
\sigma_n^{(k)}\equiv
 \left.\partial_t^k\sigma_n(t)\right|_{t=m_0^2}\, .
\end{align}
The $\sigma_1\sigma_1'$ and $\sigma_1\sigma_1''$ terms in Eqs.~\re{eq:FFdelta} and \re{eq:FFZ} are generated by evaluating the
one-loop self-energy and its derivative at the shifted pole.

Since the external virtualities move to the pole $M^2$, we find for the vertex at the physical poles 
\begin{align}
\label{eq:FFvertexTaylor}
\Gamma(s,M^2,M^2)
 &=1-2\delta+g_\epsilon^2\Gamma_1
 +g_\epsilon^4\bigl[\Gamma_2+\sigma_1\mathcal D\Gamma_1\bigr]
 +O(g_\epsilon^6)
\, ,
\end{align}
where Eq.~\re{eq:FFtree-vertex} gives
$\Gamma_0(s,M^2,M^2)=1-2\delta$, and we introduced the derivative
\begin{align}
\label{eq:FFDdef}
\mathcal D&\equiv\partial_{t_1}+\partial_{t_2}
\, .
\end{align}
It is tacitly assumed here that all unadorned $\Gamma_n$ and $\mathcal D\Gamma_n$ below are evaluated at $t_1=t_2=m_0^2$ only 
after differentiation. The derivatives hold all parameters of the model fixed; in particular, they act as on the tree-level vertex
\begin{align}
 \mathcal D\Gamma_0&=-\frac2s,\qquad
 \mathcal D^2\Gamma_0=0\, .
\end{align}

The physical form factor is the ratio of the finite matrix element of $[G]$ to the tree matrix element at the same external pole, with
renormalized parameters in its normalization. Its numerator is the matrix element defined in Eq.~\re{eq:FFme}:
\begin{align}
\label{eq:FFexp}
F &\equiv
 \frac{\mathcal F_G}
 {{\mathcal F}_G^{(0)}}
 =
1+g_R^2f_{1}+g_R^4f_{2}+O(g_R^6)
 \, ,
\end{align}
with perturbative series performed in the renormalized coupling $g_R^2$ defined by Eq.~\re{eq:coupling-renormalization} below. 
Both the numerator and denominator, i.e., the reference tree, determining $F$, are finite. 

For the intermediate diagrammatic algebra, we define, however, the ratio of the bare quantities instead
\begin{align}
\label{eq:FFexpBare}
F_B&\equiv
 \frac{\mathcal F_{G_0,B}}
 {-s(\varepsilon_1\cdot\varepsilon_2)(1-2\delta)}
 =\frac{Z\,\Gamma(s,M^2,M^2)}{1-2\delta}\notag\\*
 &=1+g_\epsilon^2 f_{1,B}+g_\epsilon^4 f_{2,B}
 +O(g_\epsilon^6)\, .
\end{align}
The second equality uses the auxiliary matrix element \re{eq:FFLSZ} and the tree vertex \re{eq:FFtree-vertex} at the shifted
pole. The label $B$ belongs only to this unsubtracted ratio and its coefficients. To obtain Eq.~\re{eq:FFexp} from Eq.~\re{eq:FFexpBare}, 
one replaces the action parameters by renormalized ones, applies the operator mixing resolved in Eq.~\re{eq:FFZGP}, and converts the 
mass-dependent tree denominator to the renormalized normalization, before finally removing the regulator; see 
Eq.~\re{eq:FFrenormalization-bridge} below for this map, and more specifically  Eq.~\re{eq:MS155} for how $f_{2}$ is defined
by $f_{2,B}$. The reference-tree conversion is power suppressed in that limit, but is retained in the exact finite-mass one-loop result 
\re{eq:FF1held}, using Eq.~\re{eq:exact-pole-shift-wy}.

Multiplying the vertex expansion \re{eq:FFvertexTaylor} by the residue \re{eq:FFZ} and expanding the denominator in 
Eq.~\re{eq:FFexpBare} gives for the first two loops
\begin{align}
\label{eq:FF1LSZ}
f_{1,B}&=\Gamma_1+\sigma_1'
\, ,\\
\label{eq:FF2LSZ}
f_{2,B}&=\Gamma_2+\sigma_1'\Gamma_1+
 \sigma_1\mathcal D\Gamma_1+\sigma_2'+\sigma_1\sigma_1''
 +(\sigma_1')^2+\frac{2\sigma_1\Gamma_1}{s}
 \, .
\end{align}
The first three terms in Eq.~\re{eq:FF2LSZ} are the two-loop vertex, the one-loop vertex times the residue, and the displacement of its
external virtualities. The next three come from the two-loop residue and its one-loop iterations. The final term comes from the tree
normalization $1/(1-2\delta)$ in Eq.~\re{eq:FFexpBare}. This specifies the normalization of all coefficients below without an additional
rescaled vertex function.

An important comment is in order regarding the off-shell derivatives in Eqs.\ \re{eq:FF1LSZ} and \re{eq:FF2LSZ}. Since the
vertex and self-energies will be reduced later on to a set of master integrals, the derivatives $\mathcal D\Gamma_1$, $\sigma_1'$, 
$\sigma_1''$ and $\sigma_2'$ act on both rational coefficients and these master integrals at independent virtualities. Terms proportional 
to $t_i-m_0^2$ or $(t_i - m_0^2)^2$ can therefore survive differentiation even though they vanish on the tree mass-shell. Obviously, a 
derivative of an already on-shell answer with respect to $m_0^2$ would also change the background charges and masses and would
never produce the correct LSZ expression. 

\subsubsection{Expansion point and Euclidean walking kinematics}
\label{sec:FFexpansion}

The LSZ formulas \re{eq:FF1LSZ}-\re{eq:FF2LSZ} are exact in the masses. The walking calculation developed here and
the branch-A calculation of the companion paper use different expansion points. We distinguish their infrared
hierarchies before specifying the Euclidean walking kinematics.

\begin{itemize}
\item{Branch A.}
In the companion calculation, the null background $z=\pm\ii$ has massless planar internal lines, $m_z^2=0$, and
massive external $W$ bosons. Their virtualities provide the infrared regulator, and the ultrasoft region is dynamical.
The branch-A master integrals and perturbative expansion are developed in that paper.
\item{Walking.}
At generic $z$, the internal mass is nonzero. We first perform the
LSZ derivatives at independent virtualities and fixed background
parameters. We then continue the external invariants to Euclidean
kinematics,
\begin{align}
 q^2&=-Q^2,\qquad t_1=t_2=-M_E^2,\qquad
 Q^2>0,\quad M_E^2>0,\quad m_z^2>0.
 \label{eq:walking-euclidean-kinematics}
\end{align}
On the tree shell, the invariant $m_0^2$ in the preceding Minkowski
formulas is continued as $m_0^2\to-M_E^2$. Thus $M_E^2$ is the
positive magnitude of the continued virtuality; substituting
$m_0^2=+M_E^2$ into those formulas would give the wrong continuation.
The background masses are held independent while the derivatives
are taken.

The ratios of Eq.~\re{eq:walkcond} specify the trajectory
$y=w^\eta$. We retain leading power for every fixed $0<\eta<1$.
Terms suppressed by positive powers of $y$ or $w/y$ then vanish.
Inverse rational coefficients in a master reduction can require
subleading master terms before this projection. The interior
expansion is not uniform at either endpoint and is not a termwise
substitute for the branch-A calculation in the companion paper.
\end{itemize}

\begin{figure}[t]
\centering
\begingroup\setlength{\tabcolsep}{14pt}
\tikzset{act/.style={circle,fill=black,inner sep=0pt,minimum size=3.6pt},
opv/.style={circle,draw,fill=white,inner sep=0pt,minimum size=8pt},
xl/.style={line width=.6pt},gh/.style={line width=.6pt,dashed},
mk/.style={densely dotted,->},lb/.style={font=\scriptsize}}
\newcommand{\opnode}[2]{\node[opv] (#1) at #2 {}; \draw[line width=.5pt] ($(#1)+(-2.6pt,-2.6pt)$)--($(#1)+(2.6pt,2.6pt)$);
\draw[line width=.5pt] ($(#1)+(-2.6pt,2.6pt)$)--($(#1)+(2.6pt,-2.6pt)$);}
\newcommand{\legs}[2]{\draw[xl] #1--(-1.0,-1.45) node[below,lb]{$p_1$}; \draw[xl] #2--(1.0,-1.45) node[below,lb]{$p_2$};}
\newcommand{\qmark}[1]{\draw[mk] ($(#1)+(0,0.75)$) node[above,lb]{$q$}--(#1);}
\begin{tabular}{ccc}
\begin{tikzpicture}[scale=.8]
\coordinate (O) at (0,0.6); \coordinate (A) at (-0.55,-0.5);
\draw[xl] (A)--(-1.0,-1.45) node[below,lb]{$p_1$}; \draw[xl] (O)--(1.0,-1.45) node[below,lb]{$p_2$};
\draw[xl] (O) to[bend right=40] (A); \draw[xl] (O) to[bend left=40] (A); \node[act] at (A) {};
\opnode{G}{(O)} \qmark{G}
\end{tikzpicture} &
\begin{tikzpicture}[scale=.8]
\coordinate (O) at (0,0.6); \coordinate (A) at (0.55,-0.5);
\draw[xl] (O)--(-1.0,-1.45) node[below,lb]{$p_1$}; \draw[xl] (A)--(1.0,-1.45) node[below,lb]{$p_2$};
\draw[xl] (O) to[bend right=40] (A); \draw[xl] (O) to[bend left=40] (A); \node[act] at (A) {};
\opnode{G}{(O)} \qmark{G}
\end{tikzpicture} &
\begin{tikzpicture}[scale=.8]
\coordinate (O) at (0,0.6); \coordinate (A) at (0,-0.55);
\draw[xl] (A)--(-1.0,-1.45) node[below,lb]{$p_1$}; \draw[xl] (A)--(1.0,-1.45) node[below,lb]{$p_2$};
\draw[xl] (O) to[bend right=55] (A); \draw[xl] (O) to[bend left=55] (A); \node[act] at (A) {};
\opnode{G}{(O)} \qmark{G}
\end{tikzpicture}
\\[-2pt]
{\small \FFgraph{1}{1}: $p_1$ bubble} & {\small \FFgraph{1}{2}: $p_2$ bubble} & {\small \FFgraph{1}{3}: $q$ bubble}\\[6pt]
\end{tabular}\par\smallskip
\begin{tabular}{cc}

\begin{tikzpicture}[scale=.8]
\coordinate (O) at (0,0.6); \coordinate (B) at (0,-0.25); \coordinate (A) at (0,-0.75);
\draw[xl] (A)--(-1.0,-1.45) node[below,lb]{$p_1$}; \draw[xl] (A)--(1.0,-1.45) node[below,lb]{$p_2$};
\draw[xl] (O) to[bend right=60] (B); \draw[xl] (O) to[bend left=60] (B); \draw[xl] (B)--(A);
\node[act] at (A) {}; \node[act] at (B) {};
\opnode{G}{(O)} \qmark{G}
\end{tikzpicture} &
\begin{tikzpicture}[scale=.8]
\coordinate (O) at (0,0.6); \coordinate (A) at (-0.6,-0.5); \coordinate (B) at (0.6,-0.5);
\draw[xl] (A)--(-1.0,-1.45) node[below,lb]{$p_1$}; \draw[xl] (B)--(1.0,-1.45) node[below,lb]{$p_2$};
\draw[xl] (O)--(A)--(B)--(O); \node[act] at (A) {}; \node[act] at (B) {};
\opnode{G}{(O)} \qmark{G}
\end{tikzpicture}
\\[-2pt]
{\small \FFgraph{1}{4}: bridge-bubble} & {\small \FFgraph{1}{5}: triangle}\\[6pt]
\end{tabular}
\endgroup
\caption{One-loop amputated vertex graphs of the form factor. The crossed circle is the nonpropagating $G$-insertion with incoming 
momentum $q$ (dotted arrow); the ordinary vertical line in \FFgraph{1}{4} is a bridge carrying $q$. Solid lines denote the uplifted field 
$X$ (vector, scalar and mixed components).}
\label{fig:FF1graphs}
\end{figure}
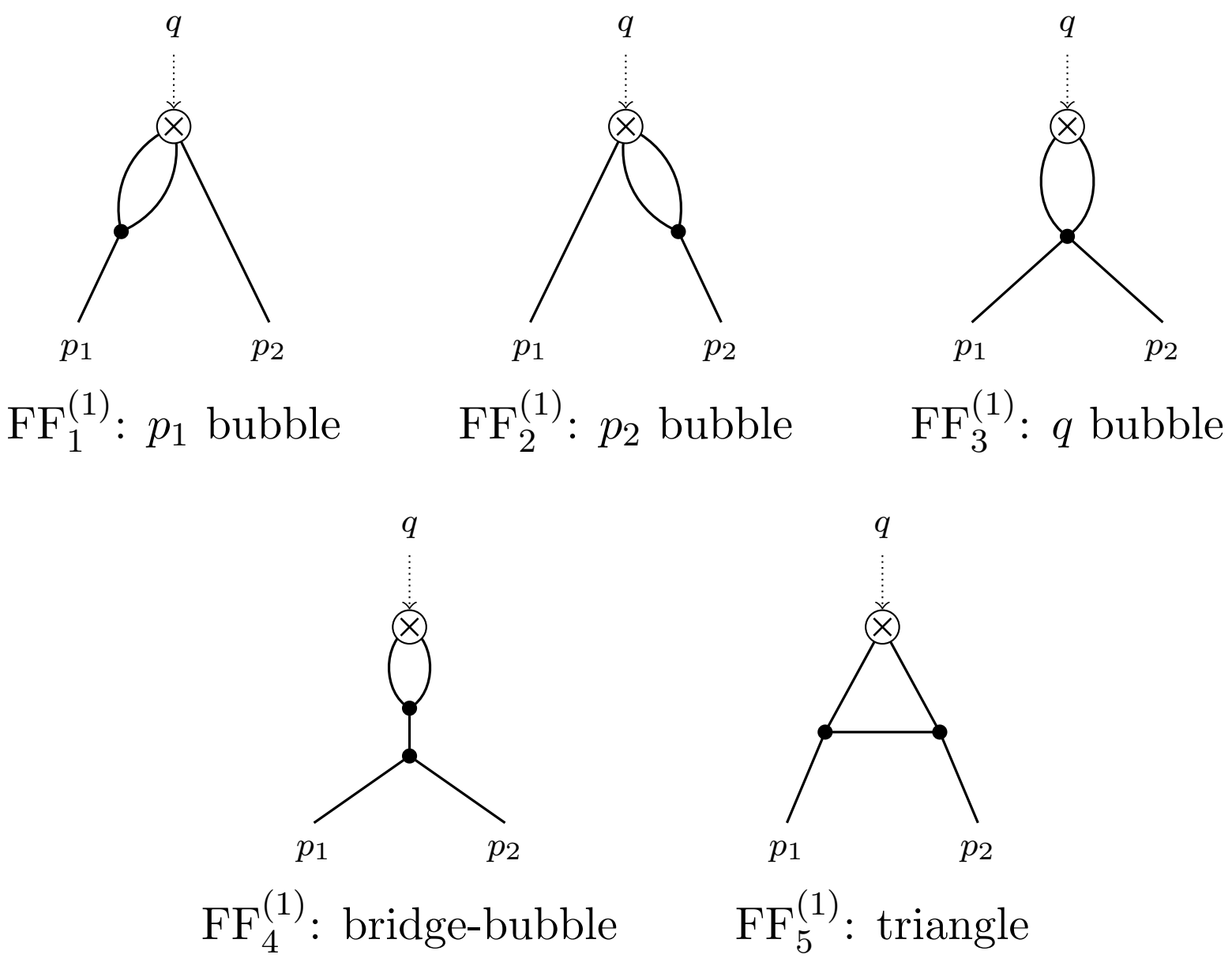

\subsection{One loop}
\label{sec:FF1}

The one-loop calculation fixes the normalization and ultraviolet subtraction before the two-loop assembly. It also supplies
the lower-loop ingredients needed there. We first assemble the vertex and residue contributions, distinguish
the two mechanisms that remove tadpoles, and renormalize the resulting pole matrix element. This gives an expression
exact in the mass ratios, from which the walking logarithms can be extracted in a controlled limit.
The graphs are generated with QGRAF using the uplifted field $X$, ghosts, and the insertion vertices of $G$, and resolved
into vector, scalar, and mixed components after the color algebra. They fall into the following three groups:
\begin{itemize}
\item \textbf{Vertex graphs}, Fig.~\ref{fig:FF1graphs}: the bubbles in the $p_1$, $p_2$ and $q$ channels, the bridge-bubble,
retained here, as we explained in Sect.\ \ref{sec:FFdef}, although it is not one-particle irreducible, and the triangle. We refer to them 
as \FFgraph{1}{1}-\FFgraph{1}{5}.
\item \textbf{Self-energy graphs}, Fig.~\ref{fig:SE1graphs}: the bosonic and ghost bubbles \SEgraph{1}{1}, \SEgraph{1}{2}, together 
with the self-energy tadpole \SEgraph{1}{3}, entering through the pole and residue \re{eq:FFpole}.
\item \textbf{Form-factor tadpoles}, Fig.~\ref{fig:TAD1graphs}: the insertion tadpole \tadgraph{1}{1}, the bridged tadpole \tadgraph{1}{2}, and 
the bosonic and ghost one-point attachments \tadgraph{1}{3}, \tadgraph{1}{4}. The genuine two-point tadpole \SEgraph{1}{3} is grouped with 
the self-energies in Fig.~\ref{fig:SE1graphs}.
\end{itemize}
The figures display the retained nonvanishing graph families; external-leg self-energy dressings are not part of the amputated vertex.
As shown in Sect.~\ref{sec:FF1tad}, the tadpole graphs do not contribute 
to the held-background form factor at leading power: the one-point attachments are cancelled exactly by the holding source, and the massive 
tadpoles are power suppressed. Anticipating what is coming at two loops, the same two mechanisms remove almost all tadpole subgraphs, 
with one exception: in \tadgraph{2}{65} a massive tadpole is inserted into the middle line of the triangle, which thereby becomes a raised-power
propagator, and the endpoint of that raised line compensates the tadpole mass factor. \tadgraph{2}{65}, therefore, contributes at leading power 
and is included among the two-loop graph families of Sect.~\ref{sec:FF2graphs}.

\begin{figure}[t]
\centering
\begingroup\setlength{\tabcolsep}{14pt}
\tikzset{act/.style={circle,fill=black,inner sep=0pt,minimum size=3.6pt},
xl/.style={line width=.6pt},gh/.style={line width=.6pt,dashed},
lb/.style={font=\scriptsize}}
\begin{tabular}{ccc}
\begin{tikzpicture}[scale=.8]
\draw[xl] (-1.3,0)--(-.55,0); \draw[xl] (.55,0)--(1.3,0);
\draw[xl] (-.55,0) to[bend left=50] (.55,0);
\draw[xl] (-.55,0) to[bend right=50] (.55,0);
\node[act] at (-.55,0) {}; \node[act] at (.55,0) {};
\node[lb] at (-1.3,.2) {$p$};
\end{tikzpicture} &
\begin{tikzpicture}[scale=.8]
\draw[xl] (-1.3,0)--(-.55,0); \draw[xl] (.55,0)--(1.3,0);
\draw[gh] (-.55,0) to[bend left=50] (.55,0);
\draw[gh] (-.55,0) to[bend right=50] (.55,0);
\node[act] at (-.55,0) {}; \node[act] at (.55,0) {};
\node[lb] at (-1.3,.2) {$p$};
\end{tikzpicture} &
\begin{tikzpicture}[scale=.8]
\draw[xl] (-1.3,0)--(1.3,0);
\node[act] at (0,0) {}; \draw[xl] (0,.4) circle (.4);
\node[lb] at (-1.3,.2) {$p$};
\end{tikzpicture}
\\[-2pt]
{\small \SEgraph{1}{1}: bubble} & {\small \SEgraph{1}{2}: ghost bubble} & {\small \SEgraph{1}{3}: tadpole}\\[6pt]
\end{tabular}
\endgroup
\caption{One-loop proper self-energies of the external $W$, entering through
the pole and residue \re{eq:FFpole}. Every panel is a genuine two-point graph
with horizontal external legs and one external momentum $p$. Dashed lines
are ghosts. The tadpole \SEgraph{1}{3} belongs to this two-point class.}
\label{fig:SE1graphs}
\end{figure}

\subsubsection{Reduction to master integrals}

The reduction must preserve the independent masses and virtualities on which the LSZ derivatives act. We therefore
perform the color and tensor algebra before specializing these invariants, using the propagators and vertices from
Appendix~\ref{App:FeynRules} at $\alpha=1$. For the normalization $\tr(T^AT^B)=\tfrac12\delta^{AB}$, the exact color contraction is
\begin{align*}
\sum_{A=1}^{(N+2)^2-1}(T^A)_{ij}(T^A)_{kl}
&=\frac12\left(\delta_{il}\delta_{jk}
-\frac{\delta_{ij}\delta_{kl}}{N+2}\right).
\end{align*}
The expression in parentheses projects onto traceless matrices. The leading planar contribution is extracted only after this contraction.
Tensor numerators of the emerging loop integrands are expressed through scalar products and then reduced by 
integration-by-parts (IBP) identities with up to three independent internal masses $m_{1,2,3}^2$ and independent virtualities $t_1,t_2,Q^2$. Only 
after the reduction are the internal masses set to their specific value $m_z^2$ of Eq.~\re{eq:mz}. Every graph produced in this
fashion is a linear combination of the following five master integrals
\begin{alignat}{2}
\label{eq:FFmasters}
&\mathcal I_1\equiv \ &&A\equiv A_0(m_z^2),\\
&\mathcal I_2\equiv \ &&B_{t_1}\equiv B_0(t_1;m_z^2,m_z^2),\\
&\mathcal I_3\equiv \ &&B_{t_2}\equiv B_0(t_2;m_z^2,m_z^2),\\
&\mathcal I_4\equiv \ &&B_Q\equiv B_0(-Q^2;m_z^2,m_z^2),\\
&\mathcal I_5\equiv \ &&C\equiv C_0(t_1,t_2,-Q^2;m_z^2,m_z^2,m_z^2)\, .
\end{alignat}
These masters are defined in Appendix~\ref{app:masters1}. Since $A$ and the bubbles have ultraviolet poles, their accompanying rational
coefficients emerging from IBPs are retained through $O(\epsilon)$. The one-loop graph families are partitioned into amputated insertion
graphs and proper self-energy graphs:
\begin{align}
 \mathcal S_1^{(1)}
 &=\{\FFgraph{1}{1},\ldots,\FFgraph{1}{5},
       \tadgraph{1}{1},\ldots,\tadgraph{1}{4}\},
 \label{eq:FFvertex-graph-set}\\
 \mathcal S_2^{(1)}
 &=\{\SEgraph{1}{1},\SEgraph{1}{2},\SEgraph{1}{3}\}.
 \label{eq:FFselfenergy-graph-set}
\end{align}
For $\gamma\in\mathcal S_1^{(1)}$, the dimensionless bare graph
coefficient, with the coupling and common polarization product
stripped, is
\begin{align}
 f_{1,\gamma,B}
 &=\frac1{Q^2}\sum_{j=1}^{5}
 \bigl[c^{(0)}_{\gamma j}+\epsilon c^{(1)}_{\gamma j}\bigr]
 \mathcal I_j.
 \label{eq:FFledger}
\end{align}
The coefficients $c^{(k)}_{\gamma j}$ are the numerator coefficients
before division by the overall vertex normalization $Q^2$.
For $\gamma\in\mathcal S_2^{(1)}$, $f_{1,\gamma,B}$ instead denotes the
virtuality derivative of that graph's contribution to $\sigma_1(t)$,
taken at fixed internal masses and evaluated at the tree shell only
after differentiation. The graph expansions therefore have the same
form at one and two loops:
\begin{align}
 \Gamma_1&=\sum_{\gamma\in\mathcal S_1^{(1)}}f_{1,\gamma,B},
 \notag\\
 \sigma_1'&=\sum_{\gamma\in\mathcal S_2^{(1)}}f_{1,\gamma,B},
 \notag\\
 f_{1,B}&=\sum_{\gamma\in\mathcal S_1^{(1)}\cup\mathcal S_2^{(1)}}
 f_{1,\gamma,B}.
 \label{eq:FFgraphmap}
\end{align}
The first set is specified before the holding-source cancellation
and leading-power projection. Its five ordinary vertex families are
drawn in Fig.~\ref{fig:FF1graphs}, and its four tadpole families in
Fig.~\ref{fig:TAD1graphs}. After the cancellations and power counting
in Sect.~\ref{sec:FF1tad}, only the five ordinary vertex families
survive at leading walking power.
The second set contains the proper two-point graphs of
Fig.~\ref{fig:SE1graphs}, including the self-energy tadpole.
They are projected onto $\Sigma_T$ by Eq.~\re{eq:FFSigma} before
the virtuality derivative is taken.

Equation~\re{eq:FFgraphmap} implements the one-loop LSZ identity
\re{eq:FF1LSZ} directly in terms of graph coefficients.
The $\epsilon c^{(1)}_{\gamma j}$ terms in
Eq.~\re{eq:FFledger} contribute to finite one-loop matching through
the poles of $A$ and the bubbles. They do not generate walking
logarithms at one loop, but are needed for the two-loop reconstruction.
The latter also uses the positive-$\epsilon$ triangle coefficient
in Eq.~\re{eq:triangle155}.

\begin{figure}[t]
\centering
\begingroup\setlength{\tabcolsep}{14pt}
\tikzset{act/.style={circle,fill=black,inner sep=0pt,minimum size=3.6pt},
opv/.style={circle,draw,fill=white,inner sep=0pt,minimum size=8pt},
xl/.style={line width=.6pt},gh/.style={line width=.6pt,dashed},
mk/.style={densely dotted,->,shorten >=4.5pt},lb/.style={font=\scriptsize}}
\newcommand{\opnode}[2]{\node[opv] (#1) at #2 {};
\draw[line width=.5pt] ($(#1)+(-2.1pt,-2.1pt)$)--($(#1)+(2.1pt,2.1pt)$);
\draw[line width=.5pt] ($(#1)+(-2.1pt,2.1pt)$)--($(#1)+(2.1pt,-2.1pt)$);}
\newcommand{\fflegs}[1]{
\draw[xl] ($(#1)+(-1.3,0)$) node[above,lb]{$p_1$} -- (#1);
\draw[xl] ($(#1)+(1.3,0)$) node[above,lb]{$p_2$} -- (#1);}
\begin{tabular}{cc}
\begin{tikzpicture}[scale=.8]
\coordinate (O) at (0,0);
\fflegs{O}
\draw[xl] (0,.4) circle (.4);
\opnode{G}{(O)}
\draw[mk] (0,-.85) node[below,lb]{$q$} -- (G);
\end{tikzpicture} &
\begin{tikzpicture}[scale=.8]
\coordinate (O) at (0,.9); \coordinate (A) at (0,0);
\fflegs{A}
\draw[xl,line width=.9pt] (O)--(A);
\node[act] at (A) {};
\draw[xl] (0,1.3) circle (.4);
\opnode{G}{(O)}
\draw[mk] (1.0,.9) node[above,lb]{$q$} -- (G);
\end{tikzpicture}
\\[-2pt]
{\small \tadgraph{1}{1}: insertion tadpole} & {\small \tadgraph{1}{2}: tadpole with bridge}\\[6pt]
\begin{tikzpicture}[scale=.8]
\coordinate (O) at (0,0); \coordinate (T) at (0,-.65);
\fflegs{O}
\draw[xl] (O)--(T); \node[act] at (T) {};
\draw[xl] (0,-.95) circle (.3);
\opnode{G}{(O)}
\draw[mk] (0,.85) node[above,lb]{$q$} -- (G);
\end{tikzpicture} &
\begin{tikzpicture}[scale=.8]
\coordinate (O) at (0,0); \coordinate (T) at (0,-.65);
\fflegs{O}
\draw[xl] (O)--(T); \node[act] at (T) {};
\draw[gh] (0,-.95) circle (.3);
\opnode{G}{(O)}
\draw[mk] (0,.85) node[above,lb]{$q$} -- (G);
\end{tikzpicture}
\\[-2pt]
{\small \tadgraph{1}{3}: one-point, bosonic} & {\small \tadgraph{1}{4}: one-point, ghost}\\[6pt]
\end{tabular}
\endgroup
\caption{One-loop tadpole graphs in the amputated form-factor vertex.
The horizontal physical legs carry $p_1,p_2$; the dotted arrow carries the
insertion momentum $q$. \tadgraph{1}{1}, \tadgraph{1}{2} are massive tadpoles, power suppressed at
leading power and scaleless at the null point. \tadgraph{1}{3}, \tadgraph{1}{4} attach the
one-point function through a zero-momentum Cartan connector and cancel
exactly against the holding source. The genuine self-energy tadpole \SEgraph{1}{3}
is displayed separately in Fig.~\ref{fig:SE1graphs}. None of the graphs
shown here contributes to the held-background form factor at leading power.}
\label{fig:TAD1graphs}
\end{figure}

\subsubsection{Bare expressions}

The master decomposition makes it possible to follow separately the vertex terms and the virtuality derivatives
entering the residue. We display the compact vertex expressions first; the longer vertex coefficients and the
self-energy contributions are organized below in the same normalization. For the compact graph families, the bare
dimensionless vertex contributions are
\begin{align}
 f_{1,\mathrm{tad}1,B}
 &=\frac{2(\Ds-1)}{Q^2}A,\notag\\
 f_{1,\mathrm{FF}3,B}
 &=\frac1{Q^2}\Biggl\{
 -\Bigl(\frac{6\Ds-11}{3}+\frac{2\epsilon}{9}\Bigr)A
 \notag\\
 &\qquad
 +\Bigl[-Q^2\Bigl(\Ds-\frac{11}{6}+\frac{\epsilon}{9}\Bigr)
 -m_z^2\Bigl(\frac23+\frac{4\epsilon}{9}\Bigr)\Bigr]B_Q
 \Biggr\},\notag\\
 f_{1,\mathrm{tad}2,B}
 &=-\frac{4m_0^2(\Ds-1)}{Q^4}A,\notag\\
 f_{1,\mathrm{FF}4,B}
 &=\frac{4m_0^2(\Ds-1)}{Q^4}A
 +\frac{2m_0^2(\Ds-3)}{Q^2}B_Q.
 \label{eq:FFbare1}
\end{align}
The numerator coefficients are retained through the evanescent order
required for the finite one-loop answer. These formulas use the
Euclidean continuation stated in
Eq.~\re{eq:walking-euclidean-kinematics}, but retain the independent
mass parameter until the tree-shell continuation is imposed.

For the proper self-energy families, the same graph convention
requires a derivative, rather than division by $Q^2$. The tadpole is
independent of the external virtuality and hence gives
\begin{align}
 f_{1,\mathrm{SE}3,B}&=0.
 \label{eq:one-loop-SE-tadpole-derivative}
\end{align}
This zero concerns its residue contribution, not its contribution to
the self-energy or the pole displacement.
The ghost bubble gives
\begin{align}
 f_{1,\mathrm{SE}2,B}
 &=\left.\frac1{4(d-1)}
 \left[B_0(t;m_z^2,m_z^2)
 +(t-4m_z^2)\partial_t B_0(t;m_z^2,m_z^2)\right]
 \right|_{t=m_0^2}.
 \label{eq:one-loop-SE-ghost-derivative}
\end{align}
All masses in this derivative are fixed. The differentiated expression
is subsequently continued to the Euclidean tree shell.
At walking leading power the individual two-point contributions
therefore reduce to
\begin{align}
 f_{1,\mathrm{SE}1,B}
 &=\left[2-\frac{(2\Ds-3)(1-\epsilon)}
 {4(3-2\epsilon)}\right]B_w+o_{\rm LP}(1),\notag\\
 f_{1,\mathrm{SE}2,B}
 &=\frac{1-\epsilon}{4(3-2\epsilon)}B_w
 +o_{\rm LP}(1).
 \label{eq:one-loop-SE-leading-coefficients}
\end{align}
Their sum reproduces $\sigma_1'$ obtained by differentiating
Eq.~\re{eq:FFsig1}. The term proportional to $t-m_0^2$ in the
undifferentiated self-energy must not be dropped before this step.

The exact numerator coefficients for the remaining vertex families
\FFgraph{1}{1}, \FFgraph{1}{2} and \FFgraph{1}{5}
contain several hundred terms and are not reproduced here.
Their leading-power tree-shell values are given in
Table~\ref{tab:FF1lead}. The table lists the $c^{(0)}_{\gamma j}$
of Eq.~\re{eq:FFledger}, so the common factor $1/Q^2$ in that
equation is still to be applied. Self-energy derivatives are given
above rather than inferred from already on-shell table entries.

\begin{table}[t]
\begin{center}
\small
\begin{tabular}{lll}
\toprule
Graph & master & $c^{(0)}$ at $t_1=t_2=m_0^2$, leading power\\
\midrule
\FFgraph{1}{1}, \FFgraph{1}{2} & $A$ & $-\frac{2\Ds}{3}-1$\\[6pt]
\FFgraph{1}{1}, \FFgraph{1}{2} & $B_t$ &
$-\frac{3Q^2}{2}+\frac{\Ds m_0^2}{3}-\frac{m_0^2}{2}
-\frac{4\Ds m_z^2}{3}+2m_z^2$\\[6pt]
\FFgraph{1}{5} & $A$ & $\frac{4\Ds}{3}+\frac13$\\[6pt]
\FFgraph{1}{5} & $B_t$ &
$5Q^2-7m_0^2+\frac{4\Ds m_0^2}{3}-4m_z^2
+\frac{8\Ds m_z^2}{3}$\\[6pt]
\FFgraph{1}{5} & $B_Q$ &
$\Ds Q^2-\frac{11Q^2}{6}-2\Ds m_0^2+6m_0^2
+\frac{2m_z^2}{3}$\\[6pt]
\FFgraph{1}{5} & $C$ &
$2Q^4+2\Ds m_z^2Q^2-8m_z^2Q^2-2\Ds m_0^4+8m_0^4$\\
\bottomrule
\end{tabular}
\end{center}
\caption{Numerator coefficients of the one-loop vertex families
with long exact expressions. The dimensionless graph coefficient
is obtained using Eq.~\re{eq:FFledger}, including its factor $1/Q^2$.
Here $B_t$ stands for $B_{t_1}$ or $B_{t_2}$ as appropriate.
The two external attachments of \FFgraph{1}{5} are combined in the
displayed entries. Graphs \FFgraph{1}{1} and \FFgraph{1}{2}
have identical coefficients.}
\label{tab:FF1lead}
\end{table}

Summing the five ordinary vertex families at $\alpha=1$ gives
the numerator coefficients
\begin{equation}
\label{eq:FFsumV}
\begin{aligned}
\sum_{\gamma\in\{\FFgraph{1}{1},\ldots,\FFgraph{1}{5}\}}
 c^{(0)}_{\gamma A}
 &=-2\Ds+2,\\
\sum_{\gamma\in\{\FFgraph{1}{1},\ldots,\FFgraph{1}{5}\}}
 c^{(0)}_{\gamma B_t}
 &=2Q^2+2\Ds m_0^2-8m_0^2,\\
\sum_{\gamma\in\{\FFgraph{1}{1},\ldots,\FFgraph{1}{5}\}}
 c^{(0)}_{\gamma B_Q}
 &=0,\\
\sum_{\gamma\in\{\FFgraph{1}{1},\ldots,\FFgraph{1}{5}\}}
 c^{(0)}_{\gamma C}
 &=2\bigl[Q^4+(\Ds-4)m_z^2Q^2-(\Ds-4)m_0^4\bigr].
\end{aligned}
\end{equation}
The LSZ residue is obtained by differentiating the self-energy at independent virtuality before expanding in the small internal mass.
In the gauge $c=\alpha=1$, the term required at leading power in the walking expansion is
\begin{align}
\label{eq:FFsig1}
\sigma_1(t)
 &=\left[2(t-m_0^2)-\frac{\Ds-2}{2(d-1)}t\right]
 B_0(t;m_z^2,m_z^2)
 +O\!\left(m_z^2\log^k\frac{m_z^2}{-t}\right).
\end{align}
Its first term vanishes at $t=m_0^2$, but contributes $2B_0$ to the residue derivative. The corresponding fixed-gauge null-point formula, 
useful as an endpoint check, is recorded in Appendix~\ref{app:masters1}. The independent-mass reductions were checked
at three generic Euclidean points against Feynman-parameter quadrature.

\subsubsection{Cancellation of the tadpoles}
\label{sec:FF1tad}

Before taking the walking limit, it is important to distinguish an exact cancellation from a power-counting statement.
One-point attachments cancel against the source that holds the background fixed, whereas ordinary massive tadpoles
drop out only after the leading-power projection. Keeping these mechanisms separate also explains why a tadpole
insertion can survive in the two-loop calculation.

\begin{itemize}
\item{One-point attachments.} The one-point graphs \tadgraph{1}{3} and \tadgraph{1}{4} attach the renormalized tadpole $T^{(1)}$ 
of Eq.~\re{eq:QSforce} to the tree $G\chi WW$ vertex through a zero-momentum Cartan line. Their sum is proportional to $T^{(1)}/\lambda^2$, 
where $\lambda^2$ regulates the massless connector, i.e., propagator. This contribution is of the same planar order as the rest of the form 
factor and is not removed by any counterterm. In the held setup, the source \re{eq:held} attaches to the same connector with $H=T^{(1)}$, 
and the two cancel identically before the connector regulator is removed. On the stationary branches A and B, $T^{(1)}=0$ and both are absent. 
Thus the held-source condition $H=T^{(1)}$ removes the one-point graphs exactly, at any value of $z$, before any expansion; this is its defining 
property. Without the source, these graphs would leave the singular term $\propto T^{(1)}/\lambda^2$ of the source-free form factor discussed in
Sect.~\ref{sec:TwoWays}. For the leading walking behavior, held-source and branch-B conditions are equivalent, so we chose the former since
it does not insist on the special choice of the mass scale $\mu$.
\item{Massive tadpoles.} The graphs \tadgraph{1}{1}, \tadgraph{1}{2} and \SEgraph{1}{3} are proportional to 
$A=A_0(m_z^2)=m_z^2[1/\epsilon+1-\log(m_z^2/\mu^2)]$ with
coefficients that carry no inverse power of $m_z^2$, Eq.~\re{eq:FFbare1}. Their contributions to $\Gamma_1$ are therefore suppressed
by $m_z^2/Q^2$, by a further $w$ in the case of \tadgraph{1}{2}, whose $A$ term is moreover cancelled exactly by the $A$ term of the
bridge-bubble \FFgraph{1}{4}. The self-energy tadpole \SEgraph{1}{3} is independent of $t$, so it does not contribute to $\sigma_1'$ and hence not to
$f_{1,B}$; it contributes only to the pole shift $\sigma_1$, which enters the normalized form factor first at two loops, through
$\sigma_1\mathcal D\Gamma_1$ and $\sigma_1\sigma_1''$, where it is suppressed by $y=m_z^2/M_E^2$. All massive tadpoles thus vanish at leading
power in the walking regime, and they are scaleless at the null point. The $A$ terms that do survive at leading power, with
coefficients $\propto1/m_z^2$, come from the tensor reduction of the vertex and self-energy bubbles, not from tadpole graphs.
\end{itemize}

To conclude, at one loop, the held-background form factor at leading power is given by the five vertex graphs of Fig.~\ref{fig:FF1graphs} and the 
self-energy graphs of Fig.~\ref{fig:SE1graphs}; all tadpole graphs of Fig.~\ref{fig:TAD1graphs} drop out, either by the holding-source cancellation 
or by power suppression. At two loops the same two mechanisms remove the tadpole subgraphs, except for the raised-power insertion 
\tadgraph{2}{65}, which survives at leading power; the complete decision matrix for every family of graphs is given in Appendix~\ref{sec:tadpole-gate}.

\subsubsection{Renormalization}
\label{sec:FF1ren}

The LSZ reduction in Eq.~\re{eq:FFLSZ} removes the external poles, but does not subtract short-distance divergences. The finite form factor of 
the renormalized operator is constructed by adding the parameter and operator counterterms to the bare pole matrix element and expressing 
its reference tree, i.e., the denominator, in the same renormalized parameters. In terms of the coefficients defined in Eqs.~\re{eq:FFexp} and 
\re{eq:FFexpBare}, this prescription is
\begin{align}
\label{eq:FFcounterterm-construction}
F=\lim_{\epsilon\to0}\Bigl\{
1
+
g_R^2\bigl[f_{1,B}+\Delta f_1\bigr]
+
g_R^4\bigl[f_{2,B}+\Delta f_2\bigr]\Bigr\}
+O(g_R^6)
\, .
\end{align}
Here the bare loop coefficients on the right are evaluated at the renormalized parameter values, before any counterterm
insertion. The symbol $\Delta f_n$ denotes the additive counterterm correction to the normalized $n$-loop coefficient;
it includes the effects of parameter replacement, renormalization of the local $G$-insertion and reference-tree conversion.
At two loops, their cross terms must also be retained. These corrections are not separately finite: only the sums in
Eq.~\re{eq:FFcounterterm-construction} have a finite regulator limit. The complete two-loop correction at walking leading power
is displayed in Eq.~\re{eq:MS155}. We first specify the separate renormalization ingredients and then combine them in the explicit one-loop construction.
No additional elementary-field renormalization factor multiplies the pole matrix element, since the external residues
have already been included by LSZ. 

\paragraph{Gauge-coupling renormalization and running.}
\label{sec:gauge-running}
The gauge coupling fixes the scale dependence of the answer and, through its counterterm, part of the subtraction
at the next loop order. We use minimal subtraction to define the renormalized, dimensionless coupling $g_R^2(\mu)$:
\begin{align}
 g_\epsilon^2&=Z_{g^2}\,g_R^2,
 \label{eq:coupling-renormalization}
\\
 Z_{g^2}&=1-\frac{b_0}{\epsilon}g_R^2
 +\left(\frac{b_0^2}{\epsilon^2}-\frac{b_1}{2\epsilon}\right)g_R^4
 +O(g_R^6).
 \label{eq:coupling-Z}
\end{align}
The one-loop gauge, ghost and scalar contributions, and the two-loop gauge coefficient, follow by specializing the
gauge-scalar result of Ref.~\cite{Machacek:1983tz} to real adjoint scalars and no fermions. For $n_s$ real adjoint scalars,
the coefficients in the planar normalization used here are
\begin{align}
 b_0&=\frac{11}{3}-\frac{n_s}{6},
\\
 b_1&=\frac{34}{3}-\frac{7n_s}{3}.
 \label{eq:coupling-general-beta}
\end{align}
Their four-dimensional values, $n_s=\Ds-4$, are
\begin{align}
 b_0&=\frac{26-\Ds}{6},
\\
 b_1&=\frac{62-7\Ds}{3}.
 \label{eq:coupling-beta-coefficients}
\end{align}
In the regulated graph algebra, the actual scalar multiplicity remains $\Ds-d=\Ds-4+2\epsilon$: its evanescent part is retained until it has 
multiplied all ultraviolet poles. Equation~\re{eq:coupling-beta-coefficients} specifies the four-dimensional subtraction coefficients, not a 
replacement for the scalar multiplicity inside individual graphs.

Since the normalized tree form factor is unity, coupling renormalization does not contribute to $\Delta f_1$. Its first nontrivial effect starts at 
two loops: it is obtained by substituting Eq.~\re{eq:coupling-renormalization} into $g_\epsilon^2 f_{1,B}$:
\begin{align}
\label{eq:FFcoupling-counterterm}
\left.\Delta f_2\right|_{\rm coupling}
=-\frac{b_0}{\epsilon}f_{1,B}.
\end{align}
This is the gauge-coupling contribution to the two-loop subtraction; the mass, scalar-potential and operator terms
are treated separately below.

Differentiating at fixed $\gym$ gives the dimensionally continued beta function,
\begin{align}
 \beta_\epsilon(g_R^2)
 &\equiv\left.\frac{\dd g_R^2}{\dd\log\mu^2}\right|_{\gym}
 =-\epsilon g_R^2-b_0g_R^4-b_1g_R^6+O(g_R^8)
  \label{eq:coupling-beta-d}
 \, ,
 \end{align}
with the four-dimensional beta function defined by
\begin{align}
\beta(g_R^2)&\equiv\lim_{\epsilon\to0}\beta_\epsilon(g_R^2)
\, . 
\label{eq:coupling-beta-4}
\end{align}
To one-loop accuracy, evolution from the subtraction scale $\mu$ to a scale $k$ gives the running coupling
\begin{align}
 g_R^2(k^2)&=\frac{g_R^2(\mu^2)}
 {1+b_0g_R^2(\mu^2)\log(k^2/\mu^2)}
 +O(g_R^6\log(k^2/\mu^2))
\, ,
\label{eq:coupling-running-one-loop}
\end{align}
expressed in terms of $g_R^2(\mu^2)$.

\paragraph{Mass renormalization and the reference tree.}
Mass renormalization must be implemented consistently in the matrix element and in its normalizing tree.
It concerns the mass parameter of the action, rather than the location of the full propagator pole.
The action parameter and its renormalized counterpart $m_R^2$ are related by
\begin{align}
\label{eq:FFmass-renormalization}
m_0^2&=Z_{m^2}m_R^2,\qquad
Z_{m^2}=1+\frac{c_m}{\epsilon}g_R^2+O(g_R^4),
\qquad c_m=\frac{\Ds-2}{6}.
\end{align}
This replacement is distinct from the pole equation for $M^2$ in Eq.~\re{eq:FFpole}.
The latter is applied after the bare parameters have been expressed in renormalized ones.
The sign in Eq.~\re{eq:FFmass-renormalization} cancels the self-energy pole
$\sigma_1(m_R^2)|_{\rm UV}=-c_m m_R^2/\epsilon$, given equivalently in Eq.~\re{eq:exact-pole-shift-wy}.

Both the matrix element and its normalizing tree must use the same renormalized mass.
After the continuation of Eq.~\re{eq:walking-euclidean-kinematics},
with $m_R^2=-wQ^2$, the bare and renormalized reference trees obey
\begin{align}
\label{eq:FFreference-tree-counterterm}
\frac{1+2(M^2-m_0^2)/Q^2}{1+2(M^2-m_R^2)/Q^2}
&=1+g_R^2\frac{2wc_m}{\epsilon}+O(g_R^4),\\
\label{eq:FFmass-counterterm}
\Delta f_1^{\rm mass}&=\frac{2wc_m}{\epsilon}.
\end{align}
Here $w$ is the existing ratio of Eq.~\re{eq:walkcond}, evaluated at the renormalized mass. This term is required at finite masses even 
though it vanishes at walking leading power. The finite fixed-tree-to-pole-tree conversion, described in Appendix~\ref{app:masters1}, is 
already part of the chosen normalization, not an additional ultraviolet counterterm to be added to $f_{1,B}$. At two loops, the same 
parameter replacement also acts on the one-loop coefficient. Its contribution is included in Eq.~\re{eq:MS155}.

\paragraph{Scalar-potential counterterms.}
The quartic couplings require their own counterterms, independently of gauge-coupling renormalization. They are the coefficients 
$h_c,h_s,h_w,h_z$ of the potential in Eq.~\re{eq:RCT}, fixed in Sect.~\ref{sec:Renormalization}. Here $h_i$ denote the pole 
counterterms of Eq.~\re{eq:Rpoles}, not the finite renormalized quartic couplings in Eq.~\re{eq:UR}. Expanding that counterterm 
potential about the background generates two- and one-point vertices as well as the quartic insertions. Let $\Delta_{\rm sc}F$ denote 
their additive contribution to the normalized form factor, including both the vertex and LSZ terms. The leading-power result with one 
such counterterm insertion is
\begin{align}
\label{eq:FFHact}
\Delta_{\rm sc}F
&=g_R^2(3h_s-h_c)\mathcal H_B
 =g_R^4\frac{\kappa_h}{\epsilon}\mathcal H_B+O(g_R^6),
\\
\label{eq:FFscalar-counterterm-coefficient}
\kappa_h&=\frac{2(\Ds-2)}3=4c_m.
\end{align}
The second equality follows by inserting Eq.~\re{eq:Rpoles}, with $g^2$ replaced by $g_R^2$ at this order:
$3h_s-h_c=\kappa_h g_R^2/\epsilon$. Consequently $\Delta_{\rm sc}F$ contributes to $f_2$, not to $f_1$.
The function $\mathcal H_B$ is the complete regulated response to the scalar-potential insertion, derived in Eq.~\re{eq:H155};
it is not the holding source $H$. Writing its expansion from Eq.~\re{eq:Hdepth155} as
$\mathcal H_B=\mathcal H_0+\epsilon\mathcal H_1+O(\epsilon^2)$, its contribution to the two-loop counterterm coefficient is
\begin{align}
\label{eq:FFscalar-counterterm-finite}
\Delta_{\rm sc}f_2
&=\frac{\kappa_h}{\epsilon}\mathcal H_B
 =\frac{\kappa_h\mathcal H_0}{\epsilon}+\kappa_h\mathcal H_1+O(\epsilon).
\end{align}
The pole cancels only after all two-loop contributions have been combined, and $\kappa_h\mathcal H_1$ is the finite contribution
of this counterterm insertion. It must, therefore, be retained when constructing $f_2$.

The holding source has a different role: $H=T^{\rm ren}$ cancels the one-point attachments before the connector regulator is removed.
As established in Sect.~\ref{sec:FF1tad}, it requires no additional one-loop counterterm in the transverse form factor.

\paragraph{Composite-operator renormalization and $G$-$P$ mixing.}
After renormalizing the action, one must also renormalize the local $G$-insertion. Gauge-invariant operators can mix with total 
derivatives, equation-of-motion operators and BRST-exact operators \cite{Joglekar:1976eb,Collins:1977um}. In the present transverse 
pole matrix element, the required gauge-invariant mixing is with
\begin{align}
\label{eq:Pdefinition}
P&\equiv\partial^2\sum_I\tr\Phi_I^2.
\end{align}
A total derivative integrates to zero, but its local insertion carries $q^2=-Q^2\ne0$ and need not vanish.
Its tree matrix element, normalized to that of $G$, is
\begin{align}
\label{eq:P-composite-definition}
\frac{\langle WW|P|0\rangle^{(0)}}{\langle WW|G|0\rangle^{(0)}}&=-2w.
\end{align}
Equation-of-motion and BRST-exact terms do not supply additional transverse pole matrix elements at this order.

The one-loop operator counterterm is
\begin{align}
\label{eq:FFZGP}
[G]&=G_0+\frac{g_R^2}{\epsilon}
 \bigl(z_{GG}G_0+z_{GP}P_0\bigr)+O(g_R^4).
\end{align}
Here $G_0$ and $P_0$ are the unsubtracted operator polynomials; finite $G$-$P$ mixing is set to zero at $\mu_*$.
Its additive contribution to the one-loop form-factor coefficient is therefore
\begin{align}
\label{eq:FFoperator-counterterm}
\Delta f_1^{\rm operator}&=\frac{z_{GG}-2wz_{GP}}{\epsilon}.
\end{align}
In particular, the mixing term is not a multiplicative rescaling of the $G$ matrix element.

The constants $z_{GG}$ and $z_{GP}$ are fixed by the ultraviolet poles of the one-loop graph sum.
Only the tadpole and bubble masters are divergent,
$A|_{\rm UV}=m_z^2/\epsilon$ and $B_\ell|_{\rm UV}=1/\epsilon$; the triangle $C$ is finite.
The contributing vertex, self-energy and tadpole graphs are shown in Figs.~\ref{fig:FF1graphs}-\ref{fig:TAD1graphs}.
Using the Euclidean kinematics of
Eq.~\re{eq:walking-euclidean-kinematics} and the ratios of
Eq.~\re{eq:walkcond} after the independent-virtuality LSZ
derivatives, their sum gives
\begin{align}
\label{eq:composite-pole-matching}
\left.f_{1,B}\right|_{\rm UV}
&=\frac{b_0+2w(c_{GP}-c_m)}{\epsilon},
&c_{GP}&=\frac{22-5\Ds}{6}.
\end{align}
The coefficient $c_{GP}$ in this equation is obtained from the bare graphs, before imposing operator renormalization.
Combining this pole with the mass/reference-tree and operator counterterms gives
\begin{align}
\label{eq:composite-pole-cancel}
0&=b_0+2w(c_{GP}-c_m)+2wc_m+z_{GG}-2wz_{GP}.
\end{align}
Matching the independent constant and linear terms in $w$ determines
\begin{align}
\label{eq:composite-z-derived}
z_{GG}&=-b_0=\frac{\Ds-26}{6},
&z_{GP}&=c_{GP}=\frac{22-5\Ds}{6}.
\end{align}
Thus the diagonal term cancels the mass-independent pole, and the $P$ counterterm cancels the remaining mass-dependent pole.
These are consequences of the displayed graph sum. The diagonal coefficient is consistent with the beta-function/trace-anomaly 
relation \cite{Collins:1977um}. However, that relation alone does not determine $z_{GP}$.

\paragraph{Finite one-loop form factor.}
The preceding ingredients now assemble into a finite one-loop coefficient. Since the normalized tree is unity,
there are no gauge-coupling or scalar-potential counterterm contributions at order $g_R^2$.
The one-loop specialization of Eq.~\re{eq:FFcounterterm-construction} is the explicit sum
\begin{align}
\label{eq:FFone-loop-finite-construction}
f_1
&=\lim_{\epsilon\to0}
 \left[f_{1,B}+\Delta f_1^{\rm mass}+\Delta f_1^{\rm operator}\right]\notag\\
&=\lim_{\epsilon\to0}
 \left[f_{1,B}+\frac{2wc_m+z_{GG}-2wz_{GP}}{\epsilon}\right].
\end{align}
Equation~\re{eq:composite-pole-cancel} proves the cancellation of its ultraviolet pole at finite $w$.
The exact finite-mass evaluation of this expression is given next.
At walking leading power the terms proportional to $w$ vanish, and the same construction reduces to
\begin{align}
\label{eq:FFone-loop-renormalization}
f_1&=\lim_{\epsilon\to0}\left[f_{1,B}-\frac{b_0}{\epsilon}\right]
\qquad\text{(walking leading power)}.
\end{align}
The bare coefficient through the positive-$\epsilon$ order required at two loops is given in Eq.~\re{eq:FFbare-one-loop-Laurent}.
Only the leading-power specialization permits the finite-mass reference-tree counterterm to be dropped.

\subsection{One-loop form factor in exact kinematics}

With the source cancellation and ultraviolet subtraction in place, the finite result can be expressed through bubble
functions and one dimensionless triangle. We keep the mass ratios exact in this subsection, so that the subsequent
walking expansion is separated from the evaluation of the scalar integrals. The variables are the ratios in
Eq.~\re{eq:walkcond} and the independent logarithms
\begin{align}
\label{eq:walklogs}
L&\equiv\log\frac1w,\qquad
\ell_y\equiv\log y,\qquad
L_Q\equiv\log\frac{Q^2}{\mu^2}.
\end{align}
In the gauge $c=\alpha=1$, the complete finite one-loop coefficient in
the normalization of Eq.~\re{eq:FFexp} is
\begin{align}
\label{eq:FF1held}
f_1 (w,y;L_Q)
&
=
\frac{26-\Ds-8(\Ds+7)w}{6(1-4w)} b_w
+
\frac{8w^2(\Ds-4)}{1-4w} b_q
\nonumber\\
&-
\frac{(\Ds-2)(1+4y)}6 d_w
-
\frac{(\Ds-4)w(1+4y)}2 
\nonumber\\
&
+
\frac{4-(\Ds+12)w}{2(1-4w)} \mathcal T_0(w,y)
+
\frac{8\Ds-34}{9},
\end{align}
The finite bubble functions are defined by separating the ultraviolet
pole of the corresponding scalar masters:
\begin{align}
 B_0(-wQ^2;wyQ^2,wyQ^2)
 &=\frac1\epsilon+b_w+O(\epsilon),\\*
 B_0(-Q^2;wyQ^2,wyQ^2)
 &=\frac1\epsilon+b_q+O(\epsilon),\\*
 d_w&=\left.-wQ^2\,\partial_t
       B_0^{\rm fin}(t;wyQ^2,wyQ^2)\right|_{t=-wQ^2},\\*
 \mathcal T_0(w,y)&=\left.Q^2
 C_0(-wQ^2,-wQ^2,-Q^2;wyQ^2,wyQ^2,wyQ^2)\right|_{\epsilon=0}.
 \label{eq:one-loop-finite-masters}
\end{align}
For the walking expansion, we also introduce the unsubtracted
massless bubbles
\begin{align}
 B_w&\equiv B_0(-wQ^2;0,0)
 =\frac1\epsilon+2+L-L_Q+O(\epsilon),\\*
 B_q&\equiv B_0(-Q^2;0,0)
 =\frac1\epsilon+2-L_Q+O(\epsilon).
 \label{eq:walking-bubble-definitions}
\end{align}
They are the leading-power limits of the massive external- and
hard-channel bubbles, respectively. Their finite parts are therefore
the walking limits of $b_w$ and $b_q$, not the exact finite-mass
functions themselves. Their higher Laurent coefficients, needed at
two loops, are given in Eq.~\re{eq:bubbles155}.

Explicit parameter and logarithmic representations of the exact functions $b_w,b_q,d_w$ and $\mathcal T_0$ are given in Eqs.~\re{eq:exact-bubbles-wy} and \re{eq:exact-triangle-wy} of
Appendix~\ref{app:masters1}. In particular, $b_w$ is the finite part of $B_{t_1}=B_{t_2}$ after equal-virtuality specialization, $b_q$ is the
finite part of $B_Q$, and $d_w$ is the virtuality derivative specified in Eq.~\re{eq:one-loop-finite-masters}. The factor $Q^2$ removes the
mass dimension of the triangle $C$ in Eq.~\re{eq:FFmasters}. Equation~\re{eq:FF1held} is {\sl exact in $w$ and $y$ at one loop}: no power 
correction in either ratio has been dropped. The holding-source one-point cancellation is imposed before this result is formed. The 
conversion from a fixed classical reference tree to the shifted, renormalized reference tree in Eq.~\re{eq:FFexp} is included.
It changes finite-mass terms, but not the walking leading-power coefficient; the self-energy terms responsible for the conversion are 
shown in Appendix~\ref{app:masters1}.

\subsection{Walking Sudakov at one loop}
\label{sec:FF1walk}

We now extract the walking behavior by taking $w$ and $y$ to zero along the correlated trajectory of
Eq.\ \re{eq:walkcond}. The triangle controls the double logarithm, while the remaining terms determine operator evolution
and finite matching. The asymptotic formulas that follow retain leading power for fixed interior $0 < \eta < 1$;
endpoint comparisons will be distinguished from this interior expansion. 

\subsubsection{Leading double logarithm}

The origin of the leading double logarithm can be isolated directly in the exact answer: only the triangle can supply
two large logarithms at this order. In the limit $w\to0$ along \re{eq:walkcond}, its coefficient becomes $2+O(w)$, while
the Euclidean triangle defined in Eq.\ \re{eq:exact-triangle-wy} has the leading double-logarithmic behavior
\begin{equation}
\label{eq:FFCDL}
\mathcal T_0\big|_{\rm DL}=-\tfrac12\bigl[L^2-2L\ell_y-\ell_y^{2}\bigr]+O(L)
\, .
\end{equation}
All other terms are at most singly logarithmic: the bubbles have no independent $L^2$ once their coefficients are
combined, the massive tadpole terms have cancelled or are power suppressed, the coefficient of $b_q$ is $O(w^2)$, and the last term in 
Eq.\ \re{eq:FF1held} is rational. Thus
\begin{equation}
\label{eq:FF1DL}
f_1\big|_{\rm DL}=-\bigl(1+2\eta-\eta^2\bigr)L^2
\, ,\qquad
\Gamma^{(1)}_{\rm walk}(\eta)=1+2\eta-\eta^2
\, ,\qquad 0\le\eta\le1
\, .
\end{equation}
All $\Ds$ dependence drops out of the double logarithm, which therefore coincides with the one-loop walking coefficient
of $\cN=4$ sYM~\cite{Alday:2026walk}. The endpoints are the cusp, $\Gamma^{(1)}_{\rm walk}(0)=\Gamma^{(1)}_{\rm cusp}/4$, and
the octagon, $\Gamma^{(1)}_{\rm walk}(1)=\Gamma^{(1)}_{\rm oct}/2$.

\subsubsection{The Sudakov plane}

\begin{figure}[t]
\centering
\includegraphics[width=\textwidth]{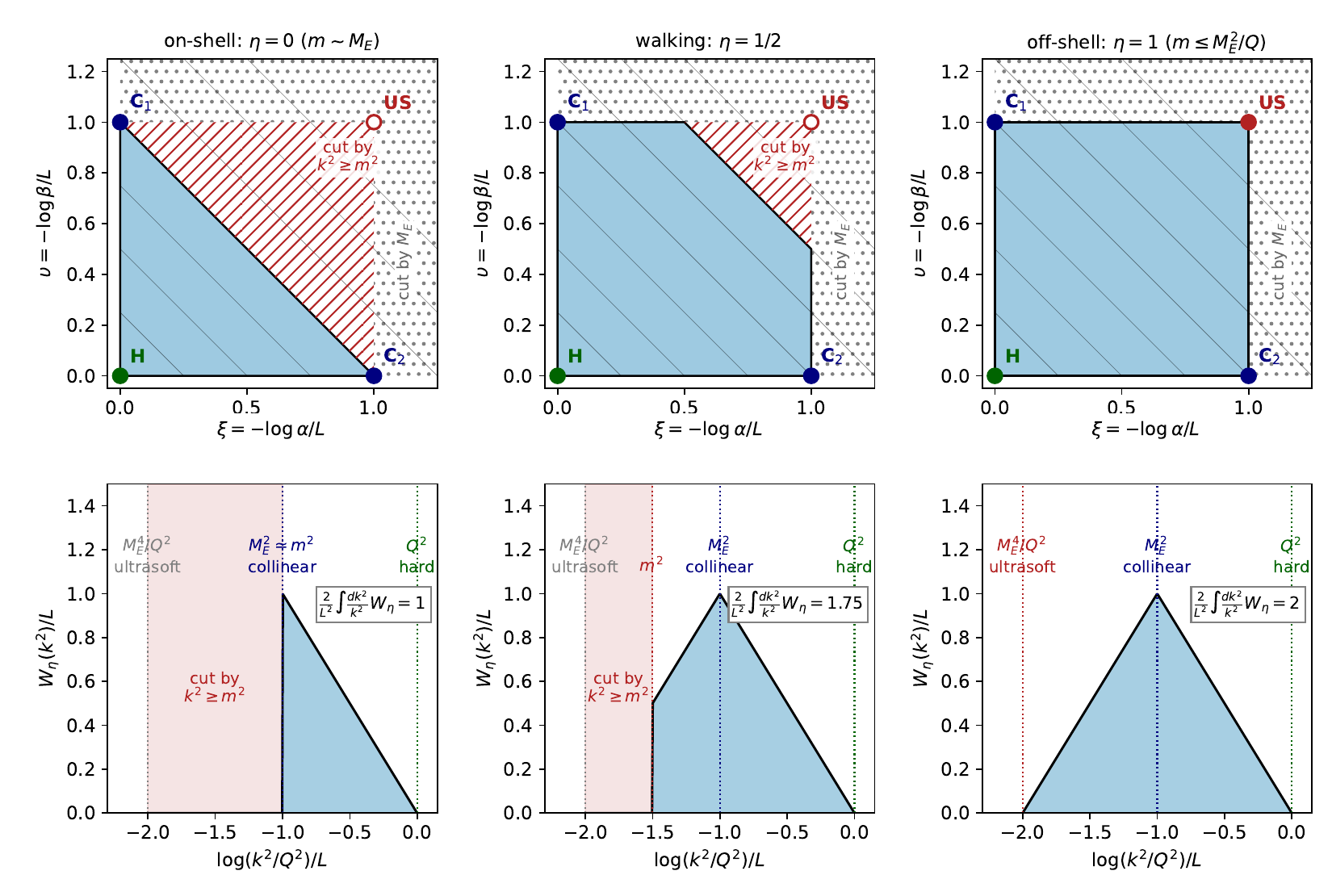}
\caption{Regions and weights for $\eta=0$ (on shell), $\eta=\tfrac12$ (walking) and $\eta=1$ (off shell). Top: the Sudakov
plane $(\xi,\upsilon)=(-\log\alpha,-\log\beta)/L$, with diagonals of fixed $k^2=\alpha\beta Q^2$. The allowed region (blue) is bounded
by the off-shellness cut $\xi,\upsilon\le1$ (dotted) and by the internal-mass cut $\xi+\upsilon\le1+\eta$ (red hatching). The modes sit at
the corners; the ultrasoft corner is excluded (open circle) unless $\eta\ge1$. Bottom: the weight $W_\eta(k^2)$, with the
part removed by the internal mass shaded red; the boxed number is $\Gamma^{(1)}_{\rm walk}=1+2\eta-\eta^2$.}
\label{fig:FFregions}
\end{figure}

The logarithmic coefficient has a geometric description as the phase space available to soft and collinear exchange.
This description both explains the interpolation between infrared regimes and supplies the weight used in the
scale-integral representation below. In the Method of Regions~\cite{Beneke:1997zp}, decompose the 
momentum of the exchanged virtual excitation as $k=\alpha p_1+\beta p_2+k_\perp$, so that $k^+=\alpha Q$, $k^-=\beta Q$. For every 
leading mode $k^+k^-$, $k_\perp^2$ and the virtuality scale homogeneously, and we label modes by the scale variable
$k^2\equiv k^+k^-=\alpha\beta Q^2$. In logarithmic variables $\xi=-\log\alpha/L$, $\upsilon=-\log\beta/L$, a line of fixed $k^2$
is the diagonal $\xi+\upsilon=\log(Q^2/k^2)/L$, along which the rapidity $\tfrac12\log(\alpha/\beta)$ runs. The four leading momentum 
modes sit at the corners of the $(\xi,\upsilon)$ plane:
\vspace{3mm}
\begin{center}
\small
\begin{tabular}{llll}
\toprule
Mode & $(k^+,k^-,k_\perp)$ & $k^+k^-\sim k_\perp^2\sim|{\rm virtuality}|$ & corner $(\xi,\upsilon)$\\
\midrule
hard H & $(Q,\,Q,\,Q)$ & $Q^2$ & $(0,0)$\\
collinear C$_1$ & $(Q,\,\lambda^2/Q,\,\lambda)$ & $\lambda^2$ & $(0,1)$\\
collinear C$_2$ & $(\lambda^2/Q,\,Q,\,\lambda)$ & $\lambda^2$ & $(1,0)$\\
ultrasoft US & $(M_E^2/Q,\,M_E^2/Q,\,M_E^2/Q)$ & $M_E^4/Q^2$ & $(1,1)$\\
\bottomrule
\end{tabular}
\end{center}
\vspace{3mm}
Here $\lambda=m_z$ in the on-shell case and $\lambda=M_E$ otherwise. The double logarithm is 
\begin{align}
\label{eq:FF1DL2}
g_R^2f_1\big|_{\rm DL}
=
- \frac12\int \frac{\dd k^2}{k^2} W(k^2)\,\Gamma_{\rm DL}
+O(g_R^4)\, , 
\end{align}
with $W(k^2)$ being the length of the diagonal inside the allowed region, i.e., the rapidity range available at fixed $k^2$, and 
$\Gamma_{\rm DL}=4g_R^2+O(g_R^4)$. The end-point regions and their interpolation are encoded in this formula and their
graphical representation displayed in Fig.\ \ref{fig:FFregions}.
\begin{itemize}
\item \textbf{On shell} ($M_E\to0$): hard and collinear modes; the internal mass cuts every line at $k^2\ge m_z^2$, and the
weight is $W_{\rm on}=\log(Q^2/k^2)$ on $[m_z^2,Q^2]$. This is the well-known form, controlled by $\Gamma_{\rm DL} = 
\Gamma_{\rm cusp}$ \cite{Korchemsky:1987wg,Korchemsky:1988hd}.
\item \textbf{Off shell} ($m_z\to0$): the off-shellness bounds the momentum fractions, $\alpha,\beta\ge M_E^2/Q^2$, so the
allowed region is the full square and the ultrasoft mode is dynamical. The weight is the ``tent'' $\log(Q^2/k^2)$ above
and $\log(k^2Q^2/M_E^4)$ below the collinear scale $M_E^2$, closing at $M_E^4/Q^2$, and the kernel is $\Gamma_{\rm DL} = 
\Gamma_{\rm oct}$. To one-loop order, the latter is identical to $\Gamma_{\rm cusp}$.
The two $Q$-dependent regions together give $-\Gamma_{\rm oct}L^2/2$, twice the on-shell
double-logarithmic contribution at one loop for the same large logarithm.
\item \textbf{Walking} ($0\le\eta\le1$): the internal mass lies between the ultrasoft and collinear scales,
$M_E^4/Q^2\le m_z^2\le M_E^2$. The ultrasoft mode is not dynamical, since its massive propagators can be Taylor expanded;
the internal mass cuts the ultrasoft corner off the square along $\xi+\upsilon=1+\eta$.
\end{itemize}
Slicing the cut square at fixed $k^2$ gives a single formula for all regimes,
\begin{equation}
\label{eq:Wwalk}
W_\eta(k^2)=\min\Bigl\{\log\frac{Q^2}{k^2},\ \log\frac{k^2Q^2}{M_E^4}\Bigr\}
\quad\text{for}\quad \max\Bigl(m_z^2,\frac{M_E^4}{Q^2}\Bigr)\le k^2\le Q^2
\, ,
\end{equation}
and $W_\eta=0$ otherwise, see Fig.~\ref{fig:FFregions}. Its moment reproduces the one-loop coefficient in each regime:
\begin{center}
\small
\begin{tabular}{llll}
\toprule
Regime & leading regions & support of $W_\eta$ & $\frac{2}{L^2}\int\frac{\dd k^2}{k^2}W_\eta$\\
\midrule
$\eta\le-1$ & none (decoupled) & empty & $0$\\
$-1\le\eta\le0$ & H + C ($k_\perp\sim m_z$) & $[m_z^2,Q^2]$, upper branch & $(1+\eta)^2$\\
$0\le\eta\le1$ & H + C ($k_\perp\sim M_E$) & $[m_z^2,Q^2]$, both branches & $1+2\eta-\eta^2$\\
$\eta\ge1$ & H + C + US & $[M_E^4/Q^2,Q^2]$ & $2$\\
\bottomrule
\end{tabular}
\end{center}

\vspace{2mm}

\begin{figure}[t]
\centering
\includegraphics[width=0.62\textwidth]{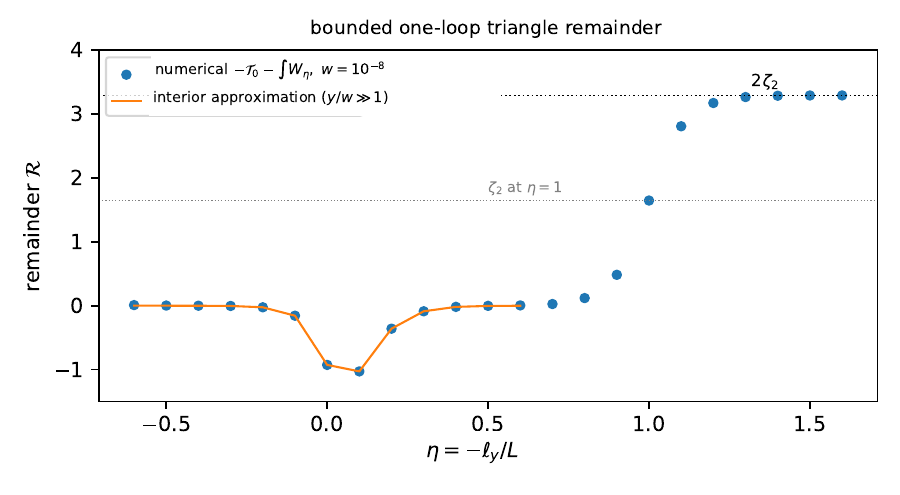}
\caption{The finite triangle remainder defined in Eq.~\re{eq:triangle-remainder-definition} at $w=10^{-8}$
(dots). The line shows the interior approximation \re{eq:FFRint} only where $y/w\gg1$; it is not extrapolated through the 
crossover. The full numerical remainder passes through $\zeta_2$ at $\eta=1$ and approaches $2\zeta_2$ on the octagon plateau.}
\label{fig:FFrem}
\end{figure}

The phase-space area accounts for the double logarithm, but the exact result also fixes the single logarithm and the
finite contribution that the area alone cannot determine. Isolating these terms in Eq.~\re{eq:FF1held} gives
\begin{equation}
\label{eq:FF1scale}
f_1=-2\int\frac{\dd k^2}{k^2}\,W_\eta(k^2)+b_0(L-L_Q)+c_1(y)-2\mathcal R(w,y)+o_{\rm LP}(1)
\, ,
\end{equation}
where
\begin{align}\label{eq:one-loop-matching-y}
c_1(y)={}&
b_0\left[2+ \bigl(\sqrt{1+4y}-1\bigr)\ell_y -2\sqrt{1+4y}\log\frac{1+\sqrt{1+4y}}2\right]\notag\\*
&
+\frac{(1+4y)(\Ds-2)}6 \left[1+\frac{2y}{\sqrt{1+4y}} \left(\ell_y-2\log\frac{1+\sqrt{1+4y}}2\right)\right] +\frac{8\Ds-34}{9}
\, ,
\end{align}
and the triangle remainder is defined by
\begin{equation}
\label{eq:triangle-remainder-definition}
\mathcal R(w,y)\equiv-\mathcal T_0(w,y)
 -\int\frac{\dd k^2}{k^2}\,W_\eta(k^2)\, ,
\end{equation}
with $\mathcal T_0$ determined exactly by the parametric integral representation in Eq.~\re{eq:exact-triangle-wy}. The support and weight of 
the scale integral is given above by Eq.~\re{eq:Wwalk}. In the limit $y/w\gg1$, its interior leading-power expression for ${\mathcal R}$ can 
be found in a closed form and reads
\begin{equation}
\label{eq:FFRint}
\mathcal R
=
\theta(1-y)\ell_y^{2}+2\Bigl[\operatorname{Li}_2\Bigl(\frac{2}{1-\sqrt{1+4y}}\Bigr)+\operatorname{Li}_2\Bigl(\frac{2}{1+\sqrt{1+4y}}\Bigr)\Bigr]
\qquad(y/w\gg1)
\, ,
\end{equation}
Equation~\re{eq:FFRint} vanishes as $y\to0$ within its stated hierarchy; it is not an expression for the crossover $y/w\sim1$ or
the octagon shoulder $y/w\to0$. Those are obtained from the exact triangle definition above. Figure~\ref{fig:FFrem} illustrates the
bounded crossover numerically along the trajectory of Eq.~\re{eq:walkcond}: the remainder passes through 
$\zeta_2$ at $\eta=1$ and approaches $2\zeta_2$ on the octagon plateau.
In the strictly off-shell limit (on the octagon plateau),
$c_1(0)-2\mathcal R=\tfrac{13\Ds+82}{18}-4\zeta_2$, in agreement with the strictly off-shell constant obtained in the companion paper. At one loop, the additional 
single logarithm is the operator-evolution term $b_0(L-L_Q)$ displayed in Eq.~\re{eq:FF1scale}. The triangle contribution is represented 
by $W_\eta$ and its bounded matching remainder.

For fixed interior $\eta$, $y\to0$ and the remainder in Eq.~\re{eq:FFRint} vanishes. The complete leading-power coefficient of the Sudakov 
form factor is
\begin{align}
f_1&=\mathcal P_{\rm DL}+b_0(L-L_Q)+c_1+o_{\rm LP}(1)\, .
 \label{eq:one-loop-walking-final}
\end{align}
with
\begin{align}
\mathcal P_{\rm DL}&\equiv-L^2+2L\ell_y+\ell_y^2,
 \label{eq:walking-double-log-polynomial}\\
c_1&\equiv\lim_{y\to0}c_1(y)=\frac{13D_s+82}{18}
 \label{eq:walking-matching-constant}
 \, .
\end{align}

\subsubsection{Scale integral representation and running coupling}

The phase-space weight separates the one-loop answer into double-logarithmic evolution, evolution of the local
operator, and a finite matching term. Promoting the coupling in these contributions to the running coupling of
Eq.~\re{eq:coupling-running-one-loop} also predicts a definite part of the next-order logarithms. In this form,
Eq.~\re{eq:FF1scale} is reproduced by
\begin{equation}
\label{eq:FFuniform}
\begin{aligned}
\log F={}&-\frac12\int\frac{\dd k^2}{k^2}\,W_\eta(k^2)\,\Gamma_{\rm DL}\bigl(g_R^2(k^2)\bigr)-\int_{\mu^2}^{M_E^2}\frac{\dd k^2}{k^2}\,
\gamma_G\bigl(g_R^2(k^2)\bigr)\\
&+g_R^2(M_E^2)\bigl[c_1(y)-2\mathcal R\bigr]+O(g_R^4)_{\text{non-running}}
\, ,
\end{aligned}
\end{equation}
where $\Gamma_{\rm DL}=4g_R^2+\dots$ denotes the common one-loop double-logarithmic kernel\footnote{At one loop
$\Gamma_{\rm cusp}$ and $\Gamma_{\rm oct}$ coincide, so this calculation alone cannot distinguish them.
The two-loop analysis below resolves their separate roles.} and $\gamma_G=b_0g_R^2+\dots$ is the anomalous dimension
of $[G]$, cf.\ Eq.~\re{eq:FFZGP}. The diagonal operator anomalous dimension is determined by the beta function
because $G$ is conjugate to the gauge coupling:
\begin{align}\label{eq:gamma-diagonal}
\gamma_G^{\rm diag}(g_R^2)
&\equiv
- g_R^2\frac{\partial}{\partial g_R^2} \left[\frac{\beta(g_R^2)}{g_R^2}\right]
=b_0g_R^2+2b_1g_R^4+O(g_R^6)\, .
\end{align}
Equation~\re{eq:FFuniform} matches the complete one-loop answer. At the next order it specifies the terms generated by
running, but not the new two-loop kernels or finite matching, which are determined in Eq.~\re{eq:integral155}.
To isolate the running contribution at order $g_R^4$, impose $\ell_y=-\eta L$ from Eq.~\re{eq:walkcond} in the
following moments of the weight:
\begin{equation}
\label{eq:FFweight-moments}
\int\frac{\dd k^2}{k^2}W_\eta=\frac{L^2}{2}\Gamma^{(1)}_{\rm walk}(\eta)
\, ,\qquad
\int\frac{\dd k^2}{k^2}W_\eta\log\frac{k^2}{Q^2}=-L^3\Bigl(\frac13+\eta-\frac{\eta^3}{3}\Bigr)
\end{equation}
Expanding $g_R^2(k^2)$ and using these moments gives the running-coupling terms
\begin{equation}
\label{eq:FFrun2}
\begin{aligned}
\log F\supset g_R^4\Bigl[&b_0\,\Gamma^{(1)}_{\rm walk}(\eta)\,L^2L_Q
-2b_0\Bigl(\frac13+\eta-\frac{\eta^3}{3}\Bigr)L^3\\
&+\frac{b_0^2}{2}(L-L_Q)^2+b_0\,c_1(y)(L-L_Q)\Bigr]
\end{aligned}
\, .
\end{equation}
At $\eta=0$ the cubic term is the standard $-\tfrac23b_0L^3$ of the on-shell cusp; at $\eta=1$ the first two terms combine
into $2b_0L^2\log(M_E^2/\mu^2)$. These cubic logarithms are therefore fixed by coupling evolution rather than by
a new double-logarithmic kernel. The complete two-loop calculation must add the genuine two-loop contributions and
the remaining renormalization and matching terms.

\subsection{Two loops}
\label{sec:FF2}

We now calculate the two-loop held-background form factor in the walking regime, retaining the complete leading-power
dependence on the independent logarithms, including the single logarithms and constants. Unlike the one-loop result,
this coefficient receives both genuine two-loop contributions and products generated by the displacement and residue of
the external poles. We therefore first identify and evaluate the graph families, then assemble the bare LSZ coefficient
of Eq.~\re{eq:FF2LSZ}, and finally apply the action and composite-operator counterterms of
Eqs.~\re{eq:MS155}-\re{eq:FFtwo-loop-counterterms}.

The graph labels follow the one-loop convention: \FFgraph{\ell}{n}, \SEgraph{\ell}{n} and \tadgraph{\ell}{n}, now with
$\ell=2$. Their numerical identifiers are inherited from the graph-family catalogue and need not be consecutive.
The FF labels denote ordinary vertex families without a tadpole subgraph; the SE labels denote proper two-point
families, including self-energy tadpoles; and the tad labels denote insertion families containing a tadpole or a
detachable one-point branch. The ancillary graph dictionary relates these physical families to the generated diagrams.

The subtraction convention matters for the finite two-loop coefficient. We use the dimensional-reduction type
scheme specified above, with scalar multiplicity $n_s=\Ds-d$ and $\Ds$ held fixed during the Laurent expansion in
$\epsilon$.
The renormalized scalar quartics obey the commutator-potential boundary conditions of Eq.~\re{eq:Rcond} at $\mu_*$. The coefficients quoted 
below are evaluated with these boundary conditions at the reference scale. At another renormalization scale, the scalar couplings generally 
depart from these commutator-potential values. The relevant finite combination $h(\mu)$, normalized by Eq.~\re{eq:finite-scalar-deformation}, 
must therefore be evolved when the result is expressed at the mass scale. Since $h=O(g_R^2)$, its contribution $g_R^2h(\mu)\mathcal H_0$ 
first enters at two loops. The transport to the mass boundary is given in Eq.~\re{eq:mass-boundary-transport}. This finite coupling is distinct from 
$h_{\rm CT}=\kappa_hg_R^2/\epsilon$ in Eq.~\re{eq:quartic-response-coefficient}: $h(\mu)$ is a parameter of the renormalized theory, whereas 
$h_{\rm CT}$ is the pole term that subtracts its ultraviolet divergence. The holding source satisfies $H=T^{\rm ren}$; it is applied to the 
one-point/source pairs before the diagonal-connector limit, and is not set to zero. However, as we already emphasized earlier, the only 
surviving tadpole insertion at two loops is \tadgraph{2}{65}. The calculation uses the gauge $c=\alpha=1$ of Sect.~\ref{sec:gauge}. We 
calculate the form factor defined in Eq.~\re{eq:FFexp} in the kinematics of Eq.~\re{eq:walkcond}. The logarithms $L$, $\ell_y$ and $L_Q$ of 
Eq.~\re{eq:walklogs} are kept independent through renormalization such that the correlated trajectory in Eq.\ \re{eq:walkcond} is imposed at 
the last step on the final result.

\subsubsection{Graphs}
\label{sec:FF2graphs}

The graph enumeration separates corrections to the operator insertion from corrections to the external propagators.
Both are required by LSZ, but their roles in the amplitude are different. The bare vertex graphs are generated with QGRAF~\cite{Nogueira:1991ex},
using the action vertices $X^3$, $X^4$, $\bar\omega\omega X$ and exactly
one insertion vertex $G_0X^2$, $G_0X^3$ or $G_0X^4$. Here $G_0$ is the
unsubtracted operator polynomial introduced in Sect.~\ref{sec:FFdef}.
The selection rules of that section exclude external-leg self-energy
chains from the amputated vertex, but retain internal self-energy
insertions and bridges in the operator channel. Proper two-point graphs
are generated separately for the pole and residue factors in LSZ.
The one-point/source pairs are included according to the held-background
prescription.

FORM~\cite{Vermaseren:2000nd} performs the color and uplifted component-index
contractions, retaining the full $d$ dependence and scalar multiplicity
$\Ds-d$. We then select the leading large-$N$ color contribution.
This is a color selection, not a criterion of planarity for the scalar
propagator drawing; in particular, it does not exclude the crossed
ladder \FFgraph{2}{70}. Contributions with the same propagator connectivity
are grouped into the paper families \FFgraph{2}{n},
\tadgraph{2}{n} and \SEgraph{2}{n} defined above.
Different mass routings, ghost orientations, symmetry factors and tensor
numerators are kept separately until the members of each family are
summed. The accompanying QGRAF-to-paper ledger identifies the constituent
graph records.

The figures below depict propagator connectivity, not complete scalar
integrands. An internal line represents an uplifted $X$ component or a
ghost, with its mass fixed by the color flow. Each drawing must therefore
be combined with its numerator and mass assignments. The transverse
projection of Sect.~\ref{sec:FFdef} removes the common external
polarization product; the subsequent tensor reduction expresses the
remaining loop-momentum contractions in terms of scalar integrals.
The individual vertex and proper self-energy families contributing at
leading power are shown separately.

\begin{figure}[!htbp]
\centering
\begin{tabular}{cccc}
\begin{tikzpicture}[x=0.40cm,y=0.40cm,baseline=(current bounding box.center)]
\coordinate (v1) at (-1.25,-1.0);
\coordinate (v2) at (1.25,-1.0);
\coordinate (v3) at (0.0,1.35);
\coordinate (p1) at (-2.05,-2.15);
\coordinate (p2) at (2.05,-2.15);
\coordinate (G) at (v3);
\draw[line width=.6pt,bend left=-28] (v1) to (v3);
\draw[line width=.6pt,bend left=28] (v1) to (v3);
\draw[line width=.6pt,bend left=-28] (v2) to (v3);
\draw[line width=.6pt,bend left=28] (v2) to (v3);
\draw[line width=.6pt] (p1) to (v1);
\draw[line width=.6pt] (p2) to (v2);
\fill (v1) circle (2.2pt);
\fill (v2) circle (2.2pt);
\node[below left,font=\scriptsize] at (p1) {$p_1$};
\node[below right,font=\scriptsize] at (p2) {$p_2$};
\draw[fill=white,line width=.4pt] (G) circle (4pt);
\draw[line width=.5pt] ($(G)+(-2.1pt,-2.1pt)$)--($(G)+(2.1pt,2.1pt)$);
\draw[line width=.5pt] ($(G)+(-2.1pt,2.1pt)$)--($(G)+(2.1pt,-2.1pt)$);
\draw[densely dotted,->,shorten >=4.5pt] ($(G)+(0,.95)$)
 node[above,font=\scriptsize]{$q$} -- (G);
\end{tikzpicture} & \begin{tikzpicture}[x=0.40cm,y=0.40cm,baseline=(current bounding box.center)]
\coordinate (v2) at (0.0,-1.15);
\coordinate (v3) at (0.0,1.35);
\coordinate (v1) at (0.0,0.25);
\coordinate (p1) at (-2.05,-2.15);
\coordinate (p2) at (2.05,-2.15);
\coordinate (G) at (v3);
\draw[line width=.6pt] (v1) to (v2);
\draw[line width=.6pt,bend left=-28] (v1) to (v3);
\draw[line width=.6pt,bend left=28] (v1) to (v3);
\draw[line width=.6pt,bend left=50] (v2) to (v3);
\draw[line width=.6pt] (p1) to (v2);
\draw[line width=.6pt] (p2) to (v2);
\fill (v2) circle (2.2pt);
\fill (v1) circle (2.2pt);
\node[below left,font=\scriptsize] at (p1) {$p_1$};
\node[below right,font=\scriptsize] at (p2) {$p_2$};
\draw[fill=white,line width=.4pt] (G) circle (4pt);
\draw[line width=.5pt] ($(G)+(-2.1pt,-2.1pt)$)--($(G)+(2.1pt,2.1pt)$);
\draw[line width=.5pt] ($(G)+(-2.1pt,2.1pt)$)--($(G)+(2.1pt,-2.1pt)$);
\draw[densely dotted,->,shorten >=4.5pt] ($(G)+(0,.95)$)
 node[above,font=\scriptsize]{$q$} -- (G);
\end{tikzpicture} & \begin{tikzpicture}[x=0.40cm,y=0.40cm,baseline=(current bounding box.center)]
\coordinate (v2) at (0.0,1.35);
\coordinate (v3) at (1.25,-1.0);
\coordinate (v1) at (0.0,0.25);
\coordinate (p1) at (-2.05,-2.15);
\coordinate (p2) at (2.05,-2.15);
\coordinate (G) at (v2);
\draw[line width=.6pt] (v1) to (v2);
\draw[line width=.6pt,bend left=-28] (v1) to (v3);
\draw[line width=.6pt,bend left=28] (v1) to (v3);
\draw[line width=.6pt] (v2) to (v3);
\draw[line width=.6pt] (p1) to (v2);
\draw[line width=.6pt] (p2) to (v3);
\fill (v3) circle (2.2pt);
\fill (v1) circle (2.2pt);
\node[below left,font=\scriptsize] at (p1) {$p_1$};
\node[below right,font=\scriptsize] at (p2) {$p_2$};
\draw[fill=white,line width=.4pt] (G) circle (4pt);
\draw[line width=.5pt] ($(G)+(-2.1pt,-2.1pt)$)--($(G)+(2.1pt,2.1pt)$);
\draw[line width=.5pt] ($(G)+(-2.1pt,2.1pt)$)--($(G)+(2.1pt,-2.1pt)$);
\draw[densely dotted,->,shorten >=4.5pt] ($(G)+(0,.95)$)
 node[above,font=\scriptsize]{$q$} -- (G);
\end{tikzpicture} & \begin{tikzpicture}[x=0.40cm,y=0.40cm,baseline=(current bounding box.center)]
\coordinate (v1) at (-1.25,-1.0);
\coordinate (v2) at (0.0,1.35);
\coordinate (v3) at (0.0,0.25);
\coordinate (p1) at (-2.05,-2.15);
\coordinate (p2) at (2.05,-2.15);
\coordinate (G) at (v2);
\draw[line width=.6pt,bend left=-28] (v1) to (v3);
\draw[line width=.6pt,bend left=28] (v1) to (v3);
\draw[line width=.6pt,bend left=-28] (v2) to (v3);
\draw[line width=.6pt,bend left=28] (v2) to (v3);
\draw[line width=.6pt] (p1) to (v1);
\draw[line width=.6pt] (p2) to (v2);
\fill (v1) circle (2.2pt);
\fill (v3) circle (2.2pt);
\node[below left,font=\scriptsize] at (p1) {$p_1$};
\node[below right,font=\scriptsize] at (p2) {$p_2$};
\draw[fill=white,line width=.4pt] (G) circle (4pt);
\draw[line width=.5pt] ($(G)+(-2.1pt,-2.1pt)$)--($(G)+(2.1pt,2.1pt)$);
\draw[line width=.5pt] ($(G)+(-2.1pt,2.1pt)$)--($(G)+(2.1pt,-2.1pt)$);
\draw[densely dotted,->,shorten >=4.5pt] ($(G)+(0,.95)$)
 node[above,font=\scriptsize]{$q$} -- (G);
\end{tikzpicture}\\[-2pt]
{\small \FFgraph{2}{12}} & {\small \FFgraph{2}{17}} & {\small \FFgraph{2}{18}} & {\small \FFgraph{2}{23}}\\[2pt]
\begin{tikzpicture}[x=0.40cm,y=0.40cm,baseline=(current bounding box.center)]
\coordinate (v1) at (0.0,1.35);
\coordinate (v2) at (0.0,-1.15);
\coordinate (v3) at (0.0,0.25);
\coordinate (p1) at (-2.05,-2.15);
\coordinate (p2) at (2.05,-2.15);
\coordinate (G) at (v1);
\draw[line width=.6pt,bend left=-28] (v1) to (v3);
\draw[line width=.6pt,bend left=28] (v1) to (v3);
\draw[line width=.6pt,bend left=-28] (v2) to (v3);
\draw[line width=.6pt,bend left=28] (v2) to (v3);
\draw[line width=.6pt] (p1) to (v2);
\draw[line width=.6pt] (p2) to (v2);
\fill (v2) circle (2.2pt);
\fill (v3) circle (2.2pt);
\node[below left,font=\scriptsize] at (p1) {$p_1$};
\node[below right,font=\scriptsize] at (p2) {$p_2$};
\draw[fill=white,line width=.4pt] (G) circle (4pt);
\draw[line width=.5pt] ($(G)+(-2.1pt,-2.1pt)$)--($(G)+(2.1pt,2.1pt)$);
\draw[line width=.5pt] ($(G)+(-2.1pt,2.1pt)$)--($(G)+(2.1pt,-2.1pt)$);
\draw[densely dotted,->,shorten >=4.5pt] ($(G)+(0,.95)$)
 node[above,font=\scriptsize]{$q$} -- (G);
\end{tikzpicture} & \begin{tikzpicture}[x=0.40cm,y=0.40cm,baseline=(current bounding box.center)]
\coordinate (v1) at (-1.25,-1.0);
\coordinate (v2) at (1.25,-1.0);
\coordinate (v3) at (0.0,1.35);
\coordinate (p1) at (-2.05,-2.15);
\coordinate (p2) at (2.05,-2.15);
\coordinate (G) at (v3);
\draw[line width=.6pt] (v1) to (v2);
\draw[line width=.6pt] (v1) to (v3);
\draw[line width=.6pt,bend left=-28] (v2) to (v3);
\draw[line width=.6pt,bend left=28] (v2) to (v3);
\draw[line width=.6pt] (p1) to (v1);
\draw[line width=.6pt] (p2) to (v2);
\fill (v1) circle (2.2pt);
\fill (v2) circle (2.2pt);
\node[below left,font=\scriptsize] at (p1) {$p_1$};
\node[below right,font=\scriptsize] at (p2) {$p_2$};
\draw[fill=white,line width=.4pt] (G) circle (4pt);
\draw[line width=.5pt] ($(G)+(-2.1pt,-2.1pt)$)--($(G)+(2.1pt,2.1pt)$);
\draw[line width=.5pt] ($(G)+(-2.1pt,2.1pt)$)--($(G)+(2.1pt,-2.1pt)$);
\draw[densely dotted,->,shorten >=4.5pt] ($(G)+(0,.95)$)
 node[above,font=\scriptsize]{$q$} -- (G);
\end{tikzpicture} & \begin{tikzpicture}[x=0.40cm,y=0.40cm,baseline=(current bounding box.center)]
\coordinate (v1) at (0.0,1.35);
\coordinate (v2) at (-1.25,-1.0);
\coordinate (v3) at (1.25,-1.0);
\coordinate (p1) at (-2.05,-2.15);
\coordinate (p2) at (2.05,-2.15);
\coordinate (G) at (v1);
\draw[line width=.6pt] (v1) to (v2);
\draw[line width=.6pt] (v1) to (v3);
\draw[line width=.6pt,bend left=-28] (v2) to (v3);
\draw[line width=.6pt,bend left=28] (v2) to (v3);
\draw[line width=.6pt] (p1) to (v2);
\draw[line width=.6pt] (p2) to (v3);
\fill (v2) circle (2.2pt);
\fill (v3) circle (2.2pt);
\node[below left,font=\scriptsize] at (p1) {$p_1$};
\node[below right,font=\scriptsize] at (p2) {$p_2$};
\draw[fill=white,line width=.4pt] (G) circle (4pt);
\draw[line width=.5pt] ($(G)+(-2.1pt,-2.1pt)$)--($(G)+(2.1pt,2.1pt)$);
\draw[line width=.5pt] ($(G)+(-2.1pt,2.1pt)$)--($(G)+(2.1pt,-2.1pt)$);
\draw[densely dotted,->,shorten >=4.5pt] ($(G)+(0,.95)$)
 node[above,font=\scriptsize]{$q$} -- (G);
\end{tikzpicture} & \begin{tikzpicture}[x=0.40cm,y=0.40cm,baseline=(current bounding box.center)]
\coordinate (v1) at (-1.25,-1.0);
\coordinate (v4) at (0.6,0.9);
\coordinate (v2) at (-1.4,0.25);
\coordinate (v3) at (-0.55,1.15);
\coordinate (p1) at (-2.05,-2.15);
\coordinate (p2) at (2.05,-2.15);
\coordinate (G) at (v4);
\draw[line width=.6pt] (v1) to (v2);
\draw[line width=.6pt] (v1) to (v4);
\draw[line width=.6pt,bend left=-28] (v2) to (v3);
\draw[line width=.6pt,bend left=28] (v2) to (v3);
\draw[line width=.6pt] (v3) to (v4);
\draw[line width=.6pt] (p1) to (v1);
\draw[line width=.6pt] (p2) to (v4);
\fill (v1) circle (2.2pt);
\fill (v2) circle (2.2pt);
\fill (v3) circle (2.2pt);
\node[below left,font=\scriptsize] at (p1) {$p_1$};
\node[below right,font=\scriptsize] at (p2) {$p_2$};
\draw[fill=white,line width=.4pt] (G) circle (4pt);
\draw[line width=.5pt] ($(G)+(-2.1pt,-2.1pt)$)--($(G)+(2.1pt,2.1pt)$);
\draw[line width=.5pt] ($(G)+(-2.1pt,2.1pt)$)--($(G)+(2.1pt,-2.1pt)$);
\draw[densely dotted,->,shorten >=4.5pt] ($(G)+(0,.95)$)
 node[above,font=\scriptsize]{$q$} -- (G);
\end{tikzpicture}\\[-2pt]
{\small \FFgraph{2}{24}} & {\small \FFgraph{2}{27}} & {\small \FFgraph{2}{28}} & {\small \FFgraph{2}{35}}\\[2pt]
\begin{tikzpicture}[x=0.40cm,y=0.40cm,baseline=(current bounding box.center)]
\coordinate (v1) at (0,1.35);
\coordinate (v4) at (0,-1.15);
\coordinate (v2) at (-1.2,0.65);
\coordinate (v3) at (-1.2,-0.5);
\coordinate (p1) at (-2.05,-2.15);
\coordinate (p2) at (2.05,-2.15);
\coordinate (G) at (v1);
\draw[line width=.6pt] (v1) to (v2);
\draw[line width=.6pt] (v1) to (v4);
\draw[line width=.6pt,bend left=-28] (v2) to (v3);
\draw[line width=.6pt,bend left=28] (v2) to (v3);
\draw[line width=.6pt] (v3) to (v4);
\draw[line width=.6pt] (p1) to (v4);
\draw[line width=.6pt] (p2) to (v4);
\fill (v4) circle (2.2pt);
\fill (v2) circle (2.2pt);
\fill (v3) circle (2.2pt);
\node[below left,font=\scriptsize] at (p1) {$p_1$};
\node[below right,font=\scriptsize] at (p2) {$p_2$};
\draw[fill=white,line width=.4pt] (G) circle (4pt);
\draw[line width=.5pt] ($(G)+(-2.1pt,-2.1pt)$)--($(G)+(2.1pt,2.1pt)$);
\draw[line width=.5pt] ($(G)+(-2.1pt,2.1pt)$)--($(G)+(2.1pt,-2.1pt)$);
\draw[densely dotted,->,shorten >=4.5pt] ($(G)+(0,.95)$)
 node[above,font=\scriptsize]{$q$} -- (G);
\end{tikzpicture} & \begin{tikzpicture}[x=0.40cm,y=0.40cm,baseline=(current bounding box.center)]
\coordinate (v1) at (-1.25,-1.0);
\coordinate (v4) at (0.6,0.7);
\coordinate (v2) at (-0.8,0.37);
\coordinate (v3) at (0.15,-0.67);
\coordinate (p1) at (-2.05,-2.15);
\coordinate (p2) at (2.05,-2.15);
\coordinate (G) at (v4);
\draw[line width=.6pt] (v1) to (v2);
\draw[line width=.6pt] (v1) to (v3);
\draw[line width=.6pt] (v2) to (v3);
\draw[line width=.6pt] (v2) to (v4);
\draw[line width=.6pt] (v3) to (v4);
\draw[line width=.6pt] (p1) to (v1);
\draw[line width=.6pt] (p2) to (v4);
\fill (v1) circle (2.2pt);
\fill (v2) circle (2.2pt);
\fill (v3) circle (2.2pt);
\node[below left,font=\scriptsize] at (p1) {$p_1$};
\node[below right,font=\scriptsize] at (p2) {$p_2$};
\draw[fill=white,line width=.4pt] (G) circle (4pt);
\draw[line width=.5pt] ($(G)+(-2.1pt,-2.1pt)$)--($(G)+(2.1pt,2.1pt)$);
\draw[line width=.5pt] ($(G)+(-2.1pt,2.1pt)$)--($(G)+(2.1pt,-2.1pt)$);
\draw[densely dotted,->,shorten >=4.5pt] ($(G)+(0,.95)$)
 node[above,font=\scriptsize]{$q$} -- (G);
\end{tikzpicture} & \begin{tikzpicture}[x=0.40cm,y=0.40cm,baseline=(current bounding box.center)]
\coordinate (v1) at (0,1.35);
\coordinate (v4) at (0,-1.15);
\coordinate (v2) at (-1.0,0.1);
\coordinate (v3) at (1.0,0.1);
\coordinate (p1) at (-2.05,-2.15);
\coordinate (p2) at (2.05,-2.15);
\coordinate (G) at (v1);
\draw[line width=.6pt] (v1) to (v2);
\draw[line width=.6pt] (v1) to (v3);
\draw[line width=.6pt] (v2) to (v3);
\draw[line width=.6pt] (v2) to (v4);
\draw[line width=.6pt] (v3) to (v4);
\draw[line width=.6pt] (p1) to (v4);
\draw[line width=.6pt] (p2) to (v4);
\fill (v4) circle (2.2pt);
\fill (v2) circle (2.2pt);
\fill (v3) circle (2.2pt);
\node[below left,font=\scriptsize] at (p1) {$p_1$};
\node[below right,font=\scriptsize] at (p2) {$p_2$};
\draw[fill=white,line width=.4pt] (G) circle (4pt);
\draw[line width=.5pt] ($(G)+(-2.1pt,-2.1pt)$)--($(G)+(2.1pt,2.1pt)$);
\draw[line width=.5pt] ($(G)+(-2.1pt,2.1pt)$)--($(G)+(2.1pt,-2.1pt)$);
\draw[densely dotted,->,shorten >=4.5pt] ($(G)+(0,.95)$)
 node[above,font=\scriptsize]{$q$} -- (G);
\end{tikzpicture} & \begin{tikzpicture}[x=0.40cm,y=0.40cm,baseline=(current bounding box.center)]
\coordinate (v1) at (-1.25,-1.0);
\coordinate (v2) at (1.25,-1.0);
\coordinate (v4) at (0.0,1.35);
\coordinate (v3) at (0.0,0.25);
\coordinate (p1) at (-2.05,-2.15);
\coordinate (p2) at (2.05,-2.15);
\coordinate (G) at (v4);
\draw[line width=.6pt] (v1) to (v3);
\draw[line width=.6pt] (v1) to (v4);
\draw[line width=.6pt] (v2) to (v3);
\draw[line width=.6pt] (v2) to (v4);
\draw[line width=.6pt] (v3) to (v4);
\draw[line width=.6pt] (p1) to (v1);
\draw[line width=.6pt] (p2) to (v2);
\fill (v1) circle (2.2pt);
\fill (v2) circle (2.2pt);
\fill (v3) circle (2.2pt);
\node[below left,font=\scriptsize] at (p1) {$p_1$};
\node[below right,font=\scriptsize] at (p2) {$p_2$};
\draw[fill=white,line width=.4pt] (G) circle (4pt);
\draw[line width=.5pt] ($(G)+(-2.1pt,-2.1pt)$)--($(G)+(2.1pt,2.1pt)$);
\draw[line width=.5pt] ($(G)+(-2.1pt,2.1pt)$)--($(G)+(2.1pt,-2.1pt)$);
\draw[densely dotted,->,shorten >=4.5pt] ($(G)+(0,.95)$)
 node[above,font=\scriptsize]{$q$} -- (G);
\end{tikzpicture}\\[-2pt]
{\small \FFgraph{2}{36}} & {\small \FFgraph{2}{37}} & {\small \FFgraph{2}{38}} & {\small \FFgraph{2}{48}}\\[2pt]
\begin{tikzpicture}[x=0.40cm,y=0.40cm,baseline=(current bounding box.center)]
\coordinate (v1) at (-1.25,-1.0);
\coordinate (v2) at (0.0,1.35);
\coordinate (v4) at (1.25,-1.0);
\coordinate (v3) at (0.0,0.25);
\coordinate (p1) at (-2.05,-2.15);
\coordinate (p2) at (2.05,-2.15);
\coordinate (G) at (v2);
\draw[line width=.6pt] (v1) to (v3);
\draw[line width=.6pt] (v1) to (v4);
\draw[line width=.6pt] (v2) to (v3);
\draw[line width=.6pt] (v2) to (v4);
\draw[line width=.6pt] (v3) to (v4);
\draw[line width=.6pt] (p1) to (v1);
\draw[line width=.6pt] (p2) to (v4);
\fill (v1) circle (2.2pt);
\fill (v4) circle (2.2pt);
\fill (v3) circle (2.2pt);
\node[below left,font=\scriptsize] at (p1) {$p_1$};
\node[below right,font=\scriptsize] at (p2) {$p_2$};
\draw[fill=white,line width=.4pt] (G) circle (4pt);
\draw[line width=.5pt] ($(G)+(-2.1pt,-2.1pt)$)--($(G)+(2.1pt,2.1pt)$);
\draw[line width=.5pt] ($(G)+(-2.1pt,2.1pt)$)--($(G)+(2.1pt,-2.1pt)$);
\draw[densely dotted,->,shorten >=4.5pt] ($(G)+(0,.95)$)
 node[above,font=\scriptsize]{$q$} -- (G);
\end{tikzpicture} & \begin{tikzpicture}[x=0.40cm,y=0.40cm,baseline=(current bounding box.center)]
\coordinate (v1) at (-1.25,-1.0);
\coordinate (v2) at (1.25,-1.0);
\coordinate (v4) at (0.0,1.35);
\coordinate (v3) at (0.0,0.25);
\coordinate (p1) at (-2.05,-2.15);
\coordinate (p2) at (2.05,-2.15);
\coordinate (G) at (v4);
\draw[line width=.6pt] (v1) to (v2);
\draw[line width=.6pt] (v1) to (v3);
\draw[line width=.6pt] (v2) to (v4);
\draw[line width=.6pt,bend left=-28] (v3) to (v4);
\draw[line width=.6pt,bend left=28] (v3) to (v4);
\draw[line width=.6pt] (p1) to (v1);
\draw[line width=.6pt] (p2) to (v2);
\fill (v1) circle (2.2pt);
\fill (v2) circle (2.2pt);
\fill (v3) circle (2.2pt);
\node[below left,font=\scriptsize] at (p1) {$p_1$};
\node[below right,font=\scriptsize] at (p2) {$p_2$};
\draw[fill=white,line width=.4pt] (G) circle (4pt);
\draw[line width=.5pt] ($(G)+(-2.1pt,-2.1pt)$)--($(G)+(2.1pt,2.1pt)$);
\draw[line width=.5pt] ($(G)+(-2.1pt,2.1pt)$)--($(G)+(2.1pt,-2.1pt)$);
\draw[densely dotted,->,shorten >=4.5pt] ($(G)+(0,.95)$)
 node[above,font=\scriptsize]{$q$} -- (G);
\end{tikzpicture} & \begin{tikzpicture}[x=0.40cm,y=0.40cm,baseline=(current bounding box.center)]
\coordinate (v1) at (-1.25,-1.0);
\coordinate (v2) at (0.0,1.35);
\coordinate (v4) at (1.25,-1.0);
\coordinate (v3) at (0.0,0.25);
\coordinate (p1) at (-2.05,-2.15);
\coordinate (p2) at (2.05,-2.15);
\coordinate (G) at (v2);
\draw[line width=.6pt] (v1) to (v2);
\draw[line width=.6pt] (v1) to (v3);
\draw[line width=.6pt] (v2) to (v4);
\draw[line width=.6pt,bend left=-28] (v3) to (v4);
\draw[line width=.6pt,bend left=28] (v3) to (v4);
\draw[line width=.6pt] (p1) to (v1);
\draw[line width=.6pt] (p2) to (v4);
\fill (v1) circle (2.2pt);
\fill (v4) circle (2.2pt);
\fill (v3) circle (2.2pt);
\node[below left,font=\scriptsize] at (p1) {$p_1$};
\node[below right,font=\scriptsize] at (p2) {$p_2$};
\draw[fill=white,line width=.4pt] (G) circle (4pt);
\draw[line width=.5pt] ($(G)+(-2.1pt,-2.1pt)$)--($(G)+(2.1pt,2.1pt)$);
\draw[line width=.5pt] ($(G)+(-2.1pt,2.1pt)$)--($(G)+(2.1pt,-2.1pt)$);
\draw[densely dotted,->,shorten >=4.5pt] ($(G)+(0,.95)$)
 node[above,font=\scriptsize]{$q$} -- (G);
\end{tikzpicture} & \begin{tikzpicture}[x=0.40cm,y=0.40cm,baseline=(current bounding box.center)]
\coordinate (v1) at (0.0,1.35);
\coordinate (v2) at (-1.25,-1.0);
\coordinate (v4) at (1.25,-1.0);
\coordinate (v3) at (0.0,0.25);
\coordinate (p1) at (-2.05,-2.15);
\coordinate (p2) at (2.05,-2.15);
\coordinate (G) at (v1);
\draw[line width=.6pt] (v1) to (v2);
\draw[line width=.6pt] (v1) to (v3);
\draw[line width=.6pt] (v2) to (v4);
\draw[line width=.6pt,bend left=-28] (v3) to (v4);
\draw[line width=.6pt,bend left=28] (v3) to (v4);
\draw[line width=.6pt] (p1) to (v2);
\draw[line width=.6pt] (p2) to (v4);
\fill (v2) circle (2.2pt);
\fill (v4) circle (2.2pt);
\fill (v3) circle (2.2pt);
\node[below left,font=\scriptsize] at (p1) {$p_1$};
\node[below right,font=\scriptsize] at (p2) {$p_2$};
\draw[fill=white,line width=.4pt] (G) circle (4pt);
\draw[line width=.5pt] ($(G)+(-2.1pt,-2.1pt)$)--($(G)+(2.1pt,2.1pt)$);
\draw[line width=.5pt] ($(G)+(-2.1pt,2.1pt)$)--($(G)+(2.1pt,-2.1pt)$);
\draw[densely dotted,->,shorten >=4.5pt] ($(G)+(0,.95)$)
 node[above,font=\scriptsize]{$q$} -- (G);
\end{tikzpicture}\\[-2pt]
{\small \FFgraph{2}{49}} & {\small \FFgraph{2}{51}} & {\small \FFgraph{2}{52}} & {\small \FFgraph{2}{53}}\\[2pt]
\begin{tikzpicture}[x=0.40cm,y=0.40cm,baseline=(current bounding box.center)]
\coordinate (v1) at (-1.25,-1.0);
\coordinate (v2) at (1.25,-1.0);
\coordinate (v3) at (0.0,1.35);
\coordinate (v4) at (0.0,0.25);
\coordinate (p1) at (-2.05,-2.15);
\coordinate (p2) at (2.05,-2.15);
\coordinate (G) at (v3);
\draw[line width=.6pt] (v1) to (v2);
\draw[line width=.6pt] (v1) to (v4);
\draw[line width=.6pt] (v2) to (v4);
\draw[line width=.6pt,bend left=-28] (v3) to (v4);
\draw[line width=.6pt,bend left=28] (v3) to (v4);
\draw[line width=.6pt] (p1) to (v1);
\draw[line width=.6pt] (p2) to (v2);
\fill (v1) circle (2.2pt);
\fill (v2) circle (2.2pt);
\fill (v4) circle (2.2pt);
\node[below left,font=\scriptsize] at (p1) {$p_1$};
\node[below right,font=\scriptsize] at (p2) {$p_2$};
\draw[fill=white,line width=.4pt] (G) circle (4pt);
\draw[line width=.5pt] ($(G)+(-2.1pt,-2.1pt)$)--($(G)+(2.1pt,2.1pt)$);
\draw[line width=.5pt] ($(G)+(-2.1pt,2.1pt)$)--($(G)+(2.1pt,-2.1pt)$);
\draw[densely dotted,->,shorten >=4.5pt] ($(G)+(0,.95)$)
 node[above,font=\scriptsize]{$q$} -- (G);
\end{tikzpicture} & \begin{tikzpicture}[x=0.40cm,y=0.40cm,baseline=(current bounding box.center)]
\coordinate (v1) at (-1.25,-1.0);
\coordinate (v2) at (0.0,1.35);
\coordinate (v3) at (1.25,-1.0);
\coordinate (v4) at (0.0,0.25);
\coordinate (p1) at (-2.05,-2.15);
\coordinate (p2) at (2.05,-2.15);
\coordinate (G) at (v2);
\draw[line width=.6pt] (v1) to (v2);
\draw[line width=.6pt] (v1) to (v4);
\draw[line width=.6pt] (v2) to (v4);
\draw[line width=.6pt,bend left=-28] (v3) to (v4);
\draw[line width=.6pt,bend left=28] (v3) to (v4);
\draw[line width=.6pt] (p1) to (v1);
\draw[line width=.6pt] (p2) to (v3);
\fill (v1) circle (2.2pt);
\fill (v3) circle (2.2pt);
\fill (v4) circle (2.2pt);
\node[below left,font=\scriptsize] at (p1) {$p_1$};
\node[below right,font=\scriptsize] at (p2) {$p_2$};
\draw[fill=white,line width=.4pt] (G) circle (4pt);
\draw[line width=.5pt] ($(G)+(-2.1pt,-2.1pt)$)--($(G)+(2.1pt,2.1pt)$);
\draw[line width=.5pt] ($(G)+(-2.1pt,2.1pt)$)--($(G)+(2.1pt,-2.1pt)$);
\draw[densely dotted,->,shorten >=4.5pt] ($(G)+(0,.95)$)
 node[above,font=\scriptsize]{$q$} -- (G);
\end{tikzpicture} & \begin{tikzpicture}[x=0.40cm,y=0.40cm,baseline=(current bounding box.center)]
\coordinate (v1) at (0,1.35);
\coordinate (v2) at (-1.25,-1.0);
\coordinate (v3) at (1.25,-1.0);
\coordinate (v4) at (0,-1.0);
\coordinate (p1) at (-2.05,-2.15);
\coordinate (p2) at (2.05,-2.15);
\coordinate (G) at (v1);
\draw[line width=.6pt] (v1) to (v2);
\draw[line width=.6pt] (v1) to (v3);
\draw[line width=.6pt] (v2) to (v4);
\draw[line width=.6pt] (v3) to (v4);
\draw[line width=.6pt] (p1) to (v2);
\draw[line width=.6pt] (p2) to (v3);
\draw[line width=.6pt] (v4) .. controls (-1.05,0.35) and (1.05,0.35) .. (v4);
\fill (v2) circle (2.2pt);
\fill (v3) circle (2.2pt);
\fill (v4) circle (2.2pt);
\node[below left,font=\scriptsize] at (p1) {$p_1$};
\node[below right,font=\scriptsize] at (p2) {$p_2$};
\draw[fill=white,line width=.4pt] (G) circle (4pt);
\draw[line width=.5pt] ($(G)+(-2.1pt,-2.1pt)$)--($(G)+(2.1pt,2.1pt)$);
\draw[line width=.5pt] ($(G)+(-2.1pt,2.1pt)$)--($(G)+(2.1pt,-2.1pt)$);
\draw[densely dotted,->,shorten >=4.5pt] ($(G)+(0,.95)$)
 node[above,font=\scriptsize]{$q$} -- (G);
\end{tikzpicture} & \begin{tikzpicture}[x=0.40cm,y=0.40cm,baseline=(current bounding box.center)]
\coordinate (v1) at (-1.0,-1.0);
\coordinate (v2) at (1.0,-1.0);
\coordinate (v3) at (0,1.35);
\coordinate (v4) at (-1.0,0.25);
\coordinate (v5) at (1.0,0.25);
\coordinate (p1) at (-2.05,-2.15);
\coordinate (p2) at (2.05,-2.15);
\coordinate (G) at (v3);
\draw[line width=.6pt] (v1) to (v4);
\draw[line width=.6pt] (v1) to (v5);
\draw[line width=.6pt] (v2) to (v4);
\draw[line width=.6pt] (v2) to (v5);
\draw[line width=.6pt] (v3) to (v4);
\draw[line width=.6pt] (v3) to (v5);
\draw[line width=.6pt] (p1) to (v1);
\draw[line width=.6pt] (p2) to (v2);
\fill (v1) circle (2.2pt);
\fill (v2) circle (2.2pt);
\fill (v4) circle (2.2pt);
\fill (v5) circle (2.2pt);
\node[below left,font=\scriptsize] at (p1) {$p_1$};
\node[below right,font=\scriptsize] at (p2) {$p_2$};
\draw[fill=white,line width=.4pt] (G) circle (4pt);
\draw[line width=.5pt] ($(G)+(-2.1pt,-2.1pt)$)--($(G)+(2.1pt,2.1pt)$);
\draw[line width=.5pt] ($(G)+(-2.1pt,2.1pt)$)--($(G)+(2.1pt,-2.1pt)$);
\draw[densely dotted,->,shorten >=4.5pt] ($(G)+(0,.95)$)
 node[above,font=\scriptsize]{$q$} -- (G);
\end{tikzpicture}\\[-2pt]
{\small \FFgraph{2}{62}} & {\small \FFgraph{2}{63}} & {\small \tadgraph{2}{65}} & {\small \FFgraph{2}{70}}\\[2pt]
\begin{tikzpicture}[x=0.40cm,y=0.40cm,baseline=(current bounding box.center)]
\coordinate (v1) at (-1.0,-1.0);
\coordinate (v2) at (1.0,-1.0);
\coordinate (v3) at (0,1.35);
\coordinate (v4) at (-1.0,0.25);
\coordinate (v5) at (1.0,0.25);
\coordinate (p1) at (-2.05,-2.15);
\coordinate (p2) at (2.05,-2.15);
\coordinate (G) at (v3);
\draw[line width=.6pt] (v1) to (v2);
\draw[line width=.6pt] (v1) to (v4);
\draw[line width=.6pt] (v2) to (v5);
\draw[line width=.6pt] (v3) to (v4);
\draw[line width=.6pt] (v3) to (v5);
\draw[line width=.6pt] (v4) to (v5);
\draw[line width=.6pt] (p1) to (v1);
\draw[line width=.6pt] (p2) to (v2);
\fill (v1) circle (2.2pt);
\fill (v2) circle (2.2pt);
\fill (v4) circle (2.2pt);
\fill (v5) circle (2.2pt);
\node[below left,font=\scriptsize] at (p1) {$p_1$};
\node[below right,font=\scriptsize] at (p2) {$p_2$};
\draw[fill=white,line width=.4pt] (G) circle (4pt);
\draw[line width=.5pt] ($(G)+(-2.1pt,-2.1pt)$)--($(G)+(2.1pt,2.1pt)$);
\draw[line width=.5pt] ($(G)+(-2.1pt,2.1pt)$)--($(G)+(2.1pt,-2.1pt)$);
\draw[densely dotted,->,shorten >=4.5pt] ($(G)+(0,.95)$)
 node[above,font=\scriptsize]{$q$} -- (G);
\end{tikzpicture} & \begin{tikzpicture}[x=0.40cm,y=0.40cm,baseline=(current bounding box.center)]
\coordinate (v1) at (-1.25,-1.0);
\coordinate (v2) at (0,1.35);
\coordinate (v3) at (1.25,-1.0);
\coordinate (v4) at (0,-1.0);
\coordinate (v5) at (0.62,0.17);
\coordinate (p1) at (-2.05,-2.15);
\coordinate (p2) at (2.05,-2.15);
\coordinate (G) at (v2);
\draw[line width=.6pt] (v1) to (v2);
\draw[line width=.6pt] (v1) to (v4);
\draw[line width=.6pt] (v2) to (v5);
\draw[line width=.6pt] (v3) to (v4);
\draw[line width=.6pt] (v3) to (v5);
\draw[line width=.6pt] (v4) to (v5);
\draw[line width=.6pt] (p1) to (v1);
\draw[line width=.6pt] (p2) to (v3);
\fill (v1) circle (2.2pt);
\fill (v3) circle (2.2pt);
\fill (v4) circle (2.2pt);
\fill (v5) circle (2.2pt);
\node[below left,font=\scriptsize] at (p1) {$p_1$};
\node[below right,font=\scriptsize] at (p2) {$p_2$};
\draw[fill=white,line width=.4pt] (G) circle (4pt);
\draw[line width=.5pt] ($(G)+(-2.1pt,-2.1pt)$)--($(G)+(2.1pt,2.1pt)$);
\draw[line width=.5pt] ($(G)+(-2.1pt,2.1pt)$)--($(G)+(2.1pt,-2.1pt)$);
\draw[densely dotted,->,shorten >=4.5pt] ($(G)+(0,.95)$)
 node[above,font=\scriptsize]{$q$} -- (G);
\end{tikzpicture} & \begin{tikzpicture}[x=0.40cm,y=0.40cm,baseline=(current bounding box.center)]
\coordinate (v1) at (-1.25,-1.0);
\coordinate (v2) at (1.25,-1.0);
\coordinate (v3) at (0,1.35);
\coordinate (v4) at (0.875,-0.295);
\coordinate (v5) at (0.375,0.645);
\coordinate (p1) at (-2.05,-2.15);
\coordinate (p2) at (2.05,-2.15);
\coordinate (G) at (v3);
\draw[line width=.6pt] (v1) to (v2);
\draw[line width=.6pt] (v1) to (v3);
\draw[line width=.6pt] (v2) to (v4);
\draw[line width=.6pt] (v3) to (v5);
\draw[line width=.6pt,bend left=-28] (v4) to (v5);
\draw[line width=.6pt,bend left=28] (v4) to (v5);
\draw[line width=.6pt] (p1) to (v1);
\draw[line width=.6pt] (p2) to (v2);
\fill (v1) circle (2.2pt);
\fill (v2) circle (2.2pt);
\fill (v4) circle (2.2pt);
\fill (v5) circle (2.2pt);
\node[below left,font=\scriptsize] at (p1) {$p_1$};
\node[below right,font=\scriptsize] at (p2) {$p_2$};
\draw[fill=white,line width=.4pt] (G) circle (4pt);
\draw[line width=.5pt] ($(G)+(-2.1pt,-2.1pt)$)--($(G)+(2.1pt,2.1pt)$);
\draw[line width=.5pt] ($(G)+(-2.1pt,2.1pt)$)--($(G)+(2.1pt,-2.1pt)$);
\draw[densely dotted,->,shorten >=4.5pt] ($(G)+(0,.95)$)
 node[above,font=\scriptsize]{$q$} -- (G);
\end{tikzpicture} & \begin{tikzpicture}[x=0.40cm,y=0.40cm,baseline=(current bounding box.center)]
\coordinate (v1) at (0,1.35);
\coordinate (v2) at (-1.25,-1.0);
\coordinate (v3) at (1.25,-1.0);
\coordinate (v4) at (-0.65,-1.0);
\coordinate (v5) at (0.65,-1.0);
\coordinate (p1) at (-2.05,-2.15);
\coordinate (p2) at (2.05,-2.15);
\coordinate (G) at (v1);
\draw[line width=.6pt] (v1) to (v2);
\draw[line width=.6pt] (v1) to (v3);
\draw[line width=.6pt] (v2) to (v4);
\draw[line width=.6pt] (v3) to (v5);
\draw[line width=.6pt,bend left=-28] (v4) to (v5);
\draw[line width=.6pt,bend left=28] (v4) to (v5);
\draw[line width=.6pt] (p1) to (v2);
\draw[line width=.6pt] (p2) to (v3);
\fill (v2) circle (2.2pt);
\fill (v3) circle (2.2pt);
\fill (v4) circle (2.2pt);
\fill (v5) circle (2.2pt);
\node[below left,font=\scriptsize] at (p1) {$p_1$};
\node[below right,font=\scriptsize] at (p2) {$p_2$};
\draw[fill=white,line width=.4pt] (G) circle (4pt);
\draw[line width=.5pt] ($(G)+(-2.1pt,-2.1pt)$)--($(G)+(2.1pt,2.1pt)$);
\draw[line width=.5pt] ($(G)+(-2.1pt,2.1pt)$)--($(G)+(2.1pt,-2.1pt)$);
\draw[densely dotted,->,shorten >=4.5pt] ($(G)+(0,.95)$)
 node[above,font=\scriptsize]{$q$} -- (G);
\end{tikzpicture}\\[-2pt]
{\small \FFgraph{2}{71}} & {\small \FFgraph{2}{72}} & {\small \FFgraph{2}{74}} & {\small \FFgraph{2}{75}}\\[2pt]
\end{tabular}
\caption{The 24 two-loop vertex graph families contributing at leading power. The crossed circle is the insertion of $G$,
and each vertical dotted arrow carries its incoming momentum $q$;
$p_{1,2}$ are the external $W$ legs. All graph families are planar and drawn without crossings, except \FFgraph{2}{70}, whose crossing
is unavoidable with the external legs and the insertion on the outer boundary.}
\label{fig:FF2graphs}
\end{figure}
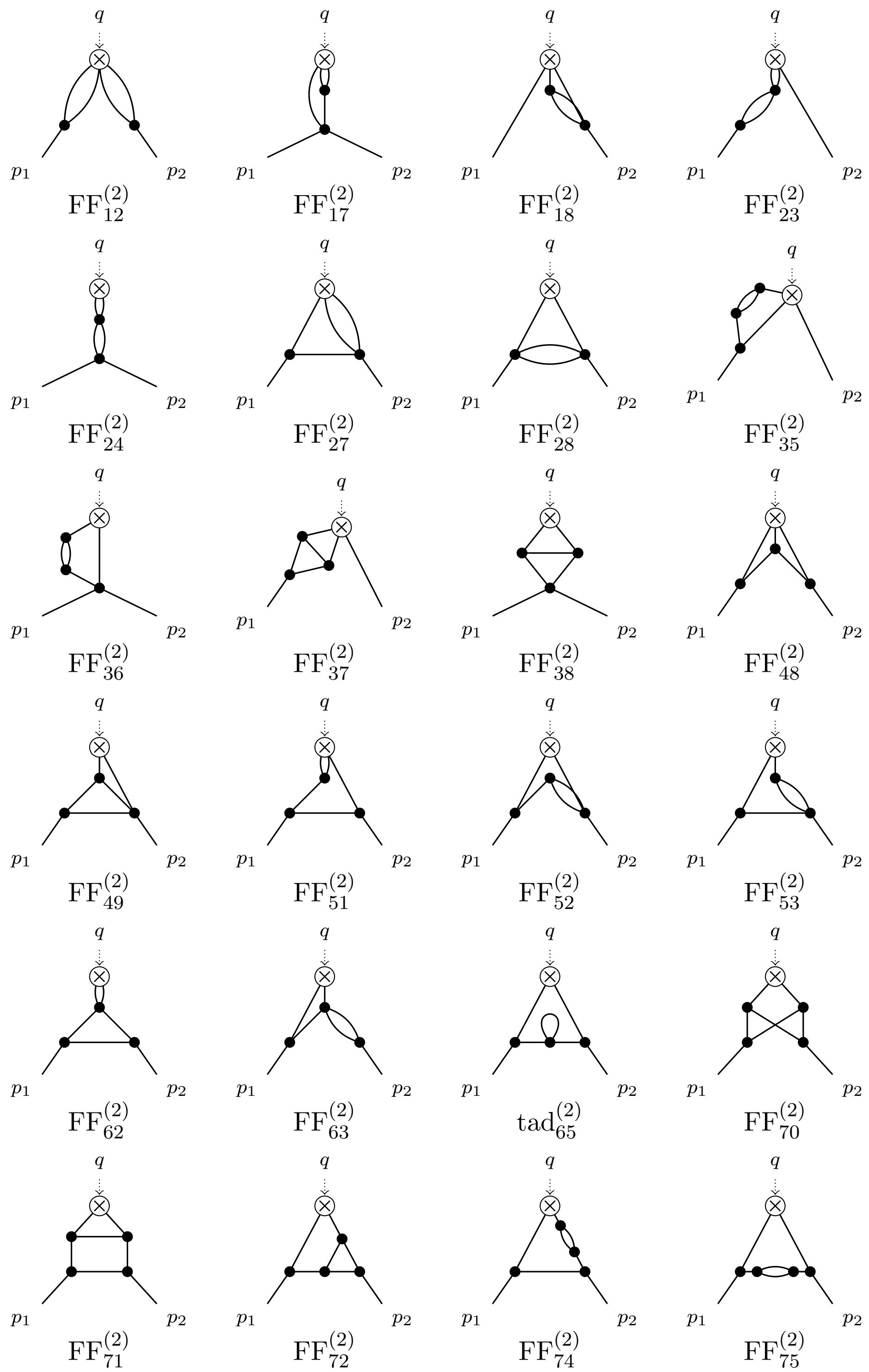

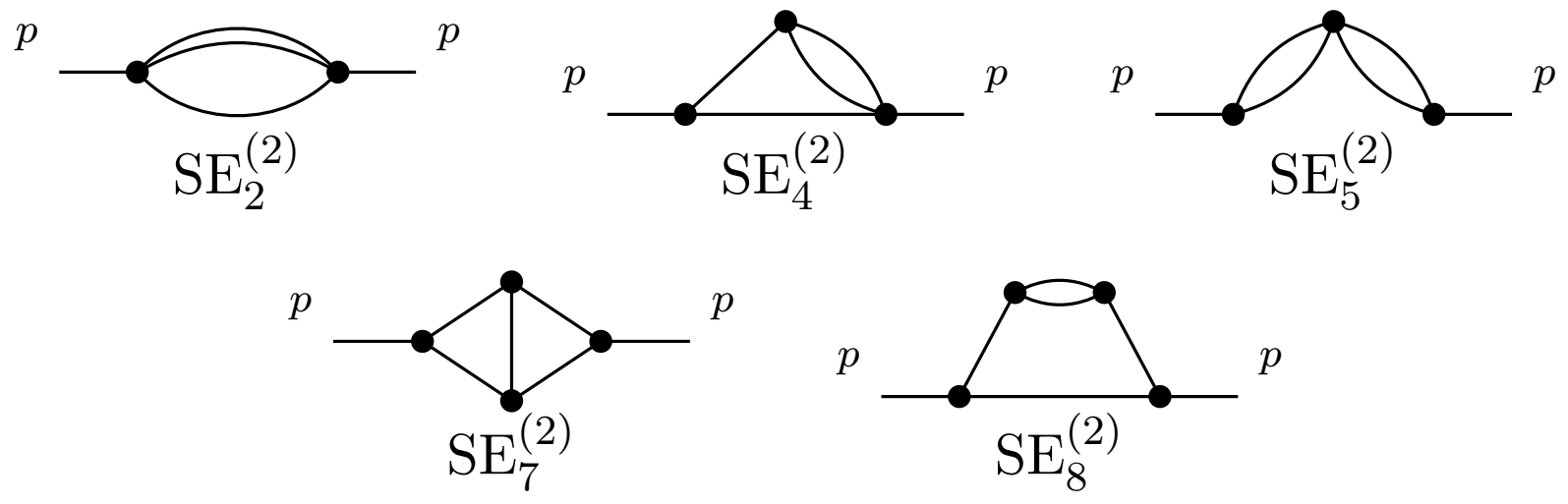
\begin{figure}[!htbp]
\centering
\begin{tabular}{ccc}
\begin{tikzpicture}[x=0.5cm,y=0.5cm,baseline=(current bounding box.center)]
\coordinate (v1) at (-1.35,-1.0);
\coordinate (v2) at (1.35,-1.0);
\coordinate (pL) at (-2.4,-1.0);
\coordinate (pR) at (2.4,-1.0);
\draw[line width=.6pt,bend left=-48] (v1) to (v2);
\draw[line width=.6pt,bend left=30] (v1) to (v2);
\draw[line width=.6pt,bend left=48] (v1) to (v2);
\draw[line width=.6pt] (pL) to (v1);
\draw[line width=.6pt] (pR) to (v2);
\fill (v1) circle (2.2pt);
\fill (v2) circle (2.2pt);
\node[above left,font=\scriptsize] at (pL) {$p$};
\node[above right,font=\scriptsize] at (pR) {$p$};
\end{tikzpicture} & \begin{tikzpicture}[x=0.5cm,y=0.5cm,baseline=(current bounding box.center)]
\coordinate (v1) at (-1.35,-1.0);
\coordinate (v3) at (1.35,-1.0);
\coordinate (v2) at (0.0,0.25);
\coordinate (pL) at (-2.4,-1.0);
\coordinate (pR) at (2.4,-1.0);
\draw[line width=.6pt] (v1) to (v2);
\draw[line width=.6pt] (v1) to (v3);
\draw[line width=.6pt,bend left=-28] (v2) to (v3);
\draw[line width=.6pt,bend left=28] (v2) to (v3);
\draw[line width=.6pt] (pL) to (v1);
\draw[line width=.6pt] (pR) to (v3);
\fill (v1) circle (2.2pt);
\fill (v3) circle (2.2pt);
\fill (v2) circle (2.2pt);
\node[above left,font=\scriptsize] at (pL) {$p$};
\node[above right,font=\scriptsize] at (pR) {$p$};
\end{tikzpicture} & \begin{tikzpicture}[x=0.5cm,y=0.5cm,baseline=(current bounding box.center)]
\coordinate (v1) at (-1.35,-1.0);
\coordinate (v2) at (1.35,-1.0);
\coordinate (v3) at (0.0,0.25);
\coordinate (pL) at (-2.4,-1.0);
\coordinate (pR) at (2.4,-1.0);
\draw[line width=.6pt,bend left=-28] (v1) to (v3);
\draw[line width=.6pt,bend left=28] (v1) to (v3);
\draw[line width=.6pt,bend left=-28] (v2) to (v3);
\draw[line width=.6pt,bend left=28] (v2) to (v3);
\draw[line width=.6pt] (pL) to (v1);
\draw[line width=.6pt] (pR) to (v2);
\fill (v1) circle (2.2pt);
\fill (v2) circle (2.2pt);
\fill (v3) circle (2.2pt);
\node[above left,font=\scriptsize] at (pL) {$p$};
\node[above right,font=\scriptsize] at (pR) {$p$};
\end{tikzpicture}\\[3pt]
{\small \SEgraph{2}{2}} & {\small \SEgraph{2}{4}} & {\small \SEgraph{2}{5}}\\[6pt]
\end{tabular}\par\smallskip
\begin{tabular}{cc}

\begin{tikzpicture}[x=0.5cm,y=0.5cm,baseline=(current bounding box.center)]
\coordinate (v1) at (-1.2,0);
\coordinate (v2) at (1.2,0);
\coordinate (v3) at (0,0.8);
\coordinate (v4) at (0,-0.8);
\coordinate (pL) at (-2.4,0);
\coordinate (pR) at (2.4,0);
\draw[line width=.6pt] (v1) to (v3);
\draw[line width=.6pt] (v1) to (v4);
\draw[line width=.6pt] (v2) to (v3);
\draw[line width=.6pt] (v2) to (v4);
\draw[line width=.6pt] (v3) to (v4);
\draw[line width=.6pt] (pL) to (v1);
\draw[line width=.6pt] (pR) to (v2);
\fill (v1) circle (2.2pt);
\fill (v2) circle (2.2pt);
\fill (v3) circle (2.2pt);
\fill (v4) circle (2.2pt);
\node[above left,font=\scriptsize] at (pL) {$p$};
\node[above right,font=\scriptsize] at (pR) {$p$};
\end{tikzpicture} & \begin{tikzpicture}[x=0.5cm,y=0.5cm,baseline=(current bounding box.center)]
\coordinate (v1) at (-1.35,-1.0);
\coordinate (v2) at (1.35,-1.0);
\coordinate (v3) at (-0.6,0.4);
\coordinate (v4) at (0.6,0.4);
\coordinate (pL) at (-2.4,-1.0);
\coordinate (pR) at (2.4,-1.0);
\draw[line width=.6pt] (v1) to (v2);
\draw[line width=.6pt] (v1) to (v3);
\draw[line width=.6pt] (v2) to (v4);
\draw[line width=.6pt,bend left=-28] (v3) to (v4);
\draw[line width=.6pt,bend left=28] (v3) to (v4);
\draw[line width=.6pt] (pL) to (v1);
\draw[line width=.6pt] (pR) to (v2);
\fill (v1) circle (2.2pt);
\fill (v2) circle (2.2pt);
\fill (v3) circle (2.2pt);
\fill (v4) circle (2.2pt);
\node[above left,font=\scriptsize] at (pL) {$p$};
\node[above right,font=\scriptsize] at (pR) {$p$};
\end{tikzpicture}\\[3pt]
{\small \SEgraph{2}{7}} & {\small \SEgraph{2}{8}}\\[6pt]
\end{tabular}
\caption{The 5 two-loop proper self-energy graph families of the external $W$ contributing at leading power. 
All are planar. The external legs are horizontal and carry the same conserved momentum $p$.}
\label{fig:SE2graphs}
\end{figure}

\begin{itemize}
\item \textbf{Contributing at leading power}: the 24 vertex graph families of Fig.~\ref{fig:FF2graphs} and the 5 self-energy
graph families of Fig.~\ref{fig:SE2graphs}. All of them are planar except \FFgraph{2}{70}, the crossed vertex, which is the scalar-YM
analog of the nonplanar triangle $T_{22}$ of the $\cN=4$ form factor. It is not suppressed by $N$ since the operator insertion
$G$ is color-neutral.
\item \textbf{Power suppressed}: completed massive tadpole products with source weights (\tadgraph{2}{1}, \tadgraph{2}{31}, 
\tadgraph{2}{32}), see Appendix~\ref{app:suppressed}.
\item \textbf{Zero or cancelling}: algebraically vanishing groups, the holding pairs, which vanish at $H=T^{\rm ren}$
before the connector limit, and graphs whose tadpole subgraphs are detachable one-point branches, as explained in
Appendix~\ref{sec:tadpole-gate}; these include the one-point and double-one-point groups \tadgraph{2}{5} and \tadgraph{2}{6}.
The internal tadpole of \tadgraph{2}{65}, inserted on a power-raised line, is the only tadpole subgraph that survives at leading 
power; it is therefore listed among the contributing graph families.
\end{itemize}

\subsubsection{Methods}
\label{sec:FF2methods}

The graph families call for different integration methods, but share a common order of operations: the tensor and
color algebra fixes the scalar integrands, the mass hierarchy determines the required asymptotic terms, and the
regulated expressions are then combined with LSZ and ultraviolet subtraction. We describe the reduction,
integration and walking expansion separately, since an analytic parameter integration is not itself an
application of the Method of Regions.

\paragraph{Algebra and reduction.}
The color and uplifted component-index contractions are performed with FORM~\cite{Vermaseren:2000nd}. Throughout the reduction, we 
retain the full dependence on $d=4-2\epsilon$ and the scalar multiplicity $\Ds-d$, including terms that contribute to finite coefficients
through multiplication by regulator poles. The external-virtuality derivatives required by LSZ are taken before imposing the
equal-virtuality and mass-shell conditions.

Gaussian moments reduce tensor integrals to scalar Feynman-parameter integrals. Algebraic cancellation of propagators
and scalar integration-by-parts identities provide further simplifications. The scalar-basis reduction for \FFgraph{2}{71} is performed with
Kira~\cite{Klappert:2020nbg}. Independent integration-by-parts reductions in the hard region of the Method of Regions also check
\FFgraph{2}{38} and \FFgraph{2}{49}. Simpler subintegrals are reduced directly, as illustrated by the raised-bubble identity in Eq.~\re{eq:I21155}.

\paragraph{Analytic integration.}
Whenever one loop can be isolated as a subintegral, its integration reduces the graph to a product of one-loop
functions or to a lower-dimensional parameter integral. The tensor coefficients and raised propagator powers
produced by the numerator reduction are retained in this step. The remaining parameter integrations use Euler
integrals and beta- and gamma-function identities.

For \FFgraph{2}{70} and \FFgraph{2}{71}, we use the scalar crossed-ladder and ladder functions of Ref.~\cite{Alday:2026walk}. Their 
scalar-YM numerator coefficients, color factors, and lower-sector contributions are determined independently. The finite
crossed-ladder constant, denoted $\kappa_{22}$ below, has a reported numerical PSLQ identification; its dependence is
kept explicit in the individual graph contribution. The Laurent depth of
each scalar integral is chosen before assembly, accounting for poles in its coefficient and in the counterterms with which it is multiplied.

\paragraph{Walking asymptotics.}
We extract the leading-power form factor in the hierarchy of Eq.~\re{eq:walkcond}, retaining all single logarithms and constants.
The logarithms $L$, $\ell_y$ and $L_Q$ remain independent through renormalization; the trajectory $y=w^\eta$ is imposed afterwards.

When an integrated analytical expression is available, its walking expansion is taken directly. Other contributions, including the direct
parameter calculations for \FFgraph{2}{17} and \FFgraph{2}{35}, are expanded before the remaining parameter integrations.
Thus, analytic parameter integration does not imply that every graph was first evaluated at arbitrary mass ratios.
If a reduction coefficient contains inverse powers of a small mass ratio, the corresponding scalar integral is expanded far
enough to retain every term contributing to the leading power of the complete graph.

The coupled families \FFgraph{2}{48} and \FFgraph{2}{72} require an explicit decomposition by the Method of
Regions~\cite{Beneke:1997zp}. Their parameter integrands are expanded in the relevant scalings, with an analytic regulator
separating singular contributions. Tropical subtraction~\cite{Salvatori:2024tropical,Belitsky:2025tropical,Giroux:2026tgd} isolates 
parameter-boundary singularities, leaving convergent coefficient integrals suitable for {\tt HyperInt}  \cite{Panzer:2014caa}.
The subtraction is implemented with one regulator, so the additional analytic shifts are chosen proportional to the dimensional
regulator, as in Ref.~\cite{Alday:2026walk}. The precise prescription is specified in Eq.~\re{eq:coupled-regulator}: after
summing the retained regions, the constant Laurent coefficient in the regulator slope is extracted for each dimensional-regulator 
coefficient. The regions and resulting expressions are given in Appendices~\ref{app:ff48} and \ref{app:ff72}.

\paragraph{Checks and assembly.}
Algebraic identities and direct quadrature of the defining parameter integrals check the reductions and analytic evaluations.
Python/SymPy is used for rational assembly and Laurent expansion, with high-precision checks using mpmath. For
\FFgraph{2}{48} and \FFgraph{2}{72}, independent numerical expansions with FIESTA~\cite{Smirnov:2021fiesta5} provide the
comparisons reported in Appendices~\ref{app:ff48} and \ref{app:ff72}, including their precision limitations.

The parameter-boundary subtractions used to evaluate an integral are distinct from ultraviolet renormalization. Proper-subgraph
forests enter the separately subtracted graph coefficients of Sect.~\ref{sec:FF2local}. The renormalized form factor is instead
constructed from the complete bare graph and LSZ sum, followed by the action and operator counterterms of
Sect.~\ref{sec:FF2ren}; the separately subtracted graph coefficients are not substituted into that bare sum.

\subsubsection{Bare graph and LSZ organization}\label{sec:assembly}

The genuine two-loop vertex and self-energy graphs do not yet constitute the two-loop form factor. Equation~\re{eq:FF2LSZ}
also contains products and derivatives of one-loop functions. We collect their normalizations and required regulator
expansions here so that all terms can be assembled before ultraviolet renormalization.

\paragraph{Graph normalization.}\label{sec:phase153}

We use the same graph convention as in
Eq.~\re{eq:FFgraphmap}. Let $\mathcal S_1^{(2)}$ contain the
24 amputated insertion families in Fig.~\ref{fig:FF2graphs},
including \tadgraph{2}{65}, and let
\begin{align}
 \mathcal S_2^{(2)}
 &=\{\SEgraph{2}{2},\SEgraph{2}{4},\SEgraph{2}{5},
       \SEgraph{2}{7},\SEgraph{2}{8}\}.
 \label{eq:two-loop-selfenergy-graph-set}
\end{align}
The latter are the five proper self-energy families in
Fig.~\ref{fig:SE2graphs}. With the vertex normalization of
Eq.~\re{eq:FFfact} and loop measure of Eq.~\re{eq:FFloop-measure},
their dimensionless graph coefficients satisfy
\begin{align}
 \Gamma_2&=\sum_{\gamma\in\mathcal S_1^{(2)}}f_{2,\gamma,B},
 \notag\\*
 \sigma_2'&=\sum_{\gamma\in\mathcal S_2^{(2)}}f_{2,\gamma,B}.
 \label{eq:phase153}
\end{align}
For a self-energy family the coefficient includes its LSZ
virtuality derivative, taken before imposing the shell.
For an identified graph the first subscript gives the loop order:
$f_{\ell,\mathrm{FF}n}$, $f_{\ell,\mathrm{SE}n}$ and
$f_{\ell,\mathrm{tad}n}$ denote the contributions of
\FFgraph{\ell}{n}, \SEgraph{\ell}{n} and \tadgraph{\ell}{n} graphs,
respectively. A further subscript $B$ has the unsubtracted meaning
defined in Eq.~\re{eq:FFexpBare}.
The same normalization applies to the proper-subgraph forests. The
source-insertion coefficients in the auxiliary derivations have the opposite
overall sign; the coefficients in this section already include that conversion.
The one-loop LSZ
products are already in the physical convention. The two coupled bare
inputs enter as
\begin{align}
f_{2,\mathrm{FF}48,B}&=-e^{-2\epsilon L_Q}G_{48},
\label{eq:bare-FF2-48}\\
f_{2,\mathrm{FF}72,B}&=2e^{-2\epsilon L_Q}G_{72}\, .
\label{eq:coupled-bare-normalization}
\end{align}
The functions $G_{48}$ and $G_{72}$, including their Laurent coefficients,
are defined in Appendices~\ref{app:ff48} and \ref{app:ff72} at
$Q^2=\mu^2=1$. The factor two includes the exchanged external attachment.
Choosing $Q^2=1$ fixes the unit in which all masses and virtualities
are measured; choosing $\mu^2=Q^2$ evaluates the dimensionless
coefficients at $L_Q=0$. Neither choice removes scale dependence.
The loop measure restores it through
$(\mu^2/Q^2)^{2\epsilon}=e^{-2\epsilon L_Q}$.
This factor multiplies the full Laurent series before expansion,
so its products with ultraviolet poles generate the required
finite $L_Q$ terms. The ratios $w,y$ are unaffected.
We retain $Q^2$ where it displays dimensions or the bounds of
a running-coupling integral.

\paragraph{The one-loop functions through the required depth.}
The positive powers of $\epsilon$ in a one-loop integral contribute to the finite two-loop answer when multiplied
by a subtraction pole. We therefore need more than its four-dimensional value. For the bubbles $B_w,B_q$,
defined in Eq.~\re{eq:walking-bubble-definitions} as the walking limits of the masters in
Appendix~\ref{app:masters1}, the exact regulator dependence and the required expansions in the same loop measure are
\begin{align}
 B_q&=e^{\gamma_E\epsilon-\epsilon L_Q}
 \frac{\Gamma(\epsilon)\Gamma(1-\epsilon)^2}{\Gamma(2-2\epsilon)},\\*
 B_w&=e^{\epsilon L}B_q,\\*
 B_q&=\frac1\epsilon+2-L_Q
 +\epsilon\left[4-2L_Q+\tfrac12L_Q^2-\tfrac12\zeta_2\right]
 +O(\epsilon^2),\\*
 B_w&=\frac1\epsilon+2+L-L_Q
 +\epsilon\left[4+2(L-L_Q)+\tfrac12(L-L_Q)^2-\tfrac12\zeta_2\right]
 +O(\epsilon^2).
 \label{eq:bubbles155}
\end{align}
Their finite terms agree with the leading-power limits of the exact
$b_w,b_q$ in Eqs.~\re{eq:one-loop-finite-masters} and
\re{eq:exact-bubbles-wy}; they are also the bubble building blocks
used in Appendix~\ref{sec:green-results}.

We use the same triangle convention at both loop orders.
The finite function $\mathcal T_0(w,y)$ was defined in
Eq.~\re{eq:one-loop-finite-masters}; it is the
$\epsilon^0$ coefficient of the dimensionless regulated master
$Q^2C_0$ with those arguments. At leading walking power its
dimensional expansion is
\begin{align}
 \mathcal T&=\mathcal T_0+\epsilon\mathcal T_1+O(\epsilon^2).
 \label{eq:regulated-walking-triangle}
\end{align}
Here and below the coefficients $\mathcal T_0,\mathcal T_1$
denote their leading-power expressions, while the exact one-loop
function retains its arguments.
The coefficients through the required order are
\begin{align}\label{eq:triangle155}
 \mathcal T_0&=-\tfrac12L^2+L\ell_y+\tfrac12\ell_y^2,\notag\\
 \mathcal T_1&=-\tfrac13L^3+\tfrac12L^2L_Q+L^2\ell_y
 -LL_Q\ell_y-\zeta_2L\notag\\*
 &\quad-\tfrac12L_Q\ell_y^2-\tfrac13\ell_y^3
 -\zeta_2\ell_y-3\zeta_3.
\end{align}
The exact-$d$ source coefficients give
\begin{equation}\label{eq:F1B155}
 f_{1,B}=2\mathcal T+
 \frac{\epsilon(D_s-4)}{1-\epsilon}B_q+
 \left[4-\frac{(D_s-2)(1-\epsilon)}{2(3-2\epsilon)}\right]B_w.
\end{equation}
Its Laurent expansion contains the finite coefficient in
Eq.~\re{eq:one-loop-walking-final}:
\begin{align}\label{eq:FFbare-one-loop-Laurent}
f_{1,B}&=\frac{b_0}{\epsilon}+f_1
+\epsilon f_{1,\epsilon}+O(\epsilon^2)\, .
\end{align}
The $O(\epsilon)$ coefficient needed by every one-loop subtraction is
\begin{align}\label{eq:f1eps155}
 f_{1,\epsilon}={}&2\mathcal T_1+(D_s-4)(3-L_Q)
 +b_0\left[\tfrac12(L-L_Q)^2+2(L-L_Q)+4-\tfrac12\zeta_2\right]\notag\\
 &+\tfrac1{18}(D_s-2)(L-L_Q+2)+\tfrac1{27}(D_s-2).
\end{align}
This retains evanescent source coefficients before they multiply poles.

\paragraph{Products generated by the one-loop pole shift and residue.}
The products of one-loop quantities in Eq.~\re{eq:FF2LSZ}
arise from expanding the external pole positions and residues.
They must be included in addition to the genuine two-loop
vertex and proper self-energy graphs.

To evaluate these products, we first differentiate the self-energy
of Eq.~\re{eq:FFsig1} at independent virtuality. Its leading-power
part can be written as
\begin{align}
 \sigma_1(t)
 &=\bigl[r_p t-2m_0^2\bigr]B_0(t;m_z^2,m_z^2)
 +O\!\left(m_z^2\log^k\frac{m_z^2}{-t}\right),\\*
 r_\Sigma&=-\frac{D_s-2}{2(3-2\epsilon)},\\*
 r_p&=2+r_\Sigma.
 \label{eq:LSZ-self-energy-coefficients}
\end{align}
The coefficient $r_p$ multiplies the independent virtuality $t$;
$r_\Sigma=r_p-2$ remains after setting $t=m_0^2$.
All derivatives hold $m_0^2$ and $m_z^2$ fixed.
The leading-power bubble satisfies
\begin{align}
 t\,\partial_t B_0(t;0,0)&=-\epsilon B_0(t;0,0).
 \label{eq:bubble-virtuality-derivative}
\end{align}
Differentiating before the tree-shell specialization therefore gives
\begin{align}
 \sigma_1&=m_0^2r_\Sigma B_w,\\*
 \sigma_1'&=(r_p-\epsilon r_\Sigma)B_w,\\*
 m_0^2\sigma_1''
 &=\bigl[-2\epsilon r_p+\epsilon(1+\epsilon)r_\Sigma\bigr]B_w.
 \label{eq:LSZ-one-loop-shell-values}
\end{align}
These equations give only the leading-power terms; the primes have
the meaning specified in Eq.~\re{eq:FFself-energy-derivatives}.

The one-loop amputated vertex entering the same products is
\begin{align}
 \Gamma_1
 &=2\mathcal T+\frac{\epsilon(D_s-4)}{1-\epsilon}B_q+2B_w.
 \label{eq:LSZ-one-loop-vertex}
\end{align}
Using the derivative $\mathcal D$ of Eq.~\re{eq:FFDdef}, define
\begin{align}
 \mathcal D_v&\equiv m_0^2\mathcal D,\\*
 \mathcal D_v\Gamma_1
 &=-2\epsilon B_w-2(\partial_L+\partial_{\ell_y})\mathcal T.
 \label{eq:LSZ-vertex-virtuality-derivative}
\end{align}
The second line is obtained from the independent-virtuality vertex,
before imposing equal virtualities. The derivatives of its rational
bubble and triangle coefficients cancel in the sum, leaving
derivatives of the scalar integrals. The logarithms remain
independent: this is not differentiation along $y=w^\eta$.

The required one-loop products are then
\begin{align}
 \label{eq:products155}
 \mathcal I_B
 &\equiv \sigma_1\sigma_1''+(\sigma_1')^2
 +\sigma_1'\Gamma_1+\sigma_1\mathcal D\Gamma_1
 \\
 &=\bigl[r_p^2-4\epsilon r_\Sigma r_p
 +\epsilon(1+2\epsilon)r_\Sigma^2\bigr]B_w^2
 +(r_p-\epsilon r_\Sigma)B_w\Gamma_1
 +r_\Sigma B_w\mathcal D_v\Gamma_1 \, . \nonumber
\end{align}
The genuine two-loop contributions $\Gamma_2$ and $\sigma_2'$
are supplied by Eq.~\re{eq:phase153}.
The remaining reference-tree term
$-2\sigma_1\Gamma_1/Q^2$ in Eq.~\re{eq:FF2LSZ} is power suppressed.

\paragraph{Raised triangle in \tadgraph{2}{65}.}
This graph is the exception to the power suppression of massive tadpoles. Inserting the tadpole on the middle
line of a triangle raises that propagator to the second power, and its parameter endpoint compensates the mass
factor supplied by the tadpole. The scalar integral that controls this enhancement is
\begin{align}
 C_{121}
 &\equiv\int[\dd k]\,
 \frac{1}
 {[k^2-m_z^2+\ii0]
  [(k-p_1)^2-m_z^2+\ii0]^2
  [(k-p_1-p_2)^2-m_z^2+\ii0]}.
 \label{eq:raised-triangle-definition}
\end{align}
The subscript records the powers of its three propagators.
Its dimensionless scalar factor has the leading-power relation
\begin{align}
 m_z^2Q^2C_{121}&=K_\epsilon+o_{\rm LP}(1),
 \label{eq:raised-triangle-walking-relation}
\end{align}
where we define the walking endpoint integral by
\begin{align}
 K_\epsilon
 &\equiv m_z^2Q^2 e^{\gamma_E\epsilon}\mu^{2\epsilon}
 \Gamma(2+\epsilon)
 \int_0^\infty
 \frac{\dd b\,\dd c}
 {[m_z^2+M_E^2(b+c)+Q^2bc]^{2+\epsilon}}.
 \label{eq:raised-triangle-corner}
\end{align}
Here $b,c$ are positive Feynman parameters, with the middle-line
parameter fixed to one. The relation to the exact finite-mass
parameter integral is derived in
Appendix~\ref{app:raised-triangle-reduction}.
The endpoint integral is sufficient for the leading-power graph;
its residual $w/y$ dependence does not include all power
corrections to $C_{121}$.

Integrating $c$ and changing the remaining parameter gives
\begin{align}\label{eq:K155}
 K_\epsilon={}&e^{\gamma_E\epsilon}\Gamma(1+\epsilon)
 e^{\epsilon(L-L_Q-\ell_y)}
 \int_0^1\frac{\tau^\epsilon\,\dd\tau}{1-(1-w/y)\tau}\notag\\
 ={}&\frac{e^{\gamma_E\epsilon}\Gamma(1+\epsilon)
 e^{\epsilon(L-L_Q-\ell_y)}}{1+\epsilon}
 {}_2F_1(1,1+\epsilon;2+\epsilon;1-w/y).
\end{align}
The Euler integral converges for $w/y>0$ and
$\Re\epsilon>-1$, and then defines a meromorphic continuation.
Its expansion as $w/y\to0$ is
\begin{align}
 \int_0^1\frac{\tau^\epsilon\,\dd\tau}{1-(1-w/y)\tau}
 &=L+\ell_y-\gamma_E-\psi(1+\epsilon)
 +O\!\left(\frac wy\log\frac wy\right),
 \label{eq:raised-triangle-euler-expansion}
\end{align}
where $\psi$ is the digamma function. Consequently,
\begin{equation}\label{eq:Kdepth155}
 K_\epsilon=L+\ell_y+\epsilon\big[(L-L_Q-\ell_y)(L+\ell_y)-\zeta_2\big]+O(\epsilon^2)
 +o_{\rm LP}(1).
\end{equation}
In the physical graph normalization of Eq.~\re{eq:phase153}, the
tadpole subintegral supplies the bare contribution
\begin{align}
 f_{2,\mathrm{tad}65,B}
 &=-(D_s-1)\frac{A_0(m_z^2)}{m_z^2}K_\epsilon+o_{\rm LP}(1).
 \label{eq:raised-triangle-bare-graph}
\end{align}
The $1/\epsilon$ pole of the tadpole explains why $K_\epsilon$ is
needed through $O(\epsilon)$. The separately subtracted finite
contribution is given in Eq.~\re{main:tad2-65}.

\paragraph{Complete bare two-loop coefficient.}
We can now combine the genuine graph contributions with the one-loop LSZ products. The ultraviolet counterterms
act on this complete bare coefficient, not on a sum in which each graph has already been replaced by its finite
part. In the graph convention fixed above, the assembly is
\begin{align}
 f_{2,B}&=\sum_{\gamma\in\mathcal S_1^{(2)}}f_{2,\gamma,B}
 +\sum_{\gamma\in\mathcal S_2^{(2)}}f_{2,\gamma,B}
 +\mathcal I_B,\label{eq:bare-graph-assembly}\\
 f_2&=\lim_{\epsilon\to0}\bigl[f_{2,B}+\Delta f_2\bigr].
 \label{eq:FFrenormalization-bridge}
\end{align}
The second line is the two-loop specialization of Eq.~\re{eq:FFcounterterm-construction}.
The counterterm coefficient $\Delta f_2$ is given explicitly in Eq.~\re{eq:FFtwo-loop-counterterms} below and is applied once to the complete bare LSZ sum.
The direct reference-tree conversion is $O(w)$ and therefore absent at this power; mass renormalization acting on the
one-loop coefficient is not power suppressed and must still be retained.
No additional graphwise subtraction is applied to this bare sum. Holding pairs are zero before the connector limit.
The bare inputs for \FFgraph{2}{48} and \FFgraph{2}{72} are given in
Appendices~\ref{app:ff48} and \ref{app:ff72}; the remaining families use
the tensor and parameter integrations described in Sect.~\ref{sec:FF2methods}.
The resulting coefficient is
\begin{equation}\label{eq:bare155}
 f_{2,B}=\frac{b_0^2}{\epsilon^2}+\frac{f_{2,-1}}{\epsilon}+f_{2,0}+O(\epsilon),
 \qquad [f_{2,B}]_{\epsilon^{-4}}=[f_{2,B}]_{\epsilon^{-3}}=0,
\end{equation}
where
\begin{align}\label{eq:A1155}
 f_{2,-1}={}&4b_0\mathcal T_0+2b_0^2(L-L_Q)-2c_m L+6c_m \ell_y+d_0,\notag\\
 d_0={}&-\frac{119}{108}D_s^2+\frac{293}{27}D_s+\frac{1153}{27}.
\end{align}
The coefficient $f_{2,0}$ is the finite part of the complete
bare two-loop form factor. Unlike the one-loop finite coefficient,
it is not yet the renormalized $f_2$: lower-loop
$O(\epsilon)$ terms multiply subtraction poles and give an
additional finite contribution, specified in Eq.~\re{eq:finite155}.
The logarithmic decomposition of $f_{2,0}$ and the individual graph
poles are given in Appendix~\ref{app:bare155}.
No graph-level overall pole has been removed before constructing
Eq.~\re{eq:bare155}.

\subsubsection{Regulated inputs for two-loop renormalization}

The two-loop subtraction depends on a small set of lower-loop matrix elements: the form factor itself, its response
to a common background-mass replacement, the scalar-potential response, and the matrix element of the operator $P$.
We evaluate these quantities through the positive powers of $\epsilon$ required by the pole counterterms, keeping
parameter renormalization distinct from the external-virtuality derivatives used in LSZ.

\paragraph{Mass-parameter replacement.}
Gauge-coupling renormalization is already fixed by Eq.~\re{eq:coupling-Z}.
For the common background mass scale, the mass replacement in Eq.~\re{eq:FFmass-renormalization} acts at fixed
dimensionless background shape, hence fixed $y$:
\begin{align}
\label{eq:common-background-scale-replacement}
M_E^2&\longmapsto Z_{m^2}M_E^2,\qquad
L\longmapsto L-\log Z_{m^2},\qquad
\ell_y\longmapsto\ell_y .
\end{align}
Its Taylor expansion on the one-loop coefficient is
\begin{align}
\label{eq:Z155}
\left.f_{1,B}\right|_{\rm mass\ replacement}
&=f_{1,B}-\frac{c_mg_R^2}{\epsilon}\partial_Lf_{1,B}+O(g_R^4).
\end{align}
This is a common-scale parameter replacement, not a universal pole-mass renormalization for every scalar mode.
Independent quartic-induced mass and contact deformations are included in the scalar-potential response below.
It must also be distinguished from the LSZ virtuality derivatives, which act at fixed internal masses before equal virtualities are imposed.

\paragraph{Finite scalar coupling and its counterterm.}
The renormalized scalar potential need not remain on the commutator-potential surface as the subtraction scale
changes. Its independent response cannot in general be absorbed into a common mass replacement. We describe
the relevant direction by the finite, dimensionless quartic deformation $h(\mu)$, normalized through
\begin{align}
\label{eq:finite-scalar-deformation}
\left.\frac{\partial F}{\partial h}\right|_{h=0}
&=g_R^2\mathcal H_0+O(g_R^4).
\end{align}
The boundary conditions in Eq.~\re{eq:Rcond} give $h(\mu_*)=0$.
The displayed coefficient $f_2$ is evaluated on the surface $h(\mu)=0$ at the reference scale.
For the fixed theory defined at $\mu_*$, this means choosing $\mu=\mu_*$ or retaining the finite running deformation at another scale.
For $h=O(g_R^2)$ the latter is restored by the already calculated response:
\begin{align}
\label{eq:finite-scalar-completion}
F(g_R^2,h)
&=1+g_R^2f_1+g_R^4f_2+g_R^2h(\mu)\mathcal H_0+O(g_R^6).
\end{align}
The associated pole counterterm, not the running coupling itself, is
\begin{align}
\label{eq:quartic-response-coefficient}
h_{\rm CT}&\equiv3h_s-h_c
=\frac{\kappa_hg_R^2}{\epsilon},\qquad
\kappa_h=\frac{2(D_s-2)}3=4c_m.
\end{align}
The $h_i$ on the right are the pole coefficients of Eq.~\re{eq:Rpoles}.
Thus the scalar contribution in Eq.~\re{eq:FFHact} is $g_R^2h_{\rm CT}\mathcal H_B$,
while a finite deformation contributes $g_R^2h(\mu)\mathcal H_0$.
On the radiatively generated trajectory $h=O(g_R^2)$, both first affect the form factor at two-loop order.

\paragraph{Direct massive scalar-counterterm response.}
We obtain the scalar-potential response $\mathcal H_B$ by differentiating the action with respect to its independent
quartic coefficients; $\mathcal H_B$ is not the external holding source $H$. On a scalar line of background charge
$v$, the test variations change the mass tensor by
\[
 \delta m^2_{IJ}=(\delta h_c+\delta h_s)v^2\delta_{IJ}
 +(3\delta h_s-\delta h_c)v_Iv_J.
\]
Choosing $\delta h_c=-h/4$ and $\delta h_s=h/4$ isolates the coefficient of the deformation $h$,
with no isotropic mass variation. These test variations are not the pole counterterms $h_i$.
The propagator insertion per unit $h$ is $i v_Iv_J/(k^2-m_z^2)^2$. The insertion is summed over the scalar lines of the one-loop vertex
and self-energy graphs. 

Mass terms in the numerators are retained before taking the walking
limit. Summing the insertion over the one-loop vertex and self-energy
graphs and performing the LSZ derivatives gives the regulated response
below. The reduction of the raised one-loop integrals is given in
Appendix~\ref{app:scalar-response-reduction}.
For the raised tadpole we use
\begin{align}
 A_2(m_z^2)&\equiv\partial_{m_z^2}A_0(m_z^2)
 =e^{\gamma_E\epsilon}\Gamma(\epsilon)
 e^{\epsilon(L-L_Q-\ell_y)}.
 \label{eq:raised-tadpole-response}
\end{align}
The resulting leading-power response is
\begin{align}\label{eq:H155}
 \mathcal H_B&=\frac{d-2}{2}\frac{A_0(m_z^2)}{m_z^2}
 +\frac{(d-6)(d-3)}2 B_w+o_{\rm LP}(1)\notag\\
 &=A_2-(1-\epsilon-2\epsilon^2)B_w+o_{\rm LP}(1).
\end{align}
The coefficients of $B_q$ and $\mathcal T$ vanish
in the interior walking limit. Terms suppressed by $y$ or $w/y$ vanish
for every fixed $0<\eta<1$. Isotropic $m_z^2\delta_{IJ}$ insertions and
the other planar scalar-potential contact directions have positive
walking degree: raised external bubbles give $y\ell_y^{r}$, hard insertion
contacts give $w$ times logarithms, and transverse scalar triangle
numerators remove an otherwise possible soft power enhancement.
The retained response is therefore
\begin{equation}\label{eq:Hdepth155}
 \mathcal H_B=\underbrace{-\ell_y-1}_{\mathcal H_0}
 +\epsilon\underbrace{\left[\ell_y^2/2-(L-L_Q)(\ell_y+1)+\zeta_2\right]}_{\mathcal H_1}
 +O(\epsilon^2)+o_{\rm LP}(1).
\end{equation}

\paragraph{Composite-operator counterterms and the \texorpdfstring{$P$}{P} matrix element.}
Renormalizing the insertion requires both the diagonal factor multiplying $G$ and its mixing with $P$.
The diagonal factor through two loops is
\begin{align}
\label{eq:ZGG-two-loop}
Z_{GG}&=1-\frac{b_0g_R^2}{\epsilon}
 +g_R^4\left(\frac{b_0^2}{\epsilon^2}-\frac{b_1}{\epsilon}\right)+O(g_R^6).
\end{align}
This is an operator renormalization constant, not $Z_{g^2}$: in particular, their two-loop simple poles differ.
The one-loop $G$-$P$ mixing coefficient remains $z_{GP}=c_{GP}$ from Eq.~\re{eq:composite-z-derived}.
For $P$ as defined in Eq.~\re{eq:Pdefinition},
$P_{\chi\chi}^{ab,IJ}=-iq^2\delta^{ab}\delta^{IJ}$.
Let $\mathcal F_{P_0,B}$ denote its unsubtracted pole matrix element, obtained by replacing $G_0$ by $P_0$ in the LSZ construction.
With the same reference-tree normalization as $F_B$, define the one-loop coefficient by
\begin{align}
\label{eq:P-bare-normalization}
\frac{\mathcal F_{P_0,B}}{\mathcal F_G^{(0)}}
&=-2w+g_\epsilon^2\bigl[f_{P,1}+o_{\rm LP}(1)\bigr]+O(g_\epsilon^4).
\end{align}
The tree term is retained here to display its power suppression; $f_{P,1}$ denotes the walking leading-power coefficient.
Its complete one-loop reduction gives
\begin{align}\label{eq:P155}
 f_{P,1}&=(D_s-d)\frac{d-4}{d-2}B_q
 =-\frac{\epsilon(D_s-4+2\epsilon)}{1-\epsilon}B_q\notag\\
 &=4-D_s
 +\epsilon\left[(4-D_s)(3-L_Q)-2\right]
 +O(\epsilon^2).
\end{align}
The $-2$ is an evanescent scalar-trace contribution. The tree matrix
element of $P$ is $O(w)$, so the two-loop part of $Z_{GP}$ is unnecessary
at this leading power. The required $g_R^4c_{GP}f_{P,1}/\epsilon$ term
is included once below.

\paragraph{Holding-source contacts.}
The source holds the background fixed by cancelling its one-point function, but it does not supply an additional
independent counterterm for the form factor. At two loops this statement must be applied to the complete
tadpole/source pairs before the diagonal connector is removed. Let $\mathcal V_{\chi,n}$ be the order-$n$ response with an open Cartan-scalar
leg attached to the two external vectors, $K_{\rm reg}$ the regulated
inverse Cartan propagator, and $\Pi_{\chi,1}$ its one-loop correction.
These are not the two-vector insertion vertices $\Gamma_n$.
The three order-two tadpole/source pairs are
$\mathcal V_{\chi,0}K_{\rm reg}^{-1}(T_2^{\rm ren}-H_2)$,
$\mathcal V_{\chi,1}K_{\rm reg}^{-1}(T_1^{\rm ren}-H_1)$ and
$\mathcal V_{\chi,0}K_{\rm reg}^{-1}\Pi_{\chi,1}K_{\rm reg}^{-1}(T_1^{\rm ren}-H_1)$.
Each is zero at $H_n=T_n^{\rm ren}$ before the connector limit.
The distinct composite $Q_{H\chi}=2\operatorname{tr}H_I\chi_I$
has $H_1\sim\gym V A_0(m_{cl}^2)$ and a hard-$q$ connector in its connected
two-vector tree response. Tree normalization gives
$M_E^2m_{cl}^2/Q^4=O(w^2)$ or $O(w^2y)$.
$Q_{HV}=2\operatorname{tr}H_IV_I$ contains no dynamical field and is a
connected vacuum zero. Here $m_{cl}^2$ denotes a tree-level mass in the
tadpole, of order $M_E^2$ or $m_z^2$; either possibility obeys the stated
power bound. The corresponding source EOM
contacts obey the same power bound. Source-free EOM/BRST matrix elements
decouple after the complete LSZ projection; ghost-number-one breaking
insertions are not added as ghost-number-zero counterterms.

\subsubsection{Ultraviolet renormalization}
\label{sec:FF2ren}

With the regulated lower-loop inputs in hand, we can subtract the ultraviolet poles of the complete bare LSZ sum.
The subtraction is local, whereas its matrix elements carry the mass logarithms of the walking hierarchy.
Their cancellation therefore tests the coefficients of these logarithms as well as the purely rational poles.

\paragraph{The complete two-loop counterterm correction.}
At walking leading power, the additive construction of Eq.~\re{eq:FFcounterterm-construction} becomes
\begin{align}
\label{eq:MS155} 
f_2&=\lim_{\epsilon\to0}\bigl[f_{2,B}+\Delta f_2\bigr],\\
\label{eq:FFtwo-loop-counterterms}
\Delta f_2={}&-\frac{b_0}{\epsilon}f_{1,B}
 -\frac{c_m}{\epsilon}\partial_L f_{1,B}
 +\frac{\kappa_h}{\epsilon}\mathcal H_B\notag\\
 &+\left[-\frac{b_0}{\epsilon}f_{1,B}
          +\frac{b_0^2}{\epsilon^2}-\frac{b_1}{\epsilon}\right]
 +\frac{c_{GP}}{\epsilon}f_{P,1}.
\end{align}
The five contributions in this order are gauge-coupling replacement, common mass-scale replacement, scalar-potential
counterterms, diagonal $G$ renormalization, and $G$-$P$ mixing.
The bracket contains both the one-loop $Z_{GG}$ multiplying the one-loop matrix element and the two-loop $Z_{GG}$ multiplying
the unit tree form factor. Thus the two terms $-b_0f_{1,B}/\epsilon$ have different origins; together they give the
$-2b_0f_{1,B}/\epsilon$ in the combined subtraction. No counterterm has been added twice.
The lower-loop quantities $f_{1,B}$, $\mathcal H_B$ and $f_{P,1}$
in this equation are the regulated expansions in
Eqs.~\re{eq:FFbare-one-loop-Laurent}, \re{eq:Hdepth155} and
\re{eq:P155}, respectively. Their $O(\epsilon)$ terms must be retained
before the pole factors are multiplied out.
Table~\ref{tab:UV155} resolves these corrections into their double poles, simple poles and finite parts.

\begin{table}[htbp]\centering\small
\begin{tabular}{p{38mm}p{20mm}p{40mm}p{40mm}}\toprule
Contribution & $\epsilon^{-2}$ & $\epsilon^{-1}$ & Finite\\\midrule
Complete bare $f_{2,B}$ & $b_0^2$ & $f_{2,-1}$ & $f_{2,0}$\\
Gauge-coupling replacement & $-b_0^2$ & $-b_0f_1$ & $-b_0f_{1,\epsilon}$\\
One-loop diagonal $G$ & $-b_0^2$ & $-b_0f_1$ & $-b_0f_{1,\epsilon}$\\
Mass replacement & $0$ & $-c_m\partial_Lf_1$ & $-c_m\partial_Lf_{1,\epsilon}$\\
Scalar potential & $0$ & $\kappa_h\mathcal H_0$ & $\kappa_h\mathcal H_1$\\
Two-loop diagonal $G$ & $b_0^2$ & $-b_1$ & $0$\\
G-P & $0$ & $c_{GP}(4-D_s)$ & $\begin{aligned}[t]c_{GP}\bigl[&(4-D_s)(3-L_Q)\\&-2\bigr]\end{aligned}$\\\bottomrule
\end{tabular}\caption{Complete physical UV subtraction, rather than a sum
of individual projected graph $R'$ residues.}\label{tab:UV155}
\end{table}

The double pole cancels as $b_0^2-2b_0^2+b_0^2=0$.
The simple-pole coefficient $f_{2,-1}$ here is the same bare
coefficient calculated explicitly in Eq.~\re{eq:A1155}, not a new
counterterm. Substituting the one-loop quantities into the
subtraction gives the identity
\begin{equation}\label{eq:UVcancel155}
 f_{2,-1}=2b_0f_1+c_m\partial_Lf_1-\kappa_h\mathcal H_0+b_1-c_{GP}(4-D_s).
\end{equation}
For example its nonlogarithmic part is the explicit identity
$d_0=2b_0c_1+c_mb_0+\kappa_h+b_1-c_{GP}(4-D_s)$.
Every coefficient of $L,\ell_y,L_Q$ cancels as well. Therefore
\begingroup\begin{equation}\label{eq:noUV155}
 [f_2]_{\epsilon^{-2}}=[f_2]_{\epsilon^{-1}}=0.
\end{equation}\endgroup

\paragraph{Comparison with the companion off-shell calculation.}
As a check on the locality of the counterterms, we compare with the
branch-A results of the companion paper. Their derivation is separate;
only the lower-loop functions and pole coefficients needed for this
comparison are quoted here. Local ultraviolet counterterms must be
common to the two infrared hierarchies, but the comparison is not a
termwise endpoint limit of the walking expansion.

Translated to the present conventions, the required lower-loop functions
from the companion paper are
\[
 \mathcal T_0^{\rm off}=-L^2-2\zeta_2,\qquad \mathcal H_0^{\rm off}=L-1,
\]
\[
 \mathcal T_1^{\rm off}=-L^3+L_Q L^2-4\zeta_2L+2\zeta_2L_Q-2\zeta_3,
\quad \mathcal H_1^{\rm off}=3L^2/2-(L_Q+1)L+L_Q+2\zeta_2.
\]
Their exact-$d$ one-loop coefficients are those in
Eq.~\eqref{eq:F1B155}. The off-shell bare double pole is $b_0^2$;
its simple pole is
\begin{align}\label{eq:Aoff155}
 f_{2,-1}^{\rm off}={}&\frac{2(D_s-26)}3L^2
 +\frac{(D_s-26)^2}{18}(L-L_Q)
 -\frac{4(D_s-2)}3L\notag\\
 &+\left(\frac{4 D_s}{3} - \frac{104}{3}\right)\zeta_2+d_0.
\end{align}
The walking and off-shell pole functions satisfy the exact identity
\begin{align}\label{eq:UVtransport155}
 f_{2,-1}-f_{2,-1}^{\rm off}
 &=2b_0\big[(L+\ell_y)^2+4\zeta_2\big]+(D_s-2)(L+\ell_y)\notag\\
 &=2b_0(f_1-f_1^{\rm off})
 +c_m\partial_L(f_1-f_1^{\rm off})
 -\kappa_h(\mathcal H_0-\mathcal H_0^{\rm off}).
\end{align}
Here $f_1^{\rm off}=2\mathcal T_0^{\rm off}+b_0(L-L_Q)+c_1$;
thus the comparison uses the same rational source coefficients as
Eq.~\re{eq:F1B155}, with the off-shell rather than the walking masters.
The difference of the subdivergences is therefore fixed by the
corresponding one-loop matrix elements, with common local ultraviolet
counterterms. Substituting the off-shell functions in
Table~\ref{tab:UV155} cancels its poles, including the scalar-potential
and $G$-$P$ contributions.

\paragraph{Finite conversion to the global subtraction.}
Graphwise forest subtraction is useful for displaying the individual contributions, but its finite sum is not
identical to the action and operator subtraction of the physical form factor. We make their relation explicit
before listing the graph results. The sum of the 29 projected graph-level $R'$ residues is
$-2b_0/\epsilon^2+(b_1-1/3)/\epsilon$. This sum excludes the LSZ products
and the action/operator replacements. It is consequently distinct from
the bare form-factor pole in Eq.~\eqref{eq:bare155} and from the operator
renormalization constant. The complete ultraviolet cancellation is
given in Table~\ref{tab:UV155}.

Taking the finite part of Eq.~\eqref{eq:MS155} gives
\begin{align}\label{eq:finite155}
 f_2&=f_{2,0}-2b_0 f_{1,\epsilon}
 -c_m\partial_Lf_{1,\epsilon}\notag\\
 &\quad+\kappa_h\mathcal H_1+c_{GP}\bigl[(4-D_s)(3-L_Q)-2\bigr].
\end{align}
Here $f_{2,0}$ is the finite bare coefficient defined by the Laurent expansion in Eq.~\re{eq:bare155}.
Its conversion to $f_2$ includes the positive-$\epsilon$ terms of the lower-loop matrix elements.
The corresponding relation to the forest-subtracted graph coefficients is
\begin{align}
\label{eq:delta155}
f_2
 -\operatorname{FP}_\epsilon
 \left[\sum_{\gamma\in\mathcal S_1^{(2)}\cup\mathcal S_2^{(2)}} R f_{2,\gamma,B}+R\mathcal I_B\right]
 &=\frac{2-L_Q}{3}\, .
\end{align}
Here the sum contains all 24 vertex and five self-energy graph families
with their physical signs and LSZ weights. The forest operation $R$
acts on each graph and one-loop product term before summation.
The notation $\operatorname{FP}_\epsilon$ selects the coefficient of
$\epsilon^0$ after these operations, and does not discard a pole before
its coefficient has been expanded to the required order.
Its finite part differs from the action and operator subtraction of
Eq.~\re{eq:MS155} by
$\operatorname{FP}_\epsilon[B_q(\epsilon)-1/\epsilon]/3=(2-L_Q)/3$.
This term changes only a single logarithm and a constant, not the
ultraviolet counterterms.

\subsubsection{Individual two-loop graph contributions}
\label{sec:FF2local}

We resolve the finite two-loop coefficient into its graph families to display how the different topologies
supply the logarithmic structures entering the final answer. Each family is first assigned the same
minimal-subtraction forest prescription, with its open subgraph tensors contracted before the kinematics
are specialized. The required operations are
\begin{align}
\label{eq:forest156main}
R'\gamma&=\gamma+\sum_{\mathcal F\ne\varnothing}
 \left[\prod_{\beta\in\mathcal F}(-K\beta)\right]\gamma/\mathcal F,\\*
R\gamma&=R'\gamma-KR'\gamma\, .
\end{align}
Here $K$ extracts the local ultraviolet pole, $\beta$ denotes a proper
ultraviolet subgraph, and $\gamma/\mathcal F$ is the contracted
graph. The sum runs over proper forests with no mutually overlapping
members.
We denote the resulting finite coefficient by $f_{2,\gamma}$, with the same
normalization and sign as $f_2$. The one-loop LSZ products are
separate terms in the assembly. \FFgraph{2}{48} and \FFgraph{2}{72} enter through their
bare expressions and the global subtraction already specified above.
More explicitly, for each graph family displayed below,
\begin{align}
 f_{2,\gamma}&=\lim_{\epsilon\to0}R f_{2,\gamma,B}
       =\operatorname{FP}_\epsilon R'f_{2,\gamma,B}.
 \label{eq:individual-finite-definition}
\end{align}
These are not the bare terms in
Eq.~\re{eq:bare-graph-assembly}; replacing those terms by the finite
numbers below would omit the lower-loop and composite subtractions.
Their relation to the physical two-loop coefficient is instead the
finite conversion in Eq.~\re{eq:delta155}. The scalar building blocks
are collected in Appendix~\ref{app:graph-calculations156}; the
graph-specific source reductions and contracted-subgraph records are
supplied with the separate audit material {\tt /anc/calculation\_audit.json}.
All formulas below use the independent logarithms $L$, $\ell_y$ and $L_Q$.

The vertex and tadpole-insertion families give
\begin{align}
f_{2,\mathrm{FF}12}&=\frac{27}{8}(2+L-L_Q)^2.\label{main:FF2-12}\\[4pt]
f_{2,\mathrm{FF}17}&=\frac{53D_s-93}{36}L_Q^2
 -\left(\frac{127 D_s}{12} - \frac{77}{4}\right)L_Q
 +\frac{28459 D_s}{1296} - \frac{17465}{432}.\label{main:FF2-17}\\[4pt]
f_{2,\mathrm{FF}18}&=\frac98(L-L_Q)^2+\frac{45}{8}(L-L_Q)+\frac{279}{32}.\label{main:FF2-18}\\[4pt]
f_{2,\mathrm{FF}23}&=\frac92(2+L-L_Q)^2.\label{main:FF2-23}\\[4pt]
f_{2,\mathrm{FF}24}&=\left(D_s^2 - 3 D_s + \frac{17}{6}\right)(2-L_Q)^2-\frac{2-L_Q}{9}.\label{main:FF2-24}\\[4pt]
f_{2,\mathrm{FF}27}&=\begin{aligned}[t]
 &\frac98L_Q^2-\frac94LL_Q
 -\frac{12D_s+27}{8}L_Q+\frac{57}{8}L\\*
 &+\frac{45}{8}D_s+\frac9{32}-\frac94\zeta_2.
\end{aligned}\label{main:FF2-27}\\[4pt]
f_{2,\mathrm{FF}28}&=\begin{aligned}[t]
 &\frac{17D_s-27}8\,L_Q^2
-\left(\frac{307 D_s}{24} - \frac{163}{8}\right)\,L_Q\\*
 &+\frac{6779 D_s}{288} - \frac{3611}{96}+\frac{17D_s-27}4\,\zeta_2.
\end{aligned}\label{main:FF2-28}\\[4pt]
f_{2,\mathrm{FF}35}&=\begin{aligned}[t]
 &\left(\frac{11 D_s}{24} - \frac{77}{12}\right)(L-L_Q)^2
 -\left(\frac{1427}{36} - \frac{221 D_s}{72}\right)(L-L_Q)\\*
 &-\left(\frac{31547}{432} - \frac{5141 D_s}{864}\right).
\end{aligned}\label{main:FF2-35}\\[4pt]
f_{2,\mathrm{FF}36}&=\begin{aligned}[t]
 &-\left(\frac{59 D_s}{36} - \frac{53}{18}\right)\,L_Q^2
+\left(\frac{395 D_s}{36} - \frac{349}{18}\right)\,L_Q\\*
 &-\left(\frac{29149 D_s}{1296} - \frac{25435}{648}\right).
\end{aligned}\label{main:FF2-36}\\[4pt]
f_{2,\mathrm{FF}37}&=\begin{aligned}[t]
 &-\frac{D_s+10}{4}(L-L_Q)^2
-\frac{3(D_s+8)}{2}(L-L_Q)\\*
 &-\left(\frac{29 D_s}{12} + \frac{52}{3}\right)-\frac{7-D_s}{2}\zeta_3.
\end{aligned}\label{main:FF2-37}\\[4pt]
f_{2,\mathrm{FF}38}&=\begin{aligned}[t]
 &-\left(\frac{529 D_s}{72} - \frac{373}{24}\right)L_Q^2
+\left(\frac{1013 D_s}{24} - \frac{6709}{72}\right)L_Q\\*
 &-\left(\frac{178553 D_s}{2592} - \frac{133433}{864}\right)-\frac{69-29D_s}{2}\zeta_3.
\end{aligned}\label{main:FF2-38}\\[4pt]
f_{2,\mathrm{FF}49}&=\begin{aligned}[t]
 &-\frac{4D_s+1}{2}L_Q^2
 +\frac98LL_Q-\frac{51}{16}L
 -\left(\frac{31}{24} - \frac{181 D_s}{12}\right)L_Q\\*
 &-\frac{7D_s+25}{4}\zeta_2-\left(\frac{4457 D_s}{144} - \frac{3353}{288}\right).
\end{aligned}\label{main:FF2-49}\\[4pt]
f_{2,\mathrm{FF}51}&=\begin{aligned}[t]
 &L^3-3L^2(L_Q+\ell_y)
 +3L_Q\ell_y(2L+\ell_y)+\ell_y^3\\*
 &+\frac94L^2-12L\ell_y-6\ell_y^2
 -\frac{6D_s-87}{4}LL_Q-\left(\frac{537}{8} - \frac{67 D_s}{12}\right)L\\*
 &-\left(\frac{55 D_s}{72} + \frac{193}{24}\right)L_Q^2+\left(\frac{431 D_s}{72} + \frac{1067}{24}\right)L_Q
 +6\zeta_2\ell_y\\*
 &-\frac{6D_s-57}{4}\zeta_2+6\zeta_3
 -\left(\frac{38015 D_s}{2592} + \frac{62069}{864}\right).
\end{aligned}\label{main:FF2-51}\\[4pt]
f_{2,\mathrm{FF}52}&=\begin{aligned}[t]
 &-\frac{3(3D_s+1)}8L_Q^2
 -\frac{1-63D_s}{8}L_Q\\*
 &-\frac{3(3D_s+1)}4\zeta_2-\left(\frac{531 D_s}{32} - \frac{515}{96}\right).
\end{aligned}\label{main:FF2-52}\\[4pt]
f_{2,\mathrm{FF}53}&=\begin{aligned}[t]
 &\frac34L^3-\frac94L^2(L_Q+\ell_y)
 +\frac94 L_Q\ell_y(2L+\ell_y)+\frac34\ell_y^3\\*
 &+\frac98L^2-9L\ell_y-\frac92\ell_y^2
 +\frac{99}{8}LL_Q-\frac{585}{16}L\\*
 &-\frac{9(D_s+1)}8L_Q^2+\frac{69D_s+12}{8}L_Q
 +\frac92\zeta_2\ell_y+\frac{45}{8}\zeta_2+\frac92\zeta_3
 -\left(\frac{621 D_s}{32} - \frac{225}{32}\right).
\end{aligned}\label{main:FF2-53}\\[4pt]
f_{2,\mathrm{FF}62}&=\begin{aligned}[t]
 &-\left[\frac{9-5D_s}{2}(2-L_Q)-\frac14\right]
 \left(\frac{L^2}{2}-L\ell_y-\frac{\ell_y^2}{2}\right)\\*
 &-\left(D_s-\frac74\right)(2-L_Q)(L+2-L_Q)
 -\left(\frac{L}{12} + \frac{1}{6} - \frac{L_Q}{12}\right)\\*
 &-\left(D_s^2-\frac52D_s+\frac{23}{12}\right)(2-L_Q)^2\\*
 &-\left(D_s^2-\frac72D_s+\frac{65}{18}\right)(2-L_Q)-\frac29\, .
\end{aligned}\label{main:FF2-62}\\[4pt]
f_{2,\mathrm{FF}63}&=\begin{aligned}[t]
 &-(L+2-L_Q)\bigg[-\frac94L^2+\frac92L\ell_y+\frac94\ell_y^2\\*
 &+\frac{81}{8}L+\frac{45}{4}(2-L_Q)-\frac{27}{8}\bigg].
\end{aligned}\label{main:FF2-63}\\[4pt]
f_{2,\mathrm{tad}65}&=-(D_s-1)(L+\ell_y)(1-L_Q+L-\ell_y).\label{main:tad2-65}
\end{align}

For the crossed-ladder and ladder families, respectively, we obtain
\begin{align}\label{main:FF2-70}
f_{2,\mathrm{FF}70}={}&\frac{L^4}{6}
 -\left(\frac23\ell_y+\frac53\right)L^3
 -\left(-\frac52\ell_y+\frac92\zeta_2-\frac{71}{8}\right)L^2\notag\\*
&-L\left[-\frac23\ell_y^3-\frac52\ell_y^2
 +\left(\frac32D_s+\frac34\right)L_Q
 +D_s\left(3\zeta_2-\frac{71}{12}\right)
 -13\zeta_3-\frac{23}{12}\zeta_2+\frac{151}{24}\right]\notag\\*
&+\frac13\ell_y^4+\frac56\ell_y^3+4\zeta_2\ell_y^2
 +(10\zeta_2+4\zeta_3)\ell_y\notag\\*
&+\left(\frac{41 D_s}{36} + \frac{17}{144}\right)L_Q^2
 -\left(\frac{33 D_s}{4} - \frac{613}{144}\right)L_Q\notag\\*
&-D_s\left(8\zeta_3-\frac{43}{9}\zeta_2-\frac{20503}{1296}\right)
 +21\zeta_4+\frac{65}{9}\zeta_2-\frac{21}{4}\zeta_3
 -\frac{98483}{5184}+\kappa_{22}\, .
\end{align}

\begin{align}\label{main:FF2-71}
f_{2,\mathrm{FF}71}={}&\frac{L^4}{6}
 -\left(\frac23\ell_y+\frac53\right)L^3\notag\\*
&-L^2\left[-\frac{5D_s+11}{4}L_Q
 +\frac72D_s-\ell_y^2-5\ell_y-5\zeta_2-\frac{17}{8}\right]\notag\\*
&-L\left[\left(\frac{5D_s+11}{2}\ell_y-\frac12D_s+\frac{93}{4}\right)L_Q
 -\frac23\ell_y^3
 -\left(7D_s+\frac{13}{4}\right)\ell_y\right.\notag\\*
&\hspace{18mm}\left.
 -2D_s\zeta_2-\frac5{12}D_s+22\zeta_3+5\zeta_2-\frac{1235}{24}\right]\notag\\*
&-\left(\frac{73}{36} - \frac{67 D_s}{9}\right)L_Q^2
 -\left[\frac{5D_s+11}{4}\ell_y^2
 +\frac{565 D_s}{12} - \frac{101}{3}\right]L_Q\notag\\*
&-\frac16\ell_y^4-\frac53\ell_y^3
 -\left(4\zeta_2-\frac{13}{8}-\frac72D_s\right)\ell_y^2
 -(10\zeta_2+12\zeta_3)\ell_y\notag\\*
&-\frac{17D_s-19}{2}\zeta_3
 -\left(\frac{203 D_s}{36} - \frac{121}{36}\right)\zeta_2-19\zeta_4
 -\left(\frac{98369}{1296} - \frac{53539 D_s}{648}\right)\, .
\end{align}

The constant $\kappa_{22}$ is the finite, scale-independent term of the
scalar crossed-ladder master. Its reported estimate is $-9\zeta_4$,
without an established analytic value or quantified fit uncertainty;
we retain it symbolically. All logarithmic coefficients of these graph contributions are
independent of this constant.

For the self-energy-inserted vertex families \FFgraph{2}{74} and
\FFgraph{2}{75}, define the cubic logarithmic polynomials
\begin{align}
\mathcal P_3&=L^3-3L^2(L_Q+\ell_y)
 +3L_Q\ell_y(2L+\ell_y)+\ell_y^3,\\*
\mathcal Q_3&=L^3-\frac32L^2L_Q-3L^2\ell_y
 +3LL_Q\ell_y+\frac32L_Q\ell_y^2+\ell_y^3\, .
\end{align}
The corresponding two contributions are
\begin{align}
f_{2,\mathrm{FF}74}&=\begin{aligned}[t]
 &\frac{D_s-14}{9}\mathcal P_3
 -\left(\frac{71}{18} - \frac{17 D_s}{36}\right)L^2-\left(\frac{16 D_s}{9} - \frac{176}{9}\right)L\ell_y\\*
 &+\frac{5(D_s-14)}{3}LL_Q-\left(\frac{55 D_s}{9} - \frac{650}{9}\right)L
 +\left(\frac{29 D_s}{36} + \frac{157}{18}\right)L_Q^2\\*
 &-\left(\frac{73 D_s}{12} + \frac{883}{18}\right)L_Q
 -\frac{8(D_s-11)}9\ell_y^2+\frac{2(D_s-14)}{3}\zeta_2\ell_y\\*
 &+\frac{5(D_s-14)}{6}\zeta_2+\frac{2(D_s-14)}{3}\zeta_3
 +\frac{19981 D_s}{1296} + \frac{51125}{648}.
\end{aligned}\label{main:FF2-74}\\[4pt]
f_{2,\mathrm{FF}75}&=\begin{aligned}[t]
 &\frac{(D_s-14)}{9}\mathcal Q_3
 -\frac{5-7D_s}{9}L^2-9LL_Q
 -\left(\frac{13 D_s}{18} - \frac{103}{9}\right)L\ell_y\\*
 &-\left(\frac{7 D_s}{4} - \frac{67}{2}\right)L-\frac{2(D_s-14)}{3}\zeta_2L
 +\left(\frac{35 D_s}{72} + \frac{115}{36}\right)L_Q^2-\frac{5D_s+8}{6}L_Q\ell_y\\*
 &-\left(\frac{221 D_s}{72} + \frac{821}{36}\right)L_Q
 -\left(\frac{43 D_s}{36} - \frac{79}{18}\right)\ell_y^2+\frac{19D_s+16}{18}\ell_y\\*
 &-\left(\frac{77}{18} - \frac{83 D_s}{36}\right)\zeta_2
 +\frac{16243 D_s}{2592} + \frac{54539}{1296}.
\end{aligned}\label{main:FF2-75}
\end{align}

The proper self-energy contributions, with the LSZ virtuality derivative
included in their graph coefficients, are
\begin{align}
f_{2,\mathrm{SE}2}&=-(D_s-1)\left[\frac34(L-L_Q)+\frac{27}{16}\right].\label{main:SE2-2}\\[4pt]
f_{2,\mathrm{SE}4}&=\begin{aligned}[t]
 &-\frac{27-6D_s}{16}(L-L_Q)^2-
   \frac{69-18D_s}{16}(L-L_Q)\\*
 &-\left(\frac{217}{64} - \frac{55 D_s}{96}\right).
\end{aligned}\label{main:SE2-4}\\[4pt]
f_{2,\mathrm{SE}5}&=-\frac{27}{8}(L-L_Q)^2-
              \frac{55}{4}(L-L_Q)-14.\label{main:SE2-5}\\[4pt]
f_{2,\mathrm{SE}7}&=\begin{aligned}[t]
 &\left(\frac{D_s}{18} + \frac{335}{144}\right)(L-L_Q)^2
              +\left(\frac{17 D_s}{36} + \frac{1415}{144}\right)(L-L_Q)\\*
 &+\frac{61 D_s}{1296} + \frac{76831}{5184}+\frac{D_s}{2}\zeta_3.
\end{aligned}\label{main:SE2-7}\\[4pt]
f_{2,\mathrm{SE}8}&=\begin{aligned}[t]
 &-\left(\frac{23 D_s}{36} - \frac{407}{72}\right)(L-L_Q)^2
       -\left(\frac{25 D_s}{9} - \frac{1969}{72}\right)(L-L_Q)\\*
 &-\left(\frac{5677 D_s}{1296} - \frac{117793}{2592}\right)
       -\frac{D_s-6}{4}\big[\ell_y^2-2(L-L_Q)\ell_y-2\ell_y\big].
\end{aligned}\label{main:SE2-8}
\end{align}

\subsubsection{Explicit renormalized form factor and its logarithm}
\label{sec:FF2result}

The preceding calculation supplies all ingredients needed to assemble the
renormalized form factor: the vertex graphs, the external-pole shifts and
residues, and the action and composite-operator counterterms. We now combine
them in the normalization of Eq.~\re{eq:FFexp}. It is useful to give the
answer first at $\mu^2=M_E^2$, or $L_Q=L$, where no logarithm separates the
subtraction scale from the external virtuality. This fixes the
renormalization scale, not a new subtraction scheme. With a superscript
$m$ denoting this choice, the two-loop coefficient is
\begin{align}\label{eq:gMass155}
 f_2^{m}(L,\ell_y)={}&\frac12\bigl(f_1^{m}\bigr)^2+\frac{b_0}{3}(L^3-3L\ell_y^2-2\ell_y^3)
 -\frac{\Gamma_{\rm cusp}^{(2)}}{4}(L-\ell_y)^2+\frac{\Gamma_{\rm c}^{(2)}}2\ell_y^2\notag\\
 &+k_s\ell_y(L+\ell_y)+J_L L+J_y\ell_y+c_2.
\end{align}

The one-loop coefficient at the same scale is
\begin{align}
 f_1^{m}&\equiv\left.f_1\right|_{L_Q=L}
 =\mathcal P_{\rm DL}+c_1\, .
 \label{eq:mass-boundary-one-loop}
\end{align}

Here $\mathcal P_{\rm DL}$ and $c_1$ are the double-logarithmic polynomial
and finite constant defined in Eqs.~\re{eq:walking-double-log-polynomial}
and \re{eq:walking-matching-constant}. Evolution away from this boundary
restores the arbitrary subtraction scale:
\begin{align}\label{eq:gGeneral155}
 f_2={}&f_2^{m}+2b_0(L-L_Q)f_1^{m}\notag\\
 &+c_m(L-L_Q)\partial_L\mathcal P_{\rm DL}
 -\kappa_h(L-L_Q)\mathcal H_0\notag\\
 &+\gamma_2(L-L_Q)+b_0^2(L-L_Q)^2+b_0c_m(L-L_Q).
\end{align}

The derivative acts at independent $L$ and $\ell_y$. The coefficients
$b_0$ and $b_1$ are given in Eq.~\re{eq:coupling-beta-coefficients},
$c_m$ in Eq.~\re{eq:FFmass-renormalization}, and $\kappa_h$ in
Eq.~\re{eq:quartic-response-coefficient}. The scalar-potential response is
$\mathcal H_0=-\ell_y-1$, Eq.~\re{eq:Hdepth155}. The remaining coefficients
of Eq.~\re{eq:gMass155} are
\begin{align}
 k_s&=\frac{D_s}{3}-\frac23,\notag\\*
 J_L&=D_s^2-\frac{92}{9}D_s+\frac{112}{9}+4\zeta_3,
 \label{eq:coeff155}\\*
 J_y&=\frac73D_s-\frac{14}{3}+\frac{28}{3}\zeta_2
 -\frac23D_s\zeta_2-4\zeta_3,\notag\\*
 c_2&=-\frac{1643}{648}D_s^2+\frac{4523}{162}D_s+\frac{7177}{162}
 +\frac23D_s\zeta_2-\frac43\zeta_2\notag\\*
 &\quad+\frac83D_s\zeta_3-\frac{184}{3}\zeta_3+8\zeta_4.\notag
\end{align}
In $c_2$ we have inserted $\kappa_{22}=-9\zeta_4$, the analytic form
suggested by the available numerical crossed-ladder evaluation. Without
assuming this identification, the general result is restored by adding
$\kappa_{22}+9\zeta_4$ to $c_2$, and hence to $f_2$. None of the logarithmic
coefficients depends on this constant. The numerical precision of the
coupled inputs is described in Appendices~\ref{app:ff48} and \ref{app:ff72}.

The coefficients $\Gamma_{\rm cusp}^{(2)}$ and $\Gamma_{\rm c}^{(2)}$
are those of the two double-logarithmic kernels specified in
Eqs.~\re{eq:scalar-YM-cusp-kernel} and \re{eq:Gc155} below.
Their different roles will follow from the two integration intervals in
Eq.~\re{eq:integral155}. Finally, the diagonal evolution of $G$,
Eq.~\re{eq:gamma-diagonal}, combines with the $P$ matrix element to give
\begin{equation}\label{eq:gamma155}
 \gamma_2=2b_1-c_{GP}(4-D_s)
 =-\frac56D_s^2+\frac73D_s+\frac{80}{3}.
\end{equation}

The form of Eq.~\re{eq:gMass155} exhibits Sudakov exponentiation:
the fourth-degree logarithms are supplied by
$\tfrac12(f_1^m)^2$. They therefore cancel in the two-loop coefficient
of the logarithm,
\begin{align}
 \log F={}&g_R^2 f_1+g_R^4\left(f_2-\tfrac12f_1^2\right)
 +O(g_R^6)+o_{\rm LP}(1).
 \label{eq:log-form-factor-two-loop}
\end{align}

At the mass boundary, the remaining cubic logarithm is proportional
to $b_0$ and is generated by the running coupling. The genuinely new
two-loop information begins in the quadratic logarithms, followed by
the single logarithms and the matching constant. The distinction
between the coefficient $f_2$ of $F$ and the connected combination
$f_2-\tfrac12f_1^2$ is thus explicit.

The scale dependence provides a further test of the assembly. With the
coupling, common-mass and scalar-potential responses included, the result
obeys
\begin{equation}\label{eq:RG155}
 -\partial_{L_Q}f_2-2b_0f_1-c_m\partial_Lf_1
 +\kappa_h\mathcal H_0-\gamma_2=0.
\end{equation}

Both the scalar-potential response and $G$-$P$ mixing are required by this
identity. The mixing contributes to operator evolution and finite
matching, but not to the cusp kernel. Equations~\re{eq:gMass155} and
\re{eq:gGeneral155} retain all leading-power single logarithms and
constants, with the treatment of the crossed-ladder constant stated
above.

\subsubsection{Local scale-integral representation and walking kernels}
\label{sec:integral}

We next reorganize the logarithm of the form factor as evolution between
its physical scales. The aim is to identify the kernels integrated over
each momentum interval and to show explicitly that their running-coupling
integrals reproduce the calculated logarithmic polynomial. Matching to
the on-shell and strictly off-shell boundaries fixes the two
double-logarithmic kernels; operator evolution, single-logarithmic
evolution and finite matching then complete the result.

\paragraph{Matching the boundary kernels.}
The gauge beta function is that of Yang-Mills theory with $\Ds-4$ real
adjoint scalars~\cite{Machacek:1983tz}, with coefficients
given in Eq.~\re{eq:coupling-beta-coefficients}. In the same coupling
normalization, the scalar-YM cusp coefficient is obtained from the
dimensional-reduction result of Ref.~\cite{Belitsky:2003ys} without
fermions:
\begin{equation}
\label{eq:scalar-YM-cusp-kernel}
 \Gamma_{\rm cusp}=4g_R^2+
 \left[\frac{320}{9} - \frac{16 D_s}{9}-8\zeta_2\right]g_R^4+O(g_R^6).
\end{equation}

The second boundary datum is supplied by the strictly off-shell
calculation in the companion paper. At $\mu^2=M_E^2$ its double-logarithmic
coefficient is
\begin{align}\label{eq:off-shell-kernel-input}
\left[\log F^{\rm off}\right]_{g_R^4 L^2}
&=8\zeta_2+\frac{5 D_s}{9} - \frac{154}{9}\, .
\end{align}

This result is an external boundary input, not a termwise endpoint limit
of the interior walking expansion. In the normalization used here, the
off-shell double logarithm is $-\Gamma_{\rm oct}L^2/2$. Its one-loop term
agrees with Eq.~\re{eq:FF1DL} at $\eta=1$, while multiplying the
two-loop coefficient in Eq.~\re{eq:off-shell-kernel-input} by $-2$ gives
\begin{equation}
\label{eq:scalar-YM-octagon-kernel}
 \Gamma_{\rm oct}=4g_R^2+
 \left[\frac{308}{9}-\frac{10D_s}{9}-16\zeta_2\right]g_R^4+O(g_R^6).
\end{equation}

To reproduce both boundaries with local scale integrals, we define the
second double-logarithmic kernel by
\begin{equation}\label{eq:Gc155}
 \Gamma_{\rm c}\equiv2\Gamma_{\rm cusp}-\Gamma_{\rm oct}
 =4g_R^2+\left(\frac{332}{9}-\frac{22D_s}{9}\right)g_R^4+O(g_R^6).
\end{equation}
The zeta terms cancel in $\Gamma_{\rm c}^{(2)}$, leaving a nonzero
rational coefficient. The quadratic logarithms of
$f_2^m-\tfrac12(f_1^m)^2$ reproduce these boundary combinations; this
agreement does not require the finite interior expansion to be uniform
at either endpoint.

\paragraph{Local scale-integral representation}
At the mass boundary let $\mu_m^2=M_E^2(\mu_m)$,
$g_m^2=g_R^2(\mu_m^2)$ and $L_m=L|_{\mu=\mu_m}$.
The ratio $y$ is defined in Eq.~\re{eq:walkcond}, and the independent logarithms in Eq.~\re{eq:walklogs}.
For fixed interior walking, $k_{\min}^2=m_z^2$; globally the lower bound
is $\max(m_z^2,M_E^4/Q^2)$. The first lower scale is the scalar mass;
the second is the invariant collinear cutoff. This lower-bound rule is
kinematic: the matching polynomial displayed here is the fixed-interior
walking result, not a uniformly valid two-loop crossover formula.
The bare graph sum fixes the cubic running term and the finite
logarithmic polynomial. The on-shell cusp coefficient and the off-shell
boundary input from the companion paper fix $\Gamma_{\rm cusp}$ and $\Gamma_{\rm c}$; the residual single-log
and constant terms fix the collinear kernel and the mass-boundary matching
function. With these boundary masses, the local scale-integral
representation through two loops is
\begingroup\begin{align}\label{eq:integral155}
 \log F={}&-\frac12\int_{k_{\min}^2}^{Q^2}\frac{dk^2}{k^2}
 \log\frac{Q^2}{k^2}\,\Gamma_{\rm cusp}[g_R^2(k^2)]\notag\\
 &+\int_{k_{\min}^2}^{\mu_m^2}\frac{dk^2}{k^2}
 \log\frac{\mu_m^2}{k^2}\,\Gamma_{\rm c}[g_R^2(k^2)]
 -\int_{\mu^2}^{\mu_m^2}\frac{dk^2}{k^2}\,\gamma_G^{\rm eff}[g_R^2(k^2)]\notag\\
 &+\int_{\mu_m^2}^{Q^2}\frac{dk^2}{k^2}\,[g_R^2(k^2)]^2
       [J_L+k_s\ell_y]+\mathcal C(g_m^2,h_m;\ell_y)+O(g_R^6),
\end{align}\endgroup
where
\begin{align}
 \gamma_G^{\rm eff}(g_R^2)&=b_0g_R^2+[2b_1-c_{GP}(4-D_s)]g_R^4,
 \label{eq:effective-operator-evolution}\\
 \mathcal C(g_m^2,h_m;\ell_y)&=g_m^2 c_1+g_m^2 h_m(-\ell_y-1)
 +g_m^4[k_s\ell_y^2+J_y\ell_y+c_2].
 \label{eq:C155}
\end{align}
The function $\mathcal C$ collects finite matching contributions
at the external-mass scale; it is not a scalar triangle integral.
Here $h_m\equiv h(\mu_m)$ is the finite running scalar deformation defined by Eq.~\re{eq:finite-scalar-deformation},
not the pole counterterm $h_{\rm CT}$ and not the external holding source.
For the reference boundary $h(\mu)=0$, the required
transport follows from the one-loop mass replacement in
Eq.~\re{eq:Z155}, the quartic pole in
Eq.~\re{eq:quartic-response-coefficient}, and the dimensionally
continued gauge beta function in Eq.~\re{eq:coupling-beta-d}:
\begin{align}
 \frac{\dd\log M_E^2}{\dd\log\mu^2}
 &=c_mg_R^2+O(g_R^4),\qquad
 \left.\frac{\dd h}{\dd\log\mu^2}\right|_{h=0}
 =\kappa_hg_R^2+O(g_R^4).
 \label{eq:boundary-mass-quartic-running}
\end{align}
In the first equation the common background mass is varied at fixed
$y$, as in Eq.~\re{eq:Z155}; the independent scalar-quartic response
is carried by $h$. Integrating between the subtraction and mass
boundaries gives
\begin{align}\label{eq:mass-boundary-transport}
 g_m^2&=g_R^2+b_0g_R^4(L-L_Q),\qquad
 L_m=L+c_mg_R^2(L-L_Q),\\
 h_m&=-\kappa_h g_R^2(L-L_Q)+O(g_R^4).
\end{align}
The mass, gauge-coupling and scalar-quartic transport in
Eq.~\re{eq:mass-boundary-transport}, applied to
Eq.~\eqref{eq:integral155}, reproduces the logarithm in
Eq.~\eqref{eq:gGeneral155}, including
the mixed $b_0c_m(L-L_Q)$ term. The representation involves only local scale
kernels.

The operator-evolution interval and its sign follow directly from the
calculated one-loop term. Using the logarithms of Eq.~\re{eq:walklogs}, the local integral is
\begin{align}\label{eq:operator-interval156}
 -\int_{\mu^2}^{M_E^2}\frac{dk^2}{k^2}\,
       \gamma_G^{\rm eff}[g_R^2(k^2)]
 ={}&b_0g_R^2(L-L_Q)\notag\\
 &+g_R^4\left[\gamma_2(L-L_Q)+\frac{b_0^2(L-L_Q)^2}{2}\right]+O(g_R^6).
\end{align}
This orientation reproduces the one-loop matrix element and satisfies
the multi-coupling RG identity, Eq.~\eqref{eq:RG155}.

For an explicit evaluation of the two double-logarithmic integrals, let
$\mathcal I_{\rm DL}$ denote the sum of the first two terms on the
right-hand side of Eq.~\re{eq:integral155}. Work
at the mass boundary and set $\tau=\log(k^2/\mu_m^2)$.
Then $g_R^2(k^2)=g_m^2-b_0g_m^4\tau+O(g_m^6)$.
The first interval is $\ell_y\le\tau\le L_m$ and the second is
$\ell_y\le\tau\le0$. Their weights are $L_m-\tau$ and $-\tau$,
respectively. Consequently their complete two-loop expansion is
\begin{align}\label{eq:localDL-expanded156}
 \mathcal I_{\rm DL}={}&g_m^2[-(L_m-\ell_y)^2+2\ell_y^2]\notag\\
 &+g_m^4\left[
 \frac{b_0}{3}(L_m^3-3L_m\ell_y^2-2\ell_y^3)
 -\frac{\Gamma_{\rm cusp}^{(2)}}{4}(L_m-\ell_y)^2
 +\frac{\Gamma_{\rm c}^{(2)}}2\ell_y^2\right]+O(g_m^6).
\end{align}
The first line equals $g_m^2\mathcal P_{\rm DL}(L_m,\ell_y)$.
The remaining local integral is $g_m^4[J_L+k_s\ell_y]L_m$ at this order;
the matching function adds $g_m^2 c_1$ and
$g_m^4[k_s\ell_y^2+J_y\ell_y+c_2]$, together with the explicitly displayed
scalar-quartic term. Thus the sum gives $f_2^{m}-\tfrac12(f_1^{m})^2$ from
Eq.~\eqref{eq:gMass155}. Applying the stated mass, coupling and scalar-quartic transport
then gives the connected combination in Eq.~\eqref{eq:gGeneral155} at general subtraction scale.

\paragraph{Double-logarithmic kernels and the full form factor}
Keeping $L$ and $\ell_y$ independent, the terms of total logarithmic
degree two in the connected coefficient at the external-mass renormalization scale are
\begin{align}\label{eq:walk155}
\left.
 \left[f_2^{m}-\tfrac12(f_1^{m})^2\right]
\right|_{\deg_{L,\ell_y}=2}
={}&-\frac{\Gamma_{\rm cusp}^{(2)}}4(L-\ell_y)^2
 +\frac{\Gamma_{\rm c}^{(2)}}2\ell_y^2\notag\\
 &+k_s\ell_y(L+\ell_y)\, .
\end{align}
The first two terms are generated by the double-logarithmic scale
kernels. The last comes from the explicitly calculated collinear
and mass-boundary matching terms and is not part of those kernels.
The cubic running term, all single logarithms, and the constant are
retained in Eqs.~\re{eq:gMass155} and \re{eq:gGeneral155}.
A correlated trajectory is imposed only after these independent-logarithm
polynomials have been assembled.

\paragraph{The contributions to the exponent.}
The first term in Eq.~\re{eq:integral155} integrates the cusp kernel
over the on-shell rapidity range. The second is the correction due
to the external virtualities. Operator evolution runs from the
subtraction scale to the external-mass scale, while the remaining
single-logarithmic integral and finite matching function complete
the result. At one loop the two double-logarithmic kernels coincide
and reproduce Eq.~\re{eq:FFuniform}.
Their roles and definitions are summarized in
Table~\ref{tab:local-kernel-summary}.
\begin{table}[htbp]
\centering\small
\begin{tabular}{p{3.7cm}p{7.0cm}p{3.9cm}}
\toprule
Quantity & Role & Definition\\
\midrule
$\Gamma_{\rm cusp}$ & Upper double-logarithmic interval &
Eq.~\re{eq:scalar-YM-cusp-kernel}\\
$\Gamma_{\rm c}$ & External-virtuality correction &
Eq.~\re{eq:Gc155}\\
$\gamma_G^{\rm eff}$ & Evolution of the operator matrix element &
Eq.~\re{eq:effective-operator-evolution}\\
$J_L+k_s\ell_y$ & Two-loop collinear single-logarithmic kernel &
Eqs.~\re{eq:coeff155}, \re{eq:integral155}\\
$\mathcal C$ & Finite matching at the external-mass scale &
Eq.~\re{eq:C155}\\
\bottomrule
\end{tabular}
\caption{The local ingredients of the two-loop scale-integral
representation. Gauge, mass and scalar-coupling running are fixed by
Eqs.~\re{eq:coupling-beta-coefficients} and
\re{eq:boundary-mass-quartic-running}.}
\label{tab:local-kernel-summary}
\end{table}

\paragraph{The walking anomalous dimension.}
Having matched the kernels and checked their scale integrals, we can
state the walking result in the same perturbative convention at both
loop orders:
\begin{align}
 \Gamma_{\rm walk}(g_R^2;\eta)
 &=g_R^2\Gamma_{\rm walk}^{(1)}(\eta)
  +g_R^4\Gamma_{\rm walk}^{(2)}(\eta)+O(g_R^6).
 \label{eq:walking-kernel-expansion}
\end{align}
On the trajectory $y=w^\eta$, the two double-logarithmic kernels give
\begin{align}
 \Gamma_{\rm walk}^{(1)}(\eta)
 &=\frac{\Gamma_{\rm cusp}^{(1)}}4(1+\eta)^2
   -\frac{\Gamma_{\rm c}^{(1)}}2\eta^2
 =1+2\eta-\eta^2,
 \label{eq:walking-kernel-one-loop}\\*
 \Gamma_{\rm walk}^{(2)}(\eta)
 &=\frac{\Gamma_{\rm cusp}^{(2)}}4(1+\eta)^2
   -\frac{\Gamma_{\rm c}^{(2)}}2\eta^2 .
 \label{eq:walking-kernel-two-loop}
\end{align}
The first line reproduces Eq.~\re{eq:FF1DL}, since
$\Gamma_{\rm cusp}^{(1)}=\Gamma_{\rm c}^{(1)}=4$. In the full connected
form factor the collinear and matching terms of Eq.~\re{eq:walk155}
also contribute $-k_s\eta(1-\eta)L^2$ at two loops. This term is
separate from the double-logarithmic scale kernels and must be retained
when reconstructing $\log F$.

Substituting the boundary coefficients yields the explicit two-loop
walking anomalous dimension,
\begin{align}
 \Gamma_{\rm walk}^{(2)}(\eta)
 &=\left(\frac{80-4D_s}{9}-2\zeta_2\right)(1+\eta)^2
   -\frac{166-11D_s}{9}\eta^2 .
 \label{eq:walking-kernel-two-loop-explicit}
\end{align}
Its term of maximal transcendental weight has the particularly simple
form
\begin{equation}
 \left.\Gamma_{\rm walk}^{(2)}(\eta)\right|_{\rm max.\,trans.}
 =-2\zeta_2(1+\eta)^2.
 \label{eq:walking-kernel-maximal-weight}
\end{equation}
It coincides with the walking kernel of $\cN=4$ sYM~\cite{Alday:2026walk},
while the rational terms display the additional structure of the
non-conformal theory.

\section{Conclusion}

In this work, we considered the Sudakov form factor in adjoint scalar Yang-Mills theory with spontaneous symmetry breaking.
The model is a prototypical set-up for off-shell infrared-sensitive observables in pure gluodynamics but in a completely 
gauge-invariant environment. We found two stationary points of the Coleman-Weinberg effective potential, dubbed branches A 
and B. The former branch is the null point in the moduli space and corresponds to the setup when only external states in the 
Sudakov form factor are massive, read off-shell. This observable is addressed in a companion paper. Here we focused on the
walking regime, motivated by branch B and evaluated with the held-background prescription specified above.

The calculation combines the complete vertex contribution with the external-pole shifts and residues required by LSZ.
The diagrams were generated with QGRAF, and their color and tensor algebra was performed with FORM, retaining the
dimensional dependence needed for ultraviolet subtraction. Scalar integration-by-parts identities, factorization into
lower-loop integrals, and direct Feynman-parameter integration reduced the simpler contributions. For the coupled
two-loop graphs, the walking expansion was applied to the parameter integrands before integration. The Method of
Regions, supplemented by tropical subtraction and hyperlogarithmic integration, then provided the required asymptotic
coefficients. The holding-source cancellations were implemented before the connector limit, and the complete bare
graph and LSZ sum were renormalized by the action counterterms and composite-operator mixing. The explicit two-loop
calculation was carried out in the gauge $c=\alpha=1$.

The resulting leading-power form factor retains the independent logarithms of the internal mass, external virtuality
and subtraction scale, including the single logarithms and finite terms with the qualifications stated above. Its
logarithm admits a local scale-integral representation with the running gauge coupling. This representation separates
the double-logarithmic kernels from operator evolution and finite matching, and displays how the running coupling
modifies the walking Sudakov exponent. The maximally transcendental part of the walking kernel agrees with that of
$\cN=4$ sYM, while the non-conformal theory supplies additional terms of lower transcendentality.

We plan to extend this analysis to QCD in follow-up work. The present calculation provides a setting in which the
interplay of a nontrivial infrared hierarchy, ultraviolet renormalization and running-coupling evolution can be
established before undertaking that extension.

\begin{acknowledgments}

OpenAI's Codex and Anthropic's Claude Code were used to create computational workflows and their execution. Details can be
found in {\tt /anc/ai\_usage.tex}. Additionally, ChatGPT and Claude were used to assist with manuscript preparation.

\end{acknowledgments}

\appendix

\section{Feynman rules for \texorpdfstring{$c=1$}{c=1}: uplifted and component forms}
\label{App:FeynRules}

This appendix supplies the propagators and interaction vertices used to construct the loop integrands. We give the
compact uplifted rules first and then resolve them into vector, scalar and ghost components, so that either form can
be used to reproduce the diagrammatic calculation. Perturbatively, at $c=1$ the reduced theory is
the $\Ds$-dimensional Yang-Mills theory in Lorenz gauge. The only differences are that loop momenta run in $d$ dimensions, while 
the extra-dimensional components of all momenta are fixed by the colors of the lines.
The uplifted field and momenta are those of Eqs.~\re{eq:X} and
\re{eq:P}; their background charges and mass invariants are fixed by
Eqs.~\re{eq:Pij} and \re{eq:invPQ}. The choice $c=1$ leaves the
covariant-gauge parameter $\alpha$ arbitrary.

\subsection{Propagators}

The background enters a propagator through its color-dependent extra-dimensional momentum, while the gauge
parameter appears in the longitudinal numerator. Keeping these structures together gives the uplifted vector and
ghost propagators
\begin{align}
\label{eq:prop}
\bigl\langle X_{M;ij}(p)X_{N;kl}(-p)\bigr\rangle_0
&
=
- 
\frac{\ii}2
\Bigl(
\delta_{il}\delta_{jk}-\frac{\delta_{ij}\delta_{kl}}{N+2}
\Bigr)
\left[
\frac{\eta_{MN}}{P^2} - (1-\alpha) \frac{P_{M;ij} P_{N;ij}}{(P^2)^2}
\right]
\, , \\
\label{eq:ghprop}
\bigl\langle\omega_{ij}(p)\bar\omega_{kl}(-p)\bigr\rangle_0
&=
+
\frac{\ii}2\Bigl(\delta_{il}\delta_{jk}-\frac{\delta_{ij}\delta_{kl}}{N+2}\Bigr) \frac{1}{P^2}
\, .
\end{align}
Here $P^2=p^2-m_{ij}^2$, with the Feynman $+\ii0$ prescription understood in every denominator. The factor $\ft12$ in the color 
structure stems from the normalization $\tr(T^AT^B)=\ft12\delta^{AB}$; for the normalized fields ($W^\pm$, $w_c$ and their scalar 
partners), it is absent. Each closed ghost loop carries a factor $-1$. Splitting $M = (\mu, \hI)$, one finds the component propagators
\begin{align}
\langle AA\rangle
:
&\quad 
-\frac{g_{\mu\nu}}{p^2-m_{ij}^2}+\frac{(1-\alpha)p_\mu p_\nu}{(p^2-m_{ij}^2)^2}
\, ,\\
\langle\chi\chi\rangle
:
&\quad 
\frac{\delta_{IJ}}{p^2-m_{ij}^2}+\frac{(1-\alpha)\kappa_{ij,I}\kappa_{ij,J}}{(p^2-m_{ij}^2)^2}
\, ,\\
\langle A_\mu\chi_I\rangle
:
&\quad 
\frac{(1-\alpha)\,p_\mu\,\kappa_{ij,I}}{(p^2-m_{ij}^2)^2}
\, ,
\end{align}
each multiplied by $\ii$ and by the color factor of Eq.~\eqref{eq:prop}. This decomposition makes the following
features of the spectrum and gauge dependence explicit:
\begin{itemize}
\item The scalars along $\kappa_{ij}$ (the would-be Goldstone bosons) and the longitudinal vectors carry the
$\alpha$-dependence. The scalars orthogonal to $\kappa_{ij}$ have the plain propagator $\ii\delta_{IJ}/(p^2-m_{ij}^2)$.
\item In the unbroken directions, $\kappa_{ij}=0$, the common-block gluons and scalars, and the Abelian $B^{(1,2)}$ and
Cartan $\sigma^{(1,2)}$, all have massless propagators with no mixing.
\item At a null point, $\kappa_{ij}^2=0$ with $\kappa_{ij}\neq0$ (the $ar$, $br$ pairs at $z=\pm\ii$), the
denominators are massless, but the mixed and scalar $(1-\alpha)$ terms keep $\kappa_{ij}$ in the numerator.
\end{itemize}

\subsection{Vertices}

The interactions retain the Yang-Mills form when ordinary and background momenta are combined into uplifted
momenta. To use this compact notation for particular charged fields, we first specify their color matrices and the
orientation of their momenta. Each field in a vertex is labelled by $\ell=1,\dots,n$. It carries an 
incoming momentum $p_\ell$, a component index $M_\ell$ and a color matrix $C_\ell$, with which it enters the expansion of $X$ 
or of the ghosts. For example,
\begin{equation}
C_{W^+}=\frac{E_{ab}}{\sqrt2},\quad C_{W^-}=\frac{E_{ba}}{\sqrt2},\quad C_{(w_c)_r}=\frac{E_{cr}}{\sqrt2},\quad
C_{(\bar w_c)^r}=\frac{E_{rc}}{\sqrt2},\quad C_{\omega_{ij}}=C_{\bar\omega_{ij}}=E_{ij},
\end{equation}
where $(E_{ij})_{kl}=\delta_{ik}\delta_{jl}$. Similar expressions hold for the normalized scalars. For $C_\ell\propto E_{i_\ell j_\ell}$, the field carries
$\kappa_\ell\equiv\kappa_{i_\ell j_\ell}$ and the uplifted momentum $P_\ell=(p_\ell,\kappa_\ell)$. At every vertex, both parts of the latter are
conserved,
\begin{equation}
\sum_\ell p_\ell=0
\, ,\qquad 
\sum_\ell\kappa_{\ell}=0
\, .
\end{equation}
The second is color conservation in the language of extra-dimensional momenta. Rules are written at finite $N$. The planar limit is taken after 
the color contraction of a complete diagram.

In the vertices below, the order of the labels corresponds to the numbering of the fields; other orderings follow by
permuting the fields together with their momenta, indices, and color matrices. The trivalent vertices are
\begin{align}
\label{eq:V3}
V_{XXX}^{M_1 M_2 M_3}(1,2,3)
&=
-2\ii\gym\tr\bigl([C_1,C_2]C_3\bigr)
\\
&
\times
\bigl[\eta^{M_1M_2}(P_1-P_2)^{M_3}+\eta^{M_2M_3}(P_2-P_3)^{M_1} +\eta^{M_3M_1}(P_3-P_1)^{M_2}\bigr]
\, , \nonumber\\
\label{eq:ghv}
V_{\bar\omega\omega X}^{M}(1,2,3)
&
=
+
2\ii\gym\tr\bigl([C_1,C_2]C_3\bigr)\,P_1^M 
 \, .
\end{align}
In the latter, the fields' order is $\bar\omega(p_1,C_1)$, $\omega(p_2,C_2)$, $X_M(p_3,C_3)$, and the antighost's uplifted 
momentum reads $P_1^M=(p_1^\mu,\ -\kappa_{1,I})$. 

The tetravalent vertex then reads
\begin{align}
\label{eq:V4}
V_{XXXX}^{M_1M_2M_3M_4}(1,2,3,4)
=
2\ii\gym^2\bigl[&\tr([C_1,C_2][C_3,C_4])\bigl(\eta^{M_1M_3}\eta^{M_2M_4}-\eta^{M_1M_4}\eta^{M_2M_3}\bigr)
\\
+&
\tr([C_1,C_3][C_2,C_4])\bigl(\eta^{M_1M_2}\eta^{M_3M_4}-\eta^{M_1M_4}\eta^{M_2M_3}\bigr)
\nonumber\\
+&
\tr([C_1,C_4][C_2,C_3])\bigl(\eta^{M_1M_2}\eta^{M_3M_4}-\eta^{M_1M_3}\eta^{M_2M_4}\bigr)\bigr]
\, . \nonumber
\end{align}

Finally, we list the component form of the above vertices. Restricting the indices to the corresponding subspaces and using
$(P_1-P_2)^{\hI}=-(\kappa_1-\kappa_2)_I$ and $\eta^{\hI\hJ}=-\delta_{IJ}$, we deduce the cubic interactions
\begin{align}
V_{AAA}^{\mu\nu\rho}&
=-2\ii\gym\tr([C_1,C_2]C_3)\bigl[g^{\mu\nu}(p_1-p_2)^\rho+g^{\nu\rho}(p_2-p_3)^\mu+g^{\rho\mu}(p_3-p_1)^\nu\bigr]
\, , \\
V_{AA\chi}^{\mu\nu;I}&
=\phantom{-}2\ii\gym\tr([C_1,C_2]C_3)\,g^{\mu\nu}(\kappa_1-\kappa_2)_I
\, ,\\
V_{A\chi\chi}^{\mu;IJ}&
=\phantom{-}2\ii\gym\tr([C_1,C_2]C_3)\,\delta_{IJ}(p_2-p_3)^\mu
\, ,\\
V_{\chi\chi\chi}^{IJK}&=
-2\ii\gym\tr([C_1,C_2]C_3)\bigl[\delta_{IJ}(\kappa_1-\kappa_2)_K+\delta_{JK}(\kappa_2-\kappa_3)_I+\delta_{KI}(\kappa_3-\kappa_1)_J\bigr]
\, , \\
V_{\bar\omega\omega A}^\mu
&
=\phantom{-}2\ii\gym\tr([C_1,C_2]C_3)\,p_1^\mu
\, , \\
V_{\bar\omega\omega\chi}^I
&
=-2\ii\gym\tr([C_1,C_2]C_3)\,\kappa_{1,I}
\, .
\end{align}
The quartic vertices $V_{AAAA}$ and $V_{\chi\chi\chi\chi}$ have the same form as Eq.~\eqref{eq:V4}, with $\eta$ replaced by
the $d$-dimensional metric ($\eta\to g$) and by the Kronecker symbol ($\eta\to\delta$, the two minus signs cancelling in each
product), respectively, while, with the vectors as fields 1, 2 and the scalars as fields 3, 4,
\begin{align}
V_{AA\chi\chi}^{\mu\nu;IJ}
&
=
-2\ii\gym^2g^{\mu\nu}\delta_{IJ}\bigl[\tr([C_1,C_3][C_2,C_4])+\tr([C_1,C_4][C_2,C_3])\bigr]
\, ,\\
V_{AAA\chi}
&
=
V_{A\chi\chi\chi}
=0 
\, .
\end{align}

\section{One-loop master integrals and coefficients}
\label{app:masters1}

The one-loop calculation provides both the first correction to the form factor and the lower-loop inputs needed for
two-loop subtraction. Here we relate the scalar basis of the graph reduction to the finite functions used in the main
text. We retain the exact mass dependence where it fixes the reference-tree normalization, and distinguish this
finite-mass information from the subsequent leading-power walking expansion.

\paragraph{Masters.}
The scalar basis consists of the tadpole, bubble and triangle, with every line carrying the Feynman prescription.
Using the measure of Eq.~\re{eq:FFloop-measure}, we define an integral by its explicit list of inverse propagators:
\begin{align}
 J_N[D_1,\ldots,D_N]
 &=\int[\dd k]\prod_{j=1}^N\frac1{D_j+\ii0}.
 \label{eq:JN-definition}
\end{align}
The tadpole, bubble and triangle in the basis \re{eq:FFmasters} are
\begin{alignat}{2}
&A_0(m^2)&{}={}&J_1[k^2-m^2]\notag\\*
&&{}={}&m^2\Bigl[\frac1\epsilon+1-\log\frac{m^2-\ii0}{\mu^2}\Bigr]
 +O(\epsilon),\notag\\
&B_0(t;m_1^2,m_2^2)&{}={}&J_2[k^2-m_1^2,(k+p)^2-m_2^2]\notag\\*
&&{}={}&\frac1\epsilon-\int_0^1\dd\tau\,
 \log\frac{\Delta_B-\ii0}{\mu^2}+O(\epsilon),\notag\\
&C_0(t_1,t_2,-Q^2;m_1^2,m_2^2,m_3^2)&{}={}&
 J_3[k^2-m_1^2,(k+p_1)^2-m_2^2, (k-p_2)^2-m_3^2]\notag\\*
&&{}={}&-\int_{\tau_i\ge0}\dd^3\tau\,\delta\Bigl(1-\sum_i\tau_i\Bigr)
 \bigl[\Delta_C-\ii0\bigr]^{-1}+O(\epsilon).
\end{alignat}
Here $\Delta_B=\tau m_1^2+(1-\tau)m_2^2-\tau(1-\tau)t$ and
$\Delta_C=\sum_i\tau_i m_i^2-\tau_1\tau_2t_1-\tau_1\tau_3t_2+\tau_2\tau_3Q^2$.
Here $\tau$ and $\tau_i$ are Feynman parameters. The ultraviolet poles of the three masters are $m^2/\epsilon$, $1/\epsilon$ and zero.

After imposing the Euclidean invariants in Eq.~\re{eq:one-loop-finite-masters}, the tadpole and finite bubbles are
\begin{align}
 A_0(wyQ^2)&=wyQ^2\left[\frac1\epsilon+1-L_Q+L-\ell_y\right]
 +O(\epsilon),\\*
 b_w&=2+L-L_Q+\bigl(\sqrt{1+4y}-1\bigr)\ell_y -2\sqrt{1+4y}
 \log\frac{1+\sqrt{1+4y}}2,\\*
 b_q&=2-L_Q+\bigl(\sqrt{1+4wy}-1\bigr)(\ell_y-L) -2\sqrt{1+4wy}
 \log\frac{1+\sqrt{1+4wy}}2,\\*
 d_w&=-1-\frac{2y}{\sqrt{1+4y}}
 \left[\ell_y-2\log\frac{1+\sqrt{1+4y}}2\right].
 \label{eq:exact-bubbles-wy}
\end{align}
Only the independent logarithms of Eq.~\re{eq:walklogs} occur here; the remaining logarithms are regular functions of the same ratios $w,y$.
Finally, the dimensionless triangle receives an exact Euclidean representation
\begin{align}
\mathcal T_0(w,y)
 &=-\int_0^1\dd\tau_1\int_0^{1-\tau_1}
       \frac{\dd\tau_2}{wy+w\tau_1(1-\tau_1)+\tau_2(1-\tau_1-\tau_2)}\notag\\*
 &=-4\int_0^1\frac{\dd\tau_1}
 {\sqrt{(1-\tau_1)^2+4w\tau_1(1-\tau_1)+4wy}}\notag\\*
 &\hspace{5mm}\times
 \operatorname{artanh}
 \frac{1-\tau_1}{\sqrt{(1-\tau_1)^2+4w\tau_1(1-\tau_1)+4wy}}.
 \label{eq:exact-triangle-wy}
\end{align}
The integration variables $\tau_1,\tau_2$ are dummy Feynman parameters, with $\tau_3=1-\tau_1-\tau_2$.
Both representations are convergent for positive $w,y$ and determine the continuation 
from the Euclidean chamber with the Feynman prescription. They retain every power correction; no region expansion has been used.

\paragraph{Finite-mass tree-normalization conversion.}
The finite-mass result requires more than the vertex correction: the reference tree in the denominator must be
converted to the same shifted pole as the matrix element in the numerator. The self-energy below supplies this
conversion and makes its power suppression in the walking limit explicit. We use the signed continuation specified in
Eq.~\re{eq:one-loop-finite-masters}, at the tree shell
$t_1=t_2=m_0^2$. Here $M_E^2$ is the positive magnitude of the Euclidean virtuality;
a positive mass magnitude from Section~\ref{sec:Realize} must not be
substituted for a continued Minkowski invariant.
At $c=\alpha=1$, write the self-energy on this shell as
\begin{align}
 \sigma_1(m_0^2)&=c_A A_0(m_z^2)
      +c_B B_0(m_0^2;m_z^2,m_z^2),\\*
 c_A&=c_A^{(0)}+\epsilon c_A^{(1)}+O(\epsilon^2),\\*
 c_B&=c_B^{(0)}+\epsilon c_B^{(1)}+O(\epsilon^2).
 \label{eq:exact-self-energy-coefficients}
\end{align}
The coefficients through the required evanescent order are
\begin{align}
 c_A^{(0)}&=-\frac{2(\Ds-2)}3,
 &c_A^{(1)}&=\frac{2(\Ds-2)}9,\\*
 c_B^{(0)}&=-\left(\frac{\Ds (m_0^2-4m_z^2)}{6} - \frac{(m_0^2-4m_z^2)}{3}\right),
 &c_B^{(1)}&=-\left(\frac{\Ds (m_0^2-4m_z^2)}{9} - \frac{2 (m_0^2-4m_z^2)}{9}\right) .
\end{align}
Thus the pole and finite part of the on-shell self-energy are
\begin{align}
 \frac{\sigma_1(-wQ^2)}{Q^2}
 &=\frac{c_mw}{\epsilon}
 +c_mw\left[(1+4y)b_w
       +4y(L_Q-L+\ell_y)+\frac23\right]+O(\epsilon).
 \label{eq:exact-pole-shift-wy}
\end{align}
The rational $2/3$ term includes the products of the evanescent
coefficients with master poles.
The auxiliary result normalized to the fixed classical tree has one-loop coefficient
$\Gamma_1+\sigma_1'+2\sigma_1/Q^2$. The shifted-tree normalization in Eq.~\re{eq:FFexpBare} instead gives
$f_{1,B}=\Gamma_1+\sigma_1'$, so conversion between these two normalizations contributes $-2\sigma_1^{\rm fin}/Q^2$ to the finite coefficient.
In the independent variables this is $-2\sigma_1^{\rm fin}(-wQ^2)/Q^2$.
It cancels the tadpole logarithm and additional mass-dependent rational terms of the fixed-tree expression.
This conversion is already included in Eq.~\re{eq:FF1held}; it is not another counterterm to add to $f_{1,B}$.
The distinct ultraviolet mass/reference-tree correction is $2wc_m/\epsilon$, derived in
Eqs.~\re{eq:FFreference-tree-counterterm} and \re{eq:FFmass-counterterm}; it enters the cancellation in
Eq.~\re{eq:composite-pole-cancel}. Both reference-tree distinctions vanish at walking leading power.

\paragraph{Null-point self-energy check.}
A useful check of the virtuality differentiation is obtained by setting the internal mass to zero while leaving the
external invariant independent. In the same 't Hooft-Feynman gauge, the resulting self-energy is
\begin{align}
 \left.\sigma_1(t)\right|_{m_z^2=0}
 &=B_0(t;0,0)\left[
 2(t-m_0^2)-\frac{\Ds-2}{2(d-1)}t\right].
 \label{eq:null-self-energy-check}
\end{align}
This is a fixed-gauge endpoint check, not the massive walking input.
It agrees with the leading term in Eq.~\re{eq:FFsig1}; the term proportional to $t-m_0^2$ confirms why
the virtuality derivative must precede the on-shell specialization.

\paragraph{Coefficients.}
The scalar functions become graph contributions only after they are combined with the numerator coefficients and
physical graph weights. The coefficients $c^{(0,1)}_{\gamma j}$ of Eq.~\re{eq:FFledger} were obtained by independent-mass
reduction with FORM and Kira~\cite{Vermaseren:2000nd,Klappert:2020nbg} from QGRAF-generated graphs~\cite{Nogueira:1991ex}.
Those of graphs \FFgraph{1}{3}, \FFgraph{1}{4}, \SEgraph{1}{2} and \tadgraph{1}{1}, \tadgraph{1}{2} and \SEgraph{1}{3} are given exactly in Eq.~\re{eq:FFbare1}; the leading-power values of the
others are collected in Table~\ref{tab:FF1lead}. Their sums reproduce the fixed-gauge coefficients in Sect.~\ref{sec:FF1} and the null-point check above.
The reductions were checked against direct Feynman-parameter quadrature at three generic Euclidean points: the bubbles,
the triangle in its dilogarithmic and one-dimensional forms, and the graph sums agree to better than $10^{-40}$.

\section{Two-loop scalar building blocks}
\label{sec:green-results}\label{app:graph-calculations156}
Several two-loop graph families can be reconstructed from lower-loop scalar functions, tensor coefficients and
integrals with a raised propagator. This appendix collects those ingredients and explains how they enter the
individual contributions of Sect.~\ref{sec:FF2local}. In particular, we give the scalar-potential response and the
raised triangle responsible for the surviving tadpole insertion. The complete bare pole sum is assembled separately
in Appendix~\ref{app:bare155}; the accompanying graph ledger identifies the diagrammatic contributions.

\subsection{Bubble, triangle and tensor coefficients}\label{sec:batches}
We first identify the one-loop functions that occur inside the two-loop graphs and list the tensor coefficients
multiplying them. This separates the scalar integration from the numerator algebra and fixes the normalization of
factorized products. The kinematic variables are those of Eq.~\eqref{eq:walkcond}; the loop
measure and regulated coupling are defined in
Eqs.~\eqref{eq:FFloop-measure} and \eqref{eq:coupling-regulated}.
The one-loop subgraph master basis is given in Eq.~\eqref{eq:FFmasters}.
The dimensionally regulated scalar multiplicity $n_s=D_s-d$ is retained
through the source contractions, ultraviolet subtractions and LSZ derivatives.

The graph results use the physical convention of
Section~\ref{sec:phase153}; one-loop product terms enter separately through
Eq.~\eqref{eq:products155}.

The external- and hard-scale bubbles, including their required Laurent
coefficients, are $B_w$ and $B_q$ in Eq.~\eqref{eq:bubbles155}.
Their finite parts agree with the leading-power limits of $b_w,b_q$
in Eq.~\eqref{eq:exact-bubbles-wy}. The physical triangle is
Eq.~\eqref{eq:triangle155}, in the same convention
as the exact finite triangle of Eq.~\re{eq:one-loop-finite-masters}.
For \FFgraph{2}{62} the remaining tensor coefficients are
\begin{align}
A_D&=(9-5D_s)/2,\\
B_D&=D_s-7/4,\\
C_D&=D_s^2-5D_s/2+23/12,\\
E_D&=D_s^2-7D_s/2+65/18.
\end{align}
These multiply, respectively, the triangle-hard-bubble product, the
mixed-scale bubble product, the hard-bubble square and the hard bubble
in Eq.~\eqref{main:FF2-62}. The remaining rational terms are displayed
explicitly there.
For \FFgraph{2}{24} the tensor coefficient is
\begin{align}
C_0^{(24)}&=\frac{D_s^2}{2} - \frac{3 D_s}{2} + \frac{17}{12}.
\end{align}
It multiplies $2(2-L_Q)^2$ in Eq.~\eqref{main:FF2-24}; the additional
term is $-(2-L_Q)/9$. It is unrelated to the triangle $C_0$ or the
constant term of the assembled logarithm. The tensor coefficients
for \FFgraph{2}{74} and \FFgraph{2}{75} are
\begin{align}
A_s&=D_s-14,\\
D_{75}&=5(7D_s+46).
\end{align}
The common coefficient of the cubic polynomial in
Eqs.~\eqref{main:FF2-74} and \eqref{main:FF2-75} is $A_s/9$;
$D_{75}/72$ multiplies $L_Q^2$ in Eq.~\eqref{main:FF2-75}.
The cubic functions $\mathcal P_3,\mathcal Q_3$ are defined directly
before those graph results.

\subsection{Reduction of the scalar-potential response}
\label{app:scalar-response-reduction}

The scalar-potential counterterm changes the masses carried by scalar lines, so its contribution can be obtained by
inserting the corresponding mass variation into the one-loop vertex and self-energy graphs. The quartic variation
of Sect.~\ref{sec:FF2} therefore produces propagators with raised powers rather than a new two-loop integral family.
We reduce these integrals at nonzero mass, before taking the virtuality derivatives and walking limit.

For propagators $D_i=(k-q_i)^2-m_z^2$, with $q_i$ fixed routing
shifts, scalar products are eliminated using
\begin{align}
 2(k-q_j)\cdot(k-q_i)
 &=D_i+D_j+2m_z^2-(q_i-q_j)^2.
 \label{eq:scalar-response-denominator-identity}
\end{align}
The routing shifts $q_i$ are not additional external invariants.
Dimensional regularization gives the integration-by-parts identity
\begin{align}
 0&=\int[\dd k]\,\frac{\partial}{\partial k^\mu}
 \left[\frac{(k-q_j)^\mu}{\prod_i(D_i+\ii0)}\right].
 \label{eq:scalar-response-IBP}
\end{align}
Tensor contractions are kept in $d$ dimensions.
The raised equal-mass bubble and its ordinary counterpart are
\begin{align}
 I_{21}(t)&\equiv
 \int[\dd k]\,
 \frac1{(k^2-m_z^2+\ii0)^2((k+p)^2-m_z^2+\ii0)},\\*
 B(t)&\equiv B_0(t;m_z^2,m_z^2),\qquad p^2=t.
 \label{eq:scalar-response-raised-bubble}
\end{align}
The raised tadpole $A_2$ is defined in
Eq.~\re{eq:raised-tadpole-response}. Applying the total-derivative
identity to the bubble gives
\begin{equation}\label{eq:I21155}
 I_{21}(t)=\frac{(d-3)B(t)-A_2(m_z^2)}{4m_z^2-t},\qquad
 A_2=(1-\epsilon)A_0(m_z^2)/m_z^2.
\end{equation}
The induced leading-power self-energy term is
$-h(m_0^2)^2I_{21}(t)$. Its virtuality derivatives are taken at
fixed $m_0^2,m_z^2$ before imposing the shell.
Combining this contribution with the insertions on the vertex
lines yields the response in Eq.~\re{eq:H155}.
The tadpole and bubble representations then give its Laurent
expansion in Eq.~\re{eq:Hdepth155}.

\subsection{Parameter representation of the raised triangle}
\label{app:raised-triangle-reduction}

The raised triangle is the exception to the usual mass suppression of tadpole insertions: its endpoint compensates
the explicit tadpole mass factor in \tadgraph{2}{65}. Its parameter representation exposes that endpoint and allows
the required finite-regulator expansion to be taken directly. For $C_{121}$ defined in
Eq.~\re{eq:raised-triangle-definition}, we set the Feynman parameter of the squared middle propagator to one
and denote the remaining parameters by $b,c$.
In the Euclidean kinematics $p_i^2=-M_E^2$,
$(p_1+p_2)^2=-Q^2$, the exact projective representation is
\begin{align}
 m_z^2Q^2C_{121}
 ={}&m_z^2Q^2 e^{\gamma_E\epsilon}\mu^{2\epsilon}
 \Gamma(2+\epsilon)\notag\\*
 &\times\int_0^\infty\dd \tau_1\,\dd \tau_2\,
 \frac{(1+\tau_1+\tau_2)^{2\epsilon}}
 {[m_z^2(1+\tau_1+\tau_2)^2+M_E^2(\tau_1+\tau_2)+Q^2\tau_1\tau_2]^{2+\epsilon}}.
 \label{eq:raised-triangle-exact-parameters}
\end{align}
The leading endpoint contribution is obtained by dropping the
$\tau_1+\tau_2$ corrections to the projective factor, giving
Eq.~\re{eq:raised-triangle-corner}. The omitted contributions
are suppressed by positive powers of $y$ or $w/y$, up to
logarithms, for every fixed $0<\eta<1$.

The $\tau_2$ integration is elementary and yields the expression
\begin{align}
 K_\epsilon
 ={}&m_z^2Q^2e^{\gamma_E\epsilon}\mu^{2\epsilon}
 \Gamma(1+\epsilon)\notag\\*
 &\times\int_0^\infty
 \frac{\dd \tau_1}
 {(M_E^2+Q^2\tau_1)(m_z^2+M_E^2\tau_1)^{1+\epsilon}}.
 \label{eq:raised-triangle-after-one-integration}
\end{align}
With the dummy parameter $\tau=m_z^2/(m_z^2+M_E^2 \tau_1)$, this becomes the Euler
integral in Eq.~\re{eq:K155}. At $\epsilon=0$ it reduces to
\begin{align}
 K_{\epsilon=0}&=\frac{L+\ell_y}{1-w/y}.
 \label{eq:raised-triangle-four-dimensional}
\end{align}
Its leading-power limit is $L+\ell_y$; the finite-regulator
expansion needed in the bare graph is Eq.~\re{eq:Kdepth155}.

\subsection{Polynomials entering the ladder contributions}
The ladder results separate naturally into a part proportional to the uplifted component multiplicity $D_s$ and a
remaining scalar part. Writing these pieces as polynomials makes their relation to the complete graph coefficients
explicit and allows the numerator reduction to be checked independently of the final sum. For the crossed ladder
\FFgraph{2}{70}, the scalar master also contributes the scale-independent constant $\kappa_{22}$.
In the physical normalization their relation to the graph results is
\begin{align}
 f_{2,\mathrm{FF}70}&=-D_sA_{70}-B_{70}+\kappa_{22},\\*
 f_{2,\mathrm{FF}71}&=-D_sA_{71}-B_{71}.
 \label{eq:scalar-template-map}
\end{align}
Thus these are the tensor-resolved forms of
Eqs.~\eqref{main:FF2-70} and \eqref{main:FF2-71}, not additional
contributions to their sum.
\begin{align}
A_{70}(L,L_Q)={}&\frac32LL_Q+
 \left(3\zeta_2-\frac{71}{12}\right)L
 -\frac{41}{36}L_Q^2+\frac{33}{4}L_Q\nonumber\\
 &+8\zeta_3-\frac{43}{9}\zeta_2-\frac{20503}{1296},\\
B_{70}(L,\ell_y,L_Q)={}&-\frac{L^4}{6}
 +\left(\frac23\ell_y+\frac53\right)L^3
 +\left(-\frac52\ell_y+\frac92\zeta_2-\frac{71}{8}\right)L^2\nonumber\\
 &+L\left[-\frac23\ell_y^3-\frac52\ell_y^2+\frac34L_Q
       -13\zeta_3-\frac{23}{12}\zeta_2+\frac{151}{24}\right]\nonumber\\
 &-\frac13\ell_y^4-\frac56\ell_y^3-4\zeta_2\ell_y^2
       -(10\zeta_2+4\zeta_3)\ell_y\nonumber\\
 &-\frac{17}{144}L_Q^2-\frac{613}{144}L_Q-21\zeta_4
       -\frac{65}{9}\zeta_2+\frac{21}{4}\zeta_3+\frac{98483}{5184}.
\end{align}

The same separation for the uncrossed ladder \FFgraph{2}{71} gives
\begin{align}
 A_{71}(L,\ell_y,L_Q)={}&
  \left(\frac72-\frac54L_Q\right)L^2
  +\left[\left(\frac52\ell_y-\frac12\right)L_Q
                 -7\ell_y-2\zeta_2-\frac5{12}\right]L\nonumber\\
 &-\frac{67}{9}L_Q^2
  +\left(\frac54\ell_y^2+\frac{565}{12}\right)L_Q
  -\frac72\ell_y^2+\frac{17}{2}\zeta_3
  +\frac{203}{36}\zeta_2-\frac{53539}{648},\\
 B_{71}(L,\ell_y,L_Q)={}&-\frac{L^4}{6}
  +\left(\frac23\ell_y+\frac53\right)L^3
  +\left(-\frac{11}{4}L_Q-\ell_y^2-5\ell_y-5\zeta_2-\frac{17}{8}\right)L^2\nonumber\\
 &+\left[\left(\frac{11}{2}\ell_y+\frac{93}{4}\right)L_Q
        -\frac23\ell_y^3-\frac{13}{4}\ell_y+22\zeta_3+5\zeta_2
        -\frac{1235}{24}\right]L\nonumber\\
 &+\frac{73}{36}L_Q^2
  +\left(\frac{11}{4}\ell_y^2-\frac{101}{3}\right)L_Q
  +\frac{\ell_y^4}{6}+\frac53\ell_y^3
  +\left(4\zeta_2-\frac{13}{8}\right)\ell_y^2\nonumber\\
 &+(10\zeta_2+12\zeta_3)\ell_y-\frac{19}{2}\zeta_3
  -\frac{121}{36}\zeta_2+19\zeta_4+\frac{98369}{1296}.
\end{align}
\section{Complete bare finite coefficient and graph pole ledger}\label{app:bare155}
Before ultraviolet subtraction, the genuine two-loop graphs must be combined with the lower-loop LSZ products.
The resulting finite bare coefficient $f_{2,0}$ is defined by the Laurent expansion in Eq.~\re{eq:bare155} and enters
the renormalized coefficient through Eq.~\re{eq:finite155}. We give its full logarithmic polynomial here, followed
by the individual graph poles needed to reconstruct the ultraviolet cancellation. Every term includes its physical
graph factor before summation. To display the polynomial in a readable order, write
$f_{2,0}=\sum_{j=0}^4 f_{2,0}^{[j]}$ by total polynomial degree
in $L,\ell_y,L_Q$; $D_s$ and zeta values have degree zero. The crossed-ladder
constant $\kappa_{22}$ retains its numerical-only status. The components are listed from the highest logarithmic
degree to the constant term, so the iterated double logarithms are separated from the lower-degree terms entering
finite subtraction and matching.

\begingroup\begin{equation}\begin{aligned}f_{2,0}^{[4]}&=\frac{L^{4}}{2} + \frac{\ell_{y}^{4}}{2} + L^{2} \ell_{y}^{2} - 2 L^{3} \ell_{y} + 2 L \ell_{y}^{3}\end{aligned}\end{equation}\endgroup

\begingroup\begin{equation}\begin{aligned}f_{2,0}^{[3]}&=- 13 L^{3} - \frac{26 \ell_{y}^{3}}{3} + \frac{D_{s} L^{3}}{2} - \frac{52 L_Q \ell_{y}^{2}}{3} + \frac{D_{s} \ell_{y}^{3}}{3}\\ & + \frac{13 L \ell_{y}^{2}}{3} + \frac{52 L^{2} L_Q}{3} + \frac{104 L^{2} \ell_{y}}{3} - \frac{104 L L_Q \ell_{y}}{3}\\ & - \frac{4 D_{s} L^{2} \ell_{y}}{3} - \frac{2 D_{s} L^{2} L_Q}{3} - \frac{D_{s} L \ell_{y}^{2}}{6} + \frac{2 D_{s} L_Q \ell_{y}^{2}}{3}\\ & + \frac{4 D_{s} L L_Q \ell_{y}}{3}\end{aligned}\end{equation}\endgroup

\begingroup\begin{equation}\begin{aligned}f_{2,0}^{[2]}&=\frac{127 \ell_{y}^{2}}{9} + \frac{229 L^{2}}{9} + \frac{338 L_Q^{2}}{9} + 4 L_Q \ell_{y} - \frac{688 L L_Q}{9}\\ & - \frac{26 D_{s} L_Q^{2}}{9} - \frac{23 D_{s} L^{2}}{6} - \frac{D_{s} \ell_{y}^{2}}{18} + \frac{\pi^{2} L^{2}}{3}\\ & + \frac{\pi^{2} \ell_{y}^{2}}{3} + \frac{D_{s}^{2} L^{2}}{18} + \frac{D_{s}^{2} L_Q^{2}}{18} + \frac{194 L \ell_{y}}{9}\\ & - 2 D_{s} L_Q \ell_{y} - \frac{2 \pi^{2} L \ell_{y}}{3} - \frac{D_{s}^{2} L L_Q}{9} + \frac{29 D_{s} L \ell_{y}}{9}\\ & + \frac{58 D_{s} L L_Q}{9}\end{aligned}\end{equation}\endgroup

\begingroup\begin{equation}\begin{aligned}f_{2,0}^{[1]}&=- \frac{2306 L_Q}{27} - \frac{14 \ell_{y}}{3} + \frac{3974 L}{27} - 4 \ell_{y} \zeta\left(3\right) + 4 L \zeta\left(3\right)\\ & - \frac{586 D_{s} L_Q}{27} - \frac{149 D_{s} L}{27} - \frac{26 \pi^{2} L}{9} - \frac{4 \pi^{2} \ell_{y}}{3} - \frac{D_{s}^{2} L}{27}\\ & + \frac{7 D_{s} \ell_{y}}{3} + \frac{119 D_{s}^{2} L_Q}{54} + \frac{\pi^{2} D_{s} L}{9}\end{aligned}\end{equation}\endgroup

\begingroup\begin{equation}\begin{aligned}f_{2,0}^{[0]}&=\frac{1588}{27} + \kappa_{22} - \frac{340 \zeta\left(3\right)}{3} - \frac{305 D_{s}^{2}}{54} - \frac{163 \pi^{2}}{54} + \frac{17 \pi^{4}}{90}\\ & + \frac{647 D_{s}}{9} - \frac{\pi^{2} D_{s}^{2}}{216} + \frac{5 \pi^{2} D_{s}}{27} + \frac{14 D_{s} \zeta\left(3\right)}{3}\end{aligned}\end{equation}\endgroup

\subsection{Bare graph poles in the physical normalization}
The pole coefficients provide a graph-by-graph check of the bare assembly before any local subtraction is made.
The table contains genuine two-loop graphs only: the one-loop products are added explicitly through
Eq.~\eqref{eq:products155}, rather than being included in an individual graph entry. Its residues can therefore be
summed independently and compared with the complete Laurent coefficient in the main text.
The columns give the poles of the dimensionless physical bare
contribution $f_{2,\gamma,B}$ of Section~\ref{sec:phase153}. For a SE graph family
this includes the virtuality derivative required by
Eq.~\eqref{eq:FF2LSZ}, not an undifferentiated self-energy integral.
\begingroup\small\setlength{\tabcolsep}{3pt}
\begin{longtable}{p{25mm}p{40mm}p{91mm}}
\caption{Genuine graph bare poles. The sum including products gives
Eqs.~\eqref{eq:bare155}-\eqref{eq:A1155}.}\label{tab:bare155}\\
\toprule Physical graph family & $[f_{2,\gamma,B}]_{\epsilon^{-2}}$ & $[f_{2,\gamma,B}]_{\epsilon^{-1}}$\\\midrule\endfirsthead
\toprule Physical graph family & $[f_{2,\gamma,B}]_{\epsilon^{-2}}$ & $[f_{2,\gamma,B}]_{\epsilon^{-1}}$\\\midrule\endhead
\Atlas{\FFgraph{2}{18}} & $\frac{9}{8}$ & $\frac{9 L}{4} - \frac{9 L_Q}{4} + \frac{81}{16}$\\\midrule
\Atlas{\FFgraph{2}{23}} & $\frac{9}{2}$ & $9 L - 9 L_Q + 18$\\\midrule
\Atlas{\FFgraph{2}{35}} & $\frac{11 D_{s}}{24} - \frac{77}{12}$ & $\frac{11 D_{s} L}{12} - \frac{11 D_{s} L_Q}{12} + \frac{329 D_{s}}{144} - \frac{77 L}{6} + \frac{77 L_Q}{6} - \frac{2327}{72}$\\\midrule
\Atlas{\FFgraph{2}{37}} & $- \frac{D_{s}}{4} - \frac{5}{2}$ & $- \frac{D_{s} L}{2} + \frac{D_{s} L_Q}{2} - \frac{5 D_{s}}{4} - 5 L + 5 L_Q - 11$\\\midrule
\Atlas{\FFgraph{2}{12}} & $\frac{27}{8}$ & $\frac{27 L}{4} - \frac{27 L_Q}{4} + \frac{27}{2}$\\\midrule
\Atlas{\tadgraph{2}{65}} & $0$ & $\left(1 - D_{s}\right) \left(L + \ell_{y}\right)$\\\midrule
\Atlas{\FFgraph{2}{17}} & $\frac{53 D_{s}}{36} - \frac{31}{12}$ & $- \frac{53 D_{s} L_Q}{18} + \frac{1775 D_{s}}{216} + \frac{31 L_Q}{6} - \frac{1045}{72}$\\\midrule
\Atlas{\FFgraph{2}{51}} & $- \frac{55 D_{s}}{72} - \frac{193}{24}$ & $\frac{3 D_{s} L}{2} + \frac{55 D_{s} L_Q}{36} - \frac{2611 D_{s}}{432} + 3 L^{2} - 6 L \ell_{y} - \frac{87 L}{4} + \frac{193 L_Q}{12} - 3 \ell_{y}^{2} - \frac{4657}{144}$\\\midrule
\Atlas{\FFgraph{2}{53}} & $- \frac{9 D_{s}}{8} - \frac{9}{8}$ & $\frac{9 D_{s} L_Q}{4} - \frac{123 D_{s}}{16} + \frac{9 L^{2}}{4} - \frac{9 L \ell_{y}}{2} - \frac{99 L}{8} + \frac{9 L_Q}{4} - \frac{9 \ell_{y}^{2}}{4} + \frac{15}{16}$\\\midrule
\Atlas{\FFgraph{2}{27}} & $\frac{9}{8}$ & $\frac{3 D_{s}}{4} + \frac{9 L}{4} - \frac{9 L_Q}{4} + \frac{63}{16}$\\\midrule
\Atlas{\FFgraph{2}{28}} & $\frac{17 D_{s}}{8} - \frac{27}{8}$ & $- \frac{17 D_{s} L_Q}{4} + \frac{169 D_{s}}{16} + \frac{27 L_Q}{4} - \frac{267}{16}$\\\midrule
\Atlas{\FFgraph{2}{38}} & $\frac{373}{24} - \frac{529 D_{s}}{72}$ & $\frac{529 D_{s} L_Q}{36} - \frac{14593 D_{s}}{432} - \frac{373 L_Q}{12} + \frac{1137}{16}$\\\midrule
\Atlas{\FFgraph{2}{52}} & $- \frac{9 D_{s}}{8} - \frac{3}{8}$ & $\frac{9 D_{s} L_Q}{4} - \frac{117 D_{s}}{16} + \frac{3 L_Q}{4} + \frac{39}{16}$\\\midrule
\Atlas{\FFgraph{2}{48}} & $- \frac{3 D_{s}}{4} - \frac{51}{16}$ & $- \frac{3 D_{s} L}{2} + \frac{3 D_{s} L_Q}{2} - \frac{49 D_{s}}{12} - \frac{51 L}{8} + \frac{51 L_Q}{8} - \frac{409}{32}$\\\midrule
\Atlas{\FFgraph{2}{24}} & $D_{s}^{2} - 3 D_{s} + \frac{17}{6}$ & $- 2 D_{s}^{2} L_Q + 4 D_{s}^{2} + 6 D_{s} L_Q - 12 D_{s} - \frac{17 L_Q}{3} + \frac{101}{9}$\\\midrule
\Atlas{\FFgraph{2}{36}} & $\frac{53}{18} - \frac{59 D_{s}}{36}$ & $\frac{59 D_{s} L_Q}{18} - \frac{1961 D_{s}}{216} - \frac{53 L_Q}{9} + \frac{1727}{108}$\\\midrule
\Atlas{\FFgraph{2}{49}} & $- 2 D_{s} - \frac{1}{2}$ & $4 D_{s} L_Q - \frac{91 D_{s}}{8} - \frac{9 L}{8} + L_Q + \frac{15}{16}$\\\midrule
\Atlas{\FFgraph{2}{62}} & $- D_{s}^{2} + \frac{3 D_{s}}{2} - \frac{1}{6}$ & $2 D_{s}^{2} L_Q - 5 D_{s}^{2} + \frac{5 D_{s} L^{2}}{4} - \frac{5 D_{s} L \ell_{y}}{2} - D_{s} L - 3 D_{s} L_Q - \frac{5 D_{s} \ell_{y}^{2}}{4} + \frac{19 D_{s}}{2} - \frac{9 L^{2}}{4} + \frac{9 L \ell_{y}}{2} + \frac{7 L}{4} + \frac{L_Q}{3} + \frac{9 \ell_{y}^{2}}{4} - \frac{40}{9}$\\\midrule
\Atlas{\FFgraph{2}{63}} & $- \frac{45}{4}$ & $\frac{9 L^{2}}{4} - \frac{9 L \ell_{y}}{2} - \frac{171 L}{8} + \frac{45 L_Q}{2} - \frac{9 \ell_{y}^{2}}{4} - \frac{333}{8}$\\\midrule
\Atlas{\FFgraph{2}{70}} & $\frac{41 D_{s}}{36} + \frac{17}{144}$ & $\frac{3 D_{s} L}{2} - \frac{41 D_{s} L_Q}{18} + \frac{1343 D_{s}}{216} + \frac{3 L}{4} - \frac{17 L_Q}{72} - \frac{1375}{864}$\\\midrule
\Atlas{\FFgraph{2}{71}} & $\frac{67 D_{s}}{9} - \frac{73}{36}$ & $- \frac{5 D_{s} L^{2}}{4} + \frac{5 D_{s} L \ell_{y}}{2} - \frac{D_{s} L}{2} - \frac{134 D_{s} L_Q}{9} + \frac{5 D_{s} \ell_{y}^{2}}{4} + \frac{2239 D_{s}}{54} - \frac{11 L^{2}}{4} + \frac{11 L \ell_{y}}{2} + \frac{93 L}{4} + \frac{73 L_Q}{18} + \frac{11 \ell_{y}^{2}}{4} - \frac{7027}{216}$\\\midrule
\Atlas{\FFgraph{2}{72}} & $\frac{22 D_{s}}{9} + \frac{275}{72}$ & $- \frac{D_{s} L^{2}}{3} + \frac{2 D_{s} L \ell_{y}}{3} + \frac{5 D_{s} L}{3} - \frac{44 D_{s} L_Q}{9} + \frac{D_{s} \ell_{y}^{2}}{3} + \frac{3187 D_{s}}{216} - \frac{11 L^{2}}{6} + \frac{11 L \ell_{y}}{3} + \frac{41 L}{3} - \frac{275 L_Q}{36} + \frac{11 \ell_{y}^{2}}{6} + \frac{3911}{432}$\\\midrule
\Atlas{\FFgraph{2}{74}} & $\frac{29 D_{s}}{36} + \frac{157}{18}$ & $\frac{D_{s} L^{2}}{3} - \frac{2 D_{s} L \ell_{y}}{3} - \frac{5 D_{s} L}{3} - \frac{29 D_{s} L_Q}{18} - \frac{D_{s} \ell_{y}^{2}}{3} + \frac{1505 D_{s}}{216} - \frac{14 L^{2}}{3} + \frac{28 L \ell_{y}}{3} + \frac{70 L}{3} - \frac{157 L_Q}{9} + \frac{14 \ell_{y}^{2}}{3} + \frac{3625}{108}$\\\midrule
\Atlas{\FFgraph{2}{75}} & $\frac{35 D_{s}}{72} + \frac{115}{36}$ & $\frac{D_{s} L^{2}}{6} - \frac{D_{s} L \ell_{y}}{3} - \frac{35 D_{s} L_Q}{36} - \frac{D_{s} \ell_{y}^{2}}{6} + \frac{5 D_{s} \ell_{y}}{6} + \frac{1583 D_{s}}{432} - \frac{7 L^{2}}{3} + \frac{14 L \ell_{y}}{3} + 9 L - \frac{115 L_Q}{18} + \frac{7 \ell_{y}^{2}}{3} + \frac{4 \ell_{y}}{3} + \frac{3367}{216}$\\\midrule
\Atlas{\SEgraph{2}{2}} & $0$ & $\frac{3}{8} - \frac{3 D_{s}}{8}$\\\midrule
\Atlas{\SEgraph{2}{4}} & $\frac{3 D_{s}}{8} - \frac{27}{16}$ & $\frac{3 D_{s} L}{4} - \frac{3 D_{s} L_Q}{4} + \frac{11 D_{s}}{16} - \frac{27 L}{8} + \frac{27 L_Q}{8} - \frac{135}{32}$\\\midrule
\Atlas{\SEgraph{2}{5}} & $- \frac{27}{8}$ & $- \frac{27 L}{4} + \frac{27 L_Q}{4} - \frac{55}{4}$\\\midrule
\Atlas{\SEgraph{2}{7}} & $\frac{D_{s}}{18} + \frac{335}{144}$ & $\frac{D_{s} L}{9} - \frac{D_{s} L_Q}{9} + \frac{143 D_{s}}{216} + \frac{335 L}{72} - \frac{335 L_Q}{72} + \frac{7739}{864}$\\\midrule
\Atlas{\SEgraph{2}{8}} & $\frac{407}{72} - \frac{23 D_{s}}{36}$ & $- \frac{23 D_{s} L}{18} + \frac{23 D_{s} L_Q}{18} + \frac{D_{s} \ell_{y}}{2} - \frac{425 D_{s}}{216} + \frac{407 L}{36} - \frac{407 L_Q}{36} - 3 \ell_{y} + \frac{9341}{432}$\\\midrule
\bottomrule\end{longtable}\endgroup

\section{Tadpole contributions: cancellation and power suppression}
\label{app:suppressed}
\label{sec:tadpole-gate}
Tadpole subgraphs disappear from the leading walking result for two distinct reasons: a detachable one-point
function is cancelled by the holding source, whereas a remaining massive tadpole is generally power suppressed.
The distinction matters because a raised propagator can compensate the explicit mass factor, as happens in
\tadgraph{2}{65}. We establish the source cancellation first and then display the power counting of the suppressed
products. Among the 77 ordinary physical graph families, 39 contain a tadpole or a detachable ordinary one-point
branch. This classification concerns complete source-weighted amplitudes; the eight SE/LSZ graph families are not
part of the ordinary-vertex census.

\subsection{Holding-source cancellation}
\label{native:tad2-5}\label{native:tad2-6}
The held-background condition cancels the renormalized force on the background. It therefore removes every
terminal ordinary one-point attachment, irrespective of the rest of the graph, provided the ordinary and source
terms are combined before their zero-momentum connector is removed. To show this explicitly, use the linear source
of Eq.~\eqref{eq:held}, retain each scalar/color connector port $a=(I,A)$, and keep the finite
zero-mode connector regulator $\lambda$. A terminal ordinary 1PI
one-point leaf of a graph has no operator insertion and no physical
external leg. Its source-held completion is, for an arbitrary common
response tensor $\mathcal C_R^a$,
\begin{equation}
\WalkingResult{
\mathcal C_R^a\Delta_{\lambda,ab}
\bigl(T_{\rm ren}^b-H^b\bigr)=0.}
\label{eq:tad-held-v137}
\end{equation}
Here $a=(I,A)$ combines a scalar-flavor index with an adjoint color
index, $\Delta_{\lambda,ab}$ is the regulated zero-momentum diagonal
scalar propagator, and $\mathcal C_R^a$ is the amputated remainder of
the connected operator matrix element attached to that port, including
its local subtractions. Repeated ports are summed. The one-point
function $T_{\rm ren}^a$ and the holding source $H^a$ are the components
of Eqs.~\eqref{eq:QSforce} and \eqref{eq:held}.
This equality holds before the zero-momentum connector limit. The
one-point function and the holding source include the same finite
subtraction prescription. Counterterms of the common response and
connector multiply the same zero difference.

For two independent leaves the exponential of the same linear action
fixes the combinatorics. Writing $T=T_{\rm ren}$ in the following two
equations gives
\begin{align}
\left[e^{i\hbar T\chi}e^{-i\hbar H\chi}\right]_{\hbar^2}
&=-\tfrac12\chi^2(T-H)^2,\\
\mathcal G_{\rm two\ leaves}
&=\tfrac12\mathcal C_R^{ab}\Delta_{\lambda,aa'}
\Delta_{\lambda,bb'}
\bigl[TT-TH-HT+HH\bigr]^{a'b'}=0.
\end{align}
Here $\mathcal G_{\rm two\ leaves}$ denotes this two-leaf contribution,
$\mathcal C_R^{ab}$ is the corresponding two-port remainder, and
$HH$ denotes two insertions of the linear source. For nested
branches, apply Eq.~\eqref{eq:tad-held-v137} recursively to terminal
one-point vertices of the 1PI effective-action tree.
In particular, the complete \tadgraph{2}{5} and \tadgraph{2}{6} groups
vanish by this identity, with one and two detachable one-point branches,
respectively. The cancellation is imposed before the diagonal-connector
regulator is removed.

\subsection{Power-suppressed tadpole products}
\label{green:tad2-1}\label{green:tad2-31}\label{green:tad2-32}
Once the detachable one-point branches have cancelled, the remaining question is whether the tadpole mass factor
survives in the complete graph or is offset by a singular integral. After the 23 held-source cancellations, fifteen
of the remaining sixteen tadpole-containing graph families are power suppressed in the walking limit of
Eq.~\eqref{eq:walkcond}.
The sole exception is \tadgraph{2}{65}, drawn in
Fig.~\ref{fig:FF2graphs}: its raised middle triangle propagator
compensates the tadpole mass factor, yielding Eq.~\eqref{main:tad2-65}.
The explicit tadpole products below illustrate the suppressed contributions.
Write
\begin{equation}
\ell_m=L_Q-L+\ell_y.
\end{equation}
Thus $\ell_m=\log(m_z^2/\mu^2)$ is a combination of the logarithms in
Eq.~\eqref{eq:walklogs}, not an independent scale variable.
The graph-local coefficients use the auxiliary source-insertion sign
convention explained in Sect.~\ref{sec:phase153}; the corresponding
physical graph contributions have the opposite overall sign.

For this subsection only, $A_\alpha(m^2)$ denotes the one-loop tadpole
with denominator $(k^2-m^2+\ii0)^\alpha$ and measure
\eqref{eq:FFloop-measure}. In particular,
$A_1(m^2)=A_0(m^2)$ in the notation of Eq.~\eqref{eq:FFmasters}.
The tadpole identities and their locally subtracted limits are
\begin{align}
A_1(m^2)&=-e^{\gamma_E\epsilon}\mu^{2\epsilon}
 (m^2)^{1-\epsilon}\Gamma(-1+\epsilon),\\
A_2(m^2)&=\partial_{m^2}A_1(m^2),\\
R(A_1A_2)&\longrightarrow m^2(\ell_m^2-\ell_m),\\
R(A_1^2)&\longrightarrow m^4(1-\ell_m)^2.
\end{align}
Here $m^2=m_z^2$, and $R$ is the local subtraction operation of
Eq.~\eqref{eq:forest156main}; the arrows denote its finite
$\epsilon\to0$ limits.
Their source-weighted finite terms are, individually,
\begin{align}
\Delta f_{2,\mathrm{tad}1}^{R}
&=(D_s-1)^2wy(\ell_m^2-\ell_m),\\
\Delta f_{2,\mathrm{tad}31}^{R}
&=2(D_s-1)^2w^2y(\ell_m^2-\ell_m),\\
\Delta f_{2,\mathrm{tad}32}^{R}
&=-4(D_s-1)^2w^3y^2(1-\ell_m)^2.
\end{align}
The corresponding walking degrees are $1+\eta$, $2+\eta$ and
$3+2\eta$, all strictly positive.

\section{The coupled graph family \FFgraph{2}{48}}
\label{app:ff48}

This family couples the two loop momenta through a massless internal line,
while the remaining propagators carry the common mass $m_z$.
We give its integral definition, the region expansion used to extract the
walking hierarchy, and the resulting Laurent coefficients. The final
paragraph distinguishes the algebraic regulator checks from the numerical
tests of the asymptotic expression.

\begin{figure}[h]
\centering
\begin{tikzpicture}[scale=.85,line width=.6pt,
 lab/.style={font=\small,fill=white,inner sep=1.2pt}]
\coordinate (O) at (0,3.0);
\coordinate (V1) at (-2.4,0); \coordinate (V2) at (2.4,0);
\coordinate (A) at (0,0);
\draw (O)--(V1) node[midway,lab]{1};
\draw (O)--(V2) node[midway,lab]{5};
\draw[dashed] (O)--(A) node[midway,lab]{3};
\draw (A)--(V1) node[midway,lab,below=2pt]{2};
\draw (A)--(V2) node[midway,lab,below=2pt]{4};
\draw (V1)--++(-.8,-.8) node[below left,font=\small]{$p_1$};
\draw (V2)--++(.8,-.8) node[below right,font=\small]{$p_2$};
\foreach \p in {V1,V2,A} \fill (\p) circle (2.2pt);
\draw[fill=white,line width=.4pt] (O) circle (4pt);
\draw[line width=.5pt] ($(O)+(-2.1pt,-2.1pt)$)--($(O)+(2.1pt,2.1pt)$);
\draw[line width=.5pt] ($(O)+(-2.1pt,2.1pt)$)--($(O)+(2.1pt,-2.1pt)$);
\draw[densely dotted,->,shorten >=4.5pt] ($(O)+(0,.9)$)
 node[above,font=\small]{$q$} -- (O);
\end{tikzpicture}
\caption{\FFgraph{2}{48}. The circled cross is the operator insertion, with incoming
momentum $q$ indicated by the dotted arrow. Line 3 (dashed) is massless;
lines 1, 2, 4, 5 carry mass squared $m_z^2=wyQ^2$.}
\label{fig:ff48}
\end{figure}

\paragraph{Integral.}
With the common coupling and color factor removed, the graph function is
\begin{align}
G_{48}
&=\int[\dd k_1][\dd k_2]\,
 \frac{\mathrm{Num}_{48}(k_1,k_2;p_1,p_2)}
 {D_1D_2D_3D_4D_5}\, .
\end{align}
The denominator contains the propagators
$D_1=(k_2-p_1)^2-m_z^2$, $D_2=k_2^2-m_z^2$, $D_3=(k_1-k_2-p_2)^2$,
$D_4=(k_1-p_2)^2-m_z^2$, $D_5=k_1^2-m_z^2$,
with the Feynman prescription understood.
The external invariants are $p_i^2=-M_E^2$ and $Q^2=\mu^2=1$.
Here $\mathrm{Num}_{48}$ is the polynomial in loop and external momenta
obtained by contracting the operator and action vertices of
Appendix~\ref{App:FeynRules}, including the component sums and the
transverse projection. It is therefore fixed by the graph's Feynman
rules, not by its propagator connectivity alone: $G_{48}$ is a
numerator-weighted graph function rather than a unit-numerator master.
The independent parameter construction gives the same Symanzik
polynomials as FIESTA. The ancillary material contains the available
reduced expressions and summaries of the checks.

\paragraph{Regions and analytic regulator.}
The hierarchy $w\ll y\ll1$ is extracted before the remaining parameter
integrations. Six regions at fixed $y$ expand into
$1+3+5+9+5+3=26$ sub-regions, in agreement with a direct census of the
joint hierarchy. Twelve contribute at leading power $y^0$; thirteen
begin at $y^1$ and one at $y^2$.

Individual regions require an auxiliary analytic regulator in addition
to the dimensional regulator $\epsilon$. Denote the analytic shift of
the power of line $j$ by $\rho_j$. We generate these shifts from one
complex parameter $\rho$ and a fixed vector of weights:
\begin{align}
 \rho_j&=\rho\nu_j,\qquad a_j=1+\rho_j,\\*
 \mathcal R_\rho(x)&=\prod_jx_j^{\rho_j},
 &\mathcal N_\rho&=\prod_j\Gamma(1+\rho_j)^{-1}\, .
\end{align}
Thus both the parameter measure and its Gamma-function normalization
inherit the same analytic shifts. Integer changes in propagator powers
from numerator reduction retain those shifts.
For this family
$\boldsymbol\nu=(2,-2,0,3,-3)$ on lines $1,\dots,5$.
The weights obey
\begin{align}
 \sum_j\nu_j&=0\, ,
\end{align}
so the insertion preserves projective rescaling.
For the parameter integrations we set
\begin{align}
 \rho&=r\epsilon .
 \label{eq:coupled-regulator}
\end{align}
The dimensionless slope $r$ labels the regulator direction.
All 36 evaluations, comprising the twelve
leading sub-regions at three slopes, were completed through
$O(\epsilon)$ with the Cheng-Wu choice $x_2=1$.
After summing all retained regions, we extract the constant Laurent
coefficient in $r$ of each $\epsilon$ coefficient. This is not the
finite part at one nonzero slope and does not presume an interchange
of regulator limits.

\paragraph{Result.}
The scalar graph coefficient with the common coupling and color
prefactor removed is
\begin{align}
\label{eq:ff48-native-result}
G_{48}
&=
{}
\frac{1}{\epsilon^2}
\Bigl(\tfrac{51}{16}+\tfrac34\Ds\Bigr)
+\frac1\epsilon
\Bigl(\tfrac{409}{32}+\tfrac{49}{12}\Ds
+
\Bigl(\tfrac{51}{8}+\tfrac32\Ds\Bigr) L
\Bigr)
\\
&
-
\tfrac32 L^3
+
\tfrac94
L^2\ell_y + \tfrac94 L\ell_y^2+\tfrac34\ell_y^3
+
\Bigl(\tfrac{265}{16}+\tfrac34\Ds\Bigr)L^2\nonumber\\
&+\Bigl(\tfrac{533}{48} + \tfrac{109}{12}\Ds- \tfrac{3\pi^2}{4}\Bigr) L
+
\tfrac{3\pi^2}{2}\ell_y
+\Ds\Bigl(\tfrac{1099}{72}-\tfrac{3\pi^2}{8}\Bigr)+\tfrac{22855}{576} + \tfrac{275 \pi^2}{96}
+O(\epsilon)
\, . \nonumber
\end{align}
There are no $1/\epsilon^4$ and $1/\epsilon^3$ poles and no $L^4$ term; only $\zeta_2$ appears through $O(\epsilon^0)$.
The function $G_{48}$ enters Eq.~\re{eq:bare-FF2-48}, which
restores the hard-scale factor and the physical graph sign before
the bare sum.

\paragraph{Regulator and numerical checks.}
Regulator independence tests the analytic continuation, whereas
comparison with an independently evaluated integral tests the walking
expansion. The coefficient of $\epsilon^n$ is a polynomial of degree $n+2$ in $r$; the poles through $1/\epsilon^2$ are $r$
independent, and the $r^0$ terms were extracted exactly. Two further regulator patterns, $(1,-1,0,2,-2)$ and
$(3,-3,0,1,-1)$, give identical $r^0$ intercepts at every order, and a fourth slope is predicted exactly by the fit from the
other three. The FIESTA expansion (58 asymptotic and 54 single-scale integrals, relative accuracy $10^{-9}$) agrees at
$y=10^{-4},10^{-5},10^{-6}$: the gaps in the $L^3$ and $L^2$ coefficients and in the poles are at the $10^{-4}$ level or below
and shrink with $y$ as expected for $O(y\ell_y^{k})$ corrections; the log-free finite part, about $-10^3$, agrees to
$10^{-4}$-$10^{-6}$ relative. The largest residual gap, $+0.29$ in that coefficient at $y=10^{-6}$, is attributed to numerical
noise but was not resolved further.

\section{The coupled graph family \FFgraph{2}{72}}
\label{app:ff72}

The second coupled family has six internal lines, one of which is
massless. We follow the same sequence as for \FFgraph{2}{48}: the
numerator-weighted integral and its mass routing are specified first,
then the region expansion and regulator prescription, followed by the
Laurent result and the checks that delimit its numerical accuracy.

\begin{figure}[h]
\centering
\begin{tikzpicture}[scale=.85,line width=.6pt,
 lab/.style={font=\small,fill=white,inner sep=1.2pt}]
\coordinate (O) at (0,3.0);
\coordinate (V1) at (-2.4,0); \coordinate (V2) at (2.4,0);
\coordinate (A) at (0,0); \coordinate (B) at (1.2,1.5);
\draw (O)--(V1) node[midway,lab]{3};
\draw (O)--(B) node[midway,lab]{1};
\draw (B)--(V2) node[midway,lab]{5};
\draw (V1)--(A) node[midway,lab]{2};
\draw (A)--(V2) node[midway,lab]{6};
\draw[dashed] (A)--(B) node[midway,lab]{4};
\draw (V1)--++(-.8,-.8) node[below left,font=\small]{$p_1$};
\draw (V2)--++(.8,-.8) node[below right,font=\small]{$p_2$};
\foreach \p in {V1,V2,A,B} \fill (\p) circle (2.2pt);
\draw[fill=white,line width=.4pt] (O) circle (4pt);
\draw[line width=.5pt] ($(O)+(-2.1pt,-2.1pt)$)--($(O)+(2.1pt,2.1pt)$);
\draw[line width=.5pt] ($(O)+(-2.1pt,2.1pt)$)--($(O)+(2.1pt,-2.1pt)$);
\draw[densely dotted,->,shorten >=4.5pt] ($(O)+(0,.9)$)
 node[above,font=\small]{$q$} -- (O);
\end{tikzpicture}
\caption{\FFgraph{2}{72}. The circled cross is the operator insertion, with incoming
momentum $q$ indicated by the dotted arrow. Line 4 (dashed) is massless;
all other lines carry mass squared $m_z^2=wyQ^2$.}
\label{fig:ff72}
\end{figure}

\paragraph{Integral.}
In the same hard-scale units and graph convention as above, write
\begin{align}
G_{72}
&=\int[\dd k_1][\dd k_2]\,
 \frac{\mathrm{Num}_{72}(k_1,k_2;p_1,p_2)}
 {D_1D_2D_3D_4D_5D_6}\, .
\end{align}
The six propagators are
$D_1=-(k_2-p_1-p_2)^2+m_z^2$, $D_2=-(k_2-p_1)^2+m_z^2$,
$D_3=-k_2^2+m_z^2$, $D_4=-(k_1-k_2+p_1)^2$,
$D_5=-(k_1-p_2)^2+m_z^2$, $D_6=-k_1^2+m_z^2$,
with $p_i^2=-M_E^2$ and $Q^2=\mu^2=1$.
The Feynman prescription is understood in this denominator convention.
The numerator $\mathrm{Num}_{72}$ contains the contractions and
component sums of the graphs belonging to this family, with the
normalization of Eq.~\re{eq:coupled-bare-normalization}.
As for $G_{48}$, replacing this numerator by unity would define a
different integral. The ancillary material supplies the available
reduced coefficients and summaries of the checks, not a complete
numerator-level reconstruction.

\paragraph{Regions and analytic regulator.}
Six regions expand into $1+9+5+9+3+4=31$ sub-regions, in agreement
with a direct census and an independent facet analysis. Fifteen
contribute at leading power; seven begin at $y^1$, seven at $y^2$,
and two at $y^3$.
We use the analytic shifts $\rho_j=\rho\nu_j$ and the normalized
parameter measure defined in Appendix~\ref{app:ff48}, with
\begin{align}
 \boldsymbol\nu&=(0,2,-2,0,3,-3),\qquad
 \rho=r\epsilon,\qquad r=\frac5{11},\frac7{13},\frac37\, .
\end{align}
The weights again sum to zero. The Cheng-Wu choice $x_2=1$ makes all
fifteen leading sub-regions linearly reducible through $O(\epsilon)$;
for three of them other natural choices fail. The analytic shifts are
confined to lines adjacent to the semi-on-shell vertices, and their
Gamma-function normalization is retained throughout.
The result is obtained from the constant Laurent coefficient in $r$
after the regions have been summed, exactly as in
Eq.~\re{eq:coupled-regulator}.

\paragraph{Result.}
In the same $L,\ell_y$ convention as the main text and \FFgraph{2}{48},
write the finite term as a sum of pieces homogeneous of degree
$k$ in $L,\ell_y$; $D_s$ and the zeta values have degree zero. Then
\begingroup\small
\begin{align}
G_{72}&
=
\frac{1}{\epsilon^2}
\Bigl(
\tfrac{275}{144}+\tfrac{11}{9}\Ds
\Bigr)
+
\frac{1}{\epsilon}\,G_{72,-1}
+
\sum_{k=0}^4G_{72,0}^{(k)}
+
O(\epsilon),
\label{eq:ff72-native-result}
\end{align}
with the following explicit coefficients
\begin{align}
G_{72,-1}
&=
\tfrac{3911}{864}+\tfrac{3187}{432}\Ds
-
\Bigl(\tfrac{11}{12}+\tfrac{\Ds}{6}\Bigr)(L^2-\ell_y^2)
+
\Bigl(\tfrac{11}{6}+\tfrac{\Ds}{3}\Bigr) L\ell_y
+
\Bigl(\tfrac{41}{6}+\tfrac{5\Ds}{6}\Bigr) L
\, ,
\notag\\
G_{72,0}^{(4)}
&=
\tfrac1{12}L^4
-
\tfrac13 L^3 \ell_y
+
\tfrac13 L \ell_y^3
+
\tfrac16\ell_y^4
\, , \notag\\
G_{72,0}^{(3)}
&=
- 
\Bigl(\tfrac{31}{12}+\tfrac{\Ds}6\Bigr) L^3
+
\Bigl(\tfrac{23}{4}+\tfrac{\Ds}{2}\Bigr)L^2\ell_y
+ 2 L \ell_y^2
-
\Bigl(\tfrac7{12}+\tfrac{\Ds}6\Bigr)\ell_y^3
\, ,\notag\\
G_{72,0}^{(2)}
&=
\Bigl(
\tfrac{151}{18}+\tfrac{13\Ds}{72}+\tfrac34\zeta_2
\Bigr)L^2
+
\Bigl(
\tfrac{29}{9} + \tfrac{8\Ds}{9} - 2\zeta_2
\Bigr)L \ell_y
+
\Bigl(
\tfrac{29}{18}+\tfrac{4\Ds}{9}+\zeta_2
\Bigr)\ell_y^2
\, ,\notag\\
G_{72,0}^{(1)}
&=
\Bigl(
\tfrac{7447}{288} - \tfrac{173\pi^2}{144}
+ 
\tfrac{13\zeta_3}{2} 
+ 
\Ds \left(\tfrac{125}{36}+\tfrac{\pi^2}{12}\right)
\Bigr) L
+
\Bigl(
\tfrac{9\pi^2+16\zeta_3}{8}-\Ds\,\tfrac{\pi^2}{6}
\Bigr) \ell_y
\, , \notag\\
G_{72,0}^{(0)}
&=
\tfrac{12845}{5184} + \tfrac{401\pi^2}{288}
+
\tfrac{\pi^4}{12}
-
\tfrac{159\zeta_3}{8}
+
\Ds\left(
\tfrac{82039}{2592}-\tfrac{155\pi^2}{432}+\tfrac{\zeta_3}{2}
\right)
\, .
\end{align}
\endgroup
There is no $1/\epsilon^3$ pole; $\zeta_2$, $\zeta_3$ and $\zeta_4$ appear through $O(\epsilon^0)$.
Here $G_{72,-1}$ is the simple-pole coefficient. The function
$G_{72}$ enters Eq.~\re{eq:coupled-bare-normalization}; it is not itself a
renormalized form-factor coefficient: the graph multiplicity and
hard-scale factor in that equation precede the full LSZ and
renormalization assembly.

\paragraph{Regulator and numerical checks.}
The continuation in $r$ is checked independently of the sector-decomposition
comparison. The poles through $1/\epsilon^2$ are $r$ independent; at $O(\epsilon^0)$ the pieces of log degree 4, 3 and 2 are linear in $r$
and only the log-free piece has a $r^2$ term. Two further regulator patterns, $(0,1,-1,0,2,-2)$ and $(0,3,-3,0,1,-1)$, give
identical $r=0$ values of the log-degree 4, 3 and 2 terms. The FIESTA expansion (252 asymptotic and 33 single-scale integrals)
was compared at $y=10^{-3},\dots,10^{-6}$: the $L^4$ and $L^3$ coefficients agree to $10^{-5}$-$10^{-4}$, the $L^2$ gap shrinks
with $y$, the poles with all their logarithms agree, and the log-free $\Ds^0$ coefficient, of size $4\times10^3$ at
$y=10^{-5}$, agrees to $1.4\times10^{-5}$ relative after the dominant integrals were raised to $5\times10^9$ evaluations. The
remaining $10^{-3}$-level gaps in the single-logarithmic and $\Ds^1$ coefficients do not clearly shrink with $y$; they could
not be separated from numerical noise in the $\sim240$ non-dominant integrals and are not used in the double-logarithmic
analysis, which depends only on the log-degree 4, 3 and 2 terms.

\providecommand{\href}[2]{#2}\begingroup\raggedright\endgroup

\end{document}